\documentclass[11pt,notitlepage]{report}
\usepackage{graphicx,vmargin,fancyhdr,amsfonts,amsmath,amsthm,amssymb,mathrsfs}
\usepackage{bibunits}
\usepackage{hyperref}
\usepackage{uwmthesis2}

\usepackage{times}

\makeatletter
\@addtoreset{equation}{section}
\makeatother

\usepackage{tensor}
\usepackage{color}
\usepackage{epsfig}
\usepackage{latexsym}

\usepackage{slashed}

\setmarginsrb{1.5in}{0.65in}{1in}{1.0in}{0pt}{0.05in}{0pt}{0mm}
\fancyhfoffset[]{0.25In}

\newcommand{\hhat}[1]{\hat {\hat{#1}}}

\newcommand{\lab}[1]{\hypertarget{lb:#1}}
\newcommand{\iref}[2]{}
\newcommand{\emp}[1]{{#1}}
\newcommand{\com}[1]{}
\newcommand{\cp}[1]{{#1}}
\newcommand{\eml}[1]{{#1}}

\newcommand{\et}[1]{{\large\begin{flushleft} \color{blue}\textbf{#1} \end{flushleft}}}
\newcommand{\sto}[1]{\begin{center} \textit{#1} \end{center}}
\newcommand{\rf}[1]{{#1}}

\newcommand{\el}[1]{\label{#1}}
\newcommand{\er}[1]{\eqref{#1}}

\newcommand{\ft}[1]{\footnote{#1}}

\newcommand{\ci}[1]{}

\newcommand{\ke}{\rangle}
\newcommand{\br}{\langle}
\newcommand{\lb}{\left(}
\newcommand{\rb}{\right)}
\newcommand{\rc}{\right.}
\newcommand{\lc}{\left.}
\newcommand{\lsb}{\left[}
\newcommand{\rsb}{\right]}

\newcommand{\nn}{\nonumber \\}
\newcommand{\p}{\partial}

\newcommand{\cd}{\nabla}

\newcommand{\ba}{\begin{eqnarray}}
\newcommand{\ea}{\end{eqnarray}}
\newcommand{\be}{\begin{equation}}
\newcommand{\ee}{\end{equation}}
\newcommand{\bay}[1]{\left(\begin{array}{#1}}
\newcommand{\eay}{\end{array}\right)}
\newcommand{\eg}{\textit{e.g.} }
\newcommand{\ie}{\textit{i.e.}, }
\newcommand{\iv}[1]{{#1}^{-1}}
\newcommand{\st}[1]{|#1\ke}
\newcommand{\at}[1]{{\Big|}_{#1}}
\newcommand{\zt}[1]{\textrm{#1}}

\def\xa{{\alpha}}

\def\xb{{\beta}}

\def\xd{{\delta}}
\def\xD{{\Delta}}
\def\xe{{\epsilon}}

\def\xve{{\varepsilon}}
\def\xg{{\gamma}}
\def\xG{{\Gamma}}
\def\xk{{\kappa}}

\def\xl{{\lambda}}
\def\xL{{\Lambda}}

\def\xO{{\Omega}}
\def\xvp{{\varphi}}
\def\xs{{\sigma}}
\def\xS{{\Sigma}}
\def\xvt{{\vartheta}}
\def\xt{{\theta}}

\def\dxa{{\dot \alpha}}
\def\dxb{{\dot \beta}}
\def\dxd{{\dot \delta}}
\def\dxg{{\dot \gamma}}

\def \Tr {{\rm Tr}}

\def\CC{{\cal C}}
\def\CD{{\cal D}}

\def\CJ{{\cal J}}
\def\CK{{\cal K}}
\def\CL{{\cal L}}
\def\CM{{\cal M}}
\def\CN{{\cal N}}
\def\CO{{\cal O}}

\def\CR{{\cal R}}
\def\CS{{\cal S}}

\def\sS{\mathscr{S}}
\def\sQ{\mathscr{Q}}

\def\sla{\slashed}

\def\hhdb{\hhat \xd_{\xa \dot \xa}{}^{\xb \dot \xb}}
\def\hhdg{\hhat \xd_{\xa \dot \xa}{}^{\xg \dot \xg}}

\def\tit{{ANOMALY PUZZLE, CURVED-SPACETIME SPINOR HAMILTONIAN, AND STRING PHENOMENOLOGY}}
\def\aut{{\bf Xing Huang}}

\newcommand{\set}[1]{\mathbb{#1}}
\newcommand{\postscript}[2]{\setlength{\epsfxsize}{#2\hsize}
   \centerline{\epsfbox{#1}}}

\def\byy{{B_{yy}}}
\def\sy{{\zt{sgn}(y)}}
\def\tree{{\clubsuit}}

\def\hf{{\frac 1 2}}
\def\ksl{{\slashed k}}
\def\ap{{\xa'}}

\newcommand{\citec}[1]{{\cite{#1}}}

\begin{document}

\title{\tit}
\author{\aut}
\majorprof{Leonard Parker}
\cmajorprof{Luis Anchordoqui}
\submitdate{May 2011}
\degree{Doctor of Philosophy}
\program{Physics}
\copyrightyear{2011}
\majordept{Physics}
\havededicationfalse
\haveminorfalse
\copyrightfalse
\doctoratetrue
\figurespagetrue
\tablespagetrue

	\pagenumbering{roman}
	\pagestyle{plain}
	\manuscriptp
	\officialaprovp
			\null\vskip1in%
	\begin{center} 
		{\Large \bf ABSTRACT}
	\end{center}
  \vskip 2em	
	\begin{center}
		{\Large\bf\uppercase\expandafter{\tit}}\\
	\vskip 2em		
		\rm By\\
	\vskip 1em	
		\aut
	\end{center}
	\vskip 4em
	\begin{center}
		\rm The University of Wisconsin--Milwaukee, \number\year\\
		Under the Supervision of Professors Leonard Parker and Luis Anchordoqui
	\end{center}
	\vskip 2em
	{\samepage 
          \paragraph{}\ 
         \setstretch{1.7} 
        The advent of the Large  Hadron Collider (LHC) and the continuing influx of
          cosmological data could inject new energy to the relatively
          quiet field of string theory. Predictions from string models
          based on large extra dimensions could be tested in the
          energy range within the reach of the LHC or other upcoming
          experiments. In the first part of this dissertation, we
          study three different aspects of string  phenomenology. 

          First,  we consider extensions of the Standard Model based on open
          strings ending on D-branes, in which gauge bosons  exist as strings attached to stacks of
          D-branes, and chiral matter as strings stretching between
          intersecting D-branes.  Under the assumptions that the
          fundamental string scale is in the TeV range and the theory
          is weakly coupled, we study the complementary signals of low
          mass superstrings at the proposed electron-positron facility
          (CLIC), in $e^+e^-$ and $\gamma\gamma$ collisions. We
          examine all relevant four-particle amplitudes evaluated at
          the center of mass energies near the mass of lightest Regge
          excitations and extract the corresponding pole terms. We
          show that, in the minimal extension of the Standard Model,
          $\gamma \gamma \to e^+ e^-$ scattering proceeds only through
          a spin-2 Regge state. We estimate that for this particular
          channel, string scales as high as 4~TeV can be discovered at
          the 11$\sigma$ level with the first fb$^{-1}$ of data
          collected at a center-of-mass energy $\approx 5$~TeV.

          Next, we consider string realizations of the Randall-Sundrum
          effective theory and explore the search for the lowest
          massive Regge excitation of the gluon and of the extra
          (color singlet) gauge boson inherent of D-brane
          constructions. In these curved backgrounds, the higher-spin
          Regge recurrences of Standard Model fields localized near
          the IR brane are warped down to close to the TeV range and
          hence can be produced at collider experiments. We make use
          of four gauge boson amplitudes evaluated near the first
          Regge pole to determine the discovery potential of LHC. We
          find that with an integrated luminosity of 100~fb$^{-1}$,
          the 5$\sigma$ discovery reach for $p p\rightarrow $ dijet
          can be as high as 4.7~TeV. We also study the ratio of dijet
          mass spectra at small and large scattering angles. We show
          that with the first~fb$^{-1}$ such a ratio can probe
          lowest-lying Regge states for masses $\sim 3$~TeV.

          Finally, we propose that the 3.2$\sigma$ excess at about $
          140~{\rm GeV}$ in the dijet mass spectrum of $W$ + jets
          reproted by the CDF Collaboration originates in the decay of
          a leptophobic $Z'$ that can be related to the $U(1)$
          symmetries inherent of D-brane models.

  In the second part, we discuss several points that may help to
  clarify some questions that remain about the anomaly puzzle in
  ${\cal N}=1$ supersymmetric Yang-Mills theory. The anomaly puzzle
  concerns the question of whether there is a consistent way in the
  quantized theory to put the $R$-current and the stress tensor in a
  single supermultiplet called the supercurrent. It was proposed that
  the classically conserved supercurrent bifurcates into two
  supercurrents having different anomalies in the quantum regime. The
  most interesting result we obtain is an explicit expression for the
  lowest component of one of the two supercurrents, namely the
  supercurrent that has the energy-momentum tensor as one of its
  components. This lowest component is an energy-dependent linear
  combination of two chiral currents, one of those being the lowest
  component of the other supercurrent, namely, the
  $R$-current. Therefore, we conclude that there is no consistent way
  to construct a single supercurrent multiplet that contains the
  $R$-current and the stress tensor in the straightforward way
  originally proposed. We also discuss and try to clarify some
  technical points in the derivations of the two supercurrents in the
  literature. These latter points concern the significance of the
  infrared contributions to the NSVZ $\beta$-function and the role of
  the equations of motion in deriving the two supercurrents.
 
  In the third part, we investigate the issue that the Dirac
  Hamiltonian of a spin-$\frac 1 2$ particle in a curved background
  appears to be non-hermitian (with respect to the conserved scalar
  product) when the metric is time-dependent. Here, we show that this
  non-hermiticity results from a time dependence of the position
  eigenstates that enter into the Schr{\"o}dinger wave function.

  In the fourth and last part of the dissertation, we proposed a new
  massive gravity theory that is free of the vDVZ discontinuity. The
  key to the absence of the discontinuity is to introduce an extra
  scalar field with negative kinetic sign.

\setstretch{1.3}}
		
	\vskip 40ex
	\profsignature 
	\vskip 4em 
	\cprofsignature
	\vskip.5in
		
\vfill	

\endabstract \newpage
	\ifcopyright\copyrightpage\fi
        \ifhavededication\dedicationpage\fi
        \newpage
\renewcommand\contentsname{\ \ \ \ \ \ \ \ \ \ \ TABLE OF CONTENTS}        
\renewcommand\listfigurename{\ \ \ \ \ \ \ \ \ \ \ \ \ LIST OF FIGURES}
\renewcommand\listtablename{\ \ \ \ \ \ \ \ \ \ \ \ \ LIST OF TABLES}
	\tableofcontents

\afterpreface

\newpage

\begin{center}
\

\vskip 2cm

{\LARGE \bf ACKNOWLEDGMENTS}
\end{center}

\vskip 2cm

I wish to thank my advisor, Distinguished Professor Leonard Parker, for suggesting
Part III and IV of this dissertation. His patience, guidance, time and 
knowledge were paramount to my work. I am also grateful to Professor Luis Anchordoqui 
for teaching me many things, sharing his ideas with me and encouraging me.
Luis graciously let me contribute to his research with other distinguished
colleagues including Haim Goldberg, Dieter L\"ust and Tomasz Taylor. This dissertation would 
not have been possible without Luis. I also appreciate his generosity and sincere 
desire to see me succeed in physics and life. 

\newpage

\begin{center}
\
\vskip 1.5cm

{\LARGE \bf PREFACE}
\end{center}

\vskip 2cm

This dissertation is based on various work I did (with collaborators) during my graduate studies. The topics range from string phenomenology, supersymmetric field theory to quantum field in curved spacetime and massive gravity theory.\\

The  part on string phenomenology (Part I) is based on material from:
\begin{itemize}
\item  L.~A.~Anchordoqui, H.~Goldberg, X.~Huang and T.~R.~Taylor,\\
{\it LHC Phenomenology of Lowest Massive Regge Recurrences in the Randall-Sundrum Orbifold},\\
  Phys.\ Rev.\  D {\bf 82}, 106010 (2010)
  [arXiv:1006.3044 [hep-ph]]. 
\item L.~A.~Anchordoqui, W.~Z.~Feng, H.~Goldberg, X.~Huang and T.~R.~Taylor,\\
  {\it Searching for string resonances in $e^+e^-$ and $\gamma \gamma$
    collisions},\\ Phys.\ Rev.\  D  (to be published)
  arXiv:1012.3466 [hep-ph]. 
\item  L.~A.~Anchordoqui, H.~Goldberg, X.~Huang, D.~L\"ust and T.~R.~Taylor,\\
{\it Stringy origin of Tevatron $Wjj$ anomaly}\\
(submitted to Phys. Lett. B)
arXiv:1104.2302 [hep-ph]. 
\end{itemize}

The part on anomaly puzzle (Part II) is based on the following paper: 
\begin{itemize}
\item  X.~Huang and L.~Parker,\\
  {\it Clarifying Some Remaining Questions in the Anomaly Puzzle},\\
  Eur.\ Phys.\ J.\  C {\bf 71}, 1570 (2011)
  [arXiv:1001.2364 [hep-th]].
\end{itemize}

Part III is based on:
\begin{itemize}
\item X.~Huang and L.~Parker,\\
  {\it Hermiticity of the Dirac Hamiltonian in Curved Spacetime},\\
  Phys.\ Rev.\  D {\bf 79}, 024020 (2009) 
  [arXiv:0811.2296 [hep-th]].
\end{itemize}

Finally, Part IV is from:
\begin{itemize}
\item X.~Huang and L.~Parker,\\
  {\it Graviton propagator in a covariant massive gravity theory},\\
  arXiv:0705.1561 [hep-th].
\end{itemize}

\newpage

\pagenumbering{arabic}

\pagestyle{uwmheadings}

\chapter{Introduction}
\pagenumbering{arabic}
\thispagestyle{fancy}
\pagestyle{fancy} 

\section{D-Brane TeV-Scale String Compactifications}

At the time of its formulation and for years thereafter, Superstring
Theory was regarded as a unifying framework for Planck-scale quantum
gravity and TeV-scale Standard Model (SM) physics.  Important advances
were fueled by the realization of the vital role played by
D-branes~\cite{joe} in connecting string theory to
phenomenology~\cite{Blumenhagen:2006ci}.  This has permitted the formulation of
string theories with compositeness setting in at TeV
scales and large extra dimensions~\cite{Antoniadis:1998ig} .

\begin{figure}[tbp]
\postscript{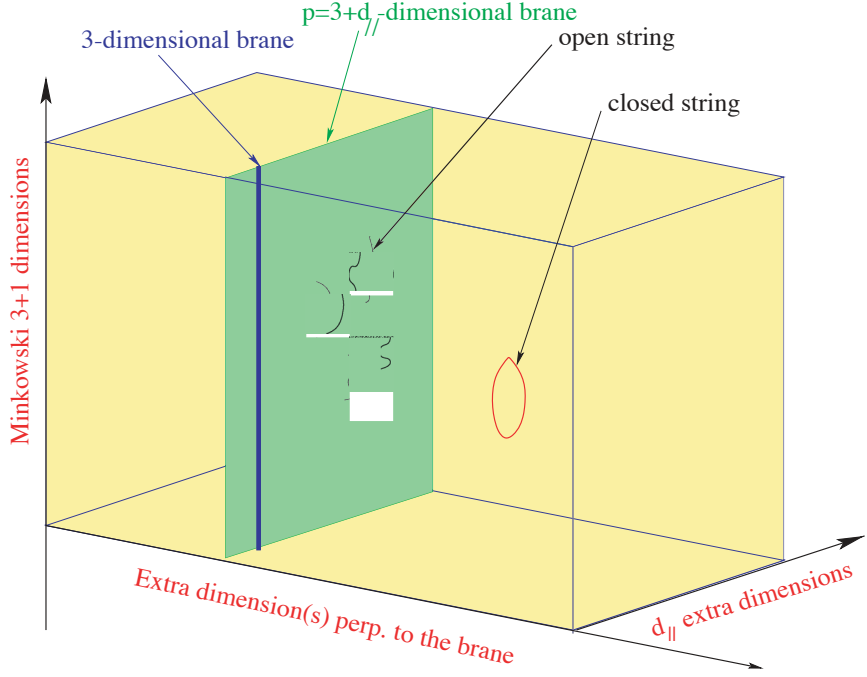}{0.8}
\caption[D-brane set-up]{D-brane
  set-up with $d_\parallel$ parallel and $d_\perp$ transverse internal
  directions. From Ref.~\cite{Antoniadis:2005aq}.}
\label{typeimodel}
\end{figure}

Conventional compactification scenarios are now widely familiar. We
imagine that in addition to the four spacetime dimensions we see, with
coordinates $x^\mu$, there are $D-4$ unseen dimensions with
coordinates $y^\mu$. The $D$-dimensional metric takes the form
\begin{equation}
ds^2 = dx^\mu dx_\mu + g_{mn} (y) dy^m dy^n \, .
\label{uno}
\end{equation}
For an illustration,  consider type II string theory compactified on a
six-dimensional torus $T^6$, which includes a D$p$-brane wrapped
around $p-3$ dimensions of $T^6$ with the remaining dimensions along
our familiar (uncompactified) three spatial dimensions. We denote the
radii of the {\it internal} longitudinal directions (of the
D$p$-brane) by $R_i^\parallel$, $i = 1, \dots p-3$ and the radii of
the  transverse directions by $R^\perp_j$, $j = 1, \dots 9-p$, see Fig.~\ref{typeimodel}. 
After dimensional reduction the effective 4-dimensional Planck scale, $M_{\rm Pl}$, is related to
the fundamental string scale,  $M_s$, according to
\begin{equation}
M_{\rm Pl}^2=8\ e^{-2\phi_{10}}\ M_s^8\
\frac{V_6}{(2\pi)^{6}}\ ,
\label{extradim}
\end{equation}
where 
\begin{equation}
V_6=(2\pi)^6\ \prod_{i=1}^{p-3} R^\parallel_i\ \prod_{j=1}^{9-p}
R^\perp_j
\end{equation}
is the volume of $T^6$ and $\phi_{10} =g_s$ is the dilaton controlling the
strength of coupling. It follows that the string scale can be
chosen hierarchically smaller than the Planck mass at the expense of
introducing $n$ extra large transverse dimensions felt only by gravity,
while keeping the string coupling  small.  Note that the coupling of
the gauge fields are not enhanced as long as  $R_i^\parallel$ remain
small,
\begin{equation}
g_{Dp}^{-2}=(2\pi)^{-1}\ M_s{}^{p-3}\ e^{-\phi_{10}}\
\prod_{i=1}^{p-3}R_i^\parallel\ .
\end{equation}
The weakness of the effective 4 dimensional gravity compared to gauge interactions is then
attributed to the largeness of the transverse space $R^\perp$
compared to the string length $M_s^{-1}$. 


\begin{figure}[tbp]
\postscript{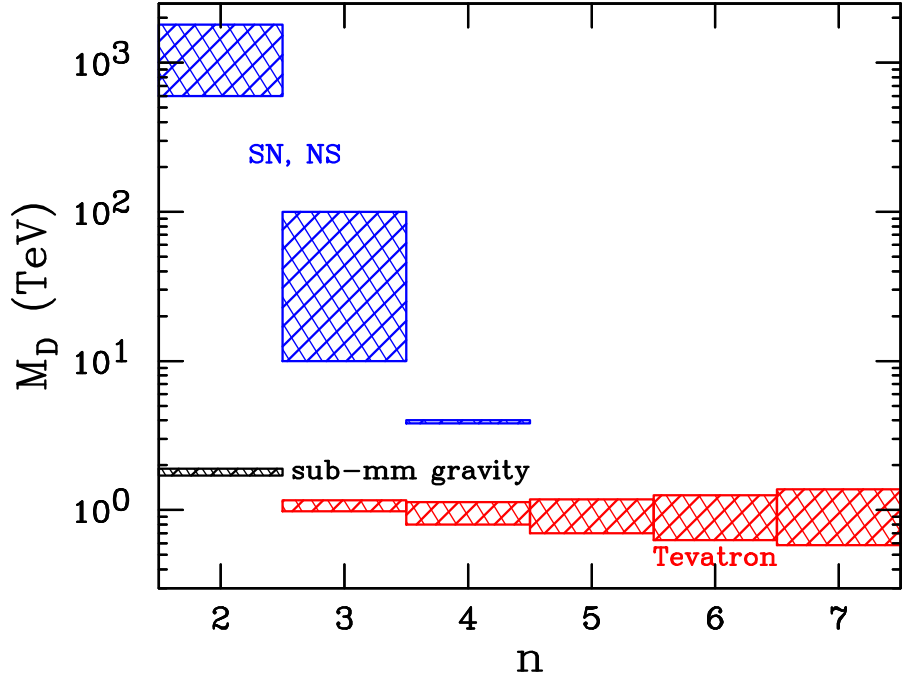}{0.8}
\caption[Bounds on TeV-scale gravity in the pre-LHC era] {Bounds on
 the fundamental Planck scale $M_D$  from: $(1)$ tests of
Newton's law on sub-millimeter scales~\cite{Hoyle:2000cv}; $(2)$ bounds on
supernova cooling (SN) and
neutron star heating (NS)~\cite{Cullen:1999hc}; $(3)$ Tevatron searches for
dielectron and diphoton production via virtual graviton exchange~\cite{Abbott:2000zb}.
The uncertainty in the Tevatron bounds corresponds to the range of
brane softening parameter; for details see
Ref.~\cite{Anchordoqui:2001cg}.}
\label{summary}
\end{figure}

A distinct property of these D-brane models is that gravity becomes
effectively $D$-dimensional with a strength comparable to those of
gauge interactions at the string
scale. Equation~(\ref{extradim}) can be
understood as a consequence of the $D$-dimensional Gauss law for
gravity, with
\begin{equation}
M_D = \left[ \frac{(2 \pi)^n}{8 \, \pi \, g_s^2}\right]^{1/(n+2)} \, M_s
\end{equation}
the fundamental scale of gravity in $D$ dimensions. Taking $M_s \sim
1$~TeV, one finds a size for the extra dimensions $R^\perp \approx
10^{30/n-19}~{\rm m}.$ This relation immediately suggests that $n=1$
is ruled out, because $R^\perp \sim 10^{11}~{\rm m}$ and the
gravitational interaction would thus be modified at the scale of our
solar system. However, already for $n=2$ one obtains $R^\perp \sim
1~{\rm mm}$. This is just the scale where our present day experimental
knowledge about gravity ends, see Fig.~\ref{summary}. All in all, in
these D-brane models gravity appears to us very weak at macroscopic
scales because its intensity is spread in the Universe's unseen
dimensions.

There are two paramount phenomenological consequences for TeV scale D-brane
string physics: the emergence of Regge recurrences at parton collision
energies $\sqrt{\hat s} \sim  M_s,$   most distinctly manifest in the
$\gamma$+jet~\citec{Anchordoqui:2007da,Anchordoqui:2008ac} and
dijet~\citec{Anchordoqui:2008di,Anchordoqui:2009mm} spectra resulting
from their decay; and the
presence of one or more additional $U(1)$ gauge symmetries, beyond the
$U(1)_Y$ of the SM. The latter follows from the property that the
gauge group for open strings terminating on a stack of $N$ identical
D-branes is $U(N)$ rather than $SU(N)$ for $N>2.$ (For $N=2$ the gauge
group can be $Sp(1)$ rather than $U(2)$.) In the first part of this Thesis we exploit
both these properties in order to obtain  ``new physics'' signals at collider experiments. 

After operating for only few months, with merely 2.9 inverse picobarns
of integrated luminosity, the LHC CMS experiment has recently ruled
out $M_s <2.5$~TeV by searching for narrow resonances in the dijet mass
spectrum~\cite{Khachatryan:2010jd}. In fact, LHC has the capacity of
discovering strongly interacting resonances in practically all range
up to $\sqrt{s}_{\rm LHC}$~\citec{Anchordoqui:2009ja}. The proper identification of Regge
recurrences, however, may not be straightforward at the LHC and
require complementary data.  We will argue that the proposed $e^+e^-$
and $\gamma\gamma$ colliders offer an excellent opportunity for
probing string physics.

In Chapter~\ref{CLIC}, we explore prospects for direct searches of
string physics at the Compact LInear Collider (CLIC). To develop our
program in the simplest way, we work within the construct of a minimal
model.  In the bosonic sector, the open strings terminating on the
(color) $U(3)$ stack of D-branes contain, in addition to the $SU(3)$
octet of gluons $g_\mu^a$, an extra $U(1)$ boson ($C_\mu$, in the
notation of~\citec{Berenstein:2006pk}), most simply the manifestation
of a gauged baryon number symmetry. The $U(1)_Y$ boson $Y_\mu$, which
gauges the usual electroweak hypercharge symmetry, is a linear
combination of $C_\mu$, the $U(1)$ boson $B_\mu$ terminating on a
separate $U(1)$ brane, and perhaps a third additional $U(1)$ field
$X_\mu$ sharing a $U(2)$ stack which is also a terminus for the
$SU(2)_L$ electroweak gauge bosons
$A_\mu^a$~\citec{Antoniadis:2000ena}.

Before proceeding, we pause to present our notation.
The first Regge excitations of the gluon $(g)$ and quarks $(q)$ will be denoted by $g^*,\ q^*$, respectively. Similarly, the first excitation of the $U(1)$ gauge bosons will be denoted by $C^*$, $B^*$, and $X^*$. Note that the $C_\mu$ ($X_\mu$) has an anomalous mass which may be less than the string scale. If that is the case, and if the mass of the $C^*$ ($X^*$) is composed (approximately) of the anomalous mass of the $C_\mu$ ($X_\mu$) and $M_s$ added in quadrature, we would expect only a minor error in our results by taking the $C^*$ ($X^*$) to be degenerate with the other resonances.

Only one assumption is necessary to build up a solid framework: the
string coupling must be small for the validity of perturbation theory
in the computations of scattering amplitudes. In this case, black hole
production and other strong gravity effects occur at energies above
the string scale, therefore at least the few lowest Regge recurrences
are available for examination, free from interference with some
complex quantum gravitational phenomena. 

We examine all relevant four-particle amplitudes evaluated at the center
of mass energies near the mass of lightest Regge excitations and
extract the corresponding pole terms.  The Regge poles of {\em all\/}
four-point amplitudes, in particular the spin content of the
resonances, are completely model independent, universal properties of
the entire landscape of string compactifications. We show that, in the
minimal extension of the SM, $\gamma
\gamma \to e^+ e^-$ scattering proceeds only through a spin-2 Regge
state. We estimate that for this particular channel, string scales as
high as 4~TeV can be discovered at the 11$\sigma$ level with the first
fb$^{-1}$ of data collected at a center-of-mass energy $\approx
5$~TeV. We also show that for $e^+e^-$ annihilation into
fermion-antifermion pairs, string theory predicts the {\em precise}
value, equal 1/3, of the relative weight of spin 2 and spin 1
contributions. This yields a dimuon angular distribution with a
pronounced forward-backward asymmetry, which will help distinguishing
between low mass strings and other beyond SM scenarios.

An interesting generalization of (\ref{uno}) that respects the (approximate) 4-dimensional
Poincar\'e invariance we observe in nature arises when the scale of the
four-dimensional metric vary depending on the location in the
extra dimension, 
\begin{equation}
ds^2 = e^{2A(y)} dx^\mu dx_\mu + g_{mn}(y) dy^m dy^n \,,
\end{equation}
for some function $A(y)$. Such a metric is referred to as {\em warped
metric}, and the factor ${\rm exp}\{2A\}$, which can be thought of as
giving a position-dependent redshift, is known as a {\em warp} factor.
In \emp{Chapter~\ref{LHCPheno}}, we explore the search for the lowest
massive Regge excitation in warped compactifications. We complement
model independent searches of top-production via $q^*$
excitation~\citec{Hassanain:2009at} by analyzing
tree-level four-point amplitudes relevant to inclusive $\gamma$ + jet
and dijet mass spectra.  We make use of four gauge boson amplitudes
evaluated near the first resonant pole to determine the discovery
potential of LHC for $g^*$ and $C^*$ excitations.  We study the
inclusive dijet mass spectrum in the central rapidity region $|y_{\rm
  jet}| < 1.0$ for dijet masses $M\geq 2.5~{\rm TeV}$. We find that
with an integrated luminosity of 100~fb$^{-1}$, the 5$\sigma$
discovery reach can be as high as 4.7~TeV. Observations of resonant
structures in $pp\rightarrow {\rm direct}\ \gamma~ +$ jet can provide
interesting corroboration for string physics up to 3.0~TeV. We also
study the ratio of dijet mass spectra at small and large
(center-of-mass) scattering angles. We show that with the
first~fb$^{-1}$ such a ratio can probe lowest-lying Regge states for
masses $\sim 3$~TeV.

New gauge bosons with SM like couplings to leptons are constrained by
collider searches to be heavier than about 1~TeV. A $Z'$ boson with
supressed couplings to leptons, however, can be much lighter and
possess substantial couplings to SM quarks. In Chapter~\ref{zprime},
we undertake a phenomenological study of the previously mentioned
$U(1)$ symmetries inherent to D-brane constructions and we show that
one of the associated $Z'$ gauge bosons  can explain the recent
excess in the $W$ + 2 jets final states reported by the CDF
Collaboration~\citec{Aaltonen:2011mk}.

\ \\

\section{Anomaly Puzzle in N = 1 Supersymmetric Gauge Theories}
The anomaly puzzle in $\CN = 1$ supersymmetric gauge theories is well known.  
Classically, a real superfield, $ \CJ_{\mu}$, called the supercurrent can be constructed \citec{Ferrara:1974pz} and is classically conserved. The lowest component of this superfield is the $R$-current.  The other components of $ \CJ_{\mu}$
are related to the supersymmetry current $J_{\xa\mu}$ (where $\xa$ is a two-component spinor index that labels the generators of the supersymmetry) and the stress tensor $\vartheta_{\mu\nu}$ through linear transformations. This construction is related to the fact that these symmetries are elements of the superconformal algebra. 

The anomaly puzzle arises as follows.
In an $\CN = 1$ SYM (supersymmetric Yang-Mills) theory, the $R$-symmetry, which is just a chiral $U(1)$ symmetry (denoted later as $U(1)_R$) has an anomaly.
This chiral anomaly is proportional to the topological invariant,
$F^{\mu\nu} {\tilde F}_{\mu\nu}$,  and can be expressed in an operator equation. One can try to generalize the operator equation
of this anomaly of the $R$-symmetry to a supersymmetric form involving $ \CJ_{\mu}$ \citec{Clark:1979te,Piguet:1981mu,Curtright:1977cg,Abbott:1977in,Lukierski:1977dr,Inagaki:1977he}.
However, this attempt led to an apparent contradiction. On the one hand, the anomaly of $R$-symmetry is known to be exactly of one-loop order because of the Adler--Bardeen theorem \citec{Adler:1969er,Jones:1982zf}. On the other hand, the trace of the stress tensor, which is another component of $D^\xa \CJ_{\xa \dot \xa}$ should be proportional to the $\xb$-function (because the trace is a measure of the breaking of scale invariance). These two components of $D^\xa \CJ_{\xa \dot \xa}$ should be proportional to the same factor, which would seem to imply that the $\xb$-function is exactly of one loop order. 
However, explicit perturbative calculations show that there are higher order corrections to the $\xb$-function \citec{Avdeev:1980bh}. Note that there are some subtleties about this formulation of the anomaly puzzle, which we shall discuss in more detail later. But the problem remains as to whether it is possible to construct a supercurrent and describe all the anomalies in a single operator equation (valid at all orders).

There have been various attempted solutions to the anomaly puzzle \citec{Grisaru:1985yk,Grisaru:1985ik,Ensign:1987wy,Shifman:1986zi}. In Grisaru et al, \citec{Grisaru:1985yk,Grisaru:1985ik}, a solution to the anomaly puzzle is given by showing that there are actually two different supercurrents $ \CJ_{\mu}$. Let us call those two different supercurrents in 4-dimensional spacetime, $ \CJ^{(1)}{}_{\mu}$ and $ \CJ^{(2)}{}_{\mu}$.  They are the same classically (meaning at tree level). One of them, $ \CJ^{(1)}{}_{\mu}$, has the $R$-current as its lowest component, but the higher components are no longer the supersymmetry current and stress tensor \ft{After the completion of the current work, we learned that there is new progress in this subject. It is proposed that \citec{Komargodski:2010rb} there is a supercurrent multiplet ($\CS$-multiplet) whose higher components contain the supersymmetry current and the stress tensor, although not in the simple way as in the original construction of the supercurrent multiplet \citec{Ferrara:1974pz}. Following this line, it has been shown that \citec{Yonekura:2010mc} the multiplet $\CJ^{(1)}{}_{\mu}$ in our notation can be identified (at least for the case of SQED)  as the $\CS$-multiplet, and the FZ-multiplet \ie $\CJ^{(2)}{}_{\mu}$ in our notation, can be obtained from $\CJ^{(1)}{}_{\mu}$ by adding a superfield. In light of the new work, when we say that the ``higher components of $\CJ^{(1)}{}_{\mu}$ are not the supersymmetry current and the stress tensor,'' the reader should interpret this as meaning that the higher components of $\CJ^{(1)}{}_{\mu}$ are not related to the supersymmetry current and the stress tensor in the straightforward way that the higher components of $\CJ^{(2)}{}_{\mu}$ are related to them.}. The anomalous non-conservation of this supercurrent is proportional to the one-loop $\xb$-function. The other supercurrent, $ \CJ^{(2)}{}_{\mu}$, has the supersymmetry current and stress tensor as its components and has an anomaly proportional the exact $\xb$-function (the so-called NSVZ $\xb$-function \citec{Novikov:1983uc}). In Ensign et al \citec{Ensign:1987wy}, they consider $\CN = 1$ supersymmetric gauge theories including matter fields and extend the construction done in \citec{Grisaru:1985yk,Grisaru:1985ik} of the two supercurrents to the case that includes matter. 

Although we believe that this ``two-supercurrent'' scenario is the correct approach to resolve the puzzle, there remains some work to be done. This approach appears to depend on a particular regularization method (the so-called superspace regularization by dimensional reduction, henceforth SRDR). The physical properties of the operators in the two supercurrents are not always easy to see. By studying the SYM with matter, we show, using conventional
dimensional regularization the different physical properties of the lowest components of the two supercurrents and provide clear evidence for the existence of two supercurrents without relying on the technique of SRDR. We find that $\CJ^{(2)}{}_{\mu}$ has as its lowest component a current which is a coupling-dependent linear combination of the $R$-current and the Konishi current \citec{Konishi:1983hf,Konishi:1985tu}. This linear combination, which we refer to as $R'_{\mu}$, does not have to satisfy the Adler--Bardeen theorem because of the coupling-dependent mixing coefficient. To avoid any ambiguity, we mention that we are using the term $R$-current (and $R$-symmetry) to describe the $U(1)$ current (denoted by $R_\mu$) that transforms the gaugino $\xl$, the matter scalar $A$ and the matter spinor $\psi$ according to the charge ratios of $1:\frac 2 3: - \frac 1 3$. The explicit expression for the lowest component of this supercurrent $\CJ^{(2)}{}_{\mu}$ had not been written earlier to our knowledge.

The anomaly equation for SYM with matter fields, as given in \citec{Shifman:1986zi}, has a term $\xg \bar D^2 (\bar \Phi e^V \Phi)$ (where $\Phi$ is a chiral superfield) that is responsible for the anomalous dimensions of the matter fields. This term is not obtained in \citec{Ensign:1987wy} because they assume that external fields are on-shell. As we shall see, it is the existence of this term that implies that the lowest component of $R'_\mu$ is not the $R$-current but a mixing (with coupling constant dependent coefficients) of the $R$-current and the Konishi current. We perform an explicit calculation (not using SRDR), which is not in the literature, to obtain the mixing. We also do the calculation using the supersymmetric background field method and SRDR. The results we obtain from either method agree and give the $\xg \bar D^2 (\bar \Phi e^V \Phi)$ term. In a word, to take into account the anomalous dimensions, the supercurrent $\CJ^{(2)}{}_{\mu}$ has to have $R'_\mu$ instead of $R_\mu$ as its lowest component. Obviously, this requirement is independent of regularization. 

As we shall see, the difference between $R'_\mu$ and $R_\mu$ is manifest in a very clear way at the infrared fixed point, where $R'_\mu$ becomes an
exact chiral symmetry current that is a linear combination of $R_\mu$ and the Konishi current. Note that the $R'_\mu$ charges of various fields also follow from the unitarity bound. In some sense, only $\CJ^{(2)}{}_{\mu}$ should be called the supercurrent as all its components are the conserved currents of
the superconformal group at the fixed point (while those of $\CJ^{(1)}{}_{\mu}$ are not). But we will continue to use the term ``two supercurrents,'' as it is widely used. 

Moreover, there are some technical issues in their construction that we discuss and attempt to clarify. In \citec{Grisaru:1985ik}, the equations of motion (EoM) are applied with the assumption that they vanish (up to contact terms). However, if one uses the expectation values of the various operators, as given in \citec{Grisaru:1985ik}, then the EoM would seem to have nonvanishing expectation values.  We show that this apparent inconsistency is resolved when one takes into account the non-local contributions.
After that, the expectation values of the bare operators are consistent with the application of the EoM. 
In particular, the expectation value, $\br\cd^\xa \CJ_{\xa \dot \xa}\ke$, of the unrenormalized operator $\cd^\xa \CJ_{\xa \dot \xa}$ vanishes as required to by the EoM. More explicitly, the non-local contribution to $\br\cd^\xa \CJ_{\xa \dot \xa}\ke$ is opposite in sign to the local contribution, which is proportional to an $\xe$ dimensional operator \ft{The
calculation is performed using dimensional reduction and the dimension is $4-\xe$ with $\xe>0$.}, and the two contributions add up to zero in the limit that $\xe\rightarrow 0$, i.e., in 4 dimensions. As a result, 
$\br\cd^\xa \CJ_{\xa \dot \xa}\ke$ does vanish.  Then, when we use the renormalization procedure of \citec{Grisaru:1985yk,Grisaru:1985ik,Ensign:1987wy}, in which the contribution proportional to an $\xe$ dimensional operator is removed by renormalization, the non-local contribution indeed gives the correct one-loop anomaly. This correct one-loop anomaly was obtained in \citec{Grisaru:1985yk,Grisaru:1985ik,Ensign:1987wy}. 
They did not explicitly discuss the role played by the non-local contributions in their derivation, so the discussion of those terms here may help clarify the consistency of the construction of the two supercurrents.

Finally, there is another version of the anomaly puzzle, which we believe is relevant but not equivalent to the one we have just discussed. The nonrenormalization \citec{Seiberg:1993vc} theorem implies that $\xb$-function of the gauge coupling $g$ will only be of one-loop, which agrees with what was found in the Wilsonian approach to the renormalization group \citec{Shifman:1986zi}. In the Wilsonian approach, effective Lagrangians (at different cutoffs) are obtained by integrating over high momenta. The renormalization group flow then implies that the $\xb$-function of the coupling constant is of one-loop order. But this is again in contradiction with the explicit calculation of \citec{Avdeev:1980bh}. 

\emp{The solution to this second version (in the context of Wilson effective action) of the anomaly puzzle certainly has nontrivial consequence on the operator form of the anomaly equation.} In Chapter~\ref{ch:APCWEA}, we will review the Wilson effective action approach to the anomaly puzzle and comment on the question of whether the higher-order terms in the $\xb$-function are the result of contributions coming from infrared modes of the fields. In \citec{Shifman:1986zi}, they show that the higher-order terms in the $\xb$-function come from the infrared modes. A different way of obtaining the same $\xb$-function is given in \citec{ArkaniHamed:1997mj}. In the latter method, the coupling constant receives its higher-order corrections from the Jacobian appearing when one rescales the measure \citec{Fujikawa:1979ay,Fujikawa:1980eg}, and as they mention in \citec{ArkaniHamed:1997mj}, the method does not appear to depend on the infrared modes. By changing the UV cutoff in the Wilson effective action, we show that the momentum modes above any arbitrary finite non-zero scale do not give a significant contribution to the Jacobian from which the multi-loop corrections to the $\beta$-function are obtained.  This shows that the method used by \citec{ArkaniHamed:1997mj} does indeed depend on the infrared modes.

In Sec~\ref{sec:RAP}, we review some basic ideas about the supercurrent and the anomaly puzzle. The supercurrent is discussed in more detail in the appendix. In Sec~\ref{sec:ABoTS}, possible solutions to the anomaly puzzle in the literature are reviewed and remaining problems are discussed. In Sec~\ref{sec:SotTA}, \eml{we perform an explicit calculation to show that the operator $R_\mu'$ in the same supermultiplet as the supersymmetry current has exactly the properties of what the anomaly equation in \citec{Shifman:1986zi} predicts but it generates a $U(1)$ transformation different from the $R$-symmetry. As a result, this superfield should be identified as $\CJ^{(2)}{}_{\mu}$
and not as $\CJ^{(1)}{}_{\mu}$ (in the notation defined above). First we do the calculation using component fields. Then in subsection~\ref{subsec:MSD}, we obtain the same result using the supersymmetric background field method. In subsection~\ref{subsec:CIFP}, we analyze the properties of the current $R_\mu'$ at the non-trivial infrared fixed point of supersymmetric QCD. We show that $R_\mu'$ does have the charge ratios to be a non-anomalous current and thus corresponds to
a true symmetry at the fixed point, as it should.} In Sec~\ref{sec:EVoEoM}, we discuss the role of non-local terms in obtaining the expectation value of the equation of motion and show how such terms enter into the construction of the two supercurrents. 

In Chapter~\ref{ch:APCWEA}, we will study the anomaly puzzle from the view point of Wilson effective action. We will investigate the significance of infrared contributions to the NSVZ $\beta$-function. In Sec~\ref{sec:ASVuWEA}, the approach by Shifman and Vainshtein, which is based on Wilson effective action, is reviewed. In Sec~\ref{sec:AoAHaM}, we also briefly review the calculation by Arkani-Hamed and Murayama of the $\xb$-function using the rescaling Jacobian. In Sec~\ref{sec:RAasaIE}, we show that the calculations of the $\xb$-function done by Shifman and Vainshtein \citec{Shifman:1986zi} and by Arkani-Hamed and Murayama \citec{ArkaniHamed:1997mj} both depend on the infrared modes.

\ \\

\section{Hermiticity of Curved-Space Spinor Hamiltonian}

In \citec{Parker:1980kw} and \citec{Parker:1980prl}, a one-electron atom was investigated as a probe of the curvature of a general spacetime.  If the curvature near the atom is sufficiently strong, then the spectrum of the atom can reveal properties of the Riemann tensor at the position of the atom. To calculate the shifts in the energy eigenvalues of the atom by means of perturbation theory, a conserved scalar product suitable to the Dirac equation in a general curved spacetime was defined in \citec{Parker:1980kw}.  This scalar product was based on a generally covariant current introduced by 
Bargmann \citec{Bargmann1932} in developing the
theory of the curved-spacetime Dirac equation obtained 
by Schr{\"o}dinger \citec{Schrodinger1932}.
The Hamiltonian for the one-electron atom was found 
in \citec{Parker:1980kw} directly from the curved-spacetime Dirac equation.  Assuming that the rate of change of the
spacetime curvature in the vicinity of the atom was negligible relative to the transition rates associated with the atom, that Hamiltonian was found to be hermitian with respect to the conserved scalar product,
and the shifts in the energy eigenvalues were obtained in 
terms of the Riemann tensor at the position of the atom.

In \citec{Parker:1980kw}, it was also found that if the time dependence of the metric can not be neglected, then the expression for the Hamiltonian coming directly from the curved-spacetime Dirac equation will violate hermiticity in a specific way. This raises the questions: Why does this non-hermiticity arise, and is there an hermitian Hamiltonian for a general curved spacetime having non-neglible time dependence?

Here, we show how to generalize the Hamiltonian of \citec{Parker:1980kw} so that it becomes exactly hermitian without neglecting the time-dependence of the metric. The key is to consider
the Hilbert space structure of the quantum mechanics of the Dirac electron.
We find that the problem with hermiticity that arises when the
metric is varying with time results from a subtle time dependence 
of the basis states $\st{x}$ (i.e., the eigenstates of position).
Once this subtlety is
taken into account, we are able to obtain an expression for the
Hamiltonian of the Dirac fermion that is exactly
Hermitian in a general curved spacetime having an arbitrary space- and time-dependent metric.

The results found in \citec{Parker:1980kw} for the perturbed spectrum of the atom are not affected, but now it
is possible to explore by means of perturbation theory in curved spacetime quantum mechanical effects on bound systems,
such as molecules and atoms, that may result from significant time-dependence of the Riemann tensor.  It would be interesting to determine if such effects could be observed.

\ \\

\section{Massive Gravity}

It has been known for more than 30 years that the linearized
theory \citec{Fierz-Pauli} of a massive graviton, 
no matter how small the graviton mass, would predict 
values for the perihelion precession
of planets, and the bending of light by the sun, that differ \citec{vanDam:1970vg, Iwa, Zakharov}
by an observable value from the confirmed predictions
of general relativity. This is the well-known Van Dam-Veltman-Zakharov (vDVZ) discontinuity. It has been suggested \citec{Vainshtein:1972sx} that the full nonlinear theory of massive gravity may overcome
this difficulty. It can be shown \citec{Karch:2001jb, Kogan:2000uy,Porrati:2000cp} and \citec{Porrati:2002cp} that in Anti-de Sitter space and in de Sitter space one can formulate a massive graviton theory that
approaches general relativity in the massless limit.

A massive linearized graviton of spin-2 has 5 degrees of
freedom. One (the scalar one) of the extra degrees of freedom affects the coupling
of the graviton to matter in a way that does not vanish
in the limit of vanishing graviton mass \citec{Boulware:1973my}.

In \citec{Arkani-Hamed:2002sp}, Arkani-Hamed \textit{et al.} constructed a covariant massive gravity theory. We will refer to their theory as AGS theory. It is shown that the vDVZ discontinuity remains in the linearized form of this theory.

Note that all the theories of massive graviton mentioned above have what is called a Fierz-Pauli mass term \citec{Fierz-Pauli}, which is the only possible form that is free of any ghost or tachyon. If this constraint is released, the vDVZ discontinuity disappears. In a theory \citec{Arkani-Hamed:2003uy} \citec{ArkaniHamed:2005gu} with GR coupled to a ghost, gravity effectively becomes massive via the coupling with the ghost. This theory has been suggested as being in agreement with observation and possibly giving the acceleration of the universe. Intuitively, the ghost cancels the extra degree of freedom from massive graviton that survives the massless limit.

Motivated by this fact, we try to construct a massive gravity that is free of vDVZ discontinuity by introducing a ghost. Here, we take a different approach. We assume that the linearized
massive graviton must be part of a larger theory involving
additional fields that bring the theory into agreement with observation
in the limit of vanishing graviton mass.  Because the observational
tests occur in the low energy limit, we can look only at the low-energy
effective theory of the additional sector of the theory.  We show
that the simplest way to achieve our goal of agreement with observation
is to introduce an effective scalar field that at low energies behaves
like a single classical scalar field.  This low-energy classical field
is able to give agreement with the classical tests of general relativity
if it couples to matter in a specified way, and if its low-energy
effective Lagrangian has a negative kinetic energy term. Such a field
has already been considered in the classical context, unrelated to
a massive graviton, by Caldwell and others \citec{Caldwell:1999ew}, as a means of
explaining the acceleration of the expansion of the universe. Its
non-standard negative kinetic energy was shown to give an
effective equation of state with $w = p/ \rho < -1$ and to lead to
an acceleration of the expansion of the universe. Here we connect
the mass of the graviton to the value of $w$. 

Our approach employs an effective Lagrangian, but purposely leaves open the question of what the fundamental
theory looks like. There is no agreement on what the complete theory actually is, so we take an agnostic approach. Therefore, we do not specify a particular magnitude of the cut-off scale for the effective field theory. As we show later, one can certainly use the results for either the phantom field sector \citec{Cline:2003gs} or the massive graviton 
sector \citec{Arkani-Hamed:2002sp} to put an upper bound on the cutoff (in energy scale). Note that the former is based on phenomenological consideration.

In Sec~\ref{sec:VDVZD}, we review the vDVZ discontinuity in a non-covariant massive gravity theory with Fierz-Pauli mass term. This Fierz-Pauli theory is formulated on a flat background spacetime, which is invariant under Lorentz transformations, but not under general coordinate transformations.  In Sec~\ref{sec:aCMGT}, we review the covariant massive gravity introduced in \citec{Arkani-Hamed:2002sp}. In subsection~\ref{subsec:GPitAGST}, we compute explicitly the gauge-fixed massive graviton propagator that is not given in the original paper \citec{Arkani-Hamed:2002sp}. With the use of this propagator, one can see that the vDVZ discontinuity persists in this theory. In Sec~\ref{sec:GatAovD}, we attempt to construct a massive gravity theory with ghost. Such a theory is free of the vDVZ discontinuity but its vacuum is unstable just like any other theories with ghost. We also summarize some of the results in the literature about the instability of ghost theories and determine an upper bound on the magnitude of the cut-off scale required for the effective theory to be in agreement with observation.

\

\newpage

\

\

\

\

\

\

\

\

\

\

\

\noindent\textbf{\Huge Part I:}

\

\noindent\textbf{\huge String Phenomenology}
\addcontentsline{toc}{chapter}{Part I - String Phenomenology}

\newpage
\thispagestyle{fancy}
\chapter{Searching for String Resonances at the Compact LInear Collider}
\label{CLIC}
\thispagestyle{fancy}
\pagestyle{fancy}

If the fundamental mass scale of superstring theory is as low as few
TeV, the massive modes of vibrating strings, Regge excitations, will
be copiously produced at the LHC.  We discuss the complementary
signals of low mass superstrings at the proposed electron-positron
facility (CLIC), in $e^+e^-$ and $\gamma\gamma$ collisions.  We
examine all relevant four-particle amplitudes evaluated at the center
of mass energies near the mass of lightest Regge excitations and
extract the corresponding pole terms.  The Regge poles of {\em all\/}
four-point amplitudes, in particular the spin content of the
resonances, are completely model independent, universal properties of
the entire landscape of string compactifications. We show that, in the
minimal extension of the SM, $\gamma
\gamma \to e^+ e^-$ scattering proceeds only through a spin-2 Regge
state. We estimate that for this particular channel, string scales as
high as 4~TeV can be discovered at the 11$\sigma$ level with the first
fb$^{-1}$ of data collected at a center-of-mass energy $\approx
5$~TeV. We also show that for $e^+e^-$ annihilation into
fermion-antifermion pairs, string theory predicts the {\em precise}
value, equal 1/3, of the relative weight of spin 2 and spin 1
contributions. This yields a dimuon angular distribution with a
pronounced forward-backward asymmetry, which will help distinguishing
between low mass strings and other beyond the standard model
scenarios. The ideas discussed in this Chapter have been published in~\citec{Anchordoqui:2010zs}.

\ \\

\section{Standard Model from intersecting D-branes}
\label{II}

The SM  is a spontaneously broken Yang-Mills theory
with gauge group 
\begin{equation}
SU(3)_C \times SU (2)_L\times U(1)_Y \, .
\end{equation}
 Matter in
the form of quarks and leptons ({\it i.e.}~$SU(3)_{C}$ triplets and
singlets, respectively) is arranged in three families ($i=1,2,3$) of
left-handed fermion doublets (of $SU(2)_L$) and right-handed fermion
singlets. Each family $i$ contains chiral gauge representations of
left-handed quarks $Q_i = (3,2)_{1/6}$ and leptons $L_i =
(1,2)_{-1/2}$ as well as right-handed up and down quarks,
$U_i~=~(3,1)_{2/3}$ and $D_i = (3,1)_{-1/3}$, respectively, and the
right-handed lepton $E_i = (1,1)_{-1}$.\footnote{The hypercharge $Y$
  is shown as a subscript of the $SU(3)_{C} \times SU(2)_L$ gauge
  representation $(A,B)$.}  The neutrino is part of the left-handed
lepton representation $L_i$ and does not have a right-handed
counterpart.

The electroweak subgroup $SU_{L}(2)\times U_{Y}(1)$ is spontaneously
broken to the electromagnetic $U(1)_{\rm em}$ by the Higgs doublet $H
= (1,2)_{1/2}$ which receives a vacuum expectation value $v\neq0$ in a
suitable potential. In the process of spontaneous symmetry breaking
quarks and leptons receive masses due to their Yukawa coupling to the
Higgs. Three of the four components of the complex Higgs are `eaten'
by the $W^\pm$ and $Z^0$ bosons, which are superpositions of the gauge
bosons $A^a_\mu$ of ${SU}(2)_{L}$ and $B_\mu$ of ${U}(1)_{Y}$,
\begin{equation}
W_\mu^\pm=\frac{1}{\sqrt{2}}A_\mu^1\mp \frac{i}{\sqrt{2}}A_\mu^2\quad {\rm and}\quad Z^0_\mu=\cos\theta_{W}\,A^3_\mu-\sin\theta_{W}\,B_\mu\,,
\end{equation}
with masses respectively $ M^2_W = \pi\alpha v^2/\sin^2\theta_{W}$, $M^2_Z = M^2_W/\cos^2\theta_{W},$ and $\alpha\simeq1/128$ at $Q^2~=~M_W^2$. The fourth vector field, $A_\mu = \sin\theta_{W}\,A^3_\mu + \cos\theta_{W}\,B_\mu$, persists massless and the
 remaining Higgs component is left as a $U(1)_{\rm em}$ neutral real
 scalar. (For further details see, {\em e.g.}~\cite{Anchordoqui:2009eg}.) 

 The study of electron positron scattering at the Large Electron
 Positron (LEP) collider, together with additional measurements from
 other experiments, in particular those at Stanford Linear Collider
 (SLC) and at the Tevatron, have allowed for tests of the SM with
 unprecedented accuracy, including some observables beyond even one
 part in a million~\cite{Nakamura:2010zzi}.  The measured values
 $M_W\simeq80.4$~GeV and $M_Z\simeq91.2$~GeV fix the weak mixing angle
 at $\sin^2\theta_{W}\simeq0.23$ and the Higgs vacuum expectation
 value at $\langle H \rangle = v \simeq 172$~GeV.

 One of the most challenging problems in high energy physics today is
 to find out what is the underlying theory that completes the
 SM. Despite its remarkable success, the SM is incomplete with many
 unsolved puzzles -- the most striking one being the huge disparity
 between the strength of gravity and of the SM forces. This hierarchy
 problem suggests that new physics could be at play at the TeV-scale,
 and is arguably {\em the} driving force behind high energy physics
 for several decades. The non-zero vacuum expectation value of the
 scalar Higgs doublet condensate sets the scale of electroweak
 interactions. However, due to the quadratic sensitivity of the Higgs
 mass to quantum corrections from an arbitrarily high mass scale, with
 no new physics between the energy scale of electroweak unification
 and the vicinity of the Planck mass ($M_{\rm Pl} \approx
 10^{19}$~GeV) the Higgs mass must be fine-tuned to an accuracy of
 ${\cal O} (10^{32})$. The traditional view is to adopt $M_{\rm Pl}$
 as {\em the} fundamental scale and attempt to derive $v$ through some
 dynamical mechanism (e.g. renormalization group evolution).  In
 recent years, however, a new framework with a diametrically opposite
 viewpoint has been proposed, in which $v$ is instead the fundamental
 scale of nature~\citec{ArkaniHamed:1998rs}. D-brane string
 compactifications with low string scale and large extra dimensions
 allow a definite representation of this innovative
 premise~\citec{Antoniadis:1998ig}.

 TeV-scale superstring theory provides a brane-world description of
 the SM, which is localized on D-branes extending in $p+3$ spatial
 dimensions~\cite{Blumenhagen:2001te,Honecker:2004kb}. Gauge
 interactions emerge as excitations of open strings with endpoints
 attached on the D-branes, whereas gravitational interactions are
 described by closed strings that can propagate in all nine spatial
 dimensions of string theory (these comprise parallel dimensions
 extended along the $(p+3)$-branes and transverse dimensions).  The
 apparent weakness of gravity at energies below a few TeV can then be
 understood as a consequence of the gravitational force ``leaking''
 into the transverse large compact dimensions of spacetime. This is
 possible only if the intrinsic scale of string excitations is also of
 order a few TeV. Should nature be so cooperative, a whole tower of
 infinite string excitations will open up at this low mass threshold,
 and new particles of spin $J$ follow the well known Regge
 trajectories of vibrating strings: $J = J_0 + \alpha' M_s^2$, where
 $\alpha'$ is the Regge slope parameter that determines the
 fundamental string mass scale
\begin{equation}
M_s={1\over \sqrt{\alpha'}}\, .
\label{Ms}
\end{equation}

The basic unit of gauge invariance for D-brane constructions is a $U(1)$ field, and so one can stack up $N$ identical D-branes to generate a $U(N)$ theory with the associated $U(N)$ gauge group.  Gauge bosons are due to strings attached to stacks of D-branes and chiral matter due to strings stretching between intersecting D-branes~\citec{Blumenhagen:2006ci}.  Each of the two strings endpoints carries a fundamental charge with respect to the stack of branes on which it terminates.  Mater fields carry quantum numbers associated with bifundamental representations.

While the existence of Regge excitations is a completely universal
feature of string theory, there are many ways of realizing SM in such
a framework.  Individual models utilize various D-brane configurations
and compactification spaces. They may lead to very different SM
extensions, but as far as the collider signatures of Regge excitations
are concerned, their differences boil down to a few parameters. The
most relevant characteristics is how the $U(1)_Y$ hypercharge is
embedded in the $U(1)$s associated to $D$-branes. One $U(1)$ (baryon
number) comes from the ``QCD'' stack of three branes, as a subgroup of
the $U(3)$ group that contains $SU(3)$ color but obviously, one needs
at least one extra $U(1)$. In D-brane compactifications, hypercharge
always appears as a linear, non-anomalous combination of the baryon
number with one, two or more $U(1)$s. The precise form of this
combination bears down on the photon couplings, however the
differences between individual models amount to numerical values of a
few parameters.  In order to develop our program in the simplest way,
we work within the construct of a minimal model in which the color
stack $a$ of three D-branes are intersected by the (weak doublet)
stack $b$ and by one (weak singlet) D-brane
$c$~\citec{Antoniadis:2000ena}. {}For the two-brane stack $b$, there is
a freedom of choosing physical state projections leading either to
$U(2)_b$ or to the symplectic $Sp(1)$ representation of Weinberg-Salam
$SU(2)_L$~\citec{Berenstein:2006pk}.

In the bosonic sector, the open strings terminating on QCD stack $a$
contain the standard $SU(3)$ octet of gluons $g_\mu^a$ and an
additional $U(1)_a$ gauge boson $C_\mu$, most simply the manifestation
of a gauged baryon number symmetry: $U(3)_a\sim SU(3)\times
U(1)_a$. On the $U(2)_b$ stack the open strings correspond to the
electroweak gauge bosons $A_\mu^a$, and again an additional $U(1)_b$
gauge field $X_\mu$.  So the associated gauge groups for these stacks
are $SU(3) \times U(1)_a,$ $SU(2)_L \times U(1)_b$, and $U(1)_c$,
respectively.  We can further simplify the model by eliminating
$X_\mu$; to this end instead we can choose the projections leading to
$Sp(1)$ instead of $U(2)_b$ \citec{Berenstein:2006pk}. The $U(1)_Y$
boson $Y_\mu$, which gauges the usual electroweak hypercharge
symmetry, is a linear combination of $C_\mu$, the $U(1)_c$ boson
$B_\mu$, and perhaps a third additional $U(1)$ gauge field, $X_\mu$.
The fermionic matter consists of open strings located at the
intersection points of the three stacks.  Concretely, the left-handed
quarks are sitting at the intersection of the $a$ and the $b$ stacks,
whereas the right-handed $u$ quarks comes from the intersection of the
$a$ and $c$ stacks and the right-handed $d$ quarks are situated at the
intersection of the $a$ stack with the $c'$ (orientifold mirror)
stack. All the scattering amplitudes between these SM particles, which
we will need in the following, essentially only depend on the local
intersection properties of these D-brane stacks. (For further details see, {\em e.g.}~\citec{MarchesanoBuznego:2003hp}.)

\begin{table}
\caption{Chiral fermion spectrum of the $U(3)_a \times Sp(1)_L \times U(1)_c$ D-brane model.}
\begin{tabular}{c|ccccc}
\hline
\hline
 Name &~~Representation~~& ~$Q_{U(3)}$~& ~$Q_{U(1)}$~ & ~$Q_Y$~&~~~$\overline {Q_Y}$~~~ \\
\hline
~~$U_i$~~ & $({\bar 3},1)$ &    $-1$ & $\phantom{-}1$ & $-\frac{2}{3}$ & $-4$  \\[1mm]
~~$D_i$~~ &  $({\bar 3},1)$&    $-1$ & $-1$ & $\phantom{-}\frac{1}{3}$ & $\phantom{-} 2$ \\[1mm]
~~$L_i$~~ & $(1,2)$&    $\phantom{-}0$ & $\phantom{-}1$ & $-\frac{1}{2}$  & $-3$ \\[1mm]
~~$E_i$~~ &  $(1,1)$&  $\phantom{-}0$ & $-2$ &  $\phantom{-}1$ & $\phantom{-} 6$ \\[1mm]
~~$Q_i$~~ & $(3,2)$& $\phantom{-}1$ & $\phantom{-}0$ & $\phantom{-}\frac{1}{6}$& $\phantom{-} 1$  \\[1mm]
\hline
\hline
\end{tabular}
\label{t1}
\end{table}

The chiral fermion spectrum of the $U(3)_a \times Sp(1) \times U(1)_c$
D-brane model is given in  Table~\ref{t1}.  In such a minimal D-brane
construction, if the coupling strength of $C_\mu$ is down by root six
when compared to the $SU(3)_C$ coupling $g_a$, the hypercharge
$Q_Y\equiv \frac{1}{6} Q_{U(3)}-\frac{1}{2}Q_{U(1)}$ is free of
anomalies.  Namely, the mixed anomaly (gauge and gravitation, with external gauge current $J_\lambda,$ and stress energy-momentum tensors $T_{\mu \nu},\, T_{\eta \rho}$) is given by
\begin{eqnarray}
{\rm Tr}\left[\overline {Q_Y}\right] & = &3 (-4)  + 3 (2) + 2 (-3)+6 +6 \nonumber \\
                                   & = & -12+6-6+6+6  \\
                                   & = & 0, \nonumber
\end{eqnarray}
whereas the chiral anomaly (with three external gauge currents, 
$J_\mu, J_\nu, J_\rho$) reads
\begin{eqnarray}
{\rm Tr}\left[(\overline {Q_Y})^3\right] & = & 3 (-4)^3 + 3 (2^3) + 2 (-3)^3 + 6^3 + 6 \nonumber \\
 & = & -192 + 24 - 54 + 216 + 6  \\
 & = & 0 \, . \nonumber
\end{eqnarray}
However, the $Q_{U(3)}$ (gauged baryon number)  is anomalous.
This anomaly is canceled by the f-D version~\citec{Witten:1984dg} of the Green-Schwarz mechanism~\citec{Green:1984sg}.  The vector boson $Y'_\mu$, orthogonal to the hypercharge, must grow a mass in order to avoid long range forces between baryons other than gravity and Coulomb forces. The anomalous mass growth allows the survival of global baryon number conservation, preventing fast proton decay~\citec{Ghilencea:2002da}.

In the $U(3)_a \times Sp(1)_L \times U(1)_c$ D-brane model, the $U(1)_a$ assignments are fixed (they
give the baryon number) and the hypercharge assignments are fixed by SM. Therefore, the mixing angle $\theta_P$ between the hypercharge and the $U(1)_a$
is obtained in a similar manner to the way the Weinberg angle is fixed
by the $SU(2)_L$ and the $U(1)_Y$ couplings ($g_b$ and $g_Y$, respectively) in the SM. The Lagrangian containing the $U(1)_a$ and $U(1)_c$ gauge fields is given by
\begin{equation}
{\cal L} = g_c \, \hat B_\mu \, J_B^\mu + \frac{g_a}{\sqrt{6}} \, \hat  C_\mu  \,  J_C^\mu
\label{apache}
\end{equation}
where 
\begin{equation}
\hat B_\mu = \cos \theta_P \, Y_\mu + \sin \theta_P \, Y'_\mu \quad
{\rm and} \quad
\hat C_\mu = -\sin \theta_P \, Y_\mu + \cos \theta_P\, Y'_\mu
\label{CBfields}
\end{equation}
are canonically normalized, and $g_c$ is the coupling strength of the $U(1)_c$ gauge field.  Substitution of these expressions into (\ref{apache}) leads to
\begin{eqnarray}
{\cal L}  =  Y_\mu \left(g_c \cos \theta_P J_B^\mu - \frac{g_a}{\sqrt{6}} \sin \theta_P J_C^\mu \right)
+  Y'_\mu \left(g_c \sin \theta_P J_B^\mu + \frac{g_a}{\sqrt{6}} \cos \theta_P J_C^\mu \right),
\end{eqnarray}
with $g_c \, \cos \theta_P\,  J_B^\mu - \frac{1}{\sqrt{6}}\,  g_a \, \sin \theta_P \,  J_C^\mu = g_Y \, J_Y^\mu$. We have seen
that the hypercharge is anomaly free if
$J_Y =  \frac{1}{6} \,  J_C^\mu - \frac{1}{2} \,  J_B^\mu$, yielding
\begin{equation}
g_c \cos \theta_P = \frac{1}{2} g_Y \quad {\rm and}  \quad\frac{g_a}{\sqrt{6}} \sin \theta_P = \frac{1}{6} g_Y \, .
\label{pipita}
\end{equation}
{}From (\ref{pipita}) we obtain the following relations
\begin{equation}
\tan \theta_P = \sqrt{\frac{2}{3}} \, \frac{g_c}{g_a}, \quad \quad
\left(\frac{g_Y}{2g_c}\right)^2 + \left(\frac{1}{\sqrt{6}} \frac{g_Y}{g_a}\right)^2 = 1, \quad {\rm and}
\quad \frac{1}{4 g_c^2} + \frac{1}{6 g_a^2} = \frac{1}{g_Y^2} \, .
\label{pellerano}
\end{equation}
We use the evolution of gauge couplings from the weak scale $M_Z$ as determined by the one-loop beta-functions of the SM with three families of quarks and leptons and one Higgs doublet,
\begin{equation}
{1\over \alpha_i(M)}={1\over \alpha_i(M_Z)}-
{b_i\over 2\pi}\ln{M \over M_Z}\ ; \quad i=a,b,Y,
\end{equation}
where $\alpha_i=g_i^2/4\pi$ and $b_a=-7$, $b_b=-19/6$, $b_Y=41/6$. We also use the measured values of the couplings at the $Z$ pole $\alpha_a(M_Z)=0.118\pm 0.003$, $\alpha_b(M_Z)=0.0338$, $\alpha_Y(M_Z)=0.01014$ (with the errors in $\alpha_{b,Y}$ less than 1\%)~\cite{Nakamura:2010zzi}.  Running couplings up to 3~TeV, which is where the phenomenology will be, we get $\kappa \equiv \sin \theta_P \sim 0.14$. When the theory undergoes electroweak symmetry breaking, because $Y'$ couples to the Higgs, one gets additional mixing. Hence $Y'$ is not exactly a mass eigenstate. The explicit form of the low energy eigenstates $A_\mu$, $Z_\mu,$ and $Z'_\mu$ is given in~\citec{Berenstein:2008xg}.

In the $U(3)_a \times U(2)_b \times U(1)_c$ D-brane model, the hypercharge is given by
\be\el{hyperchargeY} Q_Y = c_a Q_{U(3)} + c_b Q_{U(2)} + c_c Q_{U(1)}.\ee
Note that we have, in the covariant derivative $\CD_\mu$,
\be\el{covderi} \CD_\mu = \p_\mu -i g_c \, B_\mu \, Q_{U(1)}  -i \frac{g_b}{2} \, X_\mu \, Q_{U(2)}  - i \frac{g_a}{\sqrt{6}} \, C_\mu  \,  Q_{U(3)}.\ee
We can define $Y_\mu$ and two other fields $Y'{}_\mu, Y''{}_\mu$ that are related to $C_\mu, X_\mu, B_\mu$ by a orthogonal transformation $O$ defined as
\[\bay{c} Y \\ Y' \\ Y''\eay = O \bay{c} C \\ X \\ B\eay.\]
In order for $Y_\mu$ to have the hypercharge $Q_Y$ as in
Eq.~\er{hyperchargeY}, we need, \be\el{othogaugefield} C_\mu = \frac
{\sqrt{6}c_a g_Y} {g_a} Y_\mu + \dots,\quad X_\mu = \frac {2 c_b g_Y}
{g_b} Y_\mu + \dots,\quad B_\mu =\frac {c_c g_Y} {g_c} Y_\mu +
\dots.\ee where $g_Y$ is given by \be \frac{1}{g_Y^2} = \frac{6
  c_a^2}{ g_a^2} + \frac{4 c_b^2}{g_b^2} + \frac{c_c^2}{g_c^2}.\ee The
field $Y_\mu$ then appears in the covariant derivative with the
desired $Q_Y$, \be\el{covderiY}\CD_\mu = \p_\mu -i g_Y Y_\mu Q_Y +
\dots.\ee The ratio of the coefficients in Eq.~\er{othogaugefield} is
determined by the form of Eq.~\er{hyperchargeY} and
Eq.~\er{covderi}. More explicitly, only with such ratio, we can have
$Q_Y$ in Eq.~\er{covderiY}. The value of $g_Y$ is determined so that
the coefficients in Eq.~\er{othogaugefield} are components of a
normalized vector so that they can be a row vector of $O$. The rest of
the transformation (the ellipsis part) involving $Y',Y''$ is not
necessary for our calculation. The point is that we now know the first
row of the matrix $O$ and hence we can get the first column of $O^T$,
which gives the expression of $Y_\mu$ in terms of $C_\mu, X_\mu,
B_\mu$, \be Y_\mu = \frac {\sqrt{6}c_a g_Y} {g_a} C_\mu + \frac {2 c_b
  g_Y} {g_b} X_\mu +\frac {c_c g_Y} {g_c} B_\mu.\ee This is all we
need when we calculate the interaction involving $Y_\mu$; the rest of
$O$, which tells us the expression of $Y', Y''$ in terms of $C,X,B$ is
not necessary for the moment.  For later convenience, we define $\xk, \eta, \xi$ as
\be Y_\mu = \xk C_\mu + \eta X_\mu +\xi B_\mu\,;\ee therefore \be \xk
= \frac {\sqrt{6}c_a g_Y} {g_a},\quad \eta = \frac {2 c_b g_Y}
{g_b},\quad \xi = \frac {c_c g_Y} {g_c}.\label{consts}\ee

\begin{table}
\caption{Chiral fermion spectrum of the $U(3)_a \times U(2)_b \times
  U(1)_c$ D-brane model (case I).}
\begin{tabular}{c|ccccc}
\hline
\hline
 Name &~~Representation~~& ~$Q_{U(3)}$~& ~$Q_{U(2)}$~ & ~$Q_{U(1)}$~ & ~~$Q_{Y}$ ~\\
\hline
~~$U_i$~~ & $({\bar 3},1)$ &    $-1$ & $\phantom{-}0$ & $\phantom{-} 0$ & $-\frac{2}{3}$ \\[1mm]
~~$D_i$~~ &  $({\bar 3},1)$&    $-1$ & $\phantom{-}0$ & $\phantom{-} 1$ & $\phantom{-}\frac{1}{3}$ \\[1mm]
~~$L_i$~~ & $(1,2)$&    $\phantom{-}0$ & $\phantom{-}1$ & $- 1$& $-\frac{1}{2}$  \\[1mm]
~~$E_i$~~ &  $(1,1)$&  $\phantom{-}0$ & $\phantom{-} 0$ &  $\phantom{-}1$ & $\phantom{-} 1$ \\[1mm]
~~$Q_i$~~ & $(3,2)$& $\phantom{-}1$ & $\phantom{-}1 $ & $\phantom{-} 0$ & $\phantom{-}\frac{1}{6}$  \\[1mm]
\hline
\hline
\end{tabular}
\label{t:CHspectrum}
\end{table}

We pause to summarize the degree of model dependency stemming from the
multiple $U(1)$ content of the minimal model containing 3 stacks of
D-branes. First, there is an initial choice to be made for the gauge
group living on the $b$ stack. This can be either $Sp(1)$ or
$U(2)$. In the case of $Sp(1)$, the requirement that the hypercharge
remains anomaly-free was sufficient to fix its $U(1)_a$ and $U(1)_c$
content, as explicitly presented in Eqs.~(\ref{pipita}) and
(\ref{pellerano}). Consequently, the fermion couplings, as well as the
mixing angle $\theta_P$ between hypercharge and the baryon number
gauge field are wholly determined by the usual SM couplings.  The
alternative selection -- that of $U(2)$ as the gauge group tied to the
$b$ stack -- branches into some further choices. This is because the
$Q_{U(3)_a},\ Q_{U(2)_b},\ Q_{U(1)_c}$ content of the hypercharge
operator is not uniquely determined by the anomaly cancelation
requirement.  In fact, as seen in~\cite{Antoniadis:2000ena}, there are
5  possibilities. This final choice does not depend on further
symmetry considerations; in Ref.~\cite{Antoniadis:2000ena} it was
fixed ($c_a =2/3 ,\ c_b =1/2 ,\ c_c=1$) by requiring partial
unification ($g_a = g_b$) and acceptable value of $\sin^2 \theta_W$ at
string scales of 6 to 8 TeV. The chiral fermion spectrum of the $U(3)_a \times U(2)_b \times U(1)_c$ for such a choice of parameters is summarized in Table~\ref{t:CHspectrum}. In Chapter~\ref{zprime} a different
choice will be made ($c_a = -2/3 , c_b = 1 , c_c = 0$ ) to  explain the CDF
anomaly~\cite{Aaltonen:2011mk}.  Clearly the mixing possibilities within the
$U(1)_a\times U(1)_b\times U(1)_c$ serve to introduce a discrete
number of phenomenological ambiguities. This contrasts strongly with
the case where all the scattering evolves on one brane ({\em e.g.,}
the $a$ stack on the color brane, which serves as the locale for
stringy dijet processes at LHC.~\citec{Anchordoqui:2008di}).

In principle, in addition to the orthogonal field mixing induced by identifying anomalous and non-anomalous $U(1)$ sectors, there may be kinetic mixing between these sectors.  In our case, however, since there is only one $U(1)$ per stack of D-branes, the relevant
  kinetic mixing is between  $U(1)$'s  on different stacks, and hence involves
  loops with fermions at brane intersection.  Such loop  terms are typically
  down by $g_i^2/16 \pi^2 \sim 0.01$~\cite{Dienes:1996zr}. Generally, the major effect of the kinetic mixing is in communicating SUSY breaking from a hidden $U(1)$ sector to the visible sector, generally in modification of soft scalar masses.  Stability of the weak scale in various models of SUSY breaking requires the mixing to be orders of magnitude below these values~\cite{Dienes:1996zr}. For a comprehensive review of experimental limits on the mixing, see~\cite{Abel:2008ai}.  Moreover, the model discussed in the present work does not have a hidden sector-- all our $U(1)$'s (including the anomalous ones) couple to the visible sector.\footnote{We also work in the weak coupling regime. For an alternate  approach, see~\cite{Kitazawa:2009kr}.} In summary, kinetic mixing between the non-anomalous and the anomalous $U(1)$'s in our basic three stack model will be small because the fermions in the loop are all in the visible sector. In the absence of electroweak symmetry breaking, the mixing vanishes.

  The scattering amplitudes involving four gauge bosons as well as
  those with two gauge bosons plus two leptons do not depend on the
  compactification details of the transverse
  space~\citec{Lust:2008qc}. The only remnant of the compactification
  is the relation between the Yang-Mills coupling and the string
  coupling. We take this relation to reduce to field theoretical
  results in the case where they exist, e.g., $gg \to gg$. Then,
  because of the require correspondence with field theory, the
  phenomenological results are independent of the compactification of
  the transverse space. However, as we discuss in
  Chapter~\ref{LHCPheno}, a different phenomenology would result as a
  consequence of warping one or more parallel dimensions. Four gauge
  boson amplitudes will be particularly useful for testing low mass
  strings in $\gamma\gamma$ collisions.

  On the other hand, the amplitudes involving four fermions, including
  $e^+e^-\to e^+e^-$, $e^+e^-\to \mu^+\mu^-$ and $e^+e^-\to q\bar q$ (in general, $e^+e^-\to F\bar F$, where $F\bar F$ is a
  fermion-antifermion pair), which are of particular interest for the
  $e^+e^-$ collider, depend on the properties of extra dimensions and
  may include resonant contributions due to Kaluza-Klein (KK) excitations,
  string excitations of the Higgs scalar {\em etc}. However, it
  follows from Ref.~\citec{Feng:2010yx} that the three-point couplings
  of Regge excitations to fermion-antifermion pairs are
  model-independent. Furthermore, the relative weights of resonances
  with different spins $J=0,1,2$ are unambigously predicted by the
  theory. Thus the resonant contributions to these amplitudes, with
  Regge excitations propagating in the $s$-channel, are
  model-independent. $e^+e^-$ colliders can be used not only for
  discovering such resonances, but most importantly, for detailed
  studies of their spin content, therefore for distinguishing low mass
  string theory from other beyond the SM extensions predicting the
  existence of similar particles.

\ \\

\section{Regge resonances in photon-photon and electron-positron channels}
\label{III}

\subsection{Universal amplitudes for photon-photon fusion}
\noindent{\bf A:} $\gamma\gamma\to \gamma\gamma$, $\gamma\gamma\to Z^0Z^0$, $\gamma\gamma\to W^+W^-$, $\gamma\gamma\to gg$

The most direct way to compute the amplitude for the scattering of
four gauge bosons is to consider the case of polarized particles
because all non-vanishing contributions can be then generated from a
single, maximally helicity violating (MHV), amplitude -- the so-called
{\it partial\/} MHV amplitude~\cite{Parke:1986gb}.\ft{We do a brief review of this calculation
in Appendix~\ref{sec:4pTSA}.}  Assume that two vector bosons,
with the momenta $k_1$ and $k_2$, in the $U(N)$ gauge group states
corresponding to the generators $T^{a_1}$ and $T^{a_2}$ (here in the
fundamental representation), carry negative helicities while the other
two, with the momenta $k_3$ and $k_4$ and gauge group states $T^{a_3}$
and $T^{a_4}$, respectively, carry positive helicities. (All
momenta are incoming.)  Then the
partial amplitude for such an MHV configuration is given by~\cite{Stieberger:2006te}
\begin{equation}
\label{ampl}
{\cal A}(A_1^-,A_2^-,A_3^+,A_4^+) ~=~ 4\, g^2\, {\rm Tr}
  \, (\, T^{a_1}T^{a_2}T^{a_3}T^{a_4}) {\langle 12\rangle^4\over
    \langle 12\rangle\langle 23\rangle\langle 34\rangle\langle
    41\rangle}V(k_1,k_2,k_3,k_4)\ ,
\end{equation}
where $g$ is the $U(N)$ coupling constant, $\langle ij\rangle$ are the
standard spinor products written in the notation of
Ref.~\cite{Mangano:1990by,Dixon:1996wi}, and the Veneziano formfactor,
\begin{equation}
V(k_1,k_2,k_3,k_4) = V(  s,   t,   u)= \frac{s\,u}{tM_s^2}B(-s/M_s^2,-u/M_s^2)={\Gamma(1-   s/M_s^2)\ \Gamma(1-   u/M_s^2)\over
    \Gamma(1+   t/M_s^2)} \label{formf}
\end{equation}
is the function of Mandelstam variables,
$s=2k_1k_2$, $t=2  k_1k_3$, $u=2 k_1k_4$; $s+t+u=0$.  (For simplicity we drop carets for the parton subprocess.)
The physical content of the form factor becomes clear after using the
well-known expansion in terms of $s$-channel resonances~\cite{Veneziano:1968yb}
\begin{equation}
B(-s/M_s^2,-u/M_s^2)=-\sum_{n=0}^{\infty}\frac{M_s^{2-2n}}{n!}\frac{1}{s-nM_s^2}
\Bigg[\prod_{J=1}^n(u+M^2_sJ)\Bigg],\label{bexp}
\end{equation}
which exhibits $s$-channel poles associated to the propagation of
virtual Regge excitations with masses $\sqrt{n}M_s$. Thus near the
$n$th level pole $(s\to nM^2_s)$:
\begin{equation}\qquad
V(  s,   t,   u)\approx \frac{1}{s-nM^2_s}\times\frac{M_s^{2-2n}}{(n-1)!}\prod_{J=0}^{n-1}(u+M^2_sJ)\ .\label{nthpole}
\end{equation}
In specific amplitudes, the residues combine with the remaining kinematic factors, reflecting the spin content of particles exchanged in the $s$-channel, ranging from $J=0$ to $J=n+1$.\footnote{There are resonances in all the channels, i.e., there are single particle poles in the $t$ and $u$ channels which would show up as bumps if $t$ or $u$ are positive. However, for physical scattering $t$ and $u$ are negative, so we don't see the bumps.} The low-energy
  expansion reads
\begin{equation}
\label{vexp}
V(s,t,u)\approx 1-{\pi^2\over 6}s\,
    u-\zeta(3)\,s\, t\, u+\dots \, .
\end{equation}

Interestingly, because of the proximity of the 8 gluons and the photons on the D-brane, the gluon fusion into $\gamma$ + jet couples at tree level~\cite{Anchordoqui:2007da}.  This implies that there is an order $g^2$ contribution in string theory, whereas this process is not occuring until order $g^4$ (loop level) in field theory. One can write down this process projecting the gamma ray onto the hypercharge, (note that the hypercharge is a color composite state which in turn has the photon) which itself has a $\kappa$.  We discuss this next.

Consider the amplitude involving three $SU(N)$ gluons
$g_1,~g_2,~g_3$ and one $U(1)$ gauge boson $\gamma_4$ associated to
the same $U(N)$ quiver:
\begin{equation}
\label{gens}
T^{a_1}=T^a \ ,~ \ T^{a_2}=T^b\ ,~ \
  T^{a_3}=T^c \ ,~ \ T^{a_4}=QI\ ,
\end{equation}
where $I$ is the $N{\times}N$ identity matrix and $Q$ is the
$U(1)$ charge of the fundamental representation. The $U(N)$
generators are normalized according to
\begin{equation}
\label{norm}
{\rm Tr}(T^{a}T^{b})={1\over 2}\delta^{ab}.
\end{equation}
Then the color
factor \begin{equation}\label{colf}{\rm
    Tr}(T^{a_1}T^{a_2}T^{a_3}T^{a_4})=Q(d^{abc}+{i\over 4}f^{abc})\ ,
\end{equation}
where the totally symmetric symbol $d^{abc}$ is the symmetrized trace
while $f^{abc}$ is the totally antisymmetric structure constant.

The full MHV amplitude can be obtained~\cite{Stieberger:2006te} by summing
the partial amplitudes (\ref{ampl}) with the indices permuted in the
following way: \begin{equation}
\label{afull} {\cal M}(g^-_1,g^-_2,g^+_3,\gamma^+_4)
  =4\,g^{2}\langle 12\rangle^4 \sum_{\sigma } { {\rm Tr} \, (\,
    T^{a_{1_{\sigma}}}T^{a_{2_{\sigma}}}T^{a_{3_{\sigma}}}T^{a_{4}})\
    V(k_{1_{\sigma}},k_{2_{\sigma}},k_{3_{\sigma}},k_{4})\over\langle
    1_{\sigma}2_{\sigma} \rangle\langle
    2_{\sigma}3_{\sigma}\rangle\langle 3_{\sigma}4\rangle \langle
    41_{\sigma}\rangle }\ ,
\end{equation}
where the sum runs over all 6 permutations $\sigma$ of $\{1,2,3\}$ and
$i_{\sigma}\equiv\sigma(i)$. Note that in the effective field theory
of gauge bosons there are no Yang-Mills interactions that could
generate this scattering process at the tree level. Indeed, $V=1$ at
the leading order of Eq.(\ref{vexp}) and the amplitude vanishes
due to the following identity:
\begin{equation}\label{ymlimit}
{1\over\langle 12\rangle\langle
      23\rangle\langle 34\rangle\langle
      41\rangle}+{1\over\langle 23\rangle\langle
      31\rangle\langle 14\rangle\langle 42\rangle}+{1\over\langle 31\rangle\langle
      12\rangle\langle 24\rangle\langle 43\rangle} ~=~0\ .
\end{equation}
Similarly,
the antisymmetric part of
the color factor (\ref{colf}) cancels out in the full amplitude (\ref{afull}). As a result,
one obtains:
\begin{equation}\label{mhva}
{\cal
    M}(g^-_1,g^-_2,g^+_3,\gamma^+_4)=8\, Q\, d^{abc}g^{2}\langle
  12\rangle^4\left({\mu(s,t,u)\over\langle 12\rangle\langle
      23\rangle\langle 34\rangle\langle
      41\rangle}+{\mu(s,u,t)\over\langle 12\rangle\langle
      24\rangle\langle 13\rangle\langle 34\rangle}\right),
\end{equation}
 where
\begin{equation}
\label{mudef}
\mu(s,t,u)= \Gamma(1-u)\left( {\Gamma(1-s)\over
      \Gamma(1+t)}-{\Gamma(1-t)\over \Gamma(1+s)}\right) .
\end{equation}
All non-vanishing amplitudes can be obtained in a similar way. In particular,
\begin{equation}
\label{mhvb}
{\cal M}(g^-_1,g^+_2,g^-_3,\gamma^+_4)=8\, Q\,
  d^{abc}g^{2}\langle 13\rangle^4\left({\mu(t,s,u)\over\langle
      13\rangle\langle 24\rangle\langle 14\rangle\langle
      23\rangle}+{\mu(t,u,s)\over\langle 13\rangle\langle
      24\rangle\langle 12\rangle\langle 34\rangle}\right),
\end{equation}
and the remaining ones can be obtained either by appropriate
permutations or by complex conjugation~\cite{Anchordoqui:2007da}.

It is now straightforward to obtain a general expression for all string disk amplitudes with four external gauge bosons~\cite{Lust:2008qc} 
 \begin{eqnarray}
{\cal M}(A_{1}^{-},A_{2}^{-},A_{3}^{+},A_{4}^{+}) & = & 4\, g^{2}\langle12\rangle^{4}\bigg[
\frac{V_{t}}{\langle12\rangle\langle23\rangle\langle34\rangle\langle41\rangle}
\makebox{Tr}(T^{a_{1}}T^{a_{2}}T^{a_{3}}T^{a_{4}}+T^{a_{2}}T^{a_{1}}T^{a_{4}}
T^{a_{3}})\nonumber \\
 & + & \frac{V_{u}}{\langle13\rangle\langle34\rangle\langle42\rangle\langle21\rangle}
 \makebox{Tr}(T^{a_{2}}T^{a_{1}}T^{a_{3}}T^{a_{4}}+T^{a_{1}}T^{a_{2}}T^{a_{4}}
 T^{a_{3}})\nonumber \\
 & + & \frac{V_{s}}{\langle14\rangle\langle42\rangle\langle23\rangle\langle31
 \rangle}\makebox{Tr}(T^{a_{1}}T^{a_{3}}T^{a_{2}}T^{a_{4}}+T^{a_{3}}T^{a_{1}}
 T^{a_{4}}T^{a_{2}})\bigg] , \label{mhv}
\end{eqnarray}
where
\begin{equation} 
\el{formfsym}
V_t =V(  s,  t,  u) ~,\qquad V_u=V(  t,  u,  s) ~,\qquad  V_s=V(  u,   s,   t) \, .
\end{equation}
In order to
factorize amplitudes on the poles due to the lowest massive string states,
it is sufficient to consider $s=M_s^2$. In this limit, $V_s$ is
regular while
\begin{equation}
V_{t}\to \frac{u}{s-M_s^{2}}~,\qquad V_{u}\to \frac{t}{s-M_s^{2}}~.\end{equation}

Thus the $s$-channel pole term of the amplitude (\ref{mhv}), relevant
to $(--)$ decays of intermediate states, is \begin{equation}
{\cal M}(A_{1}^{-},A_{2}^{-},A_{3}^{+},A_{4}^{+})\to2\, g^{2}\,{\cal C}^{1234}\frac{\langle12\rangle^{4}}{\langle12\rangle\langle23\rangle\langle34\rangle\langle41\rangle}\frac{u}{s-M_s^{2}}\ ,\label{mhvs}\end{equation}
where \begin{equation}
\CC^{1234}=2\Tr(\{T^{a_{1}},T^{a_{2}}\}\{T^{a_{3}},T^{a_{4}}\})=
16\sum_{a=0}^{N^{2}-1}d^{a_{1}a_{2}a}d^{a_{3}a_{4}a}\,.\label{groupfacs}\end{equation}
The amplitude with the $s$-channel pole relevant to $(+-)$ decays
is \begin{equation}
{\cal M}(A_{1}^{-},A_{2}^{+},A_{3}^{+},A_{4}^{-})\to2\, g^{2}\,{\cal C}^{1234}\frac{\langle14\rangle^{4}}{\langle12\rangle\langle23\rangle
\langle34\rangle\langle41\rangle}\frac{u}{s-M_s^{2}}\ .\label{mhvo}\end{equation}
\begin{table}
\caption{Group factors and couplings for the pole terms (\ref{mhvs}) and (\ref{mhvo}).}
\begin{tabular}{l|cc}
\hline
\hline
~~ Process~~& ~~Coupling~~ &~~~~$\CC^{1234}$~~~~ \\
\hline
$CC\to gg$ & $g_a^2$  & $\frac{2}{3}\delta_{a_{3}a_{4}}$ \\[1mm]
$CC\to CC$ &  $g_a^2$ & $\frac{2}{3}$ \\ \hline
$XX\to XX$ & $g_b^2 $& 1\\
$A^{3}A^{3}\to XX$ & $g_b^2$ & 1\\
$A^{3}A^{3}\to A^{3}A^{3}$ & $g_b^2$ & 1\\
$A^{3}X\to A^{3}X$ & $g_b^2$ & 1\\ \hline
$BB\to BB$& $2g_c^2$ & 2\\
\hline
\hline
\end{tabular}
\label{t2}
\end{table}
In Table~\ref{t2}, we list the group factors and couplings [replacing $g^2$ in Eqs.(\ref{mhvs}) and (\ref{mhvo})] for the single-stack processes contributing to $\gamma\gamma$ fusion into gauge bosons, evaluated according to Eq.(\ref{groupfacs}). \footnote{As can be seen in Eq.~(\ref{covderi}) the $X_\mu$ and $C_\mu$ normalization carries a factor $1/\sqrt{2N}$,  which is absent in the $B_\mu$ field. Hence, we should recover the $\sqrt{2N}$ factor (to be $B_\mu (\sqrt 2 g_c)/\sqrt 2 Q_{U(1)}$) and use $\sqrt 2 g_c$ in any calculation that follows from a general  $N$.}

We now proceed to higher level resonances, starting from
$n=2$. Three-particle amplitudes involving one level $n$ Regge
excitation (gauge index $a$) and two massless $U(N)$ gauge bosons
(gauge indices $a_1$ and $ a_2$) are even under the world-sheet parity
(reversing the order of Chan-Paton factors) for odd $n$, and odd for
even $n$ \citec{Feng:2010yx}. As a result, the respective group factors
are the symmetric traces $d^{a_1a_2a}$ for odd $n$ and non-abelian
structure constants $f^{a_1a_2a}$ for even $n$, respectively. For all
configuration of initial particles in the processes listed in
Table~\ref{t2}, $f^{a_1a_2a}=0$, therefore the corresponding
amplitudes have no $s$-channel poles associated to Regge resonances
with even $n$.\footnote{For $n=2$, this has already been checked by
  explicit computation in Ref.\cite{Dong:2010jt}.} {}For $USp(N)$
groups, the parity assignment is reversed, however the relevant
symmetric trace $d^{33a}=0$ for $Sp(1)$, therefore the same conclusion
holds for all SM embeddings under consideration. Thus in order to
observe higher level resonances, $\gamma\gamma$ collisions would have
to reach $\sqrt{s}>\sqrt{3}M_s$, which due to the recently established
$M_s>2.5$ TeV bound~\cite{Khachatryan:2010jd} translates into $\sqrt{s}>4.3$ TeV. It is unlikely
that such high energies will be reached in the next generation of
$\gamma\gamma$ colliders, therefore from now on our discussion will be
limited to the lowest level resonances.

The $\gamma\gamma$ amplitudes are linear combinations of the amplitudes for processes listen in Table \ref{t2}, with the weights determined by the constants $\kappa$, $\eta$, $\xi$, {\em c.f}.\ Eq.(\ref{consts}), and the Weinberg angle $\theta_W$ with:
\be C_{W}=\cos\theta_{W} \quad,\quad S_{W}=\sin\theta_{W}.\ee
{}For the $U(3)_{a}\times U(2)_{b}\times U(1)_{c}$ minimal
model, they are given by:
\begin{eqnarray}
{\cal M} (\xg \xg \to gg)  & = & \kappa^2 C_W{}^2 \, {\cal M} (CC \to gg),  \\[1mm]
{\cal M} (\xg\xg \to \xg\xg) & = &  \xk^4 C_W{}^4 \, {\cal M} (C C \to CC) + 4 \eta^2 S_W{}^2 C_W{}^2 \, {\cal M} (X A^3 \to X A^3)  \nonumber \\
 & & +~\eta^4 C_W{}^4 \,  {\cal M} (X X \to XX)
 + S_W{}^4\,  {\cal M} (A^3 A^3 \to A^3 A^3)\nonumber  \\
& & +~\eta^2 S_W{}^2 C_W{}^2 \,  {\cal M} (A^3 A^3 \to XX) +~\eta^2 S_W{}^2 C_W{}^2 \,  {\cal M} (X X \to A^3 A^3)\nn
& & +~\xi^4 C_W{}^4 \,  {\cal M} (BB \to BB) \nonumber\\[1mm]
& = &  \xk^4 C_W{}^4 \, {\cal M} (C C \to CC) + 4 \eta^2 S_W{}^2 C_W{}^2 \, {\cal M} (X A^3 \to X A^3)  \nonumber \\
 & & +~(S_W{}^4+\eta^4 C_W{}^4 + 2\eta^2 S_W{}^2 C_W{}^2) \,  {\cal M} (X X \to XX)  \nonumber\\
& & +~\xi^4 C_W{}^4 \,  {\cal M} (BB \to BB) \,, 
\end{eqnarray}
\begin{eqnarray}
{\cal M} (\xg \xg \to Z^0Z^0)  & = &  \xk^4 C_W{}^2 S_W{}^2\, {\cal M} (C C \to CC) + 4 \eta^2 S_W{}^2 C_W{}^2 \, {\cal M} (X A^3 \to X A^3)  \nonumber \\
 & & +(S_W{}^2 C_W{}^2+\eta^4 C_W{}^2S_W{}^2 + \eta^2 S_W{}^4+\eta^2 C_W{}^4) \,  {\cal M} (X X \to XX) \nonumber \\
& & +~\xi^4 S_W{}^2 C_W{}^2  \,  {\cal M} (BB \to BB) \, ,\\
{\cal M} (\xg \xg \to W^+W^-)  & = & ~\eta^2 C_W{}^2 \,  {\cal M} (X X \to W^+W^-)
 + S_W{}^2\,  {\cal M} (A^3 A^3 \to W^+W^-)\nonumber  \nn
& = & ~(\eta^2 C_W{}^2+ S_W{}^2)\CM(XX\to
XX).
\end{eqnarray}
{}For the $U(3)_{a}\times Sp(1)_{L}\times U(1)_{c}$ D-brane model, $\eta=0,\xi^{2}=1-\xk^{2}$, and all amplitudes involving $X$ or
$A^{3}$ vanish. We obtain
\begin{eqnarray}
{\cal M}(\xg\xg\to gg)& = & \kappa^{2}C_{W}^{2}\,{\cal M}(CC\to gg) \, ,
\\
{\cal M}(\xg\xg\to\xg\xg)& = & \xk^{4}C_{W}^{4}{\cal M}(CC\to CC)+(1-\xk^{2})^{2}C_{W}^{4}{\cal M}(BB\to BB),\\
{\cal M} (\xg \xg \to Z^0Z^0) & = & C_W^2 S_W^2[\xk^4 {\cal M} (C C \to CC)   +(1-\xk^2)^2  {\cal M} (BB \to BB)],\quad\quad\\
{\cal M} (\xg \xg \to W^+W^-)  & = & 0 \, .
\end{eqnarray}

\noindent{\bf B:} $\gamma\gamma\to F\bar F$

Since the vertex operators creating chiral mater fermions contain boundary changing operators connecting two different stacks of intersecting D-branes, say $a$ and $b$, the disk boundary in the amplitudes involving two fermions and two gauge bosons is always attached to two stacks of D-branes. The gauge bosons can couple either to the same stack or to two different stacks. In the latter case, the amplitude with two gauge bosons in the initial state is proportional to $V_s$, which has no poles in the $s$-channel \citec{Lust:2008qc}. The only amplitudes exhibiting $s$-channel poles involve the two initial gauge bosons associated to the same stack, but carrying opposite helicities \citec{Lust:2008qc}:
\begin{equation}
{\cal M}(A_{1}^{-},A_{2}^{+},F_{3}^{-},\bar{F}_{4}^{+})=2\, g^{2}\frac{\langle13\rangle^{2}}{\langle32\rangle\langle42\rangle}
\bigg[\frac{t}{s}V_{t}(T^{a_{1}}T^{a_{2}})_{\alpha_{3}\alpha_{4}}+
\frac{u}{s}V_{u}(T^{a_{2}}T^{a_{1}})_{\alpha_{3}\alpha_{4}}\bigg]\ .\label{quarks}\end{equation}
The above equation describes the case of stack $a$, hence the (fermion) spectator indices associated to stack $b$ have been suppressed.
The lowest Regge excitations give rise to the pole term \begin{equation}
{\cal M}(A_{1}^{-},A_{2}^{+},F_{3}^{-},\bar{F}_{4}^{+})\to2\, g^{2}\ {\cal D}^{1234}\frac{\langle13\rangle^{2}}{\langle32\rangle\langle42\rangle}\frac{tu}{M_s^{2}(s-M_s^{2})}\ ,\label{qlim}\end{equation}
where the group factor
\be \CD^{1234}\equiv\{T^{a_{1}},T^{a_{2}}\}_{\xa_{3},\xa_{4}}\ .\ee
The group factors and couplings for the processes relevant to $\gamma\gamma\to F\bar F$ are listed in Table~\ref{t3}.
\begin{table}
\caption{Group factors and couplings for the pole terms (\ref{qlim}).}
\begin{tabular}{l|cc}
\hline
\hline
~~ Process~~& ~~Coupling~~ &~~~~$\CD^{1234}$~~~~ \\
\hline
$CC\to q\bar{q}$ & $g_a^2$  & $\frac{1}{3}\delta_{\alpha_{3}\alpha_{4}}$ \\ \hline
$XX\to q_{L}\bar{q}_{R}$ & $g_b^2$ &$\frac{1}{2}$\\
$A^{3}A^{3}\to q_{L}\bar{q}_{R}$ & $g_b^2$ & $\frac{1}{2}$\\
$A^{3}X\to u_{L}\bar{u}_{R}$ &$g_b^2$ & $\frac{1}{2}$\\
$A^{3}X\to d_{L}\bar{d}_{R}$ &$g_b^2$ & $-\frac{1}{2}$\\ \hline
$BB\to q_{R}\bar{q}_{L}$ & $2g_c^2$ & 1\\ \hline
$XX\to e_{R}^{+}e_{L}^{-}$ & $g_b^2$ &$\frac{1}{2}$\\
$A^{3}X\to e_{R}^{+}e_{L}^{-}$ &$g_b^2$ & $-\frac{1}{2}$\\
$A^{3}A^{3}\to e_{R}^{+}e_{L}^{-}$ & $g_b^2$ & $\frac{1}{2}$\\
$XX\to \bar{\nu}_{R}\nu_{L}$ & $g_b^2$ &$\frac{1}{2}$\\
$A^{3}X\to \bar{\nu}_{R}\nu_{L}$ &$g_b^2$ & $\frac{1}{2}$\\
$A^{3}A^{3}\to \bar{\nu}_{R}\nu_{L}$ & $g_b^2$ & $\frac{1}{2}$\\
 \hline
$BB\to e_{R}^{+}e_{L}^{-}$ & $2g_c^2$ & 1\\
$BB\to e_{L}^{+}e_{R}^{-}$ & $2g_c^2$ & 2\\
$BB\to \bar{\nu}_{R}\nu_{L}$ & $2g_c^2$ & 1\\
$BB\to\bar{\nu}_{L}\nu_{R}$ & $2g_c^2$ & 2\\
\hline
\hline
\end{tabular}
\label{t3}
\end{table}

As in the case of $\gamma\gamma$ fusion into gauge boson pairs, the higher level resonances contributing to
$\gamma\gamma\to F\bar F$ come from odd $n$ levels only, so here again, we limit our discussion to $n=1$.
{} In the  $U(3)_{a}\times U(2)_{b}\times U(1)_{c}$ case, the relevant amplitudes
are
\begin{eqnarray}
{\cal M} (\xg \xg \to q_L \bar q_R)  & = & \eta^2 C_W{}^2 \, {\cal M} (X X \to q_L \bar q_R) +S_W{}^2 \, {\cal M} (A^3 A^3 \to q_L \bar q_R)\nonumber \\
 & & + \xk^2 C_W{}^2 \, {\cal M}(CC \to q_L \bar q_R)+~2\eta C_W S_W \, {\cal M}(X A^3 \to q_L \bar q_R) \nn
 & = & (\eta^2 C_W{}^2+S_W{}^2) \, {\cal M} (X X \to q_L \bar q_R) + \xk^2 C_W{}^2 \, {\cal M}(CC \to q_L \bar q_R) \nonumber \\
 & & +~2\eta C_W S_W \, {\cal M}(X A^3 \to q_L \bar q_R) \,, \\[1mm]
{\cal M} (\xg \xg \to q_R \bar q_L)  & = &  \xi^2 C_W{}^2  \, {\cal M} (BB \to q_R \bar q_L) + \xk^2 C_W{}^2 \, {\cal M} (CC \to q_R \bar q_L) \,, \\[1mm]
{\cal M} (\xg \xg \to e_R^+ e_L^-) &  = &  \eta^2 C_W{}^2 \, {\cal M} (X X \to e_R^+ e_L^-)+S_W{}^2 \, {\cal M} (A^3 A^3 \to e_R^+ e_L^-)  \nonumber \\
 &  & + \xi^2 C_W{}^2 \,  {\cal M} (BB \to e_R^+ e_L^-)+~2\eta C_W S_W \, {\cal M}(X A^3 \to e_R^+ e_L^-) \nn
&  = &  (\eta^2 C_W{}^2+S_W{}^2) \, {\cal M} (X X \to e_R^+ e_L^-) + \xi^2 C_W{}^2 \,  {\cal M} (BB \to e_R^+ e_L^-) \nonumber \\
 &  & +~2\eta C_W S_W \, {\cal M}(X A^3 \to e_R^+ e_L^-) \, ,\label{fantif1}\\[1mm]
\CM(\xg \xg \to e_L^+ e_R^-) & = & \xi^2 C_W{}^2\CM(B B \to e_L^+ e_R^-) .\label{fantif2}
\end{eqnarray}
The amplitudes describing neutrino-antineutrino pair production can be obtained from Eqs.(\ref{fantif1}) and (\ref{fantif2}) by the replacement $e^{-}_L\to\nu_L,~ e^{+}_R\to\bar{\nu}_R$.
{}For the $U(3)_{a}\times Sp(1)_{L}\times U(1)_{c}$ D-brane model, we obtain:
\begin{eqnarray}
{\cal M}(\xg\xg\to q_{L}\bar{q}_{R}) & = & \xk^{2}C_{W}{}^{2}\,{\cal M}(CC\to q_{L}\bar{q}_{R})\,,\\[1mm]
{\cal M}(\xg\xg\to q_{R}\bar{q}_{L}) & = & (1-\xk^{2})C_{W}{}^{2}\,{\cal M}(BB\to q_{R}\bar{q}_{L})+\xk^{2}C_{W}{}^{2}\,{\cal M}(CC\to q_{R}\bar{q}_{L})\,,\quad \quad
\\[1mm]
\label{41}
{\cal M}(\xg\xg\to e^{\pm}e^{\mp}) & = & (1-\xk^{2})C_{W}{}^{2}\,{\cal M}(BB\to e^{\pm}e^{\mp})\ ,\\[1mm]
{\cal M}(\xg\xg\to \nu\bar{\nu}) & = & (1-\xk^{2})C_{W}{}^{2}\,{\cal M}(BB\to \nu\bar{\nu})\ .
\end{eqnarray}

\subsection{Various amplitudes for electron-positron annihilation}

\noindent{\bf A:} $e^+e^-\to\gamma\gamma$, $e^+e^-\to Z^0Z^0$, $e^+e^-\to Z^0\gamma$, $e^+e^-\to W^+W^-$

Leptons are decoupled from gluons at the disk level because they originate from strings ending on different D-branes. Thus $e^+e^-$ pairs can annihilate into photons and electroweak bosons only.\footnote{$e^+e^- \to \gamma \gamma$ in a toy, one-stack, stringy model has been discussed in~\citec{Cullen:2000ef}.} The corresponding resonance pole terms are obtained by crossing from Eq.(\ref{mhvs}):
 \begin{equation}
{\cal M}([e^{\pm}]_{1}^{-},[e^{\mp}]_{2}^{+},A_{3}^{-},A_{4}^{+},)\to2\, g^{2}\ {\cal D}^{1234}\frac{\langle13\rangle^{2}}
{\langle 14\rangle\langle 24\rangle}\frac{tu}{M_s^{2}(s-M_s^{2})}\ ,
\end{equation}
with the same group factors as in Table~\ref{t3}, but running in the time-reversed channels.
In the $U(3)_{a}\times U(2)_{b}\times U(1)_{c}$ case, the physical amplitudes for the processes under consideration are
\begin{align}
{\cal \mathcal{M}}(e_{R}^{+}e_{L}^{-}\rightarrow\gamma\gamma) & =\eta^{2}C_{W}{}^{2}\,{\cal M}(e_{R}^{+}e_{L}^{-}\rightarrow XX)+S_{W}{}^{2}\,{\cal M}(e_{R}^{+}e_{L}^{-}\rightarrow A^{3}A^{3})\nonumber \\
 & \quad+\xi^{2}C_{W}{}^{2}\,{\cal M}(e_{R}^{+}e_{L}^{-}\rightarrow BB)+~2\eta C_{W}S_{W}\,{\cal M}(e_{R}^{+}e_{L}^{-}\rightarrow XA^{3})\nonumber \\
 & =(\eta^{2}C_{W}{}^{2}+S_{W}{}^{2})\,{\cal M}(e_{R}^{+}e_{L}^{-}\rightarrow XX)+\xi^{2}C_{W}{}^{2}\,{\cal M}(e_{R}^{+}e_{L}^{-}\rightarrow BB)\nonumber\\
 & \quad+2\eta C_{W}S_{W}\,{\cal M}(e_{R}^{+}e_{L}^{-}\rightarrow XA^{3})\,,\\[1mm]
\mathcal{M}(e_{L}^{+}e_{R}^{-}\rightarrow\gamma\gamma) & =\xi^{2}C_{W}{}^{2}\mathcal{M}(e_{L}^{+}e_{R}^{-}\rightarrow BB) \,,\\[1mm]
\mathcal{M}(e_{R}^{+}e_{L}^{-}\rightarrow Z^0Z^0)  & =(\eta^{2}S_{W}{}^{2}+C_{W}{}^{2})\,{\cal M}(e_{R}^{+}e_{L}^{-}\rightarrow XX)+\xi^{2}S_{W}{}^{2}\,{\cal M}(e_{R}^{+}e_{L}^{-}\rightarrow BB)\nonumber\\
 & \quad+2\eta C_{W}S_{W}\,{\cal M}(e_{R}^{+}e_{L}^{-}\rightarrow XA^{3})\,,\\[1mm]
\mathcal{M}(e_{L}^{+}e_{R}^{-}\rightarrow Z^0Z^0) & =\xi^{2}S_{W}{}^{2}\mathcal{M}(e_{L}^{+}e_{R}^{-}\rightarrow BB) \,,\\[1mm]
\mathcal{M}(e_{R}^{+}e_{L}^{-}\rightarrow Z^0\gamma) & =S_{W}C_W(\eta^{2}+1)\,{\cal M}(e_{R}^{+}e_{L}^{-}\rightarrow XX)+\xi^{2}S_{W}{}C_W\,{\cal M}(e_{R}^{+}e_{L}^{-}\rightarrow BB)\nonumber\\
 & \quad+\eta (C_{W}{}^2+S_{W}{}^2)\,{\cal M}(e_{R}^{+}e_{L}^{-}\rightarrow XA^{3})\,,\\[1mm]
\mathcal{M}(e_{L}^{+}e_{R}^{-}\rightarrow Z^0\gamma) & =\xi^{2}S_{W}C_W\mathcal{M}(e_{L}^{+}e_{R}^{-}\rightarrow BB) \,,\\[1mm]
\mathcal{M}(e_{R}^{+}e_{L}^{-}\rightarrow W^+W^-) &= {\cal M}(e_{R}^{+}e_{L}^{-}\rightarrow A^{3}A^{3}) \, ,\\[1mm]
\mathcal{M}(e_{L}^{+}e_{R}^{-}\rightarrow W^+W^-) & = 0 \, .
\end{align}
{}For the $U(3)_{a}\times Sp(1)_{L}\times U(1)_{c}$ D-brane model,
we have
 \begin{align}
{\cal \mathcal{M}}(e_{R}^{+}e_{L}^{-}\rightarrow\gamma\gamma) & =\xi^{2}C_{W}{}^{2}\,{\cal M}(e_{R}^{+}e_{L}^{-}\rightarrow BB) \,,\\[1mm]
\mathcal{M}(e_{L}^{+}e_{R}^{-}\rightarrow\gamma\gamma) & =\xi^{2}C_{W}{}^{2}\mathcal{M}(e_{L}^{+}e_{R}^{-}\rightarrow BB) \,,\\[1mm]
\mathcal{M}(e_{R}^{+}e_{L}^{-}\rightarrow Z^0Z^0) & = \xi^{2}S_{W}{}^{2}{\cal M}(e_{R}^{+}e_{L}^{-}\rightarrow BB)\,,\\[1mm]
\mathcal{M}(e_{L}^{+}e_{R}^{-}\rightarrow Z^0Z^0)& =\xi^{2}S_{W}{}^{2}\mathcal{M}(e_{L}^{+}e_{R}^{-}\rightarrow BB) \,,  \\[1mm]
\mathcal{M}(e_{R}^{+}e_{L}^{-}\rightarrow Z^0\gamma) & = \xi^{2}S_{W}{}C_W\,{\cal M}(e_{R}^{+}e_{L}^{-}\rightarrow BB) \,,\\[1mm]
\mathcal{M}(e_{L}^{+}e_{R}^{-}\rightarrow Z^0\gamma) &=\xi^{2}S_{W}C_W\mathcal{M}(e_{L}^{+}e_{R}^{-}\rightarrow BB) \,, \\[1mm]
\mathcal{M}(e_{R}^{+}e_{L}^{-}\rightarrow W^+W^-) & = {\cal M}(e_{R}^{+}e_{L}^{-}\rightarrow A^{3}A^{3}) \,,\\[1mm]
\mathcal{M}(e_{L}^{+}e_{R}^{-}\rightarrow W^+W^-) & = 0 \, .
\end{align}

\noindent{\bf B:} $e^+e^-\to e^+e^-$, $e^+e^-\to \mu^+\mu^-$, $e^+e^-\to \nu\bar{\nu}$,
$e^+e^-\to q\bar{q}$
\label{eeee}

Four-fermion amplitudes \citec{Lust:2008qc} are not universal -- they depend on the internal radii and other details of extra dimensions already at the disk level. In particular, they contain resonance poles due to Kaluza-Klein excitations. More serious problems though are due to the presence of resonance poles associated to both massless and massive particles that are either unacceptable from the phenomenological point of view, or are expected to receive large mass corrections
due to quantum (anomaly) effects, see Ref.\cite{Anchordoqui:2009mm} for more details. {}For example, the same Green-Schwarz mechanism that generates non-zero masses for anomalous gauge bosons does also affect the masses of their Regge excitations. {}For the above reasons, phenomenological  analysis of $e^+e^-$ annihilation into lepton-antilepton pairs will be quite complicated, as described in more detail in the following Sec.~\ref{eplusemin}.

Here, we focus on the lowest Regge excitations of the photon and $Z^0$, remaining in the spectrum of any realistic model. Since we are considering energies far above the electroweak scale, we can replace $\gamma$ and $Z^0$ by the neutral gauge bosons of unbroken $SU(2)\times U(1)_Y$.

At the lowest, $n=1$ level, each gauge boson comes with several Regge excitations with spins ranging from 0 to 2, but only two particles couple to quark-antiquark and lepton-antilepton pairs: one spin 2 boson and one spin 1 vector particle \citec{Anchordoqui:2008hi}. All three-particle couplings involving one Regge excitation, one fermion and one antifermion have been determined in Ref.\cite{Anchordoqui:2008hi} by using the factorization methods. These S-matrix elements are completely sufficient for
reconstructing the resonance part of four-fermion amplitudes \citec{Anchordoqui:2008hi} by using the Wigner matrix techniques. In the center of mass frame, the relevant amplitudes can be written as
\begin{eqnarray}
{\cal M}(e^-_Le^+_R\to F_L\bar F_R) & \to &
\frac{M_s^2}{s-M_s^2}\frac{e^2}{4}\Big(\frac{Y_F}{C_W^2}+\frac{I_{3F}}{S_W^2}\Big)\Big[d^2_{1,1}(\theta)+
\frac{1}{3}d^1_{1,1}(\theta)\Big],\\[1mm]
{\cal M}(e^-_Le^+_R\to F_R\bar F_L) & \to & \frac{M_s^2}{s-M_s^2}\frac{e^2}{4}\frac{Y_F}{C_W^2}
\Big[d^2_{1,-1}(\theta)+
\frac{1}{3}d^1_{1,-1}(\theta)\Big],\\[1mm]
{\cal M}(e^-_Re^+_L\to F_L\bar F_R) & \to & \frac{M_s^2}{s-M_s^2}\frac{e^2}{2}\frac{Y_F}{C_W^2}
\Big[d^2_{1,-1}(\theta)+
\frac{1}{3}d^1_{1,-1}(\theta)\Big],\\[1mm]
{\cal M}(e^-_Re^+_L\to F_R\bar F_L) & \to & \frac{M_s^2}{s-M_s^2}\frac{e^2}{2}\frac{Y_F}{C_W^2}
\Big[d^2_{1,1}(\theta)+
\frac{1}{3}d^1_{1, 1}(\theta)\Big],
\end{eqnarray}
where $Y_F$ is the fermion hypercharge, $I_{3F}$ is the fermion weak isospin, and
\begin{equation}
d^2_{1,\pm 1}(\theta)=\frac{1\pm\cos\theta}{2}(2\cos\theta\mp 1)\ ,
\qquad\quad d^1_{1,\pm 1}(\theta)=\frac{1\pm\cos\theta}{2}\ ,
\end{equation}
are the spin 2 and spin 1 Wigner matrix elements~\citec{WignerGT,EdmondsAM}, respectively.
A very  interesting aspect of the above result is that string theory predicts
the {\em  precise} value, equal 1/3, of the relative weight of spin 2 and spin 1 contributions.

Here again, we would like to stress that although the full four-fermion scattering amplitudes are model-dependent, their resonance parts are universal because
the three-particle couplings involving one Regge excitation and two massless particles do not depend on the compactification space \citec{Feng:2010yx}.

\ \\

\section{CLIC phenomenology for string hunters}
\label{IV}

$e^+ e^-$ linear colliders are considered as the most desirable
facility to complement measurements at the LHC. Two alternative linear
projects are presently under consideration: the International Linear
Collider (ILC) and CLIC. The first one is based on superconducting
technology in the TeV range, whereas the second one is based on the
novel approach of two beam acceleration to extend linear colliders
into the multi-TeV range. The choice will be based on the respective
maturity of each technology and on the physics requests derived from
the LHC physics results when available.

CLIC aims at multi-TeV collisions  with high-luminosity, ${\cal
  L}_{e^+e^-} \sim 8 \times 10^{34}~{\rm cm}^{-2} \, {\rm
  s}^{-1}$~\citec{Ellis:2000iw}. The facility would be built in
phases. The initial center-of-mass energy has been arbitrarily chosen
to be $\sqrt{s} = 500$~GeV to allow direct comparison with ILC. The
collider design has been optimized for $\sqrt{s} = 3$~TeV, with a
possible upgrade path to $\sqrt{s} = 5$~TeV at constant
luminosity~\citec{Accomando:2004sz}. To keep the length (and thereby
the cost) of the machine at a reasonable level, the CLIC study
foresees a two beam accelerating scheme featuring an accelerating
gradient in the presence of a beam (loaded) in the order of 80~MV/m
and 100~MV/m, for the 500~GeV and 3~TeV options; the projected total
site lengths are 13.0~km and 48.3~km,
respectively~\citec{Ellis:2008gj}. The CLIC technology is less mature
than that of the ILC. In particular, the target accelerating gradient
is considerable higher than the ILC and requires very aggressive
performance from accelerating structures.

In addition, photon collisions that will considerably enrich the CLIC
physics program can be obtained for a relatively small incremental
cost. Recently, an exploratory study has been carried out to determine
how this facility could be turned into a collider with a high
geometric luminosity, which could be used as the basis for a $\gamma
\gamma$ collider~\citec{Telnov:2009vq}. The hard photon beam of the
$\gamma\gamma$ collider can be obtained by using the laser
back-scattering technique, i.e., the Compton scattering of laser light
on the high energy electrons~\citec{Ginzburg:1981ik}. The scattered
photons have energies close to the energy of the initial electron
beams, and the expected $\gamma \gamma$ and $\gamma e$ luminosities
can be comparable to that in $e^+e^-$ collisions, {\em e.g.,} ${\cal
  L}_{\gamma \gamma} \sim 2 \times 10^{34}~{\rm cm}^{-2} \, {\rm
  s}^{-1}$. In this section we study the distinct phenomenology of
Regge recurrences arising in the $\gamma \gamma$ and $e^+ e^-$ beam
settings.

\subsection{photon-photon collisions}

As an illustration of the CLIC potential to uncover string signals,
we focus attention on dominant $\gamma \gamma \to e^+ e^-$
scattering, within the context of the $U(3)_a \times Sp(1)_L \times
U(1)_c$ D-brane model. Let us first isolate the contribution to the
partonic cross section from the first resonant state, $B^*$. The $s$-channel
pole term of the average square amplitude can be obtained from the
formula (\ref{41}) by taking into account all possible initial
polarization configurations. However, for phenomenological purposes,
the pole needs to be softened to a Breit-Wigner form by obtaining and
utilizing the correct {\em total} widths of the resonance. After this
is done we obtain
\be
  |{\cal M} (\xg\xg \to  e^+ e^-)|^2  =  (1 + 4 ) \, (1-\xk^2)^2 \, C_W^4  \, \frac{4 g_c^4}{M_s^4}\
 \left [\frac{  u   t (   u^2+   t^2)}{( s-M_s^2)^2 +
(\Gamma_{B^*}^{J=2}\ M_s)^2} \right] \, ,
\label{gagaee}
\ee
where the factor of $(1+4)$  in the numerator accounts for the fact that the $U(1)_c$ charge of $e_R$ is twice that of $e_L$. The decay width of $B^*$ is given by \ba \xG_{B^*}^{J=2} & = &
\xG_{B^* \to l \bar l}^{J=2} + \xG_{B^* \to q_R \bar q_L}^{J=2} +
\xG_{B^* \to B B}^{J=2} \nn & = & \frac {g_c^2} {\pi} M_s \lsb \frac
{1} {40}\lb \frac 5 2 N_e+ \frac 1 2 N_\nu + \frac 1 2 N_q\rb + \frac
1 {5 N} \rsb\nn & = & \frac {13}{20} \frac {g_c^2} {4\pi} M_s, \ea
where $N_e = 3, N_\nu = 3, N_q = 18$. The first term comprises the
contribution from the left-handed ($N_e/2$) and right-handed ($2 N_e$)
electrons, the second term ($N_\nu/2$) comes from the left-handed
neutrinos, and the third term $(N_q/2)$ subsume the right-handed
quarks.

The total cross section at an $e^{+}e^{-}$ linear collider can be obtained by folding $\hat{\sigma} (\hat{s})$ with the photon distribution function~\citec{Jikia:1991hc} \begin{equation}
  \sigma_{\rm tot}(e^+e^- \Rightarrow \gamma\gamma \to  e^+ e^-) =\int^{x_{\rm max}}_{M/\sqrt{s}} dz \ \frac{d{\cal
      L}_{\gamma\gamma}}{dz} \ \hat{\sigma}(\hat{s}=z^2 s ) \,,
 \end{equation}
where  $\hat{s}$ and $s$ indicate respectively the
center-of-mass energies of the $\gamma\gamma$ and the parent $e^{+}e^{-}$
systems and
\begin{eqnarray}
\frac{d{\cal L}_{\gamma\gamma}}{dz}=2z\int_{z^2/x_{\rm max}}^{x_{\rm max}}
 \frac{dx}{x} f_{\gamma/e}(x)f_{\gamma/e}(z^2/x) \,
\end{eqnarray}
is the distribution function of photon luminosity. The energy spectrum of the back scattered photon in unpolarized incoming $e \gamma$ scattering is given by
\begin{eqnarray}
\label{structure}
f_{\gamma/e}(x)=\frac{1}{D(\xi)}\left[1-x+\frac{1}{1-x}-
\frac{4x}{\xi(1-x)}+\frac{4x^{2}}{\xi^{2}(1-x)^2}\right],~~~(x<x_{\rm max}) \,,
\end{eqnarray}
where $x=2
\omega/\sqrt{s}$ is the fraction of the energy of the incident electron carried
by the back-scattered photon  and $x_{\rm max}=2
\omega_{\rm max}/\sqrt{s}=\xi/(1+\xi)$. For $x>x_{\rm max}$,
$f_{\gamma/e}=0$. The function $D(\xi)$ is defined as
\begin{equation}
D(\xi)= \left(1-\frac{4}{\xi}-\frac{8}{\xi^{2}} \right)\ln(1+\xi)+\frac{1}{2}+\frac{8}{\xi}-\frac{1}{2(1+\xi)^{2}}.
\end{equation}
where $\xi=2 \omega_0\sqrt{s}/{m_e}^2$, $m_{e}$ and $\omega_0$ are respectively the electron mass and laser-photon energy, and (of course) the incoming electron energy is $\sqrt{s}/2$. In our evaluation, we choose $\omega_0$ such that it maximizes the backscattered photon energy without spoiling the luminosity through $e^{+}e^{-}$ pair creation, yielding ${\xi}=2(1+\sqrt{2})$, $x_{\rm
  max}\simeq 0.83$ and $D(\xi) \approx 1.84$~\citec{Cheung:1992jn}.

\begin{figure}[tbp]
\postscript{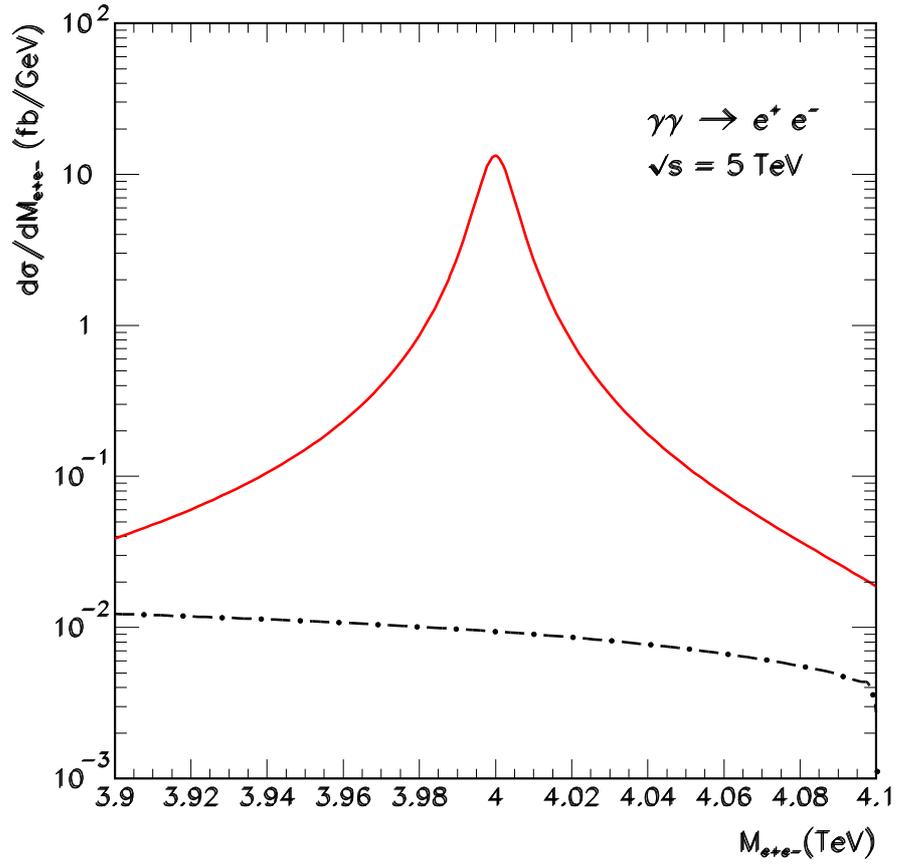}{0.8}
\caption[Invariant mass distributions for $\xg \xg \to e^+e^-$]{$d\sigma/dM_{e^+e^-}$ (units of fb/GeV) {\em vs.}
  $M_{e^+e^-}$ (TeV) is plotted for the case of SM background (dot-dashed line)
  and (first resonance) string signal + background (solid line), for  $M_s=4$~TeV and $\sqrt{s} = 5$~TeV. (We have taken $\kappa = 0.14.$)}
\label{fig:bump}
\end{figure}

We study the signal-to-noise of Regge excitations in data binned
according to the invariant mass $M_{e^+e^-}$ of the $e^+ e^-$ pair, after
setting cuts on the different electron-positron rapidities, $|y_1|, \, |y_2|
\le 2.4$ and transverse momenta $p_{\rm
  T}^{1,2}>50$~GeV.  With the definitions $Y\equiv \frac 1 2 (y_1 +
y_2)$ and $y \equiv \frac 1 2 (y_1-y_2)$, the cross section per
interval of $M_{e^+e^-}$ for $\xg\xg \rightarrow e^+ e^-$ is given
by \begin{eqnarray} \frac{d\sigma}{dM_{e^+e^-}} & = & \sqrt s z^{3}\ \left[
    \int_{-Y_{\rm max}}^{0} dY \ f_{\xg/e} (x_a) \right. \ f_{\xg/e}
  (x_b) \ \int_{-(y_{\rm max} + Y)}^{y_{\rm max} + Y} dy
  \left. \frac{d\hat \sigma}{d\hat t}\right|_{\xg\xg\rightarrow
    e^+e^-}\ \frac{1}{\cosh^2
    y} \nonumber \\
  & + &\int_{0}^{Y_{\rm max}} dY \ f_{\xg/e} (x_a) \ f_{\xg/e} (x_b) \
  \int_{-(y_{\rm max} - Y)}^{y_{\rm max} - Y} dy
  \left. \left. \frac{d\hat \sigma}{d\hat t}\right|_{\xg\xg\rightarrow
      e^+ e^-}\ \frac{1}{\cosh^2 y}
  \right] \label{longBHclic} \end{eqnarray} where $z^2 = M_{e^=e^-}^2/s$, $x_a =z
e^{Y}$, $x_b =z e^{-Y},$ and \begin{equation} |{\cal M}(\xg\xg \to
  e^+e^-) |^2 = 16 \pi \hat s^2 \, \left. \frac{d\sigma}{d\hat t}
  \right|_{\xg \xg \to e^+ e^-} \, .
\end{equation}
The string signal is
calculated using (\ref{longBHclic}) with the corresponding $\gamma \gamma \to e^+ e^-$ scattering amplitude given in Eq.~(\ref{gagaee}). The SM background is calculated using
\be
\frac {d \hat \xs}{d \hat t} =\frac {2 \pi
\xa^2}{\hat s^2} \lb\frac {\hat u}{\hat
t} + \frac {\hat t}{\hat u}\rb \, .
\label{gagaeeSM}
\ee The kinematics of the scattering also provides the relation $M_{e^+e^-} = 2p_T \cosh y$, which when combined with the standard cut $p_T \gtrsim p_{T,\rm min}$, imposes a {\em lower} bound on $y$ to be implemented in the limits of integration.  The $Y$ integration range in Eq.~(\ref{longBHclic}), $Y_{\rm max} = {\rm min} \{ \ln(x_{\rm{max}}/z),\ \ y_{\rm max}\}$, comes from requiring $x_a, \, x_b < x_{\rm{max}}$ together with the rapidity cuts $0 <|y_1|, \, |y_2| < 2.4$. Finally, the Mandelstam invariants occurring in the cross section are given by $\hat s = M_{e^+e^-}^2,$ $\hat t = -\frac 1 2 M_{e^+e^-}^2\ e^{-y}/ \cosh y,$ and $\hat u = -\frac 1 2 M_{e^+e^-}^2\ e^{+y}/ \cosh y.$ In Fig.~\ref{fig:bump} we show a representative plot of the invariant mass spectrum, for $M_s = 4$~TeV and $\sqrt{s} = 5$~TeV.

We now estimate (at the parton level)  the signal-to-noise ratio at CLIC. Standard bump-hunting methods, such as obtaining cumulative cross sections, $\sigma (M_0) = \int_{M_0}^\infty \frac{d\sigma}{dM_{e^+e^-}} \, \, dM_{e^+e^-}$, from the data and searching for regions with significant deviations from the SM background, may reveal an interval of $M_{e^+e^-}$ suspected of containing a bump.  With the establishment of such a region, one may calculate the detection significance
\begin{equation}
  S_{\rm det} = \frac{N_{\rm S}}{\sqrt{N_{\rm B} + N_{\rm S}}} \, ,
\end{equation}
with the signal rate $N_{\rm S}$ estimated in the invariant mass window $[M_s - 2 \Gamma, \, M_s + 2 \Gamma]$, and the number of background events $N_{\rm B}$ defined in the same $e^+e^-$ mass interval for the same integrated luminosity~\citec{Anchordoqui:2006pb}.  For $\sqrt{s} = 5$~TeV and $M_s = 4$~TeV we expect $S_{\rm det} \simeq 139/12 = 11\sigma$, after the first fb$^{-1}$ of data collection.  The spin-2  nature of  $\gamma \gamma \to e^+ e^-$ Regge recurrences would make them smoking guns for low mass scale D-brane string compactifications.

\subsection{electron-positron collisions}
\label{eplusemin}
\begin{figure}[tbp]
\postscript{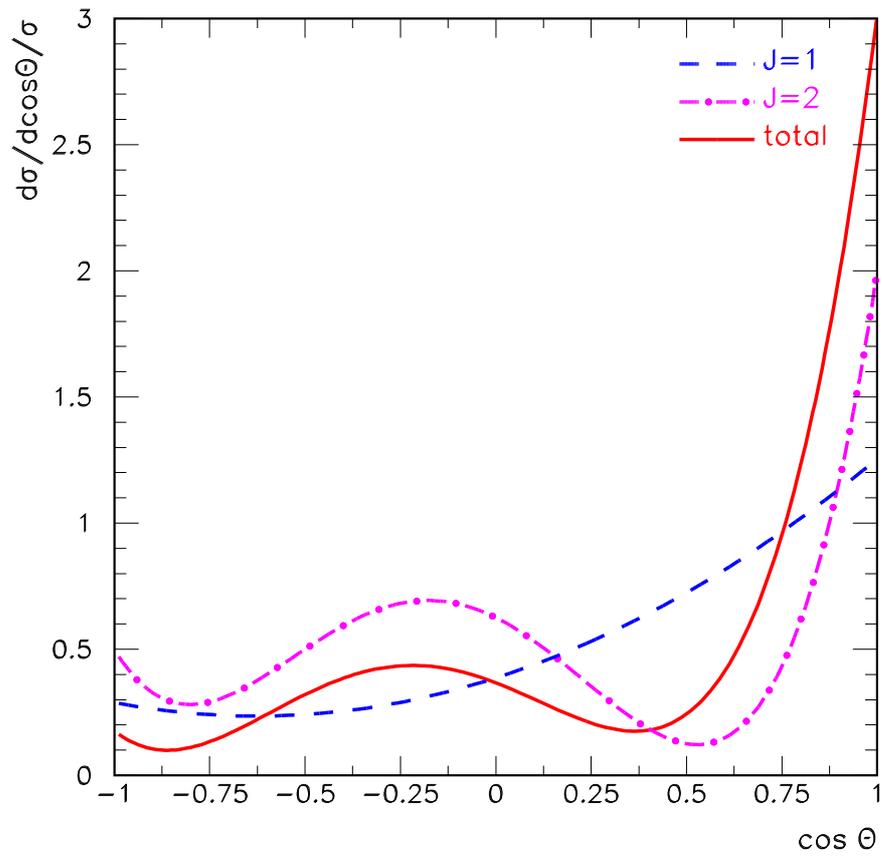}{0.8}
\caption[Normalized angular distributions of Regge recurrences]{Normalized angular distributions of Regge recurrences with spin 1, 2, and total in the $e^+ e^- \to \mu^+ \mu^-$ channel.}
\label{fig:ad}
\end{figure}

We assume that the $e^+e^-$ center-of-mass energy will be tuned to
contain the interesting range highlighted by LHC data and that the
resolution of the machine will be sufficient to probe narrow
resonances. We are interested in the $e^+e^-$ annihilation into lepton-antilepton pairs, in particular in $e^-e^+ \to \mu^-\mu^+$. Phenomenological analysis of such processes will be quite complicated, due the presence of model-dependent backgrounds of KK excitations, anomalous gauge gauge bosons and their Regge excitations. Weakly-interacting KK excitations are expected to have masses lower than the string scale \citec{Anchordoqui:2009mm}, and can appear as resonances in the $e^+e^-$ annihilation channel. Their signals will be similar to a generic $Z'$, with a unique angular momenta, commonly $J=1$ and will not provide direct evidence for the superstring substructure. The signals of gauge bosons associated to anomalous $U(1)$ gauge bosons, with masses always lower than the string scale, varying from a loop factor to a large suppression by the volume of the bulk \citec{Antoniadis:2002cs}, will have a similar character. We assume that no accidental degeneracy occurs between these particles and Regge excitations, so that the string signal discussed Sec.~\ref{eeee} can be safely isolated from the background. Even in this case, however, there is a certain amount of ambiguity due to the presence of Regge excitations of anomalous $U(1)$'s with masses shifted by radiative corrections \citec{Kitazawa:2010gh}. If this shift is large, there will be a separate resonance peak, but if it is small, it will affect the normalization of the signal.

\begin{figure}[tbp]
\postscript{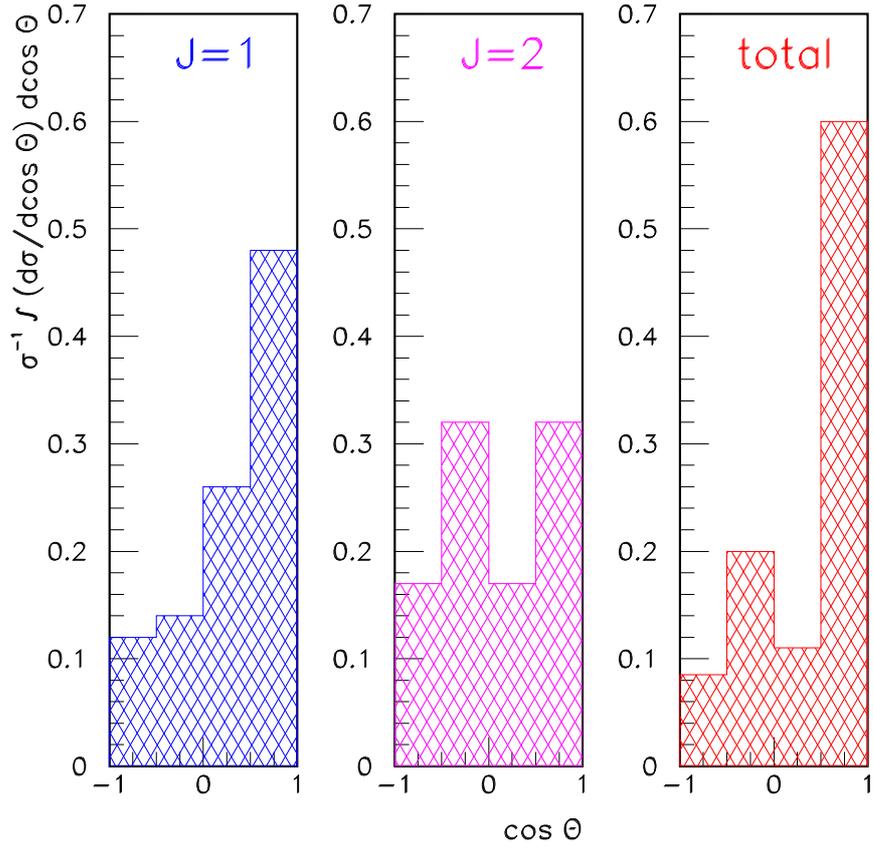}{0.8}
\caption[Binned angular distributions of Regge recurrences]{Binned angular distributions of Regge recurrences with spin 1, 2, and total in the $e^+ e^- \to \mu^+ \mu^-$ channel.}
\label{fig:3}
\end{figure}

Should a string resonance be found, a strong discriminator
between models will be the observed angular distribution. It is an interesting and exciting peculiarity
of Regge recurrences that the angular momenta content of the energy
state is more complicated. As we have shown in Sec.~\ref{eeee}, for
the lightest Regge excitation there is a specific combination of $J=1$
and $J=2$, which are access by the $e^+ e^-$ beam setting.
Specializing at this point to $e^-e^+ \to \mu^-\mu^+$, so that $I_{3F_L}
= Y_{F_L} = \frac{1}{2} Y_{F_R}= -1/2$, we obtain the normalized angular distribution
\begin{equation}
\frac{d\sigma/d\cos{\theta}}{\sigma}  =
{\cal N}\ \left\{\left[4 +\left(\frac{1}{2\ S_W^2}\right)^2 \right]
D_+(\theta)^2 + 2 \ D_-(\theta)^2\right\}\ \ ,
\end{equation}
where
\begin{equation}
D_\pm (\theta) \equiv d^2_{1,\pm 1}(\theta) + \frac{1}{3}\ d^1_{1, \pm 1}(\theta)\
\end{equation}
and
\begin{equation}
{\cal N}^{-1} =  (64/135) \left[6 + \left(\frac{1}{2\ S_W^2}\right)^2\right]
\, .
\end{equation}
For the $J=2$ piece alone, the normalization constant is
\begin{equation}
{\cal N}_2^{-1} =  (2/5) \left[6 + \left(\frac{1}{2\ S_W^2}\right)^2\right]
\end{equation}
whereas for the $J=1$ piece alone, the normalization constant is
\begin{equation}
{\cal N}_1^{-1} = (2/27) \left[6 + \left(\frac{1}{2\ S_W^2}\right)^2\right] \, .
\end{equation}
In Fig.~\ref{fig:ad} we show the resulting angular distributions. The predicted dimuon angular distribution has a pronounced forward-backward asymmetry. This is a realistic target for CLIC searches of low-mass scale string theory signals. (Note that the $e^+e^- \to e^+ e^-$ Coulomb scattering background, which peaks in the forward direction, tends to wash out the predicted string signal.) In Fig.~\ref{fig:3} we show the binned angular distributions. It is clearly seen that  it would be easy to distinguish the string excitation from single $J=2$ resonance in the dimuon angular distribution. To  completely isolate the Regge excitation from a $J=1$ resonance, one can use string predictions in alternative channels, e.g. $\gamma \gamma \to e^+ e^-$.

\newpage

\thispagestyle{fancy}
\chapter{Phenomenology of Regge Recurrences in the Randall-Sundrum Orbifold}
\label{LHCPheno}
\thispagestyle{fancy}
\pagestyle{fancy}

In this chapter, we consider string realizations of the Randall-Sundrum
  effective theory for electroweak symmetry breaking and explore the
  search for the lowest massive Regge excitation of the gluon
  and of the extra (color singlet) gauge boson  inherent of D-brane
  constructions. In these curved backgrounds, the higher-spin Regge
  recurrences of SM fields localized near the IR brane are
  warped down to close to the TeV range and hence can be produced at
  collider experiments. Assuming that the theory is weakly coupled, we
  make use of four gauge boson amplitudes evaluated near the first
  Regge pole to determine the discovery potential of LHC. We study the
  inclusive dijet mass spectrum in the central rapidity region
  $|y_{\rm jet}| < 1.0$ for dijet masses $M\geq 2.5~{\rm TeV}$. We
  find that with an integrated luminosity of 100~fb$^{-1}$, the
  5$\sigma$ discovery reach can be as high as 4.7~TeV. Observations of
  resonant structures in $pp\rightarrow {\rm direct}\ \gamma~ +$ jet
  can provide interesting corroboration for string physics up to
  3.0~TeV. We also study the ratio of dijet mass spectra at small and
  large scattering angles. We show that with the first~fb$^{-1}$ such
  a ratio can probe lowest-lying Regge states for masses $\sim 3$~TeV. The ideas discussed in this Chapter have been published in~\citec{Anchordoqui:2010vn}.

\ \\

\section{Randall-Sundrum large mass hierarchy from a small extra dimension}
\label{sec:RS}

In the canonical D-brane constructions discussed in Chapter~\ref{CLIC}
the large hierarchy between the weak scale and the fundamental scale
of gravity is eliminated through the large volume of the transverse
dimensions.  An alternative explanation to solve the gauge hierarchy
problem was suggested by Randall and Sundrum (herein
RS)~\citec{Randall:1999ee}. The RS set-up has the shape of a
gravitational condenser: two branes, which rigidly reside at
$S^1/\set{Z}_2$ orbifold fixed point boundaries $y = 0$ and $y = \pi
r_c$ (the UV and IR branes, respectively), gravitationally repel each
other and are stabilized by a slab of anti-de Sitter ($AdS$)
space. The metric satisfying this Ansatz (in horospherical
coordinates) is given by
\begin{equation}
ds^2 = e^{-2  k |y|} \,\,\eta_{\mu\nu} \,dx^\mu dx^\nu + dy^2\,,
\label{lisa-metric}
\end{equation}
where $ k$ is the $AdS$ curvature scale, which is somewhat smaller
than the fundamental 5-dimensional Planck mass $M^\star_{\rm Pl} \sim
M_{\rm Pl}$.\footnote{Greek subscripts extend over ordinary
  4-dimensional spacetime and are raised and lowered with the flat
  Minkowskian metric $\eta_{\mu \nu}$, whereas Latin subscripts span
  the full 5-dimensional space and are raised and lowered with the
  full metric $g_{MN}$.}  In this set up the distance scales get
exponentially redshifted as one moves from the UV brane towards the IR
brane. Such exponential suppression can then naturally explain why the
observed physical scales are so much smaller than the Planck scale.
For example, if the 5-dimensional Higgs condensate $v_5 \sim k$ is
IR-localized, the observed 4-dimensional value will be obtained from
$e^{-k\pi r_c} \langle H_5 \rangle$, and the observed hierarchy
between the gravitational and electroweak mass scales is reproduced if
$kr_c \approx 12$.  The most distinct signal of this set-up is the
appearance of a tower of spin-2 resonances, corresponding to the KK states of the 5-dimensional graviton, which have
masses and couplings driven by the TeV-scale. These KK gravitons
couple to all SM fields universally, yielding striking predictions for
collider experiments~\citec{Davoudiasl:1999jd}.

\begin{figure}[tbp]
\postscript{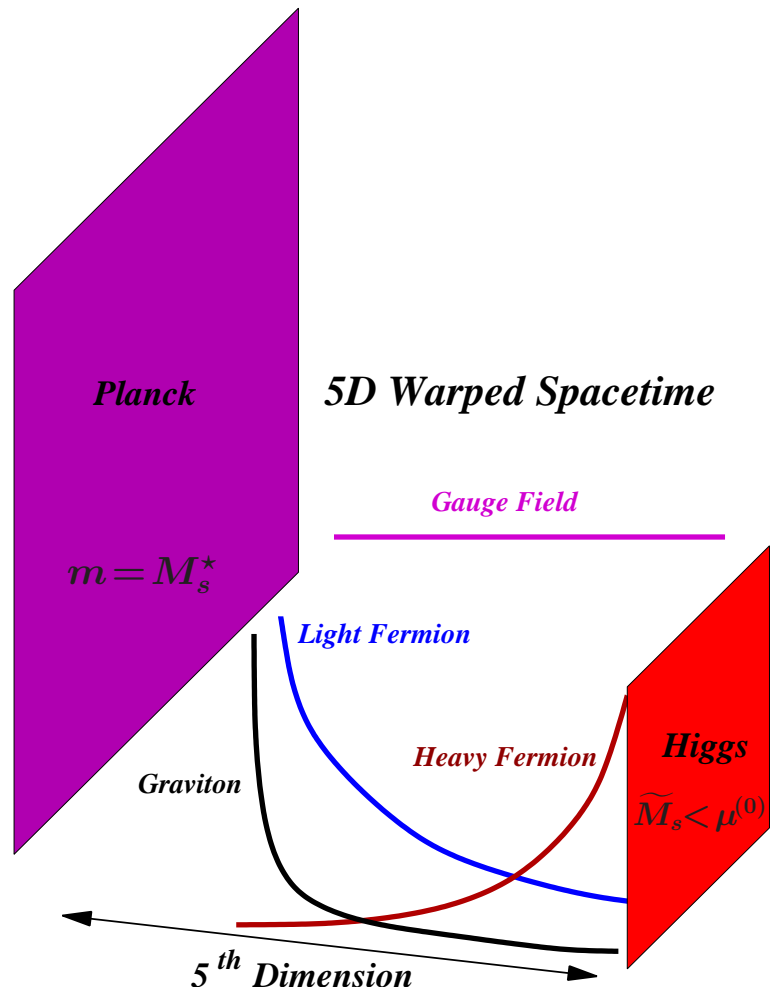}{0.8}
\caption[Randall-Sundrum brane-world]{Schematic representation of the
  RS warped model of hierarchy and flavor. From Ref.~\cite{Davoudiasl:2009sk}.}
\label{RS_cartoon}
\end{figure}

As originally noted in~\citec{Goldberger:1999wh}, to address the
hierarchy problem it is sufficient to keep the Higgs near the IR
brane.  Interestingly, if the remaining gauge bosons and fermions are
allowed to propagate into the warped  dimension, one can also
formulate an attractive mechanism to explain the flavor mass
hierarchy~\citec{Grossman:1999ra, Gherghetta:2000qt}.  The idea here is that the light
fermions are localized near the UV brane. This raises the effective
cutoff scale for operators composed of these fields far above the
TeV-regime, providing an efficient mechanism to suppress unwanted
operators, such as those mediating flavor changing  neutral currents
(FCNC) processes, related to tightly constrained light
flavors. Moreover, this results in small 4-dimensional Yukawa
couplings to the Higgs, even if there are no small 5-dimensional
Yukawa couplings. The top quark is IR-localized to obtain a large
4-dimensional top Yukawa coupling. Because the fermion profiles depend
exponentially on the bulk masses, this provides an understanding of
the hierarchy of fermion masses (and mixing) without hierarchies in
the fundamental 5-dimensional parameters, solving the SM flavor
puzzle. A schematic representation of this set-up is provided in Fig.~\ref{RS_cartoon}.

The RS set-up has also been used to construct warped Higgsless models,
where the electroweak symmetry is broken by boundary conditions on the
5-dimensional gauge fields~\citec{Csaki:2003zu}. Gauge fields are
allowed to propagate within all 5 dimensions. The electroweak gauge
structure of the minimal viable model is $SU(2)_L \times SU(2)_R
\times U(1)_{B-L},$ where $U(1)_{B-L}$ corresponds to gauging baryon
minus lepton number. Boundary conditions on the bulk gauge fields are
chosen so that the $SU(2)_L \times SU(2)_R$  symmetry is broken
on the IR brane to the diagonal subgroup $SU(2)_D$, and the $SU(2)_R
\times U(1)_{B-L}$ symmetry is broken down to the usual $U(1)_Y$
hypercharge in the UV brane to ensure that the low-energy gauge group
without electroweak symmetry breaking is $SU(2)_L \times U(1)_Y$.
The $SU(3)_C$ QCD group is unbroken everywhere, i.e., in the warped
dimension and on the branes. The spectrum of electroweak vector bosons
consists of a single massless photon along with KK towers of charged
$W_n$ and neutral $Z_n$ states. The SM massive $W$ and $Z$ vectors,
which get masses from the $SU(2)_L \times U(1)_Y$-violating boundary
condition on the IR brane, are identified with the lowest KK modes of
the $W_n$ and $Z_n$ towers.  SM fermions extend into all dimensions,
and they have explicit mass terms that are allowed by the non-chiral
structure of the theory in the bulk and on the IR brane. The most
serious challenge to construct viable models of Higgsless electroweak
symmetry breaking is satisfying the constraints from precision
electroweak measurements~\citec{Nomura:2003du}.  Mixing of the $W$ and
$Z$ with higher KK modes changes their couplings to fermions relative
to the SM. Heavier KK modes ($m_{\rm KK} \gtrsim 1~{\rm TeV}$) are
preferred to reduce these deviations to an acceptable level, but the
KK modes cannot be too heavy ($m_{\rm KK} \lesssim 1~{\rm TeV}$) if they
are to unitarize vector boson scattering. Both requirements can be
satisfied simultaneously if there are localized kinetic terms on each
of the branes~\citec{Cacciapaglia:2004jz}, and if the SM fermions (with
the exception of the right-handed top quark) have approximately flat
profiles in the extra dimension~\citec{Cacciapaglia:2004rb}.  In this
case, the first vector boson KK modes above the $Z$ and $W$ typically
have masses $\approx 0.5-1.5~{\rm TeV}$~\citec{Cacciapaglia:2004rb}.
Additional model structure is needed to generate a sufficiently large
top quark mass while not overly disrupting the measured $Z b_L \bar
b_L$ coupling. Some examples include new top-like custodial bulk
fermions~\citec{Agashe:2006at}, or a second warped bulk space on the
other side of the UV brane with its own IR
brane~\citec{Cacciapaglia:2005pa}.

In the spirit
of~\citec{Hassanain:2009at,Perelstein:2009qi,Reece:2010xj}, we {\it
  assume} that the RS orbifold arises as part of the compactification
manifold in a weakly-coupled string theory.  We further {\it assume}
that the compactification radii of the other five dimensions are
${\cal O}( {M_s^\star}^{-1})$ and therefore can be safely integrated
out.\footnote{The dearth of string constructions for a transition to
  the RS compactification~\citec{Verlinde:1999fy} makes a full
  comparison between the string scale and internal dimension radii
  difficult. A recent study~\citec{Reece:2010xj} of a range of models
  seems to indicate that $M_s^\star r_c \sim 1$ is viable.} With this
in mind, the basic relation between the curvature of the warped
internal space, the string scale (the mass of the Regge states), and
the 5-dimensional Planck mass is
 \begin{equation}   k \ll M^\star_s= \frac{1}{
    \sqrt{{\alpha'}^\star}} \ll M_{\rm Pl}^\star\,  ,
\label{Mstar}
\end{equation}
where ${\alpha'}^\star$ is the slope of the associated Regge
trajectory. The first inequality permits the warping to leave intact the
basic string properties (such as the dual resonant structure) of the perturbative
scattering amplitudes.
 The infinite tower of open string Regge
excitations have the same quantum numbers under the SM gauge group as
the gluons and the quarks, but in general higher spins, and their
masses are just square-root-of-integer multiples of the string mass
$M_s^\star$.  In string realizations of extended RS models of hierarchy and flavor,
we do expect the higher-spin Regge recurrences of the SM fields
localized near the IR brane to be redshifted close to the TeV scale
and therefore be directly produced at the
LHC~\citec{Hassanain:2009at,Perelstein:2009qi}.

\section{Four-Point Amplitudes of Gauge Bosons}
\label{sec:Fpagb}

Unfortunately, the
Veneziano amplitudes discussed in Chapter~\ref{CLIC} only apply to strings propagating on flat
Minkowski backgrounds, and their generalization to warped spaces is
presently unknown. In the absence of concrete string theory
constructions, we describe the lowest-lying Regge excitations of SM
gauge bosons following the bottom-up approach advocated
in~\citec{Perelstein:2009qi}.  In the limit where $k$ is taken to zero, this innovative
approach  reproduces the string effects encapsulated in (\ref{formf}).

Consider a free (non-interacting) massive spin-2 field $B_{MN}$ in
curved 5-dimensional spacetime,
\begin{equation}
{\cal L} =  \frac 1 4 H^{LMN}H_{LMN} - \frac 1 2  H^{LM}{}_{M} H_{LN}{}^{N} + \frac{1}{2} m^2 \left[ \left({B_M}^M \right)^2 - B^{MN}B_{MN} \right] \,,
\label{T5dL}
\end{equation}
where $H_{LMN} = \nabla_L B_{MN} - \nabla_M B_{LN}$ is the field
strength tensor and $m \equiv M_s^\star$ is the mass of the lightest Regge
excitation.  This field can be further decomposed according to its
spins ($J= 0,$ $J=1,$ and $J=2$) in 4-dimensions. The tensor, vector,
and scalar components are $B_{\mu \nu},$ $B_{\mu5}$, and $B_{55}$,
respectively. The Lagrangian (\ref{T5dL}) contains terms which mix
these components. Such mixed terms need to be canceled for a
consistent KK decomposition.  As shown
in~\citec{Perelstein:2009qi}, the
action can be factorized as
\begin{equation}
S = S_{J=2} \oplus S_{J=1, J=0} \,,
\end{equation}
where the 5-dimensional Lagrangian for $J=2$ is given by \ba\el{5dLB}
S_{J=2} & = & \int d^5 x \left\{e^{2k |y|}\left[ \frac 1 4
    H^{\xl\mu\nu}H_{\xl\mu\nu} - \frac 1 2 \lb 1 - \frac 2 \xi \rb
    H^{\xl\mu}{}_{\mu} H_{\xl\nu}{}^{\nu}\right] \right.  \nn & +&
\frac 1 2 B_\mu{}^\mu (-\p_y^2 + 4 k^2 + m^2) B_\nu{}^\nu - \frac 1 2
B^{\mu\nu} (-\p_y^2 + 4k^2 + m^2) B_{\mu\nu} \nn & + & \left. 2 k
  \phantom{\frac{1}{1}} \left[\xd(y)-\xd(y-\pi r_c) \right]
  \left[B^{\mu\nu} B_{\mu\nu} - (B_\mu{}^\mu)^2 \right]\right\} \, ,
\ea and $\xi$ is a parameter in the gauge fixing term. The field
$B_{\mu\nu}$ can be decomposed according to its wave function in the
warped dimension,
\begin{equation}
B_{\mu\nu} = \frac 1 {\sqrt{\pi r_c}} \sum_{n = 1}^\infty B_{\mu\nu}^{(n)} \  f^{(n)}(y).
\end{equation}
The equation of motion is,
\begin{equation}
e^{2k|y|} D_{\mu\nu}{}^{\xa \xb} B_{\xa \xb} + \{-\p_y^2 + 4 k^2 + m^2-4 k [\xd(y)-\xd(y-\pi r_c)]\}B_{\mu\nu} =0,
\end{equation}
where $D_{\mu\nu}{}^{\xa \xb}$ is an operator from the first line of \er{5dLB}.
A massless spin-2 field has the equation of motion of $D_{\mu\nu}{}^{\xa \xb} B_{\xa \xb} = 0$. So the masses are given by the eigenvalues of the operator,
\begin{equation}
e^{-2k|y|} \{-\p_y^2 + 4 k^2 + m^2-4 k [\xd(y)-\xd(y-\pi r_c)]\},
\end{equation}
with mode functions $f^{(n)}$ satisfying the following equation,
\begin{equation}
-f^{(n)}{}''+(4k^2 + m^2) f^{(n)} - 4 k \left[\xd(y)-\xd(y-\pi r_c) \right]f^{(n)}
= (\mu^{(n)})^2 \, e^{2k|y|} \, f^{(n)},
\label{eigeneq}
\end{equation}
and associated inner product,
\begin{equation}
\frac{1}{\pi r_c} \int_0^{\pi r_c} dy \, e^{2k|y|} f^{(n)}\, f^{(m)} = \delta^{nm},
\label{innerp}
\end{equation}
from the orthonormal condition. For this choice of $f^{(n)}$, we have
(from the second and the third line of \er{5dLB}), \ba \int d^5x \dots
&= & - \frac 1 2 \, \int d^4 x \, dy \, B^{\mu\nu} \{-\p_y^2 + 4 k^2 +
m^2-4 k [\xd(y)-\xd(y-\pi r_c)]\} B_{\mu\nu} \nn & = & - \frac 1 2 \,
\int d^4 x \, dy \, B^{(m)\mu\nu}(x) \, B^{(n)}_{\mu\nu}(x) \frac 1
{\pi r_c} \,\sum_{n=1}^\infty \sum_{m=1}^\infty (\mu^{(n)})^2 \,
e^{2k|y|} \,f^{(m)}(y) \,f^{(n)}(y) \nn & = & - \frac 1 2 \, \int d^4
x \, \sum_{n=1}^\infty (\mu^{(n)})^2 \, B^{(n)\mu\nu}(x)
B^{(n)}_{\mu\nu}(x),\ea where in the last line, we use
(\ref{innerp}). The integration of $H_{\xl \mu\nu} H^{\xl \mu\nu}$ is
trivial because there is no $y$-derivative. Hence, after the extra
dimension is integrated out, Eq.\er{5dLB} can be reduced to a
4-dimensional Lagrangian of free spin-2 fields (with different masses
$\mu^{(n)}$), \ba\el{4dLB} S_{J=2} & = & \int d^4 x \sum_{n=1}^\infty
\left\{ \frac 1 4 H^{(n)\xl\mu\nu}H^{(n)}_{\xl\mu\nu} - \frac 1 2 \lb
  1 - \frac 2 \xi \rb H^{(n)\xl\mu}{}_{\mu}
  H^{(n)}_{\xl\nu}{}^{\nu}\right. \nn & + & \left. \frac 1 2
  (\mu^{(n)})^2 [B^{(n)}_\mu{}^\mu B^{(n)}_\nu{}^\nu -B^{(n)\mu\nu}
  B^{(n)}_{\mu\nu}] \right\} ,\ea where $\xi \to \infty$ when
computing the scattering amplitude.  The general solution of
(\ref{eigeneq}) is a Bessel function~\citec{Perelstein:2009qi}
\begin{equation}
f^{(n)} (y) = \frac{1}{N} \left[J_\nu\left(\frac{\mu^{(n)}}{\Lambda_{\rm IR}} \, w \right) + c J_{-\nu}\left(\frac{\mu^{(n)}}{\Lambda_{\rm IR}} \, w \right) \right] \,,
\end{equation}
where $N$ is the normalization constant, $c$ is an integration
constant (each of these constants implicitly depends upon the level
$n$),  $\Lambda_{\rm IR} = k e^{-\pi kr_c},$ and $w = e^{k(|y| - \pi
  r_c)}, \, \in [e^{-k\pi r_c},1].$ The order of the
Bessel function is $\nu \equiv \sqrt{4+\mathfrak{m}^2}$, where
$\mathfrak{m}=m/k$ is the string scale in units of the RS
curvature. With appropriate boundary conditions, the masses
$\mu^{(n)}$ and the explicit form of $f^{(n)}$ can be obtained.

We now turn to the discussion of $J=0$. In the effective
  4-dimensional theory there is one real scalar $\Re {\rm e} (\phi)$, which comes
  from the 5-dimensional scalar and couples to the gluon strength
  $F^2$.  In addition, there is one pseudoscalar axion $A_\star^5$,
  which in 4 dimensions couples as $A_\star^5 F \, ^{^{*}\!\!}F$, with
  $^{^{*}\!\!}F = \frac{1}{2} \epsilon^{\mu \nu \rho \sigma} F_{\rho \sigma} .$  This
  pseudoscalar axion comes from the fifth component of a massive
  vector $A_\star$, with coupling $\epsilon_{\mu\nu\rho\sigma 5}
  F^{\mu\nu}F^{\rho\sigma} A_\star^5$. Then, $\Re {\rm e} (\phi)$ and
  $\Im {\rm m}
  (\phi) \equiv A_\star^5$ combine to one
  complex scalar $\phi$ which couples as $\phi
  (F+i\, ^{^{*}\!\!}F)(F+i\, ^{^{*}\!\!}F) + {\rm cc}$; this ensures that $\phi$
  and its complex conjugate $\phi^*$ couple only to the
  $++$ ($--$) helicity combinations, respectively.
  \footnote{We may trace the origin of the $J=0$ contribution to
  components of $B_{MN}$ and other fields of the 10-dimensional theory. We
discuss this decomposition in more details in Appendix~\ref{sec:spinzero}.
  Instead, we proceed by simply using the correspondence with the tree
  level string theory and identify the vertex function through comparison
  with the tree level $J=0$ pole. As described in the text this has the correct helicity structure. This approach is justified in Appendix~\ref{supermultipletRS}
using supersymmetry.} Both the scalar and the pseudoscalar will be
  affected in the same way by warping, because they sit in one SUSY
  multiplet.  Thus, to determine the $J=0$ contribution, we study the
  effect of warping on a dilaton-like scalar a with the coupling $\Re
  {\rm e}(\phi)
  F^2.$

The Klein-Gordon equation for a scalar $\phi$ in the RS spacetime is
\begin{equation}
\frac 1 {\sqrt{g}}\p_M \sqrt{g} \p^M \phi + m^2 \phi = 0 \,;
\end{equation}
more explicitly, it is,
\be
e^{2k|y|} \p^\mu \p_\mu \phi + \left[- \p_y^2 + 4 k \sy \p_y + m^2 \right] \phi = 0 \, .
\ee
The field $\phi$ can be decomposed according to its wave function in
the warped dimension,
\be
\phi(x,y) = \frac 1 {\sqrt{\pi r_c}} \sum_{n = 1}^\infty \phi^{(n)}(x) \,
h^{(n)}(y).
\ee
One can choose the mode functions $h^{(n)}$ satisfying the following equation,
\be\el{modefun} -h^{(n)}{}''+4 k\, \sy h^{(n)}{}' +  m^2 h^{(n)} = (\mu^{(n)})^2 e^{2k|y|} h^{(n)}.\ee
With a change of variable $x = \frac 1 k e^{k|y|}$, we have
\begin{equation}
\frac {d }{dy} = k x \frac {d }{dx},\quad \frac {d^2} {dy^2} =k^2 x^2
\frac {d^2 }{dy^2} + k^2 x \frac {d }{dx},
\end{equation}
so \er{modefun} can be written as
\be x^2 h^{(n)}{}'' + 3 x h^{(n)}{}' + [(\mu^{(n)})^2 x^2 - {\mathfrak m}^2] (\mu^{(n)}) = 0 .\ee
The solution to this equation is
\be
h^{(n)}(x) = \frac 1 N  \left(\mu^{(n)} x \right)^2 \Big\{J_{\nu}
\left(\mu^{(n)} x \right) + C J_{-\nu} \left(\mu^{(n)} x \right) \Big\} \equiv x^2 \tilde f^{(n)}(x),\ee
where $N$ is a normalization constant and $C$ an integration constant.
For later convenience, we also define
a new function $\tilde f^{(n)}$. The boundary conditions are
\begin{equation}
h^{(n)}{}'(0+) - h^{(n)}{}'(0-) = 0
\end{equation}
and
\begin{equation}
h^{(n)}{}'(-\pi r_c+) - h^{(n)}{}'(\pi r_c-) = 0 ,
\end{equation}
where the prime is the derivative with respect to $y$. As in the case of $B_{\mu\nu}$, the mass $\mu^{(n)}$ is determined from the second boundary condition,
\be
x^2 \tilde f^{(n)}{}'(-\pi r_c+) - 2 x k x \tilde f^{(n)}(-\pi r_c+) - x^2 \tilde f^{(n)}{}'(\pi r_c-) - 2 x k x \tilde f^{(n)}(\pi r_c-) = 0,\ee
or
\be\tilde f^{(n)}{}'(-\pi r_c+) - \tilde f^{(n)}{}'(\pi r_c-) = 4 k \tilde f^{(n)}(\pi r_c),\ee
which is essentially the boundary condition for $B_{\mu\nu}$~\citec{Perelstein:2009qi}. As a result, the mass of $\phi$ is exactly the same as that of $B_{\mu\nu}$. Note that $h^{(n)}(x)$ can be expressed as
\begin{equation}
h^{(n)} = e^{2 k |y|} f^{(n)},
\end{equation}
where $f^{(n)}$ are the mode functions for $B_{\mu\nu}$. So $h^{(n)}(x)$ are normalized as
\be
\frac{1}{\pi r_c} \int_0^{\pi r_c} dy \, e^{-2k|y|} h^{(n)}\, h^{(m)} = \delta^{nm},
\ee
This gives a canonical kinetic term for $\phi^{(n)}$ (because of the
different powers of $e^{2k|y|}$).

\emp{Here} we will restrict our calculations to incoming QCD gluons.
We then obtain the decomposition of the QCD gauge
field. Gauge freedom can be used to set $A_5
=0$~\citec{Davoudiasl:1999tf}. This is consistent with the gauge
invariant equation $\oint dx^5 A_5 =0,$ which results from the
assumption that $A_5$ is a $\set{Z}_2$-odd function of the extra
dimension. In this gauge, the 4-dimensional vector zero-mode has a constant
profile in the bulk,
\begin{equation}
A_\mu(x,y) = \frac 1 {\sqrt{\pi r_c}} A_\mu^{(0)}(x)+\dots \,,
\label{agluon}
\end{equation}
 and
the gluon field strength takes the familiar form $F_{\mu \nu}^a
= \partial_\mu A_\nu^a - \partial_\nu A_\mu^a + g_a f^{abc} A_\mu^b
A_\nu^c,$ with $a = 1, \dots, 8$.

The coupling of the 5-dimensional field $B_{MN}$ to the
gluon is given by \ft{For derivation, see
\eg Appendix~\ref{subsec:SAgggstar}.}
\begin{equation}
S_{ggg^*(C^*)} = \int d^5 x \sqrt{-g} \frac {g_5} {\sqrt 2 M_s^\star} C^{abc}\lb F^{aA C} F_{C}^{bB} - \frac 1 4 F^{a C D}F^b_{C D} g^{AB}\rb B^c_{AB}
\label{gF2}
\end{equation}
where $C^{abc} = 2 [{\rm Tr}(T^a T^b T^c) + {\rm Tr} (T^a T^b T^c)]$
is the color factor, $T^a$ are the generators of the fundamental
representation of $U(3)$ (normalized here according to ${\rm Tr}
(T^aT^b) = \frac{1}{2} \delta^{ab})$, and $F_{A B}^a = \partial_A
A_B^a - \partial_B A_A^a + g_5 f^{abc} A_A^b A_B^c$. Note that the
color indices on the field strength $F$ run from 1 to 8; on the tensor
field $B$, $U(3)$ indices ($c = 0, \dots, 8$) are permitted (with $c =0$
corresponding to the tensor excitation $C^*$).\footnote{As can be
  verified from the 4-point function~\citec{Anchordoqui:2007da} there
  is no coupling $gg \to C$, however the composite nature of $C^*$ and $g^*$
  permits respectively $gg \to C^*$ and $gC \to g^*$ couplings, with color globally preserve.}
Hence, $g_5$ is related to the Yang-Mills QCD coupling $g_a$ according
to $g_5 = g_a \sqrt {\pi r_c}$. The factor $g_5/\sqrt{2} M_s^\star$ is
determined by matching the $gg \to g^* (C^*)$ amplitude to the
s-channel pole term in the string (tree-level)
amplitude.\footnote{See Ref.\cite{Perelstein:2009qi} for some caveats
pertaining to this approach.} Thus, the 4-dimensional coupling term is
found to be \ba\el{4dintgggstar} {\cal L}_{gg \to g^* (C^*)} &=& \frac
{g^{(0)}} {\sqrt 2 \widetilde M_s} C^{abc} \left[\lb F^{\xa \xg}
  F_{\xg}{}^\xb - \frac 1 4 F^{a\xg \xd}F^b_{\xg \xd} \eta^{\xa\xb}\rb
  B^{c\,(0)}_{\xa \xb} + \frac{1}{2} \left( \phi^{c\,(0)} F^{a\mu
      \nu}F^b_{\mu \nu} \phantom{\frac{1}{2}} \right. \right. \nn & +&
\left. \left. \frac{1}{2} \, \bar \phi^{c\,(0)} F^{a\mu \nu}F^{b\rho
      \sigma} \xe_{\mu\nu\rho\sigma} \rb \right], \ea where
  $\widetilde M_s = e^{-k\pi r_c} M_s^\star \sim 1~{\rm TeV}$ is the
  redshifted string scale, $g^{(0)}$ follows from the integration of
  the zero mode $f^{(0)}(y)$ of $B_{\mu\nu}^{(0)}$, and $\bar
  \phi^{c\, (0)}$ is the zero mode for the imaginary part of the
  complex scalar.  Since each field in (\ref{gF2}) contribute to the
  integration with a factor $(\pi r_c)^{-1/2}$ we obtain,
\begin{equation}
g^{(0)} = \frac {g_a e^{-\pi k r_c}}{\pi r_c} \int_0^{\pi r_c} d y\,
e^{2 k y} \, f^{(0)}(y).
\end{equation}
The coupling \er{4dintgggstar} gives three vertices:
\be\el{gggstarvertex}i\frac {\sqrt 2  \, g^{(0)}} {\widetilde M_s}
C^{abc}\lb \xS^{\xa \xb}- \frac 1 4 \eta^{\xa \xb} \xS_\xg{}^\xg\rb
b_{\xa\xb},\,
 i\frac {g^{(0)}} {\sqrt 2  \, \widetilde M_s} C^{abc} \xS_\mu {}^\mu
 ,\,
i \frac {g^{(0)}}{\sqrt 2 \widetilde M_s} C^{abc} \ 4 \,
\xe_{\mu\nu\rho\sigma} \,
k_1^\mu \, \epsilon_1^\nu \, k_2^\rho \, \epsilon_2^\sigma   \,,
\ee
where  $b_{\xa\xb}$
is a polarization of $B_{\xa\xb}^{c\,(0)}$, $k_i^\mu$ and
$\epsilon_i^\nu$ (with $i =1,2$) are respectively the momentum
and polarization of the incoming gluons, $\xS^{\xa \xb} = (k_1^\xa \xe_1^\xg - k_1^\xg \xe_1^\xa)(k_{2\xg}
\xe_2^\xb - k_2^\xb \xe_{2\xg} ) + (\xa \leftrightarrow \xb)$,
and its trace
$\xS_\xg{}^\xg = 4(k_1 \cdot \xe_2)(k_2 \cdot \xe_1) - 4(\xe_1 \cdot \xe_2)(k_1
\cdot k_2)$~\citec{Cullen:2000ef}. As in the $J=2$ case, the coupling is determined by matching to
the $J=0$ pole term in the tree-level string amplitude.

Finally, we note that the $J=1$ resonant level exists, but is not accessible in purely gluonic
scattering~\citec{Anchordoqui:2008hi}.

The $s$-channel pole terms of the average square amplitudes contributing
to $\gamma$+ jet and  dijet production at the LHC can be obtained from the general
formulae given in Ref.~\citec{Lust:2008qc}. The 4-gluon average square amplitude is
given by
\begin{equation}
|{\cal M} (gg \to gg)| ^2  =  2 \
\left(\frac{g^{(0)}}{\widetilde M_s}\right)^4 \ \left(\frac{N^2-4+(12/N^2)}{N^2-1}\right)
 \ \frac{s^4+  t^4 +   u^4}{(  s - \mu^2)^2} \, ,
\label{ggggpole}
\end{equation}
where to simplify notation we have dropped the
superscript indicating the lowest massive Regge excitation, i.e., $\mu \equiv \mu^{(0)}$.
For phenomenological purposes, the poles
need to be softened to a Breit-Wigner form by obtaining and utilizing
the correct {\em total} widths of the
resonances~\citec{Anchordoqui:2008hi}. After this is done, the
contributions of $gg \to gg$ is as follows:
\begin{eqnarray}
|{\cal M} (gg \to gg)| ^2 & = & \frac{19}{12} \
\left(\frac{g^{(0)}}{\widetilde M_s}\right)^4 \left\{ W_{g^*}^{gg \to gg} \, \left[\frac{s^4}{(  s-\mu^2)^2
+ (\Gamma_{g^*}^{J=0}\ \mu)^2} \right. \right.
\left. +\frac{  t^4+   u^4}{(  s-\mu^{2})^2 + (\Gamma_{g^*}^{J=2}\ \mu)^2}\right] \nonumber \\
   & + &
W_{C^*}^{gg \to gg} \, \left. \left[\frac{s^4}{(  s-\mu^2)^2 + (\Gamma_{C^*}^{J=0}\ \mu)^2} \right.
\left. +\frac{  t^4+  u^4}{(  s-\mu^2)^2 + (\Gamma_{C^*}^{J=2}\ \mu)^2}\right] \right\},
\label{gggg2}
\end{eqnarray}
where
$$\Gamma_{g^*}^{J=0} = 75\,\left(\frac{g^{(0)} \mu}{g_a \widetilde M_s}\right) \, \left(\frac{\mu}{{\rm TeV}} \right)~{\rm GeV}, \quad
\Gamma_{C^*}^{J=0} = 150 \, \left(\frac{g^{(0)} \mu}{g_a \widetilde M_s}\right)
\left(\frac{\mu}{{\rm TeV}}\right)~{\rm GeV},$$
$$\Gamma_{g^*}^{J=2} = 45 \, \left(\frac{g^{(0)} \
    \mu}{g_a \widetilde M_s}\right) \left(\frac{\mu}{{\rm
      TeV}}\right)~{\rm GeV}, \quad \Gamma_{C^*}^{J=2} = 75 \,
\left(\frac{g^{(0)} \mu}{g_a \widetilde M_s}\right)
\left(\frac{\mu}{{\rm TeV}}\right)~{\rm GeV}$$ are the total decay
widths for intermediate states $g^*$, $C^*$ (with angular momentum
$J$)~\citec{Anchordoqui:2008hi,Perelstein:2009qi}. The associated
weights of these intermediate states are given in terms of the
probabilities for the various entrance and exit channels \ba
\el{totalcrossdecom} \frac{N^2-4+12/N^2}{N^2-1} & = & \frac {16}
{(N^2-1)^2}\left[\left(N^2-1\right)\left(\frac{N^2-4}{ 4N}\right)^2+
  \left(\frac{N^2-1}{2N}\right)^2\right]\nn & \propto & \frac {16} {(N^2-1)^2}
\left[(N^2-1)(\xG_{g^*\to gg})^2 + (\xG_{C^*\to gg})^2\right] \,,\ea
yielding
$$
W_{g^*}^{gg \to gg} = \frac{8 (\Gamma_{g^* \to gg})^2}{8(\Gamma_{g^* \to gg})^2 +
(\Gamma_{C^* \to gg})^2} = 0.44, \quad
W_{C^*}^{gg \to gg} = \frac{(\Gamma_{C^*
  \to gg})^2}{8(\Gamma_{g^* \to gg})^2 + (\Gamma_{C^* \to gg})^2} =
0.56 \, , $$
where superscripts $J=2$ are understood to be inserted on all the $\Gamma$'s.

As we pointed out in Chapter~\ref{CLIC}, the hypercharge is a color
composite state containing the photon. The $s$-channel pole term of
the average square amplitude contributing to $gg \to \gamma$ + jet is
given by~\citec{Anchordoqui:2007da}
\begin{eqnarray}
|{\cal M} (gg \to g \gamma)|^2  =  \frac{5}{3} Q^2\left(\frac{g^{(0)}}{\widetilde M_s}\right)^4   \Bigg[\frac{s^4}{(  s-\mu^2)^2
+ (\Gamma_{g^*}^{J=0}\ \mu)^2}  +  \left. \frac{ t^4+   u^4}{(  s-\mu^2)^2 + (\Gamma_{g^*}^{J=2} \mu)^2}\right]\quad
\label{ggggamma}
\end{eqnarray}
where $Q = \sqrt{1/6} \ \kappa \ \cos \theta_W$ is the product of the
$U(1)$ charge of the fundamental representation ($\sqrt{1/6}$)
followed by successive projections onto the hypercharge ($\kappa$) and
then onto the photon ($\cos \theta_W$).  For the phenomenological analysis
that follows we set $\kappa^2 = 0.02.$

\ \\

\section{LHC Discovery Reach}
\label{sec:LHCdr}

The most important parameter to determine the LHC discovery reach for
string recurrences is the mass of the lowest-lying Regge excitation,
which depends on $\Lambda_{\rm IR}$ and $\mathfrak{m}$. For fixed
$\mathfrak{m}$ the mass of $g^*$ and $C^*$ excitations is to a very
good approximation a linear function of $\Lambda_{\rm
  IR}$~\citec{Perelstein:2009qi}. As we already remarked in Sec.~\ref{sec:RS}, in Higgsless models $\Lambda_{\rm IR}$ is subject to
significant constraints from electroweak data. The KK excitations of
the vector gauge bosons must be near 1~TeV to simultaneously satisfy
unitarity and electroweak constraints. This leads to $\Lambda_{\rm IR}
\approx 0.5~{\rm TeV}$. Similarly, to avoid precision electroweak
constraints in scenarios where the Higgs is IR-localized the lightest
KK excitation mass \emp{(of the massless gauge boson)} is $\gtrsim 3~{\rm TeV}$~\citec{Agashe:2003zs}, yielding $\Lambda_{\rm IR} \gtrsim 1~{\rm TeV}$ \citec{Davoudiasl:1999tf}. 
From (\ref{Mstar}) we obtain the condition $\mathfrak{m} \gg 1$ for string 
propagation on a smooth geometric background. Nevertheless, as in many examples 
in various arenas of physics, $\mathfrak{m} \sim$ a few may in fact be
sufficient, depending on the behavior of the leading corrections to
the geometric limit. In our phenomenological study we will
follow~\citec{Perelstein:2009qi} and set $\mathfrak{m} \gtrsim 3$, which
leads to $\mu^{(0)} \approx 5 \, \Lambda_{\rm IR} $,  $g^{(0)}/g_a
\simeq 0.1,$ and  $\mu^{(0)} = 5 \, \mathfrak{m}^{-1}  \, \widetilde M_s
\simeq 1.7 \widetilde M_s$.

Given the particular nature of the process we are considering, the
production of a TeV Regge state and its subsequent 2-body decay, one
would hope that the resonance would be visible in data binned
according to the invariant mass $M$ of the dijet, after setting cuts
on the different jet rapidities, $|y_1|, \, |y_2| \le
1$~\citec{Bhatti:2008hz} and transverse momenta $p_{\rm T}^{1,2}>50$
GeV.  With the definitions $Y\equiv \frac{1}{2} (y_1 + y_2)$ and $y \equiv
\frac{1}{2} (y_1-y_2)$, the cross section per interval of $M$ for $p
p\rightarrow {\rm dijet}$ is given by
\begin{eqnarray}
\frac{d\sigma}{dM} & = & M\tau\ \sum_{ijkl}\left[
\int_{-Y_{\rm max}}^{0} dY \ f_i (x_a,\, M)  \right. \ f_j (x_b, \,M ) \
\int_{-(y_{\rm max} + Y)}^{y_{\rm max} + Y} dy
\left. \frac{d\sigma}{d\hat t}\right|_{ij\rightarrow kl}\ \frac{1}{\cosh^2
y} \nonumber \\
& + &\int_{0}^{Y_{\rm max}} dY \ f_i (x_a, \, M) \
f_j (x_b, M) \ \int_{-(y_{\rm max} - Y)}^{y_{\rm max} - Y} dy
\left. \left. \frac{d\sigma}{d\hat t}\right|_{ij\rightarrow kl}\
\frac{1}{\cosh^2 y} \right]
\label{longBH}
\end{eqnarray}
where $\tau = M^2/s$, $x_a =
\sqrt{\tau} e^{Y}$,  $x_b = \sqrt{\tau} e^{-Y},$
and
\begin{equation}
  |{\cal M}(ij \to kl) |^2 = 16 \pi \hat s^2 \,
  \left. \frac{d\sigma}{d\hat t} \right|_{ij \to kl} \, .
\end{equation}
In this section we reinstate the caret notation ($\hat s,\ \hat t,\
\hat u$) to specify partonic subprocesses. The $Y$ integration range
in Eq.~(\ref{longBH}), $Y_{\rm max} = {\rm min} \{
\ln(1/\sqrt{\tau}),\ \ y_{\rm max}\}$, comes from requiring $x_a, \,
x_b < 1$ together with the rapidity cuts $y_{\rm min} <|y_1|, \, |y_2|
< y_{\rm max}$. The kinematics of the scattering also provides the
relation $M = 2p_T \cosh y$, which when combined with $p_T = M/2 \
\sin \theta^* = M/2 \sqrt{1-\cos^2 \theta^*},$ yields $\cosh y = (1 -
\cos^2 \theta^*)^{-1/2},$ where $\theta^*$ is the center-of-mass
scattering angle.  Finally, the Mandelstam invariants occurring in the
cross section are given by $\hat s = M^2,$ $\hat t = -\frac{1}{2} M^2\
e^{-y}/ \cosh y,$ and $\hat u = -\frac{1}{2} M^2\ e^{+y}/ \cosh y.$

\begin{figure}[tpb]
 \postscript{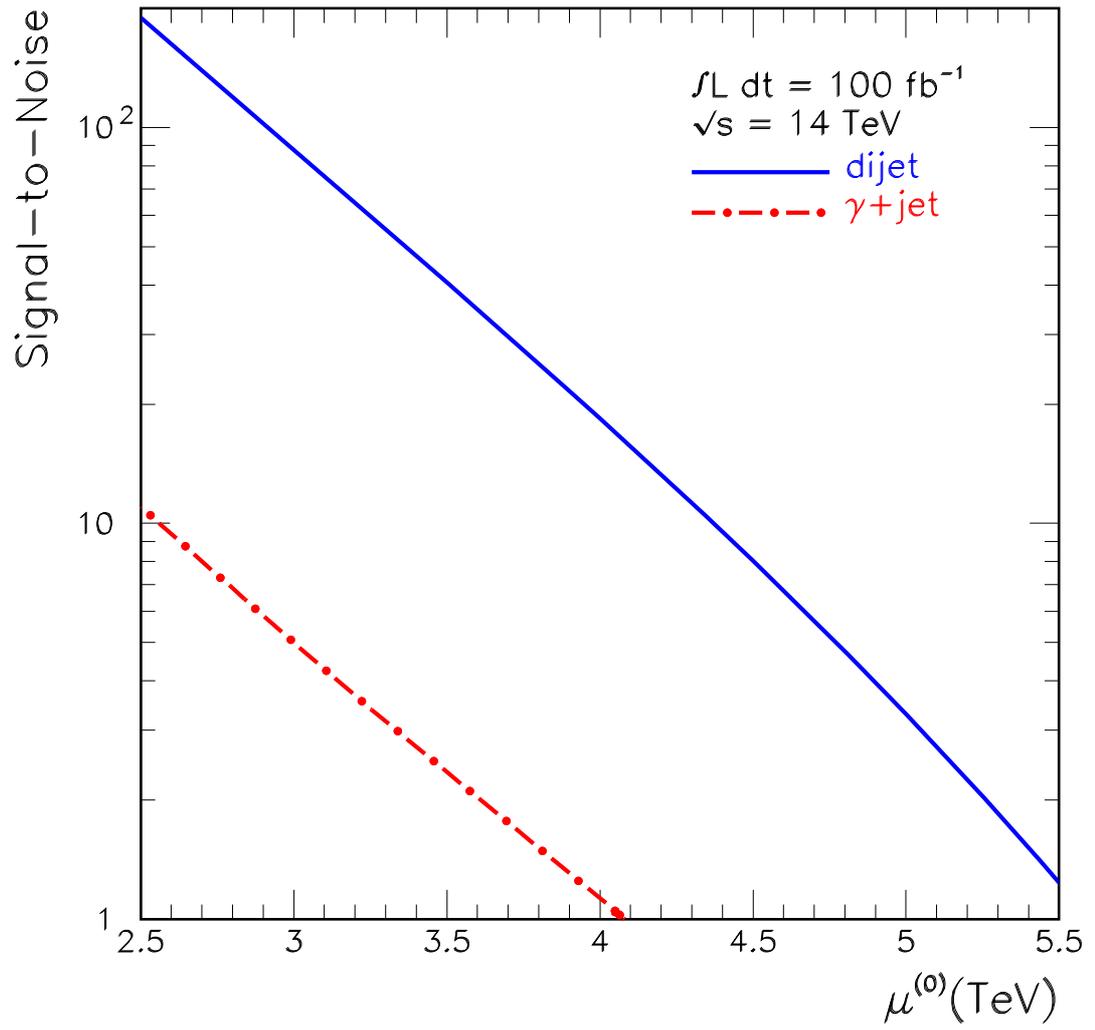}{0.98}
 \caption[$pp \to {\rm dijet}$ signal-to-noise ratio for lowest mass Regge excitation]{$pp \to {\rm
     dijet}$ and $pp \to \gamma + {\rm jet}$ signal-to-noise ratio for
    100~fb$^{-1}$ integrated luminosity.}
\label{RS_S2N}
\end{figure}

To study the feasibility of detecting the resonance we adopt the
standard bump-hunting method introduced in Chapter~\ref{CLIC}.  As
usual, the signal rate is estimated in the invariant mass window
$[\mu^{(0)} - 2 \Gamma, \, \mu^{(0)} + 2 \Gamma]$. Here the noise is
defined as the square root of the number of background events in the
same dijet mass interval for the same integrated luminosity. The QCD
background has been calculated at the partonic level considering all
SM contributions to dijet final states~\citec{Anchordoqui:2008di}.  Our
calculation, making use of the CTEQ6 parton distribution
functions~\citec{Pumplin:2002vw} agrees with that presented
in~\citec{Bhatti:2008hz}.

The top curve in Fig.~\ref{RS_S2N} shows the behavior of the
signal-to-noise (S/N) ratio as a function of the lowest massive Regge
excitation, for 100~fb$^{-1}$ of integrated luminosity and $\sqrt{s} =
14$~TeV. {\it Regge
  excitations with masses $\mu^{(0)} \lesssim 4.7$~TeV are open to discovery at
  the $\geq 5\sigma$ level.} This implies that in the Higgsless model
 discovery would be possible in a wide range of the presently
unconstrained parameter space, whereas in the model with a Higss
localized on the IR-brane the LHC discovery potential would be only marginal.
The bottom curve in Fig.~\ref{RS_S2N} shows the S/N ratio in
the $pp \to$ direct $\gamma$ + jet channel.
 To accommodate the minimal acceptance
cuts on final state photons from the CMS and ATLAS
proposals~\citec{Ball:2007zza}, we set $|y_{\rm max}|<2.4$.
The approximate equality
of the background due to misidentified $\pi^0$'s and the QCD
background~\citec{Gupta:2008zza}, across a range of large $p_T^\gamma$
as implemented in Ref.~\citec{Anchordoqui:2007da}, is maintained as an
approximate equality over a range of $\gamma$-jet invariant masses
with the rapidity cuts imposed.
Observations of resonant structures in $pp\rightarrow {\rm direct}\
\gamma~ +$ jet can provide interesting corroboration for string
physics up to 3.0~TeV. Before proceeding, we stress that the results
shown in Fig.~\ref{RS_S2N} are conservative, in the sense that we have
not included in the signal the stringy contributions of processes
containing fermions. These will be somewhat more model dependent since
they require details of the SM pattern of masses and mixings, but we
expect that these contributions can potentially increase the reach of
LHC for discovery of Regge recurrences.

\begin{figure}[tbp]
\postscript{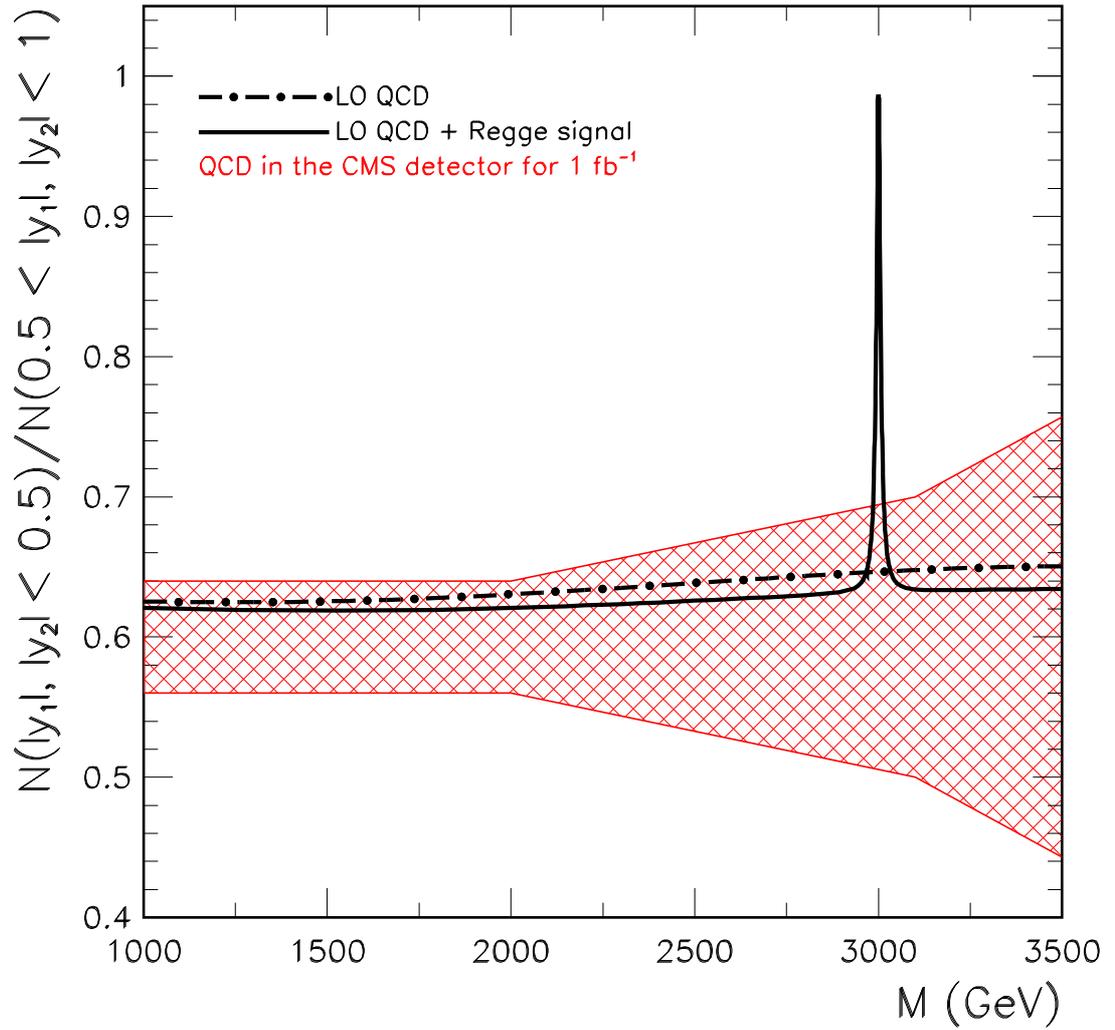}{0.99}
\caption[R-parameter at $3$~TeV]{For a luminosity of 1~fb$^{-1}$, the expected statistical error (shaded region) of the
  dijet ratio of QCD in the CMS detector~\citec{Esen} is compared with
  LO QCD (dot-dashed line) and LO QCD plus lowest massive Regge
  excitation (solid line), for $\mu^{(0)} = 3$~TeV.}
\label{RS_R}
\end{figure}

QCD parton-parton cross sections are dominated by $t$-channel
exchanges that produce dijet angular distributions which peak at small
center of mass scattering angles. In contrast,
non--standard contact interactions or excitations of resonances result
in a more isotropic distribution. In terms of rapidity variable for
standard transverse momentum cuts, dijets resulting from QCD processes
will preferentially populate the large rapidity region, while the new
processes generate events more uniformly distributed in the entire
rapidity region. To analyze the details of the rapidity space the D\O\
Collaboration introduced a new parameter~\citec{Abbott:1998wh},
\begin{equation}
R = \frac{d\sigma/dM|_ {(|y_1|,|y_2|< 0.5)}}{d\sigma/dM|_{(0.5 < |y_1|,|y_2| < 1.0)}} \, ,
\end{equation}
the ratio of the number of events, in a given dijet mass bin, for both
rapidities $|y_1|, |y_2| < 0.5$ and both rapidities $0.5 < |y_1|,
|y_2| < 1.0$.  The ratio $R$ is a genuine measure of the most
sensitive part of the angular distribution, providing a single number
that can be measure as a function of the dijet invariant
mass.\footnote{An illustration of the use of this parameter in
  a heuristic model where standard model amplitudes are modified by a
  Veneziano formfactor has been presented in~\citec{Meade:2007sz}.}

In Fig.~\ref{RS_R} we compare the results from a full CMS detector
simulation of the ratio $R$, with predictions from LO QCD and
contributions to the $g^*$ and $C^*$ excitations. The synthetic
population was generated with Pythia, passed through the full CMS
detector simulation and reconstructed with the ORCA reconstruction
package~\citec{Esen}. It is clear that with the first~fb$^{-1}$ of data
collected at the LHC, the $R$-parameter will be able to probe
lowest-lying Regge excitations for $\mu^{(0)} \sim
3$~TeV.\footnote{It should be noted that the $R$ parameter serves only
  as the crudest discriminator between QCD and stringy behavior of the
  cross section. More detailed analyses of the rapidity dependence of
  the final state jets are in order. In a recent
  paper~\citec{Kitazawa:2010gh} the behavior of the stringy amplitudes
  (for flat geometries) with respect to the rapidity difference $y$
  has been discussed. Results were presented for the separate
  contributions of the 1/2 and 3/2 resonances for the dominant
  $qg\rightarrow qg$ process, as well as for the combined cross
  sections. It remains to compare these to QCD.}

In addition to the Regge recurrences there are of course KK modes of
SM particles and gravitons propagating in the $s$-channel that we have not
yet considered. Their importance can be
gauged by their masses relative to $\mu^{(0)}$.  The ratio of string
to KK masses is model dependent, but in general there could be several
cases where the $\mu^{(0)}/m_{\rm KK}$ ratio is around a
few~\citec{Reece:2010xj}. This relation can be illustrated by comparing
with the masses of the KK states of the graviton: $m_{G}^{(n)} = x_n
\Lambda_{\rm IR}$, where the $x_n$ are the $n^{\rm th}$ roots of the
Bessel function $J_1$~\citec{Davoudiasl:1999jd}.  We find that
$\mu^{(0)}/m_{ G}^{(1)} \sim 1.25$. This implies that the KK
contribution is not significantly enhanced over the Regge
contribution~\citec{Hassanain:2009at}, and so here we have limitted our
discussion to the Regge case.

The large amount of data required for discovery may be traced to a
strong difference at the phenomenological level between the RS
scenario and the flat space result: the effective 4D coupling constant
$g^{(0)}\simeq 0.1\ g_a.$ For a given resonance mass, we also have
$\widetilde{M}_s\simeq 0.6\mu.$ The net result, following from
Eq.(\ref{gggg2}) is that for a given resonance mass, the RS cross
section is a factor of $(0.1/0.6)^4\approx 10^{-3}$ times that of the
flat case scenario. (There is also some effect from the narrowing of
the total widths.)  The drastic reduction of the effective coupling is
a direct result of permitting the gluon field to propagate in the
warped bulk.

\newpage

\chapter{Stringy Origin of the CDF Anomaly}
\label{zprime}

The invariant mass distribution of dijets produced in association with
$W$ bosons, recently observed by the CDF Collaboration at Tevatron,
reveals an excess in the dijet mass range $120-160~{\rm GeV/c}^2$,
$3\sigma$ beyond SM expectations. In this Chapter we show that such an
excess is a generic feature of low mass string theory, due to the
production and decay of a leptophobic $Z'$, a singlet partner of
$SU(3)$ gluons coupled primarily to the $U(1)$ baryon number. In this
framework, $U(1)$ and $SU(3)$ appear as subgroups of $U(3)$ associated
with open strings ending on a stack of 3 D-branes. In addition, a
minimal model contains two other stacks to accommodate the
electro-weak $SU(2)\subset U(2)$ and the hypercharge $U(1)$. Of the
three $U(1)$ gauge bosons, the two heavy $Z'$ and $Z''$ receive masses
through the Green-Schwarz mechanism. We show that for a given $Z'$
mass the model is quite constrained. Its free parameters are just
sufficient to simultaneously ensure: a small $Z-Z'$ mixing in accord
with the stringent LEP data on the $Z$ mass; very small (less than
1\%) branching ratio into leptons; and a large hierarchy between $Z''$
and $Z'$ masses. We estimate the LHC sensitivity for searches of $Z''$
in the dijet invariant mass spectrum. The ideas discussed in this
Chapter have been published in~\citec{Anchordoqui:2011ag}.

\section{Light $Z'$ boson at the Tevatron}

It appears that in the last year of the Tevatron's endeavors, it has
pierced the SM's resistant
armor~\citec{Aaltonen:2011mk,:2007qb}.  The latest foray is an excess
at $M_{jj} \simeq 140~{\rm GeV}$ in the dijet system invariant mass
distribution of the associated production of a $W$ boson with 2 jets
(hereafter $Wjj$ production)~\citec{Aaltonen:2011mk}.  The CDF
Collaboration fitted the excess to a Gaussian and estimated its
production rate to be $\sim 4$~pb. This is roughly 300 times the
SM  Higgs rate $\sigma(p\bar p \to WH) \times {\rm BR}(H
\to \bar bb)$. For a search window of $120 - 200~{\rm GeV}$, the
excess significance above SM background (including
systematic uncertainties) is $3.2 \sigma$~\citec{Aaltonen:2011mk}.

The CDF $Wjj$ anomaly has been related to the technipion of a low mass
technicolor~\citec{Eichten:2011sh}, to resonant super-partner production in a supersymmetric model with $R$-parity violation~\citec{Kilic:2011sr}, and to a leptophobic $Z'$ gauge
boson~\citec{Bai:2010dj,Buckley:2011vc,Yu:2011cw,Cheung:2011zt}. The suppressed coupling to
leptons in the latter is required to evade the strong constraints of
the Tevatron $Z'$ searches in the dilepton
mode~\citec{Acosta:2005ij}. All existing dijet-mass searches at the
Tevatron are limited to $M_{jj} > 200~{\rm GeV}$~\citec{Abe:1995jz} and
therefore cannot constrain the existence of a $Z'$ with $M_{Z'} \simeq
140~{\rm GeV}$. The strongest constraint on a light leptophobic $Z'$
comes from the dijet search by the UA2 Collaboration, which has placed
a 90\% CL upper bound on $\sigma \times {\rm BR}(Z' \to jj)$ in this
energy range~\citec{Alitti:1990kw}. In this section we show that a $Z'$
that can explain the $Wjj$ excess and is in full agreement with
exisitng limits on $Z'$ coupling to quarks and leptons can materialize
in the context of D-brane TeV-scale string compactifications.

In Chapter~\ref{CLIC} we have seen that in the minimal $U(3)_a \times
Sp(1)_L \times U(1)_c$ D-brane model, the hypercharge  is anomaly
free. However, the $Q_{U(3)}$ (gauged baryon number) is not anomaly
free and we expect this anomaly to be canceled via a Green-Schwarz
mechanism. There is an explicit mass term in the Lagrangian for the
new gauge field $-\frac{1}{2} M'^2 Y'_\mu Y'^{\mu}$ whose scale comes
from the compactification scheme. The scalar that gets eaten up to
give the longitudinal polarization of the $Y'$ is a closed string
field and there is no extra Higgs
particle~\citec{Ghilencea:2002da}. Following~\citec{Berenstein:2006pk}
we take $M'$ as a free parameter of the model and use precision
electroweak data to determine its value.  As usual, the  $U(1)$ gauge
interactions arise through the covariant derivative \be\el{covderiL}
\CD_\mu = \p_\mu -i g_c \, B_\mu \, Q_{U(1)} - i \frac{g_a}{\sqrt{6}}
\, C_\mu \, Q_{U(3)} \, .\ee
Substituting (\ref{CBfields}) into (\ref{covderiL}) we
obtain \begin{equation} g_{Y'} Q_{Y'} = \frac {g_a} {\sqrt 6}
  C_PQ_{U(3)} + g_c S_PQ_{U(1)} \, . 
\label{sura}
\end{equation} We note that a
value for $g_{Y'}$ will emerge once a normalization for $Q_{Y'}$ is
adopted. (The second relation in Eq.~(\ref{pellerano}) depends on the choice of
normalization for the hypercharge). 

For a Higgs ($Q_{U(3)} = 0$, $Q_{U(1)} = -1$, $Q_Y = -1/2$) with vacuum expectation value \begin{equation} \br H \ke = \bay{c} v \\ 0 \eay, \end{equation} the kinetic term $(D_\mu H)^* (D_\mu H)$ gives gives a mass term \begin{equation} (v, 0) \bay{cc} 
- \frac 1 2 \sqrt{g_b^2 + g_Y^2} Z -g_c S_P Y' & 0 \\ 0 & 
\frac {g_b^2 - g_Y^2}{2\sqrt{g_b^2 + g_Y^2}} Z-g_c S_P Y' 
\eay ^2 \bay{c} v 
\\ 0 \eay = (\overline M_Z Z +g_c S_P v Y')^2 , \label{massterM}\end{equation} where \be\el{covderiYY}\CD_\mu = \p_\mu -i \frac 1 {\sqrt{g_b^2 + g_Y^2}} Z_\mu (g_b^2 T^3 - g_Y^2 Y) -i g_{Y'} Y_\mu{}' Q_{Y'}  \,, \ee with $T^3 = \xs^3/ 2$ and $g_{Y'} Q_{Y'}$ given in Eq.~(\ref{sura}).  Equation~(\ref{massterM}) together with the mass term $\frac{1}{2} M'^2 Y'{}^2$ lead to a mass matrix \begin{equation}
  \frac{1}{2} (Z, Y') \bay{cc} \overline M_Z^2 & \overline M_Z g_a S_P v \\
  \overline M_Z g_a S_P v & g_a^2S_P^2 v^2 +M'^2 \eay  \bay{c} Z \\ Y' \eay = \frac{1}{2} (\overline M_Z Z + g_a v S_P Y')^2 + \frac{1}{2} M'^2 Y'^2 \, ,  \end{equation}
where $2 \overline M_Z^2 = g_b^2 v^2+ g_Y^2 v^2$ is the usual tree level formula for the mass of the $Z$ particle in the electroweak theory, before mixing~\citec{Berenstein:2006pk}. When the theory undergoes electroweak symmetry breaking, because $Y'$ couples to the Higgs, one gets additional mixing. 
However, to avoid conflict with precision measurements at LEP   we will assume negligible $Z-Z'$ mixing and consider $M' \simeq M_{Z'}$~\citec{Umeda:1998nq}.  A comprehensive study of the $M'$ parameter space has been carried out in~\citec{Berenstein:2008xg}, concluding that gauge bosons with $M_{Z'}< 700~{\rm GeV}$ are excluded by the $Z$-pole data from LEP.

On the other hand, we have seen in Chapter~\ref{CLIC} that in the
$U(3)_a \times U(2)_b \times U(1)_c$ D-brane model the $Q_a$, $Q_b$,
$Q_c$ content of the hypercharge operator, is not uniquely determined
by the anomaly cancellation requirement. Hereafter we set $c_a =
-2/3,$ $c_b = 1$, and $c_c = 0$~\citec{Lust:2008qc}.  This choice of
parameters in Eq.~(\ref{hyperchargeY}) leads to the chiral fermion
spectrum given in Table~\ref{t:spectrum}.

\begin{table}
\caption{Chiral fermion spectrum of the $U(3)_a \times U(2)_b \times
  U(1)_c$ D-brane model (case II).}
\begin{tabular}{c|ccccccc}
\hline
\hline
 Name &~~Representation~~& ~$Q_{U(3)}$~& ~$Q_{U(2)}$~ & ~$Q_{U(1)}$~ & ~~$Q_{Y}$ ~ &~~ $g_{Y'}Q_{Y'}$~ &~~$g_{Y''} Q_{Y''}$\\
\hline
~~$U_i$~~ & $({\bar 3},1)$ &    $\phantom{-} 2$ & $\phantom{-}0$ & $\phantom{-} 0$ & $-\frac{4}{3}$ & $\phantom{-} 0.265$ & $\phantom{-} 0.867$ \\[1mm]
~~$D_i$~~ &  $({\bar 3},1)$&    $-1$ & $\phantom{-}0$ & $\phantom{-} 1$ & $\phantom{-}\frac{2}{3}$ & $-0.098$ & $-0.444$ \\[1mm]
~~$L_i$~~ & $(1,2)$&    $\phantom{-}0$ &  $-1$ & $\phantom{-}1$ & $-1$ & $-0.004$ &  $-0.138$\\[1mm]
~~$E_i$~~ &  $(1,1)$&  $\phantom{-}0$ & $\phantom{-} 2$ &  $\phantom{-} 0$ & $\phantom{-} 2$ & $\phantom{-} 0.078$ & $\phantom{-} 0.255$\\[1mm]
~~$Q_i$~~ & $(3,2)$& $\phantom{-}1$ & $\phantom{-}1 $ & $\phantom{-} 0$ & $\phantom{-}\frac{1}{3}$ & $\phantom{-} 0.172$ & $\phantom{-} 0.561$ \\[1mm]
\hline
\hline
\end{tabular}
\label{t:spectrum}
\end{table}

\begin{figure}[tbp]
\begin{minipage}[t]{0.48\textwidth}
\postscript{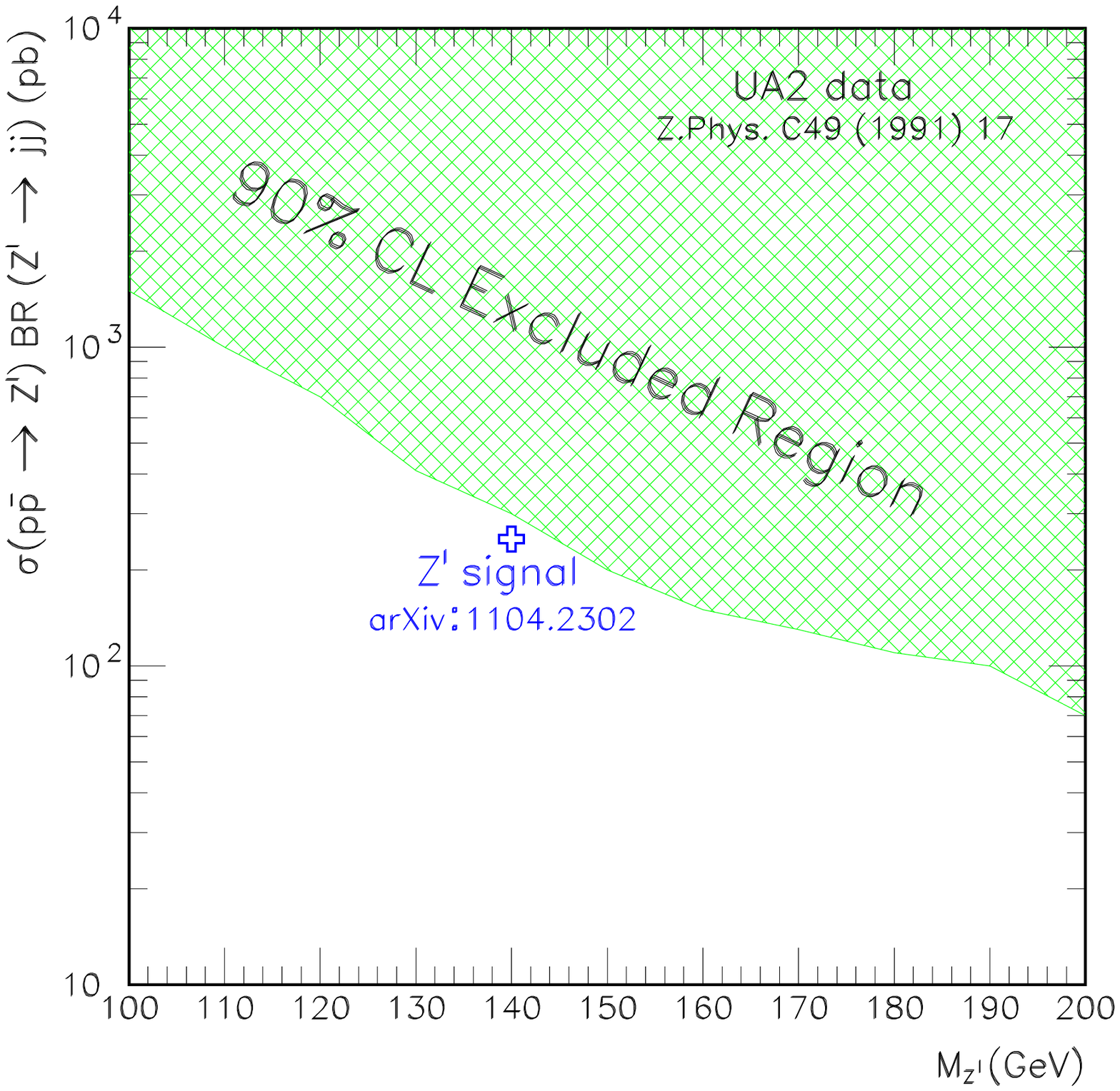}{0.99}
\end{minipage}
\hfill
\begin{minipage}[t]{0.48\textwidth}
\postscript{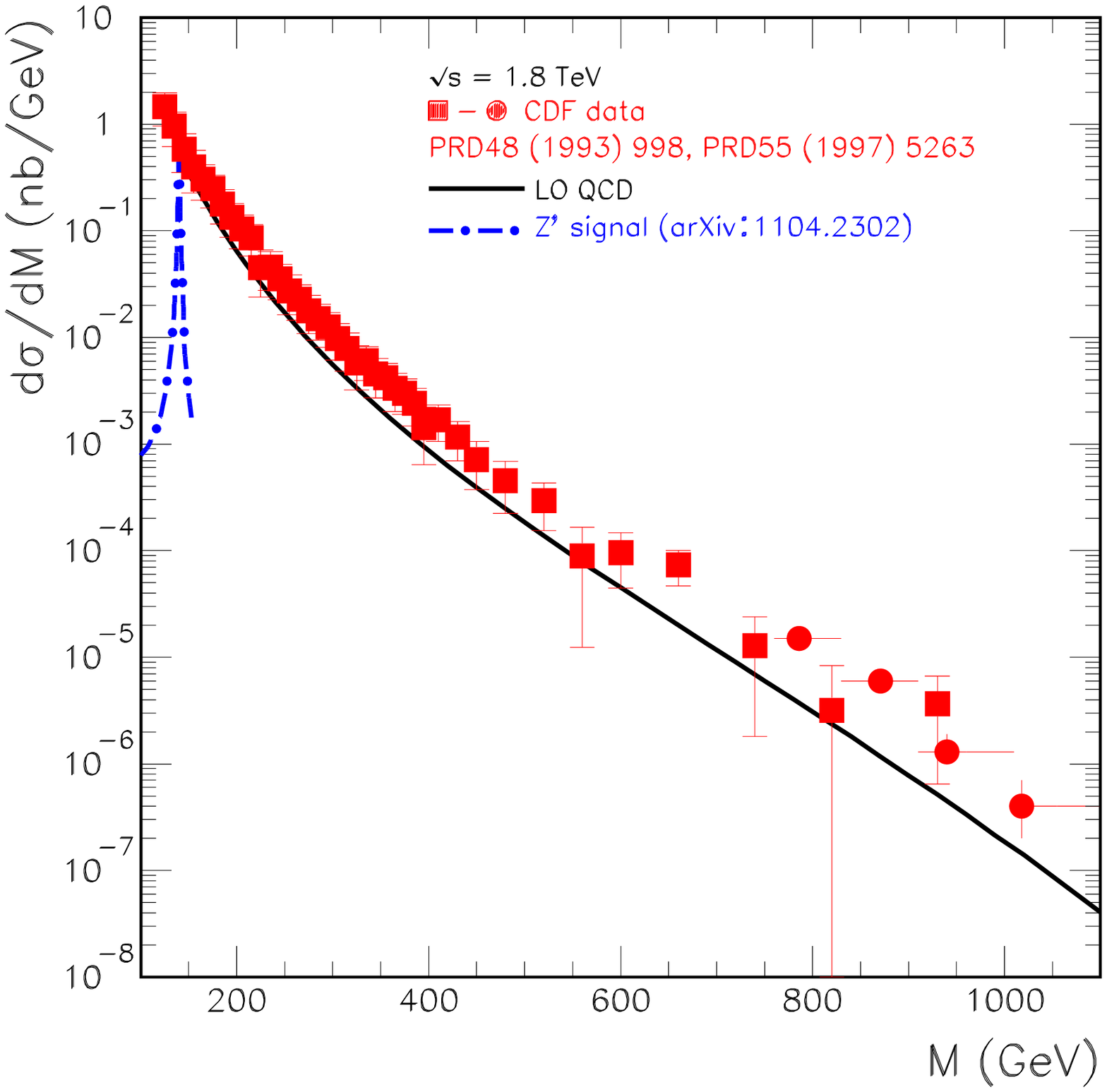}{0.99}
\end{minipage}
\caption[Bounds on $\sigma(p\bar p \to Z') \times {\rm BR} (Z' \to
jj)$]{In the left panel we show a comparisom of the total cross
  section for the production of $p \bar p \to Z' \to jj$ at $\sqrt{s}
  = 630~{\rm GeV}$ in our model and the UA2 90\% CL upper limit on the
  production of a gauge boson decaying into 2
  jets~\cite{Alitti:1990kw}. In the right panel we show the dijet
  invariant mass distribution in $p \bar p$ collisions, as measured by
  the CDF Collaboration, at $\sqrt{s} = 1.8~{\rm
    TeV}$~\cite{Abe:1995jz}. The measurement is compared to a LO QCD
  calculation and the predicted $Z'$ resonant scattering.}
\label{fig:UA2}
\end{figure}

The covariant derivative is given by~\cite{Anchordoqui:2010zs}
\be\el{covderi2} \CD_\mu = \p_\mu - i \frac{g_a}{\sqrt{6}} \, C_\mu \,
Q_{U(3)} -i \frac{g_b}{2} \, X_\mu \, Q_{U(2)} -i g_c \, B_\mu \,
Q_{U(1)} \, .\ee The fields $C_\mu, X_\mu, B_\mu$ are related to
$Y_\mu, Y_\mu{}'$ and $Y_\mu{}''$ by a rotation matrix,
\begin{equation} 
\CR=
\left(
\begin{array}{ccc}
 C_\theta C_\psi  & -C_\phi S_\psi + S_\phi S_\theta C_\psi  & S_\phi
S_\psi +  C_\phi S_\theta C_\psi  \\
 C_\theta S_\psi  & C_\phi C_\psi +  S_\phi S_\theta S_\psi  & - S_\phi
C_\psi + C_\phi S_\theta S_\psi  \\
 - S_\theta  & S_\phi C_\theta  & C_\phi C_\theta 
\end{array}
\right) \,, 
\end{equation}
with Euler angles $\theta$, $\psi,$ and $\phi$; ${\cal R} = O^{-1}$. Equation~(\ref{covderi2}) can be rewritten in terms of $Y_\mu$, $Y'_\mu$, and
$Y''_\mu$ as follows
\begin{eqnarray}
\CD_\mu & = & \partial_\mu -i Y_\mu \left(-S_\xt g_c Q_{U(1)} + \frac 1 2 C_\theta S_\psi g_b Q_{U(2)} + \frac 1 {\sqrt{6}} C_\theta C_\psi g_a Q_{U(3)} \right)  \\  
 & - & i Y'_\mu \left[ C_\theta S_\phi  g_c Q_{U(1)} + \frac{1}{2} \left( C_\phi C_\psi + S_\theta S_\phi S_\psi \right) g_b Q_{U(2)} + \frac{1}{\sqrt{6}} (C_\psi S_\theta S_\phi - C_\phi S_\psi) g_a Q_{U(3)} \right] \nonumber \\
& - & i Y''_\mu \left[ C_\theta C_\phi g_c Q_{U(1)} + \frac 1 2 \left(-C_\psi S_\phi + C_\phi S_\theta S_\psi \right) g_b Q_{U(2)} + \frac 1 {\sqrt{6}} \left( C_\phi C_\psi S_\theta + S_\phi S_\psi\right) g_a Q_{U(3)} \right]   \, .  \nonumber 
\label{linda}
\end{eqnarray}
Now, by demanding that $Y_\mu$ has the 
hypercharge $Q_Y$ given in Eq.~\er{hyperchargeY}  we  fix the first column of the rotation matrix $\CR$
\begin{equation}
\bay{c} C_\mu \\ X_\mu \\ B_\mu
\eay = \left(
\begin{array}{lr}
  Y_\mu \, \sqrt{6}c_ag_Y /g_a& \dots \\
  Y_\mu \, 2c_bg_Y/g_b & \dots\\
   Y_\mu \, c_cg_Y/g_c & \dots
\end{array}
\right) \, ,
\end{equation}
and we determine the value of the two associated Euler angles 
\begin{equation}
\theta = {\rm arcsin} [c_c g_Y/g_c] = 0
\end{equation}
and 
\begin{equation}
\psi = {\rm arcsin}  [2 c_b g_Y/ (g_b \, C_\theta)] = 1.99 \,, 
\end{equation}
where we have taken $M_Z = 91.1876$, $g_b = 0.6596$, $g_a =
1.215$. The third Euler angle $\phi$ and the coupling $g_c$ are
determined by requiring sufficient suppression ($\lesssim 1\%$) to
leptons and compatibility with the 90\%CL upper limit reported by the
UA2 Collaboration on $\sigma (p\bar p \to Z') \times {\rm BR} (Z' \to
jj)$ at $\sqrt{s} = 630~{\rm GeV}$. The decay width of $Z' \to f\bar
f$ is given by~\cite{Barger:1996kr}
\begin{equation}
\Gamma(Z' \to f \bar f) = \frac{G_F M_Z^2}{6 \pi \sqrt{2}}  N_c C(M_{Z'}^2) M_{Z'} \sqrt{1 -4x} \left[v_f^2 (1+2x) + a_f^2 (1-4x) \right] \, ,
\end{equation}
where $G_F$ is the Fermi coupling constant, $C(M_{Z'}^2) = 1 + \alpha_s/\pi + 1.409 (\alpha_s/\pi)^2 - 12.77 (\alpha_s/\pi)^3$, $\alpha_s = \alpha_s(M_{Z'})$ is the strong coupling constant at the scale $M_{Z'}$, $x = m_f^2/M_{Z'}^2$, $v_f$ and $a_f$ are the vector and axial couplings, and $N_c =3$ or 1 if $f$ is a quark or a lepton, respectively. The parton-parton
cross section in the narrow $Z'$ width approximation is given
by
\begin{equation}
\hat \sigma (q \bar q \to Z') =  K \frac{2 \pi}{3} \, \frac{G_F \, M_Z^2}{\sqrt{2}}  \left[v_q^2 (\phi, g_1)+ a_q^2 (\phi, g_1) \right] \, \delta \left(\hat s - m_{Z'}^2 \right) \,,
\end{equation}
where the $K$-factor represents the enhancement from higher order
QCD processes estimated to be $K \simeq 1.3$~\cite{Barger}. After
folding $\hat \sigma$ with the CTEQ6 parton distribution
functions~\cite{Pumplin:2002vw}, taking $M_{Z'} = 140~{\rm GeV}$,  the branching ratio of electrons to quarks is minimized within the $\phi-g_c$ parameter space, subject to saturation of the 90\%CL upper limit~\cite{Alitti:1990kw},
\begin{equation}
\sigma (p\bar p \to Z') \times {\rm BR} (Z' \to jj) \approx 250~{\rm pb} \, ,
\end{equation}
see Fig.~\ref{fig:UA2}. This occurs for
for $\phi = 1.87$ and $g_c = 0.036$, corresponding to a suppression $\Gamma_{Z' \to e^+ e^-}/\Gamma_{Z'\to q \bar q} \sim 0.5\%$. (This also corresponds to $v_u^2 + a_u^2 = 0.355$, and $v_d^2 + a_d^2 = 0.139$.) The UA2 data has a dijet mass resolution $\Delta M_{jj}/M_{jj} \sim 10\%$~\cite{Alitti:1990kw}. Therefore, at 140 GeV the dijet mass resolution is about 15~GeV. This is much larger than the resonance width, which is calculated to be $\Gamma(Z' \to f \bar f)  \simeq 2~{\rm GeV}.$  All the couplings of the $Y'$ boson are now detemined and contained in Eq.~(\ref{linda}). Numerical values are given in Table~\ref{t:spectrum} under the heading of $g_{Y'} Q_{Y'}$.  The corresponding $Wjj$ production rate at the Tevatron ($\sqrt{s} = 1.96~{\rm TeV}$) mediated through $t$ and $u$ channel quark exchange is found to be $\approx 4~{\rm pb}$, in agreement with observation~\cite{Aaltonen:2011mk} and with the recent estimate of~\cite{Cheung:2011zt}.
The rate for the 
associated production channels $ZZ'$, $\gamma Z'$, and  $Z'Z'$ is down by  factors of approximately 3,  5, and 9, respectively~\cite{Cheung:2011zt}. 

The second strong constraint on the model derives from the mixing of the $Z$ and the $Y'$ through their coupling to the two Higgs doublets $H$ and $H'$, with $Q_{U(3)} = -3,$ $Q_{U(1)} = -1$, $Q_{U(2)} = 0$, $Q_Y = 1$ and $Q_{U(3)} = 0,$ $Q_{U(1)} = 1$, $Q_{U(2)} = 1$, $Q_Y = 1$, respectively. Here, $\br H \ke = (^{v_u}_{0})$, $\br H' \ke = (^{ v_d}_{0}),$
   $v = \sqrt{v_u^2 + v_d^2} = 172~{\rm GeV}$, and $\tan \beta \equiv
v_u/v_d$~\cite{Antoniadis:2002qm}. To account for $Y''$ we introduced a second term in (\ref{covderiY}), $\CD_\mu = \p_\mu ... -i g_{Y'} Y_\mu{}' Q_{Y'} -i g_{Y''})
Y_\mu{}'' Q_{Y''}$, which is convenient to write as 
\begin{equation}
-i \frac {x_H}{ v_u} \overline M_Z Y_\mu{}' - i \frac{y_H} {v_u} \overline M_Z Y_\mu{}'' + H \to H' \ ,
\end{equation}
where for the two Higgs doublets 
\begin{equation}
x_{H} = -0.252 C_\phi + 1.886 \, g_c \, S_\phi, \quad  \quad x_{H'} = 2.817 C_\phi
\end{equation}
and 
\begin{equation}
y_{H} = 1.886 \, g_c \, C_\phi + 0.252 S_\phi, \quad \quad y_{H'} = -2.817 S_\phi \, . 
\end{equation}
The Higgs field  kinetic term together with the Green-Schwarz mass terms  ($-\frac{1}{2} M'^2 Y'_\mu Y'^\mu - \frac{1}{2} M''^2 Y''_\mu Y''^\mu$) yield the following mass square matrix
$$ \bay{ccc} \overline M_Z^2 & \overline M_Z^2 (x_{H} C_\beta^2 + x_{H'} S_\beta^2) & \overline M_Z^2 (y_{H}  C_\beta^2 + y_{H'} S_\beta^2) \\ \overline M_Z^2 (x_{H} C_\beta^2 + x_{H'} S_\beta^2) &
M_Z^2 (C_\beta^2 x_{H}^2 + S_\beta^2 x_{H'}^2) + M'^2 & \overline M_Z^2
(C_\beta^2 x_{H} y_{H} + S_\beta^2  x_{H'} y_{H'})  \\
\overline M_Z^2 (y_{H} C_\beta^2 + y_{H'} S_\beta^2) & \overline M_Z^2 (C_\beta^2 x_{H} y_{H} + S_\beta^2 x_{H'} y_{H'}) & \overline M_Z^2 (y_{H}^2 C_\beta^2 + y_{H'}^2 S_\beta^2) + M''^2\eay \, ,$$ where $x_{H} = 0.139$, $x_{H'} = -0.824,$ $y_{H} = 0.221$, and $y_{H'} = -2.694$.  The free parameters are $\tan \beta$, $M_{Z'},$ and $M_{Z''}$ which will be fixed  by requiring  the shift of the $Z$ mass to lie within 1 standard deviation of the experimental value and $M_{Z'} = 140 \pm 2~{\rm GeV}$. We are also minimizing $M_{Z''}$  to ascertain whether it can be detected at existing colliders.
This leads to $\tan \beta =0.4$, $M_{Z'} \simeq M' \simeq 140~{\rm GeV},$ and $M_{Z''} \simeq M'' \geq 3~{\rm TeV}$.

\section{LHC sensitivity to $Z''$}

We now explore (at the parton level) prospects for searches of $Z''$
signals at the LHC.   All the couplings of the $Y''$ boson are given
in Table~\ref{t:spectrum} under the heading of $g_{Y''} Q_{Y''}$.
Using these figures we determine  $\Gamma_{Z'' \to e^+ e^-}/\Gamma_{Z''\to q \bar q} \sim 0.7\%$. We therefore consider the standard bump-hunting procedure for dijet searches.
 
The cross section (for
incoming quark $q$ and outgoing quark $q'$) is given by, 
\ba 
  |{\cal M} (q\bar q \stackrel{Z''} {\to} q'\bar q {}')|^2  &= &\frac 1 4 \lsb g_{Y''}^2 Q_{Y''}^2(q_L)  + g_{Y''}^2 Q_{Y''}^2(q_R) \rsb \lsb g_{Y''}^2 Q_{Y''}^2(q_L{}')  +  g_{Y''}^2 Q_{Y''}^2(q_R{}') \rsb \nn
& \times & \left [\frac{2(  u^2+   t^2)}{( s-M_{Z''}^2)^2 + (\Gamma_{Z''}\ M_{Z''})^2} \right],
\label{Zdprimecross} 
\ea where $g_{Y''}Q_{Y''}(q_L)$ and $g_{Y''} Q_{Y''}(q_R)$ are the
couplings of $Z''$ to quarks. Note that we have not summed over the flavors,
but we did average and sum the colors. We calculate a signal-to-noise ratio, with the signal
rate estimated in the invariant mass window $[M_{Z''} - 2 \Gamma, \, M_{Z''} +
2 \Gamma]$; ;  we set the rapidity cut to be $|y_{\rm max}|<1.0$. The noise is defined as the square root of the number of QCD 
background events in the same dijet mass interval for the same
integrated luminosity. 

\begin{figure}[tpb]
 \postscript{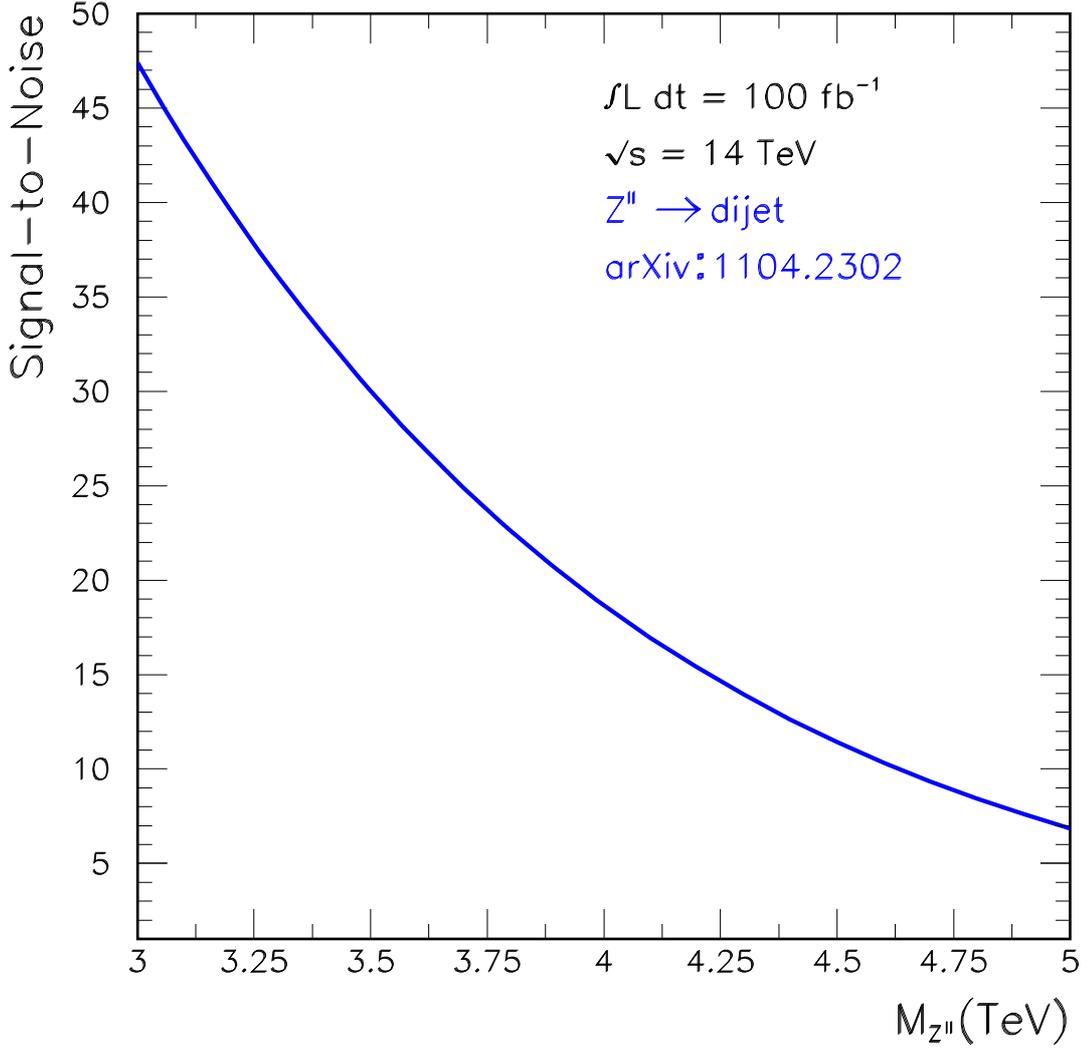}{0.98}
 \caption[$pp \to {\rm dijet}$ signal-to-noise ratio for $Z''$]{$pp \to {\rm
     dijet}$ signal-to-noise ratio for
    100~fb$^{-1}$ integrated luminosity.}
\label{zprime_S2N}
\end{figure}

The curve in Fig.~(\ref{zprime_S2N}) shows the behavior of the
signal-to-noise ratio as a function of the mass of $Z''$, for
100~fb$^{-1}$ of integrated luminosity and $\sqrt{s} = 14$~TeV. As an
illustration, we take $M_{Z''} = 3~{\rm TeV}$, for which $\Gamma(Z''
\to f \bar f) = 493~{\rm GeV}$.  For 10~fb$^{-1}$ of data collected at
$\sqrt{s} = 14~{\rm TeV}$, we obtain a signal-to-noise ratio of
$15\sigma$.

An obvious question is whether the existing data allow determination of the string mass scale. The anomalous mass contributions to $M_{Z'}$ and $M_{Z''}$ are  proportional (with computable coefficients~\cite{Antoniadis:2002cs}) to $g_{Y'} M_s$ and $g_{Y''} M_s$, respectively. However,  existing data can only determine the products  $g_{Y'} Q_{Y'}$ and $g_{Y''} Q_{Y''}$, see Table~\ref{t:spectrum}. Therefore, a separate measurement of the different quark flavor charges (e.g., by tagging on $b$'s and $t$'s in $Z''$ decays) is necessary to determine the absolute normalization of the couplings and predict the string mass scale.

In closing, we note that there are some aspects of the model which can lead to observable consequences even in the absence of a light resonant signal. {\em (1)}~The chiral nature of the couplings in Table~\ref{t:spectrum} implies substantial parity violation. Hence, for $M_{Z'} \gtrsim 400~{\rm GeV}$, the parity violating couplings of the $Z'$ to fermions can generate a $t\bar t$ forward-backward asymmetry in $p \bar p$ collisions. {\em (2)} It was noted in~\cite{Buckley:2011vc} that both the $Wjj$ anomaly and the forward-backward asymmetry observed at the Tevatron can be simultaneously explained by a $Z'$ of $M_{Z'} \simeq 140~{\rm GeV}$ with flavor-violating coupling $g_{utZ'} \sim 0.45.$ In principle  these two conditions can be accommodated in  D-brane constructions by introducing two quark families originating  from strings stretching between two stacks of D-branes, and one family  looping  with both ends of a string attached to the color stack~\cite{Blumenhagen:2001te,Lust:2008qc}.  This can give different charges to $u$ and $t$ quarks. 

\vspace{1cm}

\noindent \underline{Note added:} Related models, including theories
involving $Z'$ gauge bosons, have been recently proposed to explain the $Wjj$ anomaly~\cite{Wang:2011uq}.

\newpage

\noindent\textbf{\Huge Part II:}

\

\noindent\textbf{\huge Anomaly Puzzle in $\CN = 1$}
\addcontentsline{toc}{chapter}{Part II - Anomaly Puzzle in N = 1 Supersymmetric Yang-Mills Theory}

\

\noindent\textbf{\huge Supersymmetric Yang-Mills Theory}

\

\newpage
\thispagestyle{fancy}
\chapter{Two-Supercurrent Scenario as the Solution of the Anomaly Puzzle}
\label{ch:TSSasSAP}
\thispagestyle{fancy}
\pagestyle{fancy}
We discuss several points that may help to clarify some questions that remain about the anomaly puzzle in supersymmetric theories. In particular, we consider a general ${\cal N}=1$ supersymmetric Yang-Mills theory. The anomaly puzzle concerns the question of whether there is a consistent way in the quantized
theory to put the $R$-current and the stress tensor in a single supermultiplet called the supercurrent, even though in the classical theory they are in the same supermultiplet. It was proposed that the classically conserved supercurrent bifurcates into two supercurrents having different anomalies in the quantum regime. The most interesting result we obtain is an explicit expression for the lowest component of one of the two supercurrents in 4-dimensional spacetime, namely the supercurrent that has the energy-momentum tensor as one of its components. This expression for the lowest component is an energy-dependent linear combination of two chiral currents, which itself does not correspond to a classically conserved chiral current. The lowest component of the other supercurrent, namely, the $R$-current, satisfies the Adler--Bardeen theorem. The lowest component of the first supercurrent has an anomaly that we show is consistent with the anomaly of the trace of the energy-momentum tensor. Therefore, we conclude that there is no consistent way to construct a single supercurrent multiplet that contains the $R$-current and the stress tensor in the straightforward way originally proposed. We also discuss and try to clarify some technical points in the derivations of the two supercurrents in the literature. These latter points concern the role of the equations of motion in deriving the two supercurrents.

\ \\

\section{Review of Anomaly Puzzle}
\label{sec:RAP}
First of all, let us review the basics of the supercurrent and the anomaly puzzle. $\CN = 1$ gauge theory is described by a Lagrangian, 
\ba \el{APN1SYM:L}\CL & = & \frac{1}{8g^2T(R)}\int d^2\xt \Tr W^2 + \zt{h.c.} \nn
& = & \frac{1}{4g^2}\int d^2\xt \Tr W^2 + \zt{h.c.}.\ea
$T(R)$ denotes one half of the Dynkin index for the representation $R$, $\Tr(T^a T^b) = T(R) \xd^{ab}$.
The superfield $W_\xa \equiv W_\xa^a T^a$ in components is \ft{We mostly use the conventions in Wess and Bagger \citec{Wess:1992cp} including the choice of $\xs^\mu$ matrices and superderivatives. Our conventions differ only with regard to the normalization of the vector superfield and the integration of Grassmann variables.}
\[W_\xa = -\frac 1 8 \bar D^2 e^{-V} D_\xa e^V = -i (\xl_\xa + i\xt_\xa D +
\xt^\xb f_{\xa\xb} + i\xt \xt \CD_{\xa \dot \xa} \bar \xl^{\dot \xa}),\]
where $f_{\xa \xb}$ is the field strength in the spinor coordinate, $f_{\xa \xb} = -\frac 1 2 (\xs^\mu \bar \xs^\nu)_{\xa \xb} F_{\mu\nu} = -\xs^{\mu\nu}F_{\mu\nu}$.
The vector superfield $V$ in the Wess-Zumino gauge is,
$V = -2\xt^\xa \bar \xt^{\dot \xa} v_{\xa \dot \xa} + 2i \xt \xt \bar \xt
\bar \xl - 2i \bar \xt \bar \xt \xt \xl + \xt \xt \bar \xt \bar
\xt (D + i \p^\mu v_\mu)$. The integration over Grassmann numbers is defined by $\int \xt^2 d^2\xt = 2$.
Recall that the anomaly puzzle can be stated in terms of the absence of a supersymmetric anomaly equation. Such a possible equation is described by a supercurrent
$\CJ_\mu$, which is a superfield and can be defined as \citec{Shifman:1986zi}
\be\el{APN1SYM:1} \CJ_{\xa \dot \xa} \equiv -\frac{4}{g^2} \Tr[e^V W_\xa e^{-V} {\bar W}_{\dot \xa}] = -\frac{2}{g^2 T(R)} \Tr[e^V W_\xa e^{-V} {\bar W}_{\dot \xa}].\ee
Generally, the components of the supercurrent superfield are related to the $R$-current $R_\mu$, the supercurrent $J_{\xa\mu}$ and the stress tensor $\vartheta_{\mu\nu}$ respectively,
\ba\el{APN1SYM:supercurrent}
\CJ_{\xa\dot \xa} & = & C_{\xa\dot \xa} + \{\xt^\xb \chi_{\xb \xa \dot \xa} + h.c.\} + 2\xt^\xb \bar \xt^{\dot \xb} \tau_{\xa \dot \xa \xb \dot \xb}\nn
& & - \frac{1}{2} \{\xt_\xa \bar \xt_{\dot \xb} i \p^{\xg \dot \xb} C_{\xg\dot \xa} + h.c.\} + \{\frac 1 2 \xt^2 M_{\xa\dot \xa} + H.c\}\nn
& &  + \{\frac 1 2 \xt^2 \bar \xt^{\dot \xb} \bar \xl_{\dot \xb \xa \dot \xa} + H.c\} + \frac 1 4 \xt^2\bar
\xt^2 D_{\xa \dot \xa}.
\ea
$C_\mu, \chi_\mu$ and $\tau_{\mu\nu}$ are related to $R_\mu, J_\mu$ and $\xvt_{\mu\nu}$
as we shall see in the appendix.

For the supercurrent defined by \er{APN1SYM:1}, we have,
\be\el{MuSu2:rcurrent} C_{a\dot a} = R_{a\dot a} = -\frac{4}{g^2} \Tr(\xl_\xa \bar \xl_{\dot \xa}),\ee

The $\xt \bar \xt$ component
of $\CJ_{\xa \dot \xa}$, \er{APN1SYM:supercurrent} corresponds to the stress tensor and the exterior derivative of $R_\mu$. Note that $\tau_{\mu\nu}$ is not really the stress tensor as we shall see in the appendix. However,
the trace $\tau_\mu{}^\mu$ is proportional to that of the stress tensor $\xvt_\mu{}^\mu$.
So the operator $\tau_\mu{}^\mu$ also gives the trace anomaly.

As usual, this $R$-symmetry is broken at the quantum level because 
it is a chiral $U(1)$ symmetry.  The anomaly equation is,
\be\el{chiano}\p^\mu R_\mu = - \frac{T(G)}{16 \pi^2}F^a_{\mu\nu} \tilde F^{a\mu\nu},\ee
as follows from the Adler--Bardeen theorem \citec{Adler:1969er}. $T(G)$ is the $T(R)$ of the adjoint representation.

One can lift \er{chiano} to the supersymmetric form,
\be\el{APN1SYM:0a} \p_\mu \CJ^\mu = \frac i 2 {\cal C}\, \Tr [D^2 W^2 - \bar D^2 \bar W^2],\ee 
where $\cal C$ is some coefficient to be determined. The lowest component of \er{APN1SYM:0a} is the chiral anomaly equation, \er{chiano}. Equivalently,
we have,
\be\el{APN1SYM:0} \bar D^{\dot \xa} \CJ_{\xa \dot \xa} = {\cal C}\, D_\xa \Tr W^2.\ee

The real part of the $\xt$ component $\bar D^{\dot \xa} \CJ_{\xa \dot \xa}$ in \er{APN1SYM:0} corresponds to $\tau_\mu{}^\mu$ while the imaginary part corresponds to $-\p^\mu R_\mu$. This matches the $\xt$ component of $D_\xa \Tr W^2$, whose real and imaginary parts are $-\xe_{\xa\xb}\Tr (F F)$ and $-\xe_{\xa\xb}\Tr (F \tilde F)$ respectively.

To be consistent with the prediction of the Adler--Bardeen theorem, both sides should be bare operator and $\CC$ should be of one-loop just like \er{chiano}. However, to get the correct trace anomaly, $\xb$-function, which has higher-loop contributions should appear on the right hand side (henceforth rhs, and similarly lhs for left hand side) . So $\CC$ has to be proportional to $\xb$. Now we get the anomaly puzzle. At least, this is how this puzzle was
originally stated. There are quite a few subtleties as we shall see.

The situation becomes more complicated when matter
is introduced. The Lagrangian becomes,
\be\el{APN1SYM:Lmat} \CL = \frac{1}{4g^2}\int d^2\xt \Tr W^2 + \zt{h.c.} + \frac 1 4 \int d^4 \xt \sum_f \bar \Phi^f e^V \Phi_f.\ee
where $\Phi_f$ are chiral superfields describing matter. The supercurrent is defined as
\ba \CJ_{\xa \dot \xa} & = & -\frac{4}{g^2} \Tr[e^V W_\xa e^{-V} {\bar W}_{\dot \xa}] + \frac 1 3 \sum_f \bar \Phi^f \lb \overleftarrow{\bar \nabla}_{\dot \xa} e^V \nabla_\xa \right.\nn
& & \left.- e^V \bar D_{\dot \xa}\nabla_\xa + \overleftarrow{\bar \nabla}_{\dot \xa}\overleftarrow{D}_\xa e^V\rb \Phi_f, \ea
where covariant derivative is introduced $\nabla_\xa \Phi_f = e^{-V} D_\xa e^{V} \Phi_f$. The $R$-current has the form,
\be \el{RcurrentFull}R_\mu = \frac 2 {g^2} \Tr (\xl \xs_\mu \bar \xl)- \frac 1 3 \sum_f
\lb \psi_f \xs_\mu \bar \psi_f - 2 i A_f \overleftrightarrow{\CD}_\mu
A_f^*\rb,\ee
where $A_f$ is the scalar component of the chiral superfield $\Phi_f$ and
$\psi_f$ is the spinor component.

With the introduction of matter, there is another $U(1)$
symmetry $\Phi_f \to e^{i\xa} \Phi_f$. The corresponding current is the so-called Konishi current (denoted by $R_\mu^f$). This symmetry is certainly chiral and its anomaly, the Konishi anomaly is given by,
\be\el{KonishiAn}\bar D^2 \CJ^f = \bar D^2 (\bar \Phi^f e^V \Phi_f) = \frac{T(R_f)}{2\pi^2} \Tr W^2.\ee

Equivalently, we can define a superfield $Q_{\xa \dot \xa}$ as in \er{Konishicurrent},
which has $R_\mu^f$ as its lowest component.

\section{Possible Solutions to the Puzzle}
\label{sec:ABoTS}
As mentioned before, we can use Wilson effective action to describe the anomaly
puzzle \citec{Shifman:1986zi} \ft{This approach will be further discussed in Sec~\ref{sec:ASVuWEA}.}. In this scenario, the theory has a large but finite cutoff. The Wilson effective action at scale $\xL$ is denoted by $S_W(\xL)$. Higher momentum modes can be integrated out to provide the running of the coupling constant. It can be shown that the new $S_W(\xL-\xd\xL)$ obtained by this renormalization group flow will only have a one-loop correction to the coefficient of $\Tr W^2$ in the Lagrangian \er{APN1SYM:Lmat}. This
agrees with the conclusion based on the non-renormalization theorem \citec{Seiberg:1993vc}. However, this result appears to be in contradiction to the multi-loop $\xb$
function. Note that the coefficient of the Wilson effective action (at scale $\xL$) can be related to the 1PI amplitude with an infrared cutoff $\xL$ (see \eg \citec{Kim:1998wz}). As noted by Shifman and Vainshtein, the absence of the infrared modes is the reason for the absence of multi-loop corrections in the $\xb$-function. Shifman and Vainshtein distinguish between the physical coupling constant and the corresponding coefficient in the Wilson effective action $S_W$. The latter is renormalized only at one-loop level as predicted by the nonrenormalization theorem. On the other hand, the physical coupling
can be obtained by evaluating the matrix elements (or the effective action). To do that, all the infrared modes have to be included and the higher-order corrections emerge.

Shifman and Vainshtein then proceed to propose that a single supercurrent
can contain both the stress tensor and the $R$-current. The anomaly equation for this bare supercurrent $\CJ_{\xa \dot \xa}$ is of the form of \er{APN1SYM:0} with a one-loop coefficient
$\CC$ (see \eg Eq.(19) in \citec{Shifman:1986zi}),
\ba\el{superanomalySV}
\bar D^{\dot \xa} \CJ_{\xa \dot \xa} & = & \frac 2 3 D_\xa\Big[ \frac {\xb_g^{(1)}(g_0)} {g_0^{3}} \Tr W^2 - \frac 1 8 \sum_f \xg_f \bar D^2 (\bar \Phi^f e^V \Phi_f)\Big]\nn
& = & \frac 2 3 D_\xa\Big[-\frac {3 T(G) - \sum_f T(R_f) } {16 \pi^2}  \Tr W^2 \nn
& & - \frac 1 8 \sum_f \xg_f \bar D^2 (\bar \Phi^f e^V \Phi_f) \Big]
\ea
The $\xb_g^{(1)}(g_0)$
is the one-loop $\xb$-function and $\xb_g^{(1)}(g_0)/g_0^3$ is a $g_0$-independent
number. Note that the operators in the equation are bare operators.
To obtain the physical coupling constant one needs to take the matrix elements of the operators on the rhs. The matrix of element of $W^2$ is
shown to have finite multi-loop contribution that exactly reproduce the correct
NSVZ $\xb$-function \er{betafunctionmatter}. More explicitly, when the operators on the right are expressed in terms of renormalized operators, \er{superanomalySV} becomes the anomaly equation that has the correct multi-loop $\xb$-function,
\be\el{APN1SYM:2} \bar D^{\dot \xa} \CJ_{\xa \dot \xa} = \frac 2 3 D_\xa \Big\{\frac {\xb_g(g)} {g^3} [\Tr\, W^2] - \frac 1 8 \sum_f \xg_f \bar D^2 [(\bar \Phi^f e^V \Phi_f)]\Big\},\ee
where $[\ ]$ indicates renormalized operators, and 
\be\el{betafunctionmatter} \xb_{g}(g) = - \frac {g^3}{16 \pi^2} \frac {3 T(G)-\sum_f
T(R_f)(1-\xg_f)}{1- T(G) \xa/ 2\pi}.\ee
The term proportional to $\xg_f$ comes from the second term in \er{superanomalySV} because of the Konishi anomaly \er{KonishiAn}. More explicitly, the contribution (proportional to $\xg_f$) from the second term in \er{superanomalySV}
to the $\xb$-function follows when the operators on the rhs are diagonalized.

Note that the trace of the stress tensor, $\xvt_\mu{}^\mu$ should be equal to $\sum \xb_a (M)O_a(M)$, in which $O_a (M)$ are renormalized operators
(at scale $M$). Moreover, \citec{Adler:1976zt} the coefficients of these operators $O_a(M)$ can be considered as $\xb$-functions (with $g_a(M)$ as variables) only when the operators $O_a (M)$ are ``orthonormal" at scale $M$. More explicitly, the operators $O_a (M)$ are chosen so that the corresponding matrix element
\ft{Without the insertion, this matrix element corresponds to a certain amplitude that defines the coupling constant at scale $M$.} of every coupling constant
$g_a$ only receives contribution from a single operator (on the rhs of the trace anomaly) and the matrix element should be exactly unity (up to some power of $M$'s). Only in this case we can take the coefficients of the operators on the rhs of the trace anomaly to be the $\xb$-functions.

However, there are some subtleties about \er{superanomalySV} \eml{that imply a contradiction with the Adler--Bardeen theorem.} Let us look at it more carefully. For the example of pure SYM, the $\xt^2$ component of an operator $W^2$ in fact has an imaginary part equal to (where $\xa_0 \equiv g^2_0/4\pi$),
\[-\frac 1 2 F\tilde F - \p_\mu (\xl \xs^\mu \bar \xl) = -\frac 1 2 F \tilde F - 4 \pi \xa_0 \p_\mu R^\mu,\]
where we used the fact that the second term on the lhs is proportional to $\p^\mu R_\mu$ (see \er{MuSu2:rcurrent}).
After this term is moved to the left side in \er{superanomalySV}, it is clear that \er{superanomalySV} does not reproduce the Adler--Bardeen theorem; namely that the anomaly of the $R$-current $R_\mu$ is no longer proportional (with a coupling-constant-independent proportionality factor) to the topological term $F \tilde F$ as in the non-supersymmetry gauge theory. In other words, \er{chiano} no longer
holds as an operator equation of bare operators. Moreover, unlike it was previously claimed in the literature \citec{Jones:1983ip}, \er{APN1SYM:2} does not agree with the Adler--Bardeen theorem. Even if \er{APN1SYM:2} is not obtained from the one-loop equation (\ref{superanomalySV}) but is taken as the starting point, what appears on the right are just renormalized operators and can not be moved to the left side, which only contains bare operators. On the other hand, it has been shown \citec{Jones:1984mi} that if the lowest component of the supercurrent on the left of \er{superanomalySV} is taken as a renormalized operator, which is different from the bare $R$-current by a multiplicative renormalization factor, correct anomaly equations (for both trace anomaly and chiral anomaly) can be obtained.

There is another way to show the inconsistency between \er{superanomalySV} and the Adler--Bardeen theorem.
Together with the proposed expectation value of $W^2$ ((46) in \citec{Shifman:1986zi}),
\[\br W^2\ke = \lb 1 + \frac {T(G)\xa}{2\pi}+\dots \rb W^2_{\zt{ext}},\]
equation \er{superanomalySV} predicts a nonvanishing expectation value for the bare chiral current $\br \p^\mu R_\mu \ke$ at two-loop level. More explicitly, the two-loop value is ${T(G)\xa}/{(2\pi)}$ times
the one-loop value. This conclusion however, is in contradiction with the combination of the
Adler--Bardeen theorem and the proposed expectation value of $F \tilde F$ ((57) in \citec{Shifman:1986zi}),
\[\br F\tilde F \ke = (F\tilde F)_{\zt{ext}} \lb 1 + \frac {T(G)\xa}{\pi}\rb,\]
which implies a two-loop expectation value being ${T(G)\xa}/{(2\pi)}$ times
the one-loop value. Note that the Adler--Bardeen theorem states that \er{chiano} is an operator equation of bare operators and the expectation values of both sides should have the same quantum corrections \ft{Higher order quantum corrections to $F\tilde F$ in QED are discussed in \citec{Anselm:1989gi}.}. Such an agreement is confirmed up to two-loop in \citec{Jones:1982zf}.

So eventually we have no choice but to construct two supercurrents. One of them, $\CJ^{(1)}{}_{\mu}$ has
the $R$-current as its lowest component, but does not have the stress tensor
among its components, while the other $\CJ^{(2)}{}_{\mu}$ has the stress tensor but not the $R$-current. As a result, there is no reason to have a single operator equation to describe both chiral anomaly and trace anomaly. The construction of two supercurrents using the background field method and dimensional reduction is first proposed by Grisaru et al \citec{Grisaru:1985yk}
\cite{Grisaru:1985ik} for the pure SYM and is further developed by Ensign and Mahanthappa \citec{Ensign:1987wy} for the coupled SYM. As we shall see later, there is an inconsistency in their calculation. However, we show by careful calculation that their results for the two currents are indeed correct.

Let us briefly review their results. In this approach, two different renormalized currents (both superfields) are defined. Each satisfies an anomaly equation with the anomaly expressed in terms of renormalized operators. One of the anomaly equations is similar to \er{APN1SYM:2} with the renormalized coupling constant and operators on the rhs. The other has an one-loop coefficient for the $W^2$ term, which agrees with the Adler--Bardeen theorem
\ft{\eml{In \citec{Grisaru:1985ik} and \citec{Ensign:1987wy}, only the divergent contribution to the expectation value is considered and therefore the Adler--Bardeen
theorem implies the absence of any two-loop contribution to the coefficient $\CC$}.}. The explicit form is (Eq.(4.29) in \citec{Ensign:1987wy} after a change in convention),
\be\el{twosupABanomaly} \cd^{\xa \dot \xa} [J_{\xa \dot \xa} ] =  -\frac i 3 \frac {\xb_g^{(1)}} {g^3} \lb [\cd^\xa W^\xb \cd_\xa W_\xb] - [\bar \cd^{\dot \xa} \bar W^{\dot
\xb} \bar \cd_{\dot \xa} \bar W_{\dot \xb}]\rb.\ee
Both currents are renormalized operators whose expectation values are finite.

We shall see how these two supercurrents are constructed. For later convenience, we will give the operators that are involved,
\ba
W_{\xa \dot \xa} & = & \frac 4 {g^2_0}\Tr [e^{-V}\overline W_{\dot \xa} e^V
W_\xa] \\
K_{\xa \dot \xa} & = & \hhat W_{\xa \dot \xa} - \frac 1 {g^2} \Tr \lb i \hhat \xG_{\xb \dot \xa} \cd_\xa W^\xb - i \hhat \xG_{\xa \dot \xb} \bar \cd_{\dot \xa} \overline W^{\dot \xb} \rb\\
P_{\xa \dot \xa} & = & i (\bar \Phi e^V {\cd}_{\xa \dot \xa} \Phi - \bar \Phi {\overleftarrow {\cd}}_{\xa \dot \xa}  e^V \Phi)\\
\el{Konishicurrent} Q_{\xa \dot \xa} & = & -\frac 1 2 [\cd_\xa, \bar \cd_{\dot \xa}] (\bar \Phi e^V \Phi).
\ea
To be consistent with the expressions given above in Sec~\ref{sec:RAP}, we use
the covariant derivatives in the gauge chiral representation. Note that $\hhat \xG_{\xa \dot \xa}$ introduced in \citec{Grisaru:1985ik} is the $\xe$-dimensional projection of the gauge connection and is gauge covariant under the $K$ gauge transformation (and invariant under the $\xL$ gauge transformation). Now in the gauge chiral representation, it is covariant under the $\xL$ gauge transformation. The expectation values of various operators are given by (3.3) (3.4) (3.5) in \citec{Ensign:1987wy}. They are obtained by the background field method and dimensional reduction. The dimension is $d = 4 -2 \xe$. Some of the expectation values given below are different from
from those in \citec{Ensign:1987wy}. This is due to different conventions.
For example, one of the equation given by (3.4) in \citec{Ensign:1987wy} is,
\be\el{twoloopQr} \br Q^r_{\xa \dot \xa}\ke^{(2)} =  0\times K_{\xa
\dot \xa}^e +\dots,\ee
The superscript $e$ denotes (renormalized) external fields. The superscript $r$ denotes renormalized fields ($\Phi = Z_\Phi^{1/2} \Phi^r$) with the field strength $Z_\Phi$
given by,
\ba\el{fieldstrengthphi} Z_\Phi & = & 1 + 2\frac {g^2} \xe C(R) + g^4\lb \frac 1 {\xe^2} -\frac {1} \xe \rb(-3 T(G) C(R)\nn 
& & + C(R) T(R) + 2 C(R)^2 ),\ea
and $Z_V$ given by
\ba Z_V & = & 1+ \frac {g^2} \xe [3T(G) - T(R)] + \frac {g^4} \xe [3 T(G)^2 - T(G)T(R)\nn
& & -2C(R)T(R)].\ea
However, $\br Q^r_{\xa \dot \xa}\ke^{(2)}$ are not the two-loop expectation values of the $Q^r_{\xa \dot \xa}$. Instead, it is the two-loop expectation value of the operator renormalized to one-loop order. In \emp{this dissertation}, we use the symbol $\br \CO \ke^{(n)}$ for the expectation value of an operator $\CO$ without subtracting
any subdivergence due to renormalization of this operator. In this convention,
\er{twoloopQr} is expressed as 
\be\el{twoloopQ} \br Q^r_{\xa \dot \xa} + \frac {g^2} \xe (2C (R) Q^r_{\xa \dot \xa} + T (R) K^r_{\xa \dot \xa}) \ke^{(2)} = 0 \times K_{\xa \dot \xa}^e + \dots,\ee
where $C(R)$ is the quadratic Casimir operator of representation $R$. Note that only the two-loop contribution proportional to $K_{\xa \dot \xa}$ is evaluated
and the rest is unknown. With field strength renormalization $Z_\Phi$ ($Q_{\xa
\dot \xa} \equiv Z_\Phi Q_{\xa \dot \xa}^r$) given by \er{fieldstrengthphi}
and $\br K^r_{\xa \dot \xa} \ke^{(1)} = 0$, \er{twoloopQ} can be rewritten as,
\[\br Q_{\xa \dot \xa} \ke^{(2)} = 0 \times K_{\xa \dot \xa}^e + \dots.\]
Let us work with SQED, in which the corrections to the expectation value of $F \tilde F$ start at two-loop and the corrections to the rhs of the Adler--Bardeen theorem start at three-loop. Naively, one can speculate that the $U(1)$ current in $Q_{\xa \dot \xa}$ satisfies the Adler--Bardeen theorem in the sense that there is no anomaly at two-loop level. Note that $\br Q_{\xa \dot \xa} \ke^{(1)}$ is nonvanishing and leads to a nonvanishing expectation value of $\p^\mu C_\mu^Q$ ($C_\mu^Q$ being the lowest component of $Q_\mu$).

However, $Q_{\xa \dot \xa}$ is not the correct superfield containing the anomalous $U(1)$ current. In the approach used by \citec{Ensign:1987wy}, $Q_\mu$ has to be renormalized
and the anomaly is described by a renormalized operator $[Q_\mu]$. However,
in the usual anomaly calculation of non-supersymmetric gauge theories (with
matter), the expectation value of
a bare chiral current $\p^\mu j^5_\mu$ is proportional to $F\tilde F$. Anyway, \eml{if we ignore this difference
and just apply the equations of motion on those bare fields from which $[Q_\mu]$
is constructed, the correct anomaly equation follows.}
 
Following Eq.(3.8) in \citec{Ensign:1987wy}, we can find out the relationship between the bare operators and the renormalized
operators as,
\ba \el{AnomalyN1SYM:1}
W_{\xa \dot \xa} & = &[W_{\xa \dot \xa}] - \frac {g^2} \xe (T(R) [W_{\xa \dot \xa}] -T(G) [K] - 3 C(R) [P_{\xa \dot \xa}] \nn
& & - C(R) [Q_{\xa \dot \xa}]) - \Big[\frac {g^4 } {4\xe} (3 T(G) T(R) + C(R) T(R)) \nn
& & + \frac {g^4}{\xe^2} (\frac 1 2 T(G) T(R) + \frac 1 2 C(R) T(R)) \Big][K_{\xa \dot \xa}] \\
\el{AnomalyN1SYM:1a}
P_{\xa \dot \xa} & = & [P_{\xa \dot \xa}] - \frac {g^2} \xe (3 C(R) [P_{\xa \dot \xa}] + C(R)[Q_{\xa \dot \xa}] - T(R) [W_{\xa \dot \xa}]) \nn
& & + \Big[\frac {g^4 } {4\xe} (3 T(G) T(R) + C(R) T(R)) -  \frac {g^4}{2\xe^2} (T(G) T(R) \nn
& & - C(R) T(R)) \Big] [K_{\xa \dot \xa}]\ea
\ba
\el{AnomalyN1SYM:1b}
Q_{\xa \dot \xa} & = & [Q_{\xa \dot \xa}] - \frac {g^2} \xe T(R) [K_{\xa \dot \xa}]
\ea
Renormalized operators like $[W_{\xa \dot \xa}]$ are defined to have the expectation values of the background fields. With \er{AnomalyN1SYM:1}, \er{AnomalyN1SYM:1a} and \er{AnomalyN1SYM:1b}, we get to the conclusion that a current
\[\tilde \CJ_{\xa \dot \xa} \equiv W_{\xa \dot \xa} + P_{\xa \dot \xa} + \frac 1 3 Q_{\xa \dot \xa},\]
has no anomaly at two-loop level because of the lack of $g^4 [K_{\xa \dot \xa}]$ in $\br \tilde \CJ_{\xa \dot \xa} \ke$. However, this bare operator $\tilde \CJ_{\xa \dot \xa}$
contains neither $R_\mu$ nor $\xt_\mu{}^\mu$. Instead, two different renormalized operators need to be constructed to describe the two anomalies (trace and chiral). One of them is the supercurrent ($ \CJ^{(2)}{}_{\mu}$ in our notation),
\be \el{defsuprcurrent} \hat J_{\xa \dot \xa} \equiv \hat W_{\xa \dot \xa} -\frac 1 3 \hat K_{\xa \dot \xa}+ \hat P_{\xa \dot \xa} + \frac 1 3 \hat
Q_{\xa \dot \xa} + \CO(\xe),\ee
where the $\CO(\xe)$ terms do not affect
the $\xb$-function. The other is the Adler--Bardeen current ($\CJ^{(1)}{}_{\mu}$ in our notation),
\[[J_{\xa \dot \xa}] \equiv [W_{\xa \dot \xa}] + [P_{\xa \dot \xa}] + \frac 1 3 [Q_{\xa \dot \xa}].\]
As shown in \citec{Ensign:1987wy}, with the use of the equations of motion,
the desired anomaly equations \er{APN1SYM:2} and \er{twosupABanomaly} can be obtained.

A major problem this approach has is about the use of equation of motion. For example, the trace anomaly is described by \er{traceano}. The lhs is a renormalized operator in the sense that expectation value of the operator $[\hat J_{\xa \dot \xa}]$ is given by the background
fields. There is no way this expectation value can give what appears on the
rhs under a derivative $\bar \cd^{\dot \xa}$. In fact, the rhs is obtained
by taking the expectation value of another operator $W^2$ which is obtained
by the equation of motion of $\hat K_{\xa \dot \xa}$. $\hat K_{\xa \dot \xa}$ is in the definition of the renormalized operator $[\hat \CJ_{\xa \dot \xa}]$ (see \er{defsuprcurrent}). The rest of $\hat \CJ_{\xa \dot \xa}$, the operator $\hat W_{\xa \dot \xa} + \hat P_{\xa \dot \xa} + \frac 1 3 \hat Q_{\xa \dot \xa}$ gives no anomalous contribution because of the EoM. This is very confusing. As shown in \citec{Ensign:1987wy},
the two sides of an equation of motion generally do not have the same expectation values. For example, $\bar \cd^{\dot \xa} W_{\xa \dot \xa} = 0$ (for pure SYM) follows from the equations of motion but apparently $\bar \cd^{\dot \xa} \br W_{\xa \dot \xa} \ke \ne 0$ following from \er{AnomalyN1SYM:1}
(see also Eq.(3.3) in \citec{Ensign:1987wy}). \eml{This problem will be further discussed in Sec~\ref{sec:EVoEoM} and a possible solution will be proposed.}

\section{Superpartner of the Trace Anomaly}
\label{sec:SotTA}
Before we move on to talk about the solution to the problem about the equation
of motion in the construction of the two supercurrents, let us give a supporting
argument for this approach. As explained in Sec~\ref{sec:ABoTS}, the one-loop anomaly equation \er{superanomalySV}
implies that the operator $R_\mu'$, defined as the lowest component of $\hat
\CJ_\mu$, does not satisfy the Adler--Bardeen theorem. So it is unlikely that $R_\mu'$ is the the $R$-current $R_\mu$ \ft{Note that we define $R$-current as the $U(1)$ current associated with the (anomalous) symmetry that transforms the gaugino $\xl$, the matter scalar $A$ and the matter spinor $\psi$ according to the charge ratios of $1:\frac 2 3: - \frac 1 3$, which is determined by the classical supercurrent.}. Here we try to use some explicit calculation to show that for a general coupling $g$, this operator $R_\mu'$ is a mixing of the current $R_\mu$ and the Konishi current $R_\mu^f$ ($R_\mu^f$ being the $U(1)$ current in $Q_{\xa \dot \xa}$). This result clearly supports the approach
of two supercurrents. It also gives a clear physical interpretation
of the lowest component of $\hat J_{\xa \dot \xa}$ (the one having $\xvt_{\mu\nu}$),
which is not given before. This validity of this interpretation is particularly clear at the infrared fixed point where the superconformal symmetry is restored. At this
point, the charges of those fields $\xl$, $A$ and $\psi$ under the $R$-symmetry are different from their classical values. This new $U(1)$ symmetry is also a classical symmetry whose current $R_\mu'$ is a linear combination of $R_\mu$ and the Konishi current $R_\mu^f$. The latter assigns charge $+1$ to both $A$ and $\psi$. Moreover, as we shall explain, this property of $R_\mu'$ agrees with the last term in the anomaly equation \er{superanomalySV}, which is actually not obtained in \citec{Ensign:1987wy}. So the $\CJ_{\xa \dot \xa}$
in the Shifman-Vainshtein scenario should be identified as the supercurrent
with $\xvt_{\mu\nu}$.
 
To study $R_\mu'$, we compute the Green's functions of this operator (or
rather $\p^\mu R_\mu'$) and various other fields. More explicitly, we compute the Green's functions with
an insertion of the operator obtained via supersymmetry transformation of the gamma trace (see below) of the supersymmetry current. This operator has a term $\p^\mu R_\mu'$ according to the superconformal algebra,
\be\el{supconfalgebra} \{S_{\xa},Q_{\xb}\} = 4 M_{\xa \xb} - 2 i D \xe_{\xa \xb} - 3 R' \xe_{\xa \xb}.\ee
where $M_{\xa\xb}$ are the Lorentz generators and $S_{\xa}$ is the generator corresponding to the gamma trace of $\bar J_\mu$. We will now show that the contact terms of the Green's function, which are the changes of the other fields under the transformation generated by $R_\mu'$, can be described by the transformation of a combination of the original $R$-symmetry and the Konishi $U(1)$ symmetry.

Let us start with the computation of $\xs^\mu \bar J_\mu$. The gamma trace of the supersymmetry current $\bar J_\mu$ in the Wess-Zumino model is,
\ba
\el{gammatrace}
\xs^\mu \bar J_\mu & = & - i \xs^\mu (\bar \chi_\mu + \bar \xs_\mu \xs^\nu \bar \chi_\nu)
=  3i\xs^\mu \bar \chi_\mu = -2\sqrt 2\xs^\mu \p_\mu \bar \psi A \nn
& & \to -2\sqrt 2\xs^\mu \CD_\mu \bar \psi A
\ea
In the last step, we include the effect of the gauge field by covariantizing the derivative. \er{gammatrace} corresponds to the contribution from the matter multiplet to the gamma trace of the full supersymmetry current.

However, \er{gammatrace} does not vanish on-shell because of the interaction
with the gauge field. Let us consider SQED for simplicity. The equation of motion of $\bar \psi$ is,
\[i\CD_\mu\xs^\mu \bar \psi = - \sqrt 2i e A^* \xl.\]
So the gamma trace in 4d is
\[\xs^\mu \bar J_\mu = -2\sqrt 2\xs^\mu \CD_\mu \bar \psi A -4e AA^*\xl,\]
or equivalently, we should add $-\frac i 3 e\bar \xs_\mu \xl
A A^*$ to the definition of $\bar \chi_\mu$ in \er{gammatrace}. Of course, this term can also be obtained from explicit calculation (from a $ \xl \xs_\mu D$ term in the supercurrent
of the gauge multiplet). 
In $4+\xe$ dimension, the gamma trace becomes,
\ba
\xs^\mu \bar J_\mu & = & \xs^\mu \sqrt 2 \Big[\CD_\nu A \bar \xs^\nu \xs_\mu \bar \psi + \frac {4} 3 \bar \xs_{\mu \nu} \p^\nu (A \bar \psi)\Big] \nn
& & - i \xs^\mu [-\frac i 3 e\bar \xs_\mu \xl
A A^* + \bar \xs_\mu \xs^\nu (-\frac i 3 e\bar \xs_\mu \xl
A A^*)] \nn
& = & \sqrt 2\Big\{ (2+\xe)\CD_\mu A \xs^\mu \bar \psi+ \frac 1 3 \Big[-\frac
{(4+\xe)} 2 \nn
& & - (2+\xe) \Big]\p_\mu (A \xs^\mu \bar \psi) \Big\} -(4+ \xe)(1+\frac \xe 3) e AA^*\xl \nn
& = & \frac {\sqrt 2} 3 \xe \CD_\mu A\xs^\mu \bar \psi - \frac {2\sqrt 2}
3 \xe A\xs^\mu \CD_\mu \bar \psi -(\xe + \frac {4\xe} 3 )e AA^*\xl
\nn
& = & \frac {\sqrt 2} 3 \xe \CD_\mu A\xs^\mu \bar \psi -\xe\, e AA^*\xl.
\ea

The supersymmetry transformation (parametrized by $\xi_\xa$)  of the gamma trace is (note that the indices in $\bar \xs^\nu \xi$ are $(\bar \xs^\nu)^{\dot \xb \xb} \xi_\xb$),
\ba
\xd_\xi (\xs^\mu \bar J_\mu) & = & \xe\Big[-\frac 1 3 (\CD_\mu \psi \xs^\mu \bar \psi -2\sqrt{2} e A \bar \psi \bar \xl \nn
& &+ 2i \CD_\mu A^* \CD^\mu A)+\frac 1 {2\sqrt 2} e A^*\psi \xl \Big]\xe_{\xa \xb} \xi^\xb
\nn
& = & \Big[ \xe \sqrt 2 e  A \bar \psi \bar \xl+\xe\frac {\sqrt 2 e} {2} A^* \psi \xl + \dots\Big]\xe_{\xa \xb} \xi^\xb  \nn
\Rightarrow \mathop{\rm Re} [\xd_\xi (\xs^\mu \bar J_\mu)] & = & \Big[\xe\frac {3\sqrt 2 e} {4 } A^* \psi \xl + \zt{h.c.}\Big]\xe_{\xa \xb} \xi^\xb\ea
We use the equation of motion in the middle step. This is justified because \[(\CD_\mu \psi \xs^\mu + \sqrt 2 e A \bar \xl) \bar \psi,\]
is the the counting operator \citec{Lowenstein:1971jk}, which has a finite expectation value. In the last step, we just keep the
real part that is needed (and drop the imaginary term $\CD_\mu A^* \CD^\mu A$). Note that this is because the imaginary part should be proportional to the trace anomaly $\xvt_\mu{}^\mu$ following the superconformal algebra.

The Green's functions with an insertion of $\xd_\xi (\xs^\mu \bar J_\mu)$
can be evaluated. Alternately, the same results can be obtained by calculating the expectation value of
$\xd_\xi (\xs^\mu \bar J_\mu)$ in a certain background. The result bilinear in the external gaugino field $\xl$ is 
\ba \el{expxlxl}
\br -\xe {\sqrt 2} e A^* \psi \xl\ke_{\xl\bar\xl} & = & i \cdot (-i) \cdot \lb-{2e^2} \rb\xe\, \int \frac {d^4 p}{(2\pi)^4} \frac {\xl^e \sla p \bar \xl^e}{p^2(p+k)^2} \nn
& = &- \frac {2ie^2}{(4\pi)^2} \xl^e \sla k \bar \xl^e.
\ea
where $\xl^e$ is understood as Fourier transformation (with momentum suppressed)
of the external field $\xl^e(x)$. We have momentum $k$ flow into the vertex $-\xe {\sqrt 2} e A^* \psi \xl$
and for simplicity, we set the momentum exchange through the external field
$\xl^e$ to be $0$ and that through $\bar \xl^e$ to be $k$. Similarly we have scalar
contribution,   
\ba
\br -\xe {\sqrt 2} e A^* \psi \xl\ke_{AA^*} & = &  \lb 2
e^2 \rb \xe\,\int \frac {d^4 p}{(2\pi)^4} \frac {\Tr[(\sla p + \sla k)\sla p]}{p^2(p+k)^2}
A^e (A^e)^* \nn
& = & -\frac {4 i} {(4\pi)^2} e^2 k^2 A^e (A^e)^*.
\ea
Again, the momentum exchange through $(A^e)^*$ is set to be $0$ for simplicity.
Note that the expectation value proportional to $\psi^e \bar \psi^e$ is the same as \er{expxlxl} with the replacement of $\xl^e \to \psi^e$. Combine these results, the total expectation value is
\ba \el{mixRkonishi}
\br -\xe {\sqrt 2} e  A^* \psi \xl\ke & = & \frac {2ie^2}{(4\pi)^2} \xl^e \sla k \bar \xl^e + \frac {2ie^2}{(4\pi)^2} \psi^e \sla k \bar \psi^e
\nn
& & +\frac {4i} {(4\pi)^2} e^2 k^2 A^e (A^e)^* \nn
\el{expoverall} & = & -\frac {2ie^2}{(4\pi)^2}\Big[ (\xl^e \sla k \bar \xl^e- \frac 1 3 \psi^e \sla k \bar \psi^e + \frac 2 3 k^2 A^e (A^e)^*) \nn
& &+ \frac 4 3 (\psi^e \sla k \bar \psi^e + k^2 A^e (A^e)^*)\Big]
\ea
In the convention we are using, an operator with a momentum inflow $k$ is given by
\[\int \frac {d^4 x}{(2\pi)^4} e^{-i k \cdot x} O(x).\]
A current with the same charge $+1$ assigned to $\psi$ and $A$ (\ie Konishi current) is
\[i A \overleftrightarrow {\p} A^* + \psi \xs_\mu \bar \psi \to -i k^2 A A^* - i \psi \sla k \bar \psi.\]
So it is clear that \er{expoverall} is a linear combination of $R_\mu$ and
the Konishi current $R_\mu^f$.

\er{mixRkonishi} is also consistent with \er{superanomalySV}, at
least loosely.
So the overall contribution to $\mathop{\rm Re} [\br \xd_\xi (\xs^\mu \bar J_\mu)\ke]$
is
\be\el{JconKonishi}\br\xe \frac {3e} {4\sqrt 2} A^* \psi \xl\ke \to- \frac {2 e^2}{(4\pi)^2} \p_\mu (\psi \xs^\mu \bar \psi).\ee
On the other hand, \er{superanomalySV} predicts a correction to the $U(1)$ current $R_\mu'$,
\ba \frac {\xb_g^{(1)}(g_0)} {g_0^{3}} \Tr W^2 - \frac 1 8 \sum_f \xg_f \bar D^2 (\bar \Phi e^V \Phi)\nn 
\to \xt \xt\Big[-\frac {\xb_g^{(1)}(g)} {g^{3}} \frac 1 2 F^2 + \frac i 4 \xg\p^\mu (\bar \psi \bar \xs_\mu \psi) + \dots \Big].\ea
The first term is part of the trace anomaly $\xvt_\mu{}^\mu$. Compare this with the superconformal algebra \er{supconfalgebra},
we know that the correction to $R_\mu'$ is
\[-\frac \xg 2 \p^\mu (\bar \psi \bar \xs_\mu \psi) =- \frac {2e^{2}}{(4\pi)^2}
\p^\mu (\psi \xs_\mu \bar \psi).\]
which agrees with \er{JconKonishi}. So in some sense, we correctly calculate the $- \frac 1 8 \sum_f \xg_f \bar D^2 (\bar \Phi^f e^V \Phi_f)$ term, though
only in a way that is not manifestly supersymmetric.

However, our calculation is not without flaw. We use $- i \xs^\mu (\bar \chi_\mu + \bar \xs_\mu \xs^\nu \bar \chi_\nu)$ with $\bar \chi_\mu$ from $W_{\xa \dot \xa}$, the gauge field part of the supercurrent to get contribution
to the supersymmetry
current $\bar J_\mu$ from the gauge sector while the contribution from the
matter sector is obtained from a direct generalization of \er{MuSu2:supcur} to $4+\xe$
dimension. Without this double standard, the coefficient in front of $\p_\mu (\psi \xs^\mu \bar \psi)$ in \er{JconKonishi} will be different and does
not match that predicted by \er{superanomalySV}. The physical
implication remains the same; namely that $R_\mu'$ receives correction proportional
to the Konishi current and is no longer the $R$-current that satisfies the
Adler--Bardeen theorem.  

It is not clear whether the operator defined by \er{RcurrentFull} generates a new $R$-symmetry at the fixed point. The terminology we use may be a little confusing. By saying ``new $R$-symmetry", we refer to the $U(1)$ symmetry that forms the superconformal algebra together with supersymmetry and the scaling, instead of the one that transforms the gaugino and matter fields according to the charge ratios of $1:\frac 2 3: - \frac 1 3$. The latter is referred to as $R$-symmetry. Anyway, it is not out of question that the operator of \er{RcurrentFull} can be the right current to generate the new $R$-symmetry. Note that at the fixed point, the trace $\xt_\mu{}^\mu$ scales various fields according to their ``quantum dimensions" instead of their canonical dimensions though the operator form of this dilatation current is defined according to the canonical dimensions.

In any explicit calculation, it is hard to see how the charges associated
with the operator defined by \er{RcurrentFull} can receive quantum corrections. So this operator, after some renormalization, is likely to be the $R_\mu$ that generates the anomalous chiral $U(1)$ symmetry as it is the case in QED. Anyway, in our opinion, the point is that there should be an anomalous current that transforms the fields according to the charge $1:\frac 2 3: - \frac 1 3$ and it satisfies the Adler--Bardeen theorem. Moreover, the latter is definitely not in the same multiplet as the stress tensor. The former may or may not be generated by the bare operator defined by \er{RcurrentFull} but this could depend on the calculation scheme and is hardly physically relevant.

\subsection{A Manifestly Supersymmetric Derivation}
\label{subsec:MSD}
The calculation above can be done using the dimensional
reduction and the background field method. In \citec{Ensign:1987wy}, the anomalous
dimension term is argued to be zero because of the assumption of on-shell external fields. The assumption of on-shell external fields is in general not justified and in this particular
case, leads to the missing of a term that
has physical meaning. In this subsection we recover the anomalous term so
that the super-anomaly equation of the current $\hat \CJ_{\xa \dot \xa}$ is exactly of the form of \er{APN1SYM:2}.

In this subsection, we are going to use the convention in \citec{Ensign:1987wy}.
For simplicity, we only consider a single matter (chiral) superfield and drop the indices
of $f$. So we need to determine the corresponding form (in this new convention) of
\er{APN1SYM:2}. According to \er{EMA28}, and Eq.(C39) in \citec{Ensign:1987wy},
\[\bar \cd^2 [\hhat \xG \cdot \hhat \xG] = - \xe W^\xa W_\xa.\]
we have the Konishi anomaly,
\be\el{Konishianomaly} \bar \cd^2 \br \bar \Phi \Phi \ke = - T(R)[W^2],\ee

Let $[\hat \CJ_{\xa \dot \xa}] $ be the supercurrent, renormalized so that its expectation value is finite and is exactly equal to what one would get by putting into $\hat \CJ_{\xa \dot \xa}$ the external fields alone \ft{Here, certain non-local contributions are ignored in the previous literature.}.  Note that $[\hat \CJ_{\xa \dot \xa}] $ is the supercurrent whose $\theta {\bar\theta}$ component is the energy-momentum tensor.
Then the trace anomaly is the $\theta$ component of its super-trace, which
is given by Eq.(3.15) in \citec{Ensign:1987wy},
\be\el{traceano} \bar \cd^{\dot \xa} [\hat \CJ_{\xa \dot \xa}] = \frac 1 3 \frac {\xb_g} {g^3} \cd_\xa [W^2].\ee
The matter contribution to the $\xb$-function, up to two-loop, is (from \er{betafunctionmatter})
\[\frac {\xb_g} {g^3} = - 3 T(G) + T(R)(1-\xg) + \dots,\]
where $\xg$ is the anomalous dimension (defined from the anomalous scaling of the renormalized operators) The scaling dimension of the renormalized field $\Phi$ is $1 + \gamma/2$. 
In order for the missing (missing on the rhs of the anomaly equation \er{traceano})
term, $\bar \cd^2[\bar \Phi \Phi]$ to give the correct contribution to the $\xb$ function following \er{Konishianomaly} (see \citec{Shifman:1986zi}), we require \er{traceano} to be modified as,
\be\el{tanomwanod} \bar \cd^{\dot \xa} [\hat \CJ_{\xa \dot \xa}] = \frac 1 3 \lb\frac {\xb_g} {g^3} \cd_\xa
[W^2] + \xg \cd_\xa \bar \cd^2[\bar \Phi \Phi]\rb.\ee
Note that the numerical factors in this form are slightly different from
those in \er{APN1SYM:2}. Now both terms on the rhs follow from the vev of \[-\frac \xe {g^2} \cd_\xa W^2.\]
As explained in Sec~\ref{sec:ABoTS}, the anomaly is determined by $\bar \cd^{\dot \xa} K_{\xa \dot \xa}$ and from
\[\bar \cd^{\dot \xa} K_{\xa \dot \xa} =  \frac \xe {g^2} \cd_\xa W^2 + \frac {3 \xe} 2  \bar \cd^2\cd_\xa
\bar \Phi e^V \Phi + 4 i \xe \bar \Phi e^{V} W_\xa \Phi,\]
we get $-\xe \cd_\xa W^2/{g^2}$. Let us now show the expectation value of the latter can indeed give
the correct super-anomaly equation \er{tanomwanod}. The contribution proportional to $\bar \cd^2 \bar \Phi^e \Phi^e$ in the vev of $W^2$ can be obtained in a similar
way as that of $W_{\xa \dot \xa}$. Now
instead of
\[W_{\xa\dot \xa} \to (\cd^2 \bar \cd_{\dot \xa} V) (\bar \cd^2 \cd_{\xa} V) + \dots,\]
we have the expansion,
\[W^2 \to \frac 1 2 (\bar \cd^2 \cd^{\xa} V) (\bar \cd^2 \cd_{\xa} V) + \dots,\]
as the vertex, where $V$ is the quantum fluctuation of the gauge field. The relevant diagrams are 1(d) (from two vertices of $\bar \Phi V \Phi$) and 1(e) (one vertex of $\frac 1 2 \bar \Phi V^2 \Phi$) in \citec{Ensign:1987wy}. It is not hard to see only the latter gives nonvanishing contribution. We have,
following the Feynman rule,
\[\br V V \ke = 2 g^2 \iv {\hat \Box},\]
that the vev of $W^2$ is
\ba
\br W^2 \ke_{\bar \Phi \Phi} = 1(e) & = & 2 g^2 \bar \Phi^e \iv \Box \overleftarrow{\cd}^\xa \overleftarrow {\bar \cd^2} \bar \cd^2 \cd_\xa \iv \Box \Phi^{e} \nn
& = & 4 g^2 \bar \cd^2 \bar \Phi^e \iv \Box \bar \cd^2 \cd^2 \iv \Box \Phi^{e} \nn
& = & \frac 4 \xe C(R) g^{2}\bar \cd^2 \bar \Phi^e \Phi^e \ea
In other words, the last term on the rhs of  \er{tanomwanod} is supposed to be
\be\el{vevW2Phi} \br -\frac {\xe} {g^2}\cd_\xa  W^2 \ke_{\bar \Phi \Phi}  = -4 C(R)g^2 \cd_\xa \bar \cd^2 \bar \Phi^e \Phi^e.\ee
The anomalous dimension $\xg$ is given in \citec{Shifman:1986zi} as,
\[\xg = - C(R) \frac \xa \pi \to - 4 C(R) g^2,\]
where we recall that $\alpha= g^2/(4\pi)$.
In the last step, we use the convention $(4\pi)^2 = 1$ in \citec{Ensign:1987wy}.
One can see that \er{vevW2Phi} exactly agrees with \er{tanomwanod}.

\subsection{Charges at the Infrared Fixed Point}
\label{subsec:CIFP}
Previously, we have shown that the current $R_\mu'$ is a linear combination of the $R$-current $R_\mu$ and $R_\mu^f$. In this subsection, we apply our result to study the current $R_\mu'$ at the infrared fixed point of an
$\mathop{\rm SU}(N)$ SYM that has $N_f$ matter fields $Q_f$ in the fundamental representation and
$N_f$ matter fields $\tilde Q_f$ in the anti-fundamental representation \ft{For a review on supersymmetric QCD and especially the properties at the infrared non-trivial fixed points, see \eg \citec{Intriligator:1995au}.}. The
current $R_\mu'$ is shown to be the anomaly-free
current, whose charge ratios for $\xl, A, \psi$ is,
\be\el{anomfreeratio} 1: \frac{N_f - N}{N_f}: -\frac{N}{N_f}.\ee
We then argue for the advantage of our method compared to the argument  \citec{Shifman:1999mv}
based on the approach in \citec{Shifman:1986zi}.

At the infrared fixed point, we have the current $R_\mu'$ as,
\[R_\mu' = R_\mu + \frac 1 3\sum_f \xg_f R_\mu^f.\]
This follows from the coefficient $3T(G) - \sum_f (1-\xg_f) T(R_f)$ in \er{APN1SYM:2} (and \er{betafunctionmatter}). For later convenience, let us rewrite \er{APN1SYM:2}
in the form,
\ba\el{APN1SYM:2d} \p^{\xa\dot \xa} \CJ_{\xa \dot \xa} & = & -\frac i 3 D^2 \Big(-\frac
1 {16\pi^2} \left[ \frac{3T(G)- \sum_f (1-\xg_f) T(R_f)}{1- T(G)\xa/2\pi}\right]  \nn
& &  \times [\Tr\, W^2] - \frac 1 8 \sum_f \xg_f \bar D^2 [(\bar \Phi^f e^V \Phi_f)]\Big).
\ea
From the lhs of \er{APN1SYM:2d}, we have an operator $2\p^\mu R_\mu'$ \ft{In fact, we don't really need to know the factor in front of
$\p^\mu R_\mu'$.}. Taking its expectation value, the lowest order term is $2\p^\mu R_\mu^e$ (\ie \er{RcurrentFull} with all fields replaced by their
external counterparts). In the context of Slavnov-Taylor identity in the background field method \citec{Clark:1979te}, this term corresponds to the contact term and tells us the $R$-charges. Moreover, although the Adler--Bardeen theorem does not hold for this current $R_\mu'$, the lowest component of \er{APN1SYM:2d} still gives the chiral anomaly equation up to one-loop. So a connection between the factor in front of $\p^\mu R_\mu^e$ and the coefficient of $[\Tr W^2]$ can still be established. From \er{chiano} and the ratios of $3T(G)/\xg_f T(R_f)$ in \er{APN1SYM:2d}, we can infer that there is another renormalized operator $\frac 2 3 \sum_f\xg_f [\p^\mu R_\mu^f]$ coming out of the term
$- \frac 1 8 \sum_f \xg_f \bar D^2 [(\bar \Phi^f e^V \Phi_f)]$. The reason
is that $R_\mu$ assigns charge $+1$ to $\xl$ while $R_\mu^f$ assigns charge $+1$ to $\psi_f$ and a combination of $R_\mu + \frac 1 3 \sum_f\xg_f R_\mu^f$ gives the correct coefficient in the chiral anomaly equation (coefficient of $[\Tr W^2]$). Note that according
to \er{KonishiAn} the vev of the operator $\bar D^2 (\bar \Phi^f e^V \Phi_f)$ is going to give a term proportional to $\Tr[W^2]$ and a term proportional
to $\bar D^2 [(\bar \Phi^f e^V \Phi_f)]$. The $\xt \xt$ component of these two superfields have $f^2$ and $\p^\mu R_\mu^f$ respectively with the appropriate
coefficient determined by \er{chiano}.

Anyway, at the fixed point, there is no other contribution (from the anomaly)
of the form of $\p^{\xa \dot \xa} (\xl_\xa \bar \xl_{\dot \xa})$ \ft{As explained
above, the anomaly term $[\Tr W^2]$ has a contribution to the gaugino $U(1)$
current.} and we only have $\p^\mu R_\mu^e + \frac 1 3 \sum_f\xg_f \p^\mu(R_\mu^f)^e$ (up to a factor) in the vev of $\bar D^{\dot \xa}\CJ_{\xa \dot \xa}$. As a result, the $R'$-charge at the fixed point for $\Phi_f$ is $\frac 2 3 (1 + \frac 1 2 \xg_f )$. At a different scales, the $R'$-charge is different. Naively this operator $R_\mu'$ behaves as a current that has different charges when it acts on states of different energy scales.

With anomalous dimension at the fixed point being \ft{Physically, we have $N_f$ flavors
and they are all in the fundamental represetation.},
\[\xg_{f}=\xg  = 1 - \frac {3 N} {N_f},\]
which is necessary for the NSVZ $\xb$-function to vanish, the charge of $\Phi$
is,
\[\frac 2 3 (1 + \frac 1 2 \xg ) = \frac{N_f - N}{N_f}.\]
This result agrees with \er{anomfreeratio}.

The $R'$-charges at the fixed point can be obtained in
a very different way \citec{Shifman:1999mv}. In this case, a conserved current
is defined for every different $\xg_f$ (see Eq.(2.114) in \citec{Shifman:1999mv}),
\[\tilde \CJ_{\xa \dot \xa} \equiv \CJ_{\xa \dot \xa} - \frac {3T(G)- \sum_f (1-\xg_f) T(R_f)}{3 \sum_f T(R_f)} Q_{\xa \dot \xa},\]
At the infrared fixed point, the second term vanishes and this current is
just the supercurrent $\CJ_{\xa \dot \xa}$. However, the $R'$-charges are obtained from the form of $\tilde \CJ_{\xa \dot \xa}$ at the UV fixed point,
where $\xg_f = 0$. It is not clear why this works because $\tilde \CJ_{\xa
\dot \xa}$ with different $\xg_f$ are different operators. In other words,
$R'$-symmetries for different $\tilde \CJ_{\xa \dot \xa}$ are different. Why the charges of one symmetry is determined by that of the other needs to be explained. In our approach, the values of the charges for $R_\mu'$ come out naturally.

\ \\

\section{Problems and Solutions in the Two-Supercurrent Approach}
\label{sec:EVoEoM}
In this section, we justify the use of the equation of motion in \citec{Grisaru:1985ik} and \citec{Ensign:1987wy}. An equation of motion, when inserted
into the n-point functions serves like a functional derivative. For example, the expectation
value of the equation of motion of a field $\phi$, denoted as $\sS_{,\, \phi}$
satisfies
\[\br \sS_{,\, \phi}(x) X \ke = i\frac \xd {\xd \phi(x)} \br X \ke,\]
where $X$ denotes other operators. In the background field method, the expectation value of the equation of motion of $\phi$ gives something like
\be\el{trivialexp} i\frac {\xd \xG[\phi, \dots]} {\xd \phi(x)},\ee
where $\xG$ is the effective action.
\eml{In the standard non-supersymmetric calculation of the chiral anomaly using dimensional regularization, the chiral current is no longer conserved.
In other word, $\p^\mu j_\mu^5 \ne 0$ by the equation of motion. Instead, we have ($\psi$ being a Dirac spinor),}
\[\p^\mu j_\mu^5 = -\bar \psi \xg^5 \sla D \psi + \zt{h.c.} + \frac 1
4 \bar \psi \{\overleftrightarrow{\sla D}, \xg^5\} \psi\]
The first term on the rhs and its Hermitian conjugate are proportional to the equation of motion and both vanish on-shell. The insertion of these terms in a Green's function only gives contact
terms (or in the context of expectation value, only trivial terms like \er{trivialexp})
but not any anomalous contributions.

However, in \citec{Grisaru:1985ik} and \citec{Ensign:1987wy} expectation values of the equations of motion apparently do have contributions from the anomaly terms.
This can be seen as follows.
From \er{AnomalyN1SYM:1}, we have 
\be\el{EoMnvanexp} \br \bar \cd^{\dot \xa} W_{\xa \dot \xa}\ke \sim T(G) W^2_{\zt{ext}} \ne 0.\ee
The rhs is the anomalous contribution (given by the external
fields). Note that assuming no matter fields, the equation of motion of the gauge field implies, 
\be\el{EoMWnomatter} \bar \cd^{\dot \xa} W_{\xa \dot \xa} = 0.\ee
In \citec{Grisaru:1985ik} and \citec{Ensign:1987wy}, \eml{the lowest order contribution to the expectation value vanishes because of the
on-shell assumption} and a trivial expectation value (no anomaly) means a vanishing expectation value. Therefore, one expects
\be\el{EoMWnomatterexp} \br \bar \cd^{\dot \xa} W_{\xa \dot \xa} \ke = 0,\ee
which is in contradiction with \er{EoMnvanexp}. Moreover, when the equation of motion is used assuming that the expectation value of the equation of motion is trivial, which is necessary in the calculation of anomaly, the
nontrivial expectation value leads to an inconsistency. For example, \er{EoMWnomatter} is used on an operator 
\[\bar \cd^{\dot \xa} J_{\xa \dot \xa} = \bar \cd^{\dot \xa} \lb W_{\xa \dot \xa} - \frac 1 3 K_{\xa \dot \xa} \rb,\]
to get $-\frac \xe 3 \cd_\xa W^2$ (coming from the second term $-\frac 1 3
\bar
\cd^{\dot \xa} K_{\xa \dot \xa}$), which is then taken
to be the expectation value of $\bar \cd^{\dot \xa} J_{\xa \dot \xa}$. 
In other words, \er{EoMWnomatterexp} is assumed, which is apparently inconsistent
with \er{EoMnvanexp}.

In fact, in the scheme of dimensional reduction, the $U(1)_R$ current is actually conserved (by the equation of motion) and one may expect trivial expectation
values for $\p^\mu R_\mu$ and the supertrace of the supercurrent that has $R_\mu$ as its lowest component. This is possible only when we consider the non-local contributions to the expectation values.

In \citec{Grisaru:1985tc}, non-local contributions to the expectation values are considered but do not give any divergent contributions. The point is that it is necessary to include the non-local contributions in this calculation
scheme (background field method and dimensional reduction). For simplicity, the chiral anomaly in (non-supersymmetric)
QED is considered, as it was the same example that was discussed in \citec{Grisaru:1985ik}.  The non-local contribution to the expectation value of the chiral current $\bar \psi_{\dot \xa} \psi_\xa$ can be evaluated (in this calculation the superscript $e$ for the external field is dropped),
\ba
\br \bar \psi_{\dot \xa} \psi_\xa \ke & = &  i\cd^\xb{}_{\dot\xa} (\Box - i f)^{-1} {}_\xb{}^\xg \xe_{\xg\xa} \nn
& = &  i\cd^\xb{}_{\dot\xa} \iv \Box_0 (i\xe_{\xb \xg} A^{\mu}\p_{\mu} + f_{\xb\xg}) \iv \Box_0 (-iA^{\mu}\p_{\mu} \xd^{\xg}{}_{\xa} \nn
& & +i f^\xg{}_{\xa}) \iv \Box_0 \nn
& = & -ip{}^\xb{}_{\dot\xa} \iiint \frac 1 {p^2} f_{\xb\xg}(q) \frac 1 {(p-q)^2} f^\xg{}_{\xa}(k-q) \nn
& &
\times\, \frac 1 {(p-k)^2} e^{-i k x} d^4 q d^4 p d^4 k\nn
& = & \int_0^1 2 dy \int_0^{1-y} dz \iint d^4 q d^4 k\, (-iyq -i z k){}^\xb{}_{\dot\xa} \nn
& & \times\, f_{\xb\xg}(q)f^\xg{}_{\xa}(k-q)  \int d^4 l \frac 1 {(l^2 - \xD)^3} e^{-i k x}\nn
& = & - \frac i {2} \int_0^1 2 dy \int_0^{1-y} dz \iint d^4 q d^4 k \nn
& & \times \, \frac {(-i z k){}^\xb{}_{\dot\xa} f_{\xb\xg}(q)f^\xg{}_{\xa}(k-q)}{z^2 k^2 - z k^2} e^{-i k x}\nn
& = &  -\frac i 2 \iint d^4 q d^4 k\, \frac {i k_{\xa \dot\xa} f_{\xg\rho}(q)f^{\xg \rho}(k-q)}{k^2} e^{-i k x}
\ea
where $\xD = y^2 q^2 - yq^2 + z^2 k^2 - z k^2 + 2 y z q \cdot k$. In the
sixth line, we remove the $q$ dependence. Anyway, in the position space, the non-local contribution reads,
\[\br \p^{\xa \dot \xa} (\bar \psi_{\dot \xa} \psi_\xa) \ke = - \frac i 2 (f^2
- \bar f^2).\]
Note that the $A\p$ term, which we did not consider, has contribution too. But it appears that this contribution is proportional to $k_{\xa \dot \xa}F^2$, so after a procedure to make $\br \bar \psi_{\dot \xa} \psi_\xa\ke$ real, this contribution drops.

The non-local contribution is opposite to the contribution from the $\xe$-dimension operators, which is given in \citec{Grisaru:1985ik}. So an operator equation like,
\[\br \p^\mu j^5_\mu \ke = 0,\]
is valid. A renormalized operator $[j^5_\mu]$ defined in \citec{Grisaru:1985ik} has a nonvanishing expectation value that gives the correct chiral anomaly.

This result can be generalized to supersymmetric theories. In principle, one can compute the non-local contributions to the supercurrent $\CJ_\mu$ and show that its supertrace vanishes. Such contributions to some of the operators have actually been worked out in the literature. For example, the expectation value of the operator $\bar \Phi e^V \Phi$ has a (one-loop) non-local contribution (see \eg Eq.(6.7.10) in \citec{Gates:1983nr}). The contribution from the $\xe$-dimension operator can be found in Eq.(A28) in \citec{Ensign:1987wy},
\be\el{EMA28} \br \bar \Phi^r \Phi^r \ke^{(1)} = -2 C(R) \frac 1 \xe \bar \Phi^e \Phi^e - \frac 1 2 T(R) \frac 1 \xe \hhat \xG^e \cdot \hhat \xG^e,\ee
These two contributions are opposite to each other. So we have
\[ \bar D^2 \br \bar \Phi e^V \Phi \ke = 0.\]
On the other hand, a renormalized operator $[\bar \Phi e^{V} \Phi]$,
\[[\bar \Phi e^{V} \Phi] \equiv \bar \Phi e^{V} \Phi - \frac 1 \xe T(R) \hhat
\xG \cdot \hhat \xG.\]
can be defined to provide the correct Konishi anomaly \er{KonishiAn}. Here
we use a bare field $\Phi$ and therefore the first term on the right of \er{EMA28} in \citec{Ensign:1987wy} is removed.

The calculations of the non-local contribution that have be given in the literature
were not without flaw. The non-local contribution to $\br \bar \Phi e^V \Phi \ke$ is given by the same graphs that contributes
to the effective action (Eq.(8) in \citec{Grisaru:1985tc}, more explicitly $I_1,I_2,I_3$). In \citec{Gates:1983nr}, it appears that only part of the contribution ($I_3$) is considered. Moreover, $I_2$ is infrared divergent. It is not clear whether one can just drop this term.

However, non-local contributions to the expectation values of operators are infrared finite because the 4-momentum injected into the operators becomes an infrared cutoff. As shown by explicit calculation, the expectation value of operator $\bar \Phi e^V \Phi$ has terms containing $\cd^\xa W_\xa$
($W_\xa$ being the background field). These terms are finite and remain non-local after a differentiation by $\bar D^2$. We would like to see whether they have an impact on the anomaly. In \citec{Grisaru:1985ik}, they are discarded with the assumption of on-shell external fields.

Let us take a look at the $\xt \bar \xt$ component of $\cd^\xa W_\xa \cd^\xb W_\xb$. Note that all the fields are functions of $y^\mu = x^\mu + i \xt
\xs^\mu \bar \xt$. The expansion around
$x$ gives the $\xt^2 \bar \xt^2$ component. These two components ($\xt \bar \xt$, $\xt^2 \bar \xt^2$) are all that
are relevant in the chiral anomaly equation. First of all, let us take a
look at the bosonic contribution, which is of the form of $\p f^2$, from $\cd^\xa W_\xa \cd^\xb W_\xb$. To get
a nonvanishing $\xt \bar \xt$ component, we need one $\xt$ and one
$\bar \xt$. The lowest component of $\cd^\xa W_\xa$ vanishes identically.
Therefore both $\xt$ and $\bar \xt$ can not come from one of the $\cd^\xa W_\xa$. The $\xt$ (or $\bar \xt$) component of any scalar superfield like $\cd^\xa W_\xa$ is certainly fermionic. So there is no bosonic contribution
to the $\xt \bar \xt$ component (and the $\xt^2 \bar \xt^2$ component) of $\cd^\xa W_\xa \cd^\xb W_\xb$. A similar
argument can be made for $(\cd_\xa \cd^\xb W_\xb) W^\xa$.

In summary, we found that the terms containing $\cd^\xa W_\xa$ do not give contributions to the chiral anomaly. Nor do they give any
contributions proportional to $F^2$, which would appear in the trace anomaly. Note
that the derivative expansion does not apply to those $\cd^\xa W_\xa$ terms because of the infrared divergence in the expansion coefficients. As a result those $\cd^\xa W_\xa$ terms don't give local contributions even after being
acted on by $\bar D^2$. These terms have contributions quadratic in the spinor fields. It is not clear what their physical meanings are. Naively there can be such contributions
to the one-loop expectation value of the chiral current because of the Yukawa
coupling. Even though these contributions are just contact terms, the equation of motion is still spoiled. So in order to use the dimensional reduction, it appears that we have to assume that the external fields are on-shell, satisfying the classical equations of motion.

Note that other relevant contributions to the expectation value of $\bar \Phi e^V \Phi$ are proportional to $\cd^{(\xa} W^{\xb)} \cd_{(\xa} W_{\xb)}$. Let us also consider the $\xt \bar \xt$ component of $\cd^{(\xa} W^{\xb)} \cd_{(\xa} W_{\xb)}$, which is
\[\cd^{(\xa} W^{\xb)} \cd_{(\xa} W_{\xb)} \to f^{\xa \xb} i \bar \xt^{\dot \xa}\p_{\xa \dot \xa}(\xt^\xg
f_{\xg \xb}) \propto i\xt^\xa \bar \xt^{\dot \xa}\p_{\xa \dot \xa}f^2.\] So we have
\[\cd^{(\xa} W^{\xb)} \cd_{(\xa} W_{\xb)} + \zt{h.c} \to \xt^\xa \bar \xt^{\dot \xa}\p_{\xa \dot \xa} (if^2 - i\bar f^2).\]
With the assumption of $D^\xa W_\xa = 0$, $\cd^{(\xa} W^{\xb)} \cd_{(\xa} W_{\xb)}$ can be expressed as $D^2 W^2$ and we can get
to the usual form of the Konishi anomaly \er{KonishiAn}.

Despite these technical difficulties, it is quite tempting to expect similar non-local contributions to the expectation values of $W_{\xa \dot \xa}$, $P_{\xa \dot \xa}$ and $Q_{\xa \dot \xa}$. These non-local contributions cancel the contributions proportional to $K_{\xa \dot \xa}$ and make the use of the equations of motion justified.

\newpage
\thispagestyle{fancy}
\chapter{Anomaly Puzzle in the Context of Wilson Effective Action}
\label{ch:APCWEA}
\thispagestyle{fancy}
\pagestyle{fancy}
Shifman and Vainshtein \citec{Shifman:1986zi} considered the anomaly puzzle in the context of Wilson effective action and found a solution. They distinguish between the coefficient in the Wilson effective action and the physical coupling constant. The former is the coefficient appearing in front of the $W^2$ term in the Wilsonian effective action. Its running, obtained by integrating out higher momentum modes, is only of one-loop order as predicted by the nonrenormalization theorem. However, the physical coupling, defined from the physical amplitudes, includes all the higher-order contributions. So the problem of violation of the nonrenormalization theorem is successfully resolved. In a word, all the higher-order effects simply come from the infrared modes.

However, the calculation \citec{ArkaniHamed:1997mj} of the $\xb$-function using Jacobian appears to be independent of the infrared behavior. We study this problem and find by changing the UV cutoff that the momentum modes above an arbitrary scale do not appear to give a significant contribution to the Jacobian from which the multi-loop corrections to the $\xb$-function are obtained. 

\ \\

\section{Multi-loop beta-Function from Matrix Elements}
\label{sec:ASVuWEA}
In this section, the work \citec{Shifman:1986zi} by Shifman and Vainshtein
is reviewed. We also comment on some technical details that are skipped
in the original paper. Shifman and Vainshtein distinguish between the physical coupling constant and the corresponding coefficient in the Wilson effective action $S_W$. The latter is renormalized
only at one-loop level as predicted by the nonrenormalization theorem. An anomaly equation \er{superanomalySV} (see also \eg Eq.(19) in \citec{Shifman:1986zi}) is proposed to describe both one-loop anomalies (trace, chiral). The operators in the equation are bare operators. On the other hand, to obtain the physical coupling constant one needs to take the matrix elements of the operators on the rhs. Equivalently,
we need to evaluate the effective action $\xG$ to obtain the physical coupling
constant. This can be done from the expectation value of $e^{i S_W}$. Note that $\br e^{i S_W}\ke$ is understood as path integral with some external sources. 
 
However, there is a subtlety. For a theory whose Lagrangian is taken to be of the classical form but with a cutoff $\xL_0$, there are non-renormalizable terms appearing on the rhs of the anomaly equation \er{superanomalySV}. This is obvious for the trace anomaly $\vartheta_\mu^\mu$ since we know that the renormalization group flow produces non-renormalizable terms. The case for chiral anomaly needs some explanation. The method (change of measure)
to derive chiral anomaly \citec{Fujikawa:1979ay} \citec{Fujikawa:1980eg} certainly allows some extra non-renormalizable terms on the rhs of \er{APN1SYM:2} if we keep those terms proportional to the negative powers of the cutoff $\xL_0$.
Of course, when the cutoff is sent to infinity, we are back to \er{APN1SYM:2}.
After the cutoff is lowered, the Wilson effective action certainly has non-renormalizable
terms. The non-renormalizable terms, which violate both the $R$-symmetry and the scaling symmetry, lead to extra terms on the rhs, which are understood as classical contributions.

The role played by these non-renormalizable terms in the physical $\xb$-function will be discussed later. Let us first review the basic idea of the Shifman-Vainshtein approach. When ``shells" of the higher momentum modes (between $\xL < p < \xL_0$) are integrated, new Wilson effective actions $S_W$ are obtained. As we shall see, in each step, the trace anomaly equation holds and there is no contribution to the gauge coupling $g$ other than that of one-loop,
\be\el{xboneloop} \frac {2 \pi}{\xa(\xL)} = \frac {2 \pi}{\xa_0} + 2 \ln \frac {\xL_0} \xL.\ee

To get the physical $\xb$-function, which contains the multi-loop corrections, we need to evaluate the effective action. Note that the calculation of the effective action in the background field method is not much different from the calculation of the Wilsonian renormalization group. So we will discuss
both together. In the latter calculation, we take the modes with $k < \xL$ as external background field and only integrate the modes between $\xL < p < \xL_0$. They do not have contribution to the coupling constant (coefficient of $W^2$) other than at one-loop level. This result can be shown from the perturbative calculation (using Feynman diagrams). The Feynman diagram contains vertex with integration of $d^4\xt$. To go from $d^4\xt$ to $d^2 \xt$, which is necessary to have an $F$-term like the superpotential or the kinetic term $W^2$, a $\bar D^2$ will also appear through the identity,
\be \int d^4 \xt = -\int d^2 \xt \bar D^2.\ee
As a result, the multi-loop $F$-terms we get from renormalization group flow do not include terms like the superpotential in the Wess-Zumino model or $W^2$ in the present SQED theory. This is the basic idea in the perturbative proof of the non-renormalization theorem for the superpotential. However, there is an exception that the argument above does not apply: a $D$-term of the form of $\int d^4 \xt W \frac {D^{2}} {\p^2}
W$,
\[\int d^4 k d^4 \xt f(k^2) W(-k) D^2 W(k),\]
where $f(k^2)$ is a loop integration (over some momentum $p$) depending on the momentum exchange $k$ with the external $W$ field, can produce the F-term $W^2$. If the integration of $f(k^2)$
is proportional to some negative power of $k^2$, then it is possible to cancel the $k^2$ from $\bar D^2  D^2 = \Box$. However, with the range of integration restricted to $\xL < p < \xL_0$, there is no way to get this term. This is essentially why we only get the multi-loop contribution in the $\xb$-function of the physical coupling (but not in the Wilsonian renormalization group flow). In a word, we need to take into account all the momentum modes.

Before we move on to the explicit calculation of the physical $\xb$-function,
let us go back to the issue of the non-renormalizable terms. In the calculation of the exact $\xb$-function, all of them are actually ignored because none
of the non-renormalizable terms give any contributions. To get a nonvanishing contribution to the term $W^2$, we need a loop integration dominated by the infrared contribution. However, non-renormalizable terms introduce couplings proportional to negative powers of $\xL$ (cutoff), so from dimensional analysis, the Feynman diagrams will not have loop integrations proportional to negative powers of momentum.  Therefore we don't have any cancellation of the $\bar D^2$ and hence
no contribution to the $W^2$.  So, for any Lagrangian at the cutoff scale $\xL$, we can just ignore all the non-renormalizable terms. This is essentially what is done in \citec{Shifman:1986zi} and \citec{ArkaniHamed:1997mj}.

Let us now take a look at the explicit calculation (using the background field
method) of the effective action of SQED.  After we integrate the momentum modes between $\xL$ and $\xL_0$, the Wilson effective action at the new lower cutoff scale $\xL$ is
\be\el{APN1SYM:3a} S_{W} = \frac 1 {8 e^2(\xL)} \int d^4x d^2\xt W^2+\frac
1 4 Z(\xL) \int d^4 x d^4 \xt (\bar T e^V T + \bar U e^{-V} U),\ee
where $T$ and $U$ are differently charged chiral superfields describing matter. We can rewrite \er{APN1SYM:3a} so that it contains the perturbative term
\be\el{APN1SYM:3} \frac 1 4 [Z(\xL)-1] \int d^4 x d^4 \xt (\bar T e^V T + \bar U e^{-V} U),\ee
The effective action $\xG$ can be obtained from \er{APN1SYM:3a}. We classify
all the graphs (proportional to the external field $W^2$) into two categories: type (a) with or type (b) without the vertex of \er{APN1SYM:3}.
The sum, $\Gamma_b$, of graphs of type (b) just gives us the effective action of the original theory (having bare coupling $e(\xL_0)\equiv e_0$ and cutoff $\xL_0$), but with a different bare coupling $e(\xL)$ and a different cutoff $\xL$.  The $e^2(\xL)$'s in the multi-loop contribution to $\Gamma_b$ are accompanied by powers of $\log \frac{\xL}{\mu}$ ($\mu$ being the momentum
exchange with the external field $W_\xa$) \ft{We introduce a different scale $\mu < \xL$ to avoid the technical difficulty in evaluating momentum integral with the external momenta being at the cutoff scale.}.
For the effective action $\xG$ corresponding to the action in \er{APN1SYM:3a}, there are also other multi-loop corrections in the form of $\log Z(\Lambda)$ (from the graphs with the vertices of \er{APN1SYM:3}, explained below), which have a cutoff dependence of $\log \frac{\xL_0} \xL$. Since $\Lambda$ is an arbitrary intermediate scale between the momentum exchange $\mu$ and the cutoff $\Lambda_0$, the scale $\Lambda$ should not appear in the result. So all the graphs of type (b) should combine with those of type (a) to give a $\log \frac{\xL_0}{\mu}$ factor. In other words, we never need to consider the graphs of type (b). All they do is to help changing the $\log Z(\xL)$ obtained later and also the lowest order contribution (in $\frac {8\pi^2} {e^2(\xL)}$) $2 \log \frac {\xL_0}{\xL}$ to $\log Z(\mu)$ and $2 \log \frac {\xL_0}{\mu}$.

The multi-loop (geometrically) graphs of type (a) are a little harder to
analyze. However, as we shall see the one-loop graphs of type (a)
correctly reproduce the $\log Z(\xL)$. So multi-loop graphs of type (a) essentially serve as bridge to combine the $\log (\frac {\xL_0} {\xL})$ and the $\log (\frac {\xL} {\mu})$ together. In other words, we can also ignore the multi-loop graphs of type (a). 

The connected one-loop graph with a single vertex of \er{APN1SYM:3} has been shown to give a term
\be \el{APN1SYM:4} \sim (Z-1) \int d^4 x d^2 \xt W^2.\ee
The disconnected graphs with multiple such vertices just promote \er{APN1SYM:4} to the exponent. We consider the connected one-loop graphs with multiple such vertices. Each vertex provides one more internal lines (with $D^2 \bar D^2$) to the one-loop graph. So the graph is proportional to ($n$ being the number of vertices),
\[\int \frac{k^{2 n}}{p^{2n+2}} d p^2 \]
Note that this provides an $\frac 1 n$ factor from the integration. So we will get a $\log Z$ term after we sum all the connected graphs, which is exactly what we want. The result, as given in \citec{Shifman:1986zi}, is
\[\frac {8\pi^2}{e^2(\mu)} = \frac {8\pi^2}{e_0^2} + 2 \log \frac {\xL_0}{\mu Z(\mu)}.\]

\ \\

\section{Rescaling Jacobian}
\label{sec:AoAHaM}
As discussed in Sec~\ref{sec:ASVuWEA}, the coefficient of the Wilson effective Lagrangian, $g_W(\xL)$ should be distinguished from the physical coupling at $\xL$, $g_P(\xL)$. The former is protected by nonrenormalization theorem and has corrections only up to one-loop. The latter can have multi-loop corrections. These two coupling constants can be related via an exact formula. As a result, the $\xb$-function defined by the running of $g_P(\xL)$ with respect to $\xL$ can be related to that of $g_W(\xL)$ to all orders in perturbation theory. This exact $\xb$-function of $g_P(\xL)$, which is usually called the NSVZ $\xb$-function, can also be computed via instanton method \citec{Novikov:1983uc}.

Arkani-Hamed and Murayama present a different way to compute this NSVZ $\xb$-function. In this calculation, they don't use the effective action to define the coupling constant. Instead, they define the coupling constant $g_c (\xL)$ as the one in the canonically renormalized Lagrangian,
\ba\el{APN1SYM:cLmat} \CL_c & = & \frac 1 4 \lb \frac{1}{g_c^2} + i \frac \xt {8 \pi^2} \rb\int d^2\xt \Tr \Big[W^\xa(g_c V_c) W_\xa(g_c V_c)\Big] + \zt{H.c.} \nn
& & + \frac 1 4 \int d^4 \xt \sum_f {\bar \Phi_c}^f e^V \Phi_{c f}.\ea
To discuss the renormalization group flow, it is convenient to restore the $\xt$ angle in the Lagrangian \er{APN1SYM:Lmat},
\ba
\el{APN1SYM:hLmat} \CL & = & \frac 1 4 \lb\frac{1}{g_W^2} + i \frac \xt {8 \pi^2} \rb\int d^2\xt \Tr \Big[W^\xa (V)W_\xa (V)\Big] + \zt{H.c.} \nn
& & + \frac 1 4 Z \int d^4 \xt \sum_f \bar \Phi^f e^{V} \Phi_f.
\ea
We also restore the $Z$ which appears under the renormalization group flow. To get to the canonically renormalized Lagrangian from \er{APN1SYM:hLmat}, one need to rescale the fields. The measure is not invariant under this scaling ($\CD g_c V_c \ne \CD V_c$) and the Jacobian provide multi-loop corrections to the coefficient of the $W^2$ term. Before we discuss this part in more details, let us digress slightly to review the nonrenormalization theorem, which constrain the $\xb$-function of $g_W$ to be of one-loop.

\subsection{Nonrenormalization Theorem}
Supersymmetry like any other symmetries, imposes constraint on the possible quantum corrections. More supersymmetries (larger $\CN$), more constraints. For $\CN =1$, there is no correction to the superpotential. For $\CN =2$, the correction (and $\xb$ function) is only to one-loop order. For $\CN =4$, we have no quantum correction.

The basic principle is holomorphicity and symmetry. Symmetry is not hard to understood. Note that the R-symmetry plays an important role. Holomorphicity means that the $F$ terms (like superpotential) are local holomorphic functions of the coupling constants and the chiral superfields. 

The key is that the coupling constants are treated as fields and they have R-sym charge too. For example the $\xl$ in superpotential $\xl \Phi^3$ has charge $-1$ under $U(1)$ while $\Phi$ has charge $+1$,
(which is different from the commonly assigned R-charge of $+2/3$). Coupling constants are defined at a particular scale. So we need to use a Wilsonian renormalization group picture and the exact statement is like this: The $F$ terms in the effective Lagrangian at scale $\xL$ depend holomorphically on the chiral superfields and the coupling constants at the scale $\xL_0$.

The idea of coupling constants being fields is not easy to understand. The simplest argument for this is that the coupling constants are just vacuum expectation values of some scalars, which we assume to be the lowest components of some superfields. So the superpotential is just a holomorphic function of all these superfields (with some numerical constants).

For the $\Phi^3$ superpotential, it is obvious that with the assignment of R-charges, the superpotential cannot have any higher powers of $\xl$ multiplied by $\Phi^3$. In other words, the superpotential $\xl \Phi^3$ does not receive any corrections. To study the nonrenormalization, we can take the factor in front of $\Tr W^2$ as the coupling constant $\tau$,
\[\frac{1}{g_W^2} + i \frac \xt {8 \pi^2}.\]
The kinetic term should be a holomorphic function of $\tau$. Classically it is just $\frac 1 4 \tau \Tr W^2$. We use the shift symmetry $\tau \to \tau + i \xvp$ to restrict the possible corrections. This shift symmetry follows from the R-symmetry. Note that with an $R$ charge $+1$ for $W$ field, the classical R-charge of $\tau$ is $0$ and R-symmetry does not give any constraint classically. However, R-symmetry is not a quantum symmetry. Its anomaly implies that a combination of chiral transformation and a shift to the $\xt$ angle is a true symmetry. So we have correction only up to one loop (the constant in $\tau+\zt{constant}$),
\[\frac 1 4 (\tau + \zt{constant})\Tr W^2.\]
This essentially gives the one-loop $\xb$-function \er{xboneloop}.

\subsection{Beta-Function from Rescaling Anomaly}
We haven seen that the $\xb$ function for $g_W$ is exhausted at one-loop. What about $g_c$? The $\xb$-function for $g_c$ is defined in a conventional way. After the cutoff is lower from $\xL_0$ to $\xL$, we try to find a $g_c(\xL)$ that gives the same partition function. More explicitly, 
\[\int \CD V_c e^{i S_c[V_c, g_c(\xL), \xL]} = \int \CD V_c e^{i S_c[V_c, g_c(\xL_0), \xL_0]} = \int \CD V e^{i S[V, g_W(\xL), \xL]}.\]
By comparing \er{APN1SYM:cLmat} and \er{APN1SYM:hLmat}, we can see that classically $S_c[V_c, g_c(\xL),\xL] = S[g_c V_c, g_W (\xL), \xL]$. This implies $g_c = g_W$. However, what we actually have from the invariance of partition function is
\[\int \CD V e^{i S[V, g_W(\xL), \xL]} = \int \CD (g_c V) e^{i S[g_c V, g_W(\xL), \xL]}.\]
The measure is not invariant $\CD (g_c V_c) \ne \CD V_c$. For example, for pure SYM, they are related by
\[\CD (g_c V_c) = \CD V_c \exp \Big\{-\frac i 4 \int d^4 x \int d^2\xt \frac {2T(G)}{8 \pi^2}\log g_c \Tr \Big[W^\xa (g_c V_c)W_\xa (g_c V_c)\Big] + \zt{H.c.}\Big\}.\]
As a result, $g_c (\xL) \ne g_W(\xL)$ and they are related in a nontrivial way.
\[\frac 1 {g_c^2} = \frac 1 {g_W^2} - \frac {2T(G)}{8 \pi^2} \log g_c.\]
Since the $\xb$ function of $g_W$ is known, the $\xb$-function defined by the running of $g_c(\xL)$ can be evaluated easily,
\ba \Big[\frac 1 {g_c^2 (\xL)} + \frac {2T(G)}{8 \pi^2} \log g_c (\xL) \Big]- \Big[\frac 1 {g_c^2 (\xL_0)} + \frac {2T(G)}{8 \pi^2} \log g_c (\xL_0) \Big] & = & \frac 1 {g_W^2(\xL)} - \frac 1 {g_W^2(\xL_0)} \nn
& = & -\frac {3T(G)}{8 \pi^2} \log \frac {\xL_0}{\xL} \nonumber
\ea
It turns out that 
\[\xb_c(g_c) \equiv \frac {d g_c(\xL)} {d\log \xL}\]
agrees with the NSVZ $\xb$-function \er{betafunctionmatter} (with all $T(R_f) = 0$).

This implies, at least loosely, that $g_c (\xL)$ of the effective Lagrangian with canonical normalization is the physical coupling $g_P(\xL)$ at $\xL$.

\ \\

\section{Rescaling Anomaly as an Infrared Effect}
\label{sec:RAasaIE}
The $\xb$-function obtained by the method introduced in \citec{Shifman:1986zi}, which is reviewed in Sec~\ref{sec:ASVuWEA}, agrees with the one obtained by a different method in \citec{ArkaniHamed:1997mj}. Moreover, the role played by the field strength renormalization is almost the same. Take SQED as an example.
As pointed out in \citec{ArkaniHamed:1997mj} the rescaling of the field strength will change the coupling constant and the new theory (at cutoff $\xL$) is canonically normalized, i.e., has $Z = 1$. For
such a theory with $Z = 1$ both calculations predict that there is no quantum correction for process at the scale $\xL$. The effective Lagrangian after the rescaling is exactly the effective action for the external fields at the scale $\xL$.

The calculation (of the Jacobian) in \citec{ArkaniHamed:1997mj} involves UV
regularization and infrared effects seem to be irrelevant. However, it can be shown that if the momentum modes below an
arbitrary scale $\xL$ are ignored, there will be no contribution except for some non-renormalizable terms. The idea is to separate the contribution to the Jacobian from modes above $\xL$. To do this, one can consider two Jacobians under the rescaling of field strengths
at cutoffs $\xL$ and $\xL'$ (assuming $\xL < \xL'$) respectively. The Jacobian
for the scaling of a chiral superfield $\Phi_f$ (in a representation of $R_f$) can be computed (Eq.(A.20) in \citec{ArkaniHamed:1997mj}),
\be\log J(\xL, e^{\xa(\xL)}) = -\frac 1 {16} \int d^2 \xt \frac {2 T(R_f)}{8\pi^2}\log (e^{\xa(\xL)}) W^2 + O(\frac 1 {\xL^4}).\ee
Note that $J(\xL, e^{\xa(\xL)})$ can be understood as the Jacobian from the rescaling (by a factor of $e^{\xa(\xL)}$) of the momentum modes $k \le \xL$. We can set the scaling factor to be the same, then $J(\xL, e^{\xa(\xL')})$
is the contribution to $J(\xL', e^{\xa(\xL')})$ by modes $k \le \xL$. The difference $\log J(\xL', e^{\xa(\xL')})  - \log J(\xL, e^{\xa(\xL')})$ starts from the $\frac 1 {\xL^4} - \frac 1 {{\xL'}^4}$, which is the difference of the coefficients of a certain non-renormalizable term. This is consistent with conclusion (in Sec~\ref{sec:ASVuWEA}) about the contributions from the modes between $\xL$ and $\xL'$ in the analysis of the Wilsonian renormalization group flow in the sense that the effective theory with a cutoff $\xL$ generally has non-renormalizable terms proportional to the negative power of $\xL$.

The point is that the multi-loop contributions proportional to $\log \frac{\xL_0}{\xL'}$ do not come from the rescaling of modes between $\xL$ and $\xL'$. The choice of $\xL$ is arbitrary. In other words, the multi-loop contributions to the $\xb$-function do not come from any modes $k>\xL$. Therefore, the result in \citec{ArkaniHamed:1997mj} must also come from the infrared modes as in \citec{Shifman:1986zi}.

\newpage

\

\

\

\

\

\

\

\

\

\

\

\noindent\textbf{\Huge Part III:}

\

\noindent\textbf{\huge Hermiticity of the Dirac}
\addcontentsline{toc}{chapter}{Part III - Hermiticity of the Dirac Hamiltonian in Curved Spacetime}

\

\noindent\textbf{\huge Hamiltonian in Curved Spacetime}

\newpage
\thispagestyle{fancy}
\chapter{Restoration of Hermiticity}
\thispagestyle{fancy}
\pagestyle{fancy}

In previous work on the quantum mechanics of an atom freely falling in a general curved background spacetime, the metric was taken to be sufficiently slowly varying on time scales relevant to atomic transitions that time derivatives of the metric in the vicinity of the atom could be neglected.  However, when the time-dependence of the metric cannot be neglected, it was shown that the Hamiltonian used there was not Hermitian with respect to the conserved scalar product.  This Hamiltonian was obtained directly from the Dirac equation in curved spacetime. This raises the paradox of how it is possible for this Hamiltonian to be non-hermitian. Here, we 
show that this non-hermiticity results from a time dependence of the position eigenstates that enter into the Schr{\"o}dinger wave function, and we write the expression for the Hamiltonian that is Hermitian
for a general metric when the time-dependence of the metric is not neglected. 

\ \\

\section{Hamiltonian of a Spin-1/2 Particle in a Curved Background}

The Dirac equation in curved spacetime is
\be\el{DiracE} ({\underline \xg}^\mu (x) \nabla_\mu + m )\psi(x) = 0,\ee
where the  ${\underline \xg}^\mu (x)$ matrices satisfy
\be\el{gammabarE} {\underline \xg}^\mu{\underline \xg}^\nu
+{\underline \xg}^\nu{\underline \xg}^\mu = 2g^{\mu\nu}.\ee
The covariant derivative of the spinor $\psi(x)$ is
\be
\el{covaspin} \nabla_\mu \psi(x) \equiv (\p_{\mu} - \Gamma_{\mu})\psi(x),
\ee
where $\Gamma_{\mu}$ is the spinor affine connection.
The spinor covariant derivative of ${\underline \xg}{}_{\nu} (x)$ is
\be \el{derigamma}
 \nabla_\mu \underline {\xg}{}_{\nu} = 
 \p_{\mu} {\underline \xg}{}_{\nu}
 - \Gamma^{\xl}{}_{\mu\nu}{\underline\xg}{}_{\xl}
 -\Gamma_{\mu} {\underline \xg}{}_{\nu}
 +{\underline \xg}{}_{\nu}\Gamma_{\mu} = 0,
\ee
which must vanish so that the covariant derivative of the metric
will be $0$.

A convenient representation of the matrices 
${\underline \xg}^{\mu} (x)$ is
\be {\underline \xg}^{\mu} (x) \equiv {b_a}^{\mu}(x) \xg^a,\ee
where ${b_a}^{\mu}$ is the vierbein (often denoted by
${e_a}^{\mu}$) defined by 
$g^{\mu\nu} = {b_a}^{\mu}{b_b}^{\nu} \eta^{ab}$,
and the $\xg^a$ are the flat spacetime 
gamma-matrices, satisfying
${\xg}^a { \xg}^b
+{ \xg}^b{ \xg}^a = 2\eta^{a b}.$ 
We use the conventions that the metric in Minkowski space is 
$\eta_{ab} = {\rm diag}(-1, 1, 1, 1)$ and therefore
$\xg_0^{\dagger} = -\xg_0$ and
$\xg_i^{\dagger} = \xg_i$. 
The corresponding representation of the spinor affine connection 
$\Gamma_{\mu}$ is
\be
\xG_\mu = -\frac 1 4 \xg_a \xg_b {b^a}_\nu g^{\nu \xl} {b^b}_{\xl;\mu} + i q A_\mu.
\ee
The ``;" here acts on the vierbein as a curved-spacetime vector
\be
{b^b}_{\xl;\mu} \equiv \p_\mu {b^b}_{\xl} - \xG^\rho_{\mu \xl} {b^b}_{\rho}.
\ee
Here, $A_\mu$ is the electromagnetic vector potential. For the atom $A_\mu$ is important, but in considering the Hermiticity of the Hamiltonian we can set $A_\mu = 0$ because it does not contribute to the non-Hermiticity. Therefore, in the following discussion, we will set $A_\mu = 0$.\\ 

It is possible to interpret $\psi(x)$ as the wave function of a spin-1/2 particle moving in curved spacetime.  In Dirac notation \footnote{We will suppress the spinor index.},  it is
$\br x | \psi \ke$. We will take a closer look at that later. The scalar product for the wave function is defined to be \citec{Parker:1980kw},
\be\label{eq:scalarProduct}(\phi, \psi) = -\int d^3 x \sqrt{-g} 
\phi^\dagger(x) \xg^0 {\underline \xg}^0 (x) \psi(x).\ee

It is straightforward to rewrite the Dirac equation \er{DiracE} in the form of a Schr\"odinger equation,
\be\el{Shro} i\frac \p {\p t} \psi(x) = \hat H \psi(x),\ee
where $\hat H$ is given by 
\be\el{Hami} \hat H \equiv - i \iv {g^{00}} {\underline \xg}^0 {\underline \xg}^i \nabla_i + i \xG_0 - i \iv {g^{00}} {\underline \xg}^0 m .\ee
However, as mentioned in \citec{Parker:1980kw},
the Hamiltonian defined in this way is not hermitian when
the metric explicitly depends on the time $t$. One finds that
\goodbreak
\ba 
(\phi, \hat H\psi) - (\hat H \phi, \psi) & = & - \int d^3 x \sqrt{-g} \phi^\dagger \xg^0 {\underline \xg}^0  (-i \iv {(g^{00})} \underline \xg^0 \underline \xg^i \nabla_i + i \xG_0 \nn 
& & - i \iv {(g^{00})} \underline \xg^0 m )\psi + \int d^3 x \sqrt{-g} \Big[{(\nabla_i \phi)}^\dagger i \iv {(g^{00})} {\underline \xg^i}^\dagger {\underline \xg^0}^\dagger) \nn
& & - \phi^\dagger  i \xG_0^\dagger + \phi^\dagger i \iv {(g^{00})} m {\underline \xg}^{0\dagger} \Big]\xg^0 {\underline \xg}^0 \psi  \nn
& = & \int d^3 x \sqrt{-g} \Big[{(\nabla_i \phi)}^\dagger (i \xg^0 \underline \xg^i )\psi - i\phi^\dagger \xG_0^\dagger\xg^0 {\underline \xg}^0 \psi + i\phi^\dagger\xg^0 \underline \xg^i \nabla_i \psi \nn 
& & - i \phi^\dagger \xg^0 {\underline \xg}^0  \xG_0\psi \Big]\nn
& = & \int d^3 x \sqrt{-g} \Big[-i \phi^\dagger \frac 1 {\sqrt{-g}} \p_i(\sqrt{-g}\xg^0 \underline \xg^i )\psi - i \phi^\dagger \xG_i^\dagger\xg^0 \underline \xg^i \psi \nn 
& & - i\phi^\dagger \xG_0^\dagger\xg^0 {\underline \xg}^0 \psi - i\phi^\dagger\xg^0 \underline \xg^i \xG_i \psi - i \phi^\dagger \xg^0 {\underline \xg}^0  \xG_0\psi \Big]\nn
& = & \int d^3 x \sqrt{-g} \Big[-i \phi^\dagger (\p_\mu+\xG^\nu_{\nu \mu}) (\xg^0 \underline \xg^\mu ) \psi + i \phi^\dagger \frac 1 {\sqrt{-g}} \p_0 (\sqrt{-g}\xg^0 \underline \xg^0 )\psi \nn 
& & + i \phi^\dagger \xg^0 \xG_i \underline \xg^i \psi + i\phi^\dagger \xg^0 \xG_0 {\underline \xg}^0 \psi - i\phi^\dagger\xg^0 \underline \xg^i \xG_i \psi  - i \phi^\dagger \xg^0 {\underline \xg}^0  \xG_0\psi \Big]\nn
& = & \int d^3 x \Big[+ i \phi^\dagger \frac 1 {\sqrt{-g}} \p_0 (\sqrt{-g}\xg^0 \underline \xg^0 )\psi - i \phi^\dagger \xg^0 \nabla_\mu (\underline \xg^\mu ) \psi \Big]\nn
\el{nonHermi1}& = & i\int d^3 x \phi^\dagger \xg^0 \frac \p {\p t} \lb \sqrt{-g}  {\underline \xg}^0 \rb \psi.
\ea
In obtaining the 3rd equality we used \er{covaspin}, and to obtain the 4th and 5th equalities we used \er{derigamma}. 
In summary,
\be
\el{nonHermi}
(\phi, \hat H\psi) - (\hat H \phi, \psi) = i\int d^3 x \phi^\dagger \xg^0 \frac \p {\p t} \lb \sqrt{-g} {\underline \xg}^0 \rb \psi.
\ee
The rhs of \er{nonHermi} is generally nonzero.

\ \\

\section{Definition of the Wave Function}
This apparent paradox concerning the non-hermiticity of $\hat H$ in fact comes from the definition of $\psi(x)$. It is defined as $\br \vec x | \psi\ke$ (where $\vec x$ denotes the spatial coordinates). We must require that
\be
\br \phi | \psi \ke = (\phi, \psi),
\ee
where $(\phi, \psi)$ is the conserved scalar product defined in
Eq.~(\ref{eq:scalarProduct}). It follows that
the complete basis $\{\st {\vec x}\}$ actually satisfies
\be\el{comple} \int d^3x \st {\vec x} \sqrt{-g}\xg^0 {\underline \xg}^0 (x) \br \vec{x}| = 1.\ee
Therefore, when 
$\sqrt{-g}$ depends on time, 
so does $\st {\vec x} \equiv \st {\vec x, t}$. 
As a result,
\be 
\el{Shro1}
i \frac \p {\p t} \br \vec{x},t | \psi \ke \ne i \br \vec{x},t | \lb \frac \p {\p t} \st{\psi}\rb = \br \vec{x},t | {\cal H} |\psi\ke,
\ee
where ${\cal H}$ is the hermitian Hamiltonian in the Schr\"odinger dynamical picture in the abstract Hilbert space. It is the operator that satisfies
\be\el{ShroH} i\frac \p {\p t} \st{\psi} = {\cal H} \st{\psi}.\ee
Note that the lhs of \er{Shro1} is what appears on the 
left of \er{Shro}. 
In other word, the $\hat H$ (defined in \er{Hami}), which is on the 
rhs of \er{Shro}, is not quite the Hamiltonian in the
Schr{\"o}dinger or configuration-space representation when the
metric depends on $t$. 

Let us find the matrix elements
$\br{\vec x, t}| \mathcal{H} \st{\vec x', t}$.
One can show from \er{comple} that
\be\frac \p {\p t} \st{\vec x, t} = 
-\frac 1 2 \st{\vec x, t}  \frac \p {\p t} \lb \sqrt{-g} \xg^0 {\underline \xg}^0 (x)\rb {(\sqrt{-g}\xg^0 {\underline \xg}^0 (x))}^{-1}.\ee
Taking the conjugate, we have 
\ba\frac \p {\p t} \br\vec x, t| & = &
-\frac 1 2  {{(\sqrt{-g}\xg^0 {\underline \xg}^0 (x))}^\dagger}^{-1} \frac \p {\p t} {\lb \sqrt{-g} \xg^0 {\underline \xg}^0 (x)\rb}^\dagger \br\vec x, t|\nn
& = & -\frac 1 2  {(\sqrt{-g}\xg^0 {\underline \xg}^0 (x))}^{-1} \frac \p {\p t} \lb \sqrt{-g}\xg^0 {\underline \xg}^0 (x) \rb \br\vec x, t|.
\ea
Here we used the fact that $\xg^0 {\underline \xg}^0 (x)$ is Hermitian. It is then easy to see that the completeness relation of \er{comple} is independent of time.
 We also have
\be{(\xg^0 {\underline \xg}^0 (x))}^{-1} = \frac {{\underline \xg}^0 (x)\xg^0} {-g^{00}}.\ee
So the Hamiltonian $H$ satisfying the condition
$(\psi, H \phi) = (H \psi, \phi)$
is
\ba
\el{Hami1} H & \equiv & -i \frac 1 2  \frac {{\underline \xg}^0 (x)\xg^0} {g^{00}\sqrt{-g}} \frac \p {\p t} \lb \sqrt{-g} \xg^0 {\underline \xg}^0 (x)\rb - i \iv {g^{00}} {\underline \xg}^0 {\underline \xg}^i \nabla_i + i \xG_0 - i \iv {g^{00}} {\underline \xg}^0 m \nn 
&  = & -i \frac 1 2  \frac {{\underline \xg}^0 (x)\xg^0} {g^{00}\sqrt{-g}} \frac \p {\p t} \lb \sqrt{-g} \xg^0 {\underline \xg}^0 (x)\rb +\hat H.\ea
It thus follows that $H$ is Hermitian (with the use of \er{nonHermi})
for a general metric.
The matrix elements of ${\cal H}$ satisfy the relation
\be\el{xrep}\br{\vec x, t}| \mathcal{H} \st{\vec x', t} 
=  H  \delta(\vec{x} - \vec{x}')(\sqrt{-g})^{-1}, \ee
with $H$ being the operator given in Eq.~(\ref{Hami1}).

If we are dealing with a one-electron atom, the spinor
affine connections will contain the vector potential of the
electromagnetic field, but the derivation is unchanged. Let us
consider the effect of the time dependence of the 
Riemann tensor on the spectrum of the atom.
The Hamiltonian $H$ of Eq.~(\ref{Hami1}) reduces to $\hat H$ 
if the time dependence
of the metric can be neglected, as was the case 
in \citec{Parker:1980kw}.  For rapidly changing gravitational fields
the additional term is needed to enforce Hermiticity.

In the Fermi normal coordinates along the geodesic of
a bound system such as an atom, the difference, $H-\hat H$,
given by \er{Hami1} must vanish on the geodesic because
it involves only the first time-derivative of the metric. 
Furthermore, for
small distances from the geodesic, this difference, $H-\hat H$, is not vanishing, but it is of higher
order in $\frac {a_0}{r}$ (where $a_0$ is the atomic size and $r$ is a
characteristic length or time scale of the background spacetime) compared to the other terms in $H$. This can be seen by dimensional analysis from the Hamiltonian $\hat H$ that is
given in Fermi normal coordinates in \citec{Parker:1980kw}.
Therefore, when $\frac {a_0}{r} \ll 1$ the difference between
$H$ and $\hat H$ can be neglected.
Similarly, when $\frac {a_0}{r} \ll 1$, 
it is also possible to use $\hat H$  with
time-dependent perturbation theory to calculate transition
rates induced by the Riemann tensor along the path of the atom.

\newpage

\

\

\

\

\

\

\

\

\

\

\

\noindent\textbf{\Huge Part IV:}

\

\noindent\textbf{\Huge Massive Gravitons}
\addcontentsline{toc}{chapter}{Part IV - Massive Gravitons}

\newpage
\thispagestyle{fancy}
\chapter{Extended Theory of Massive Gravitons}
\thispagestyle{fancy}
\pagestyle{fancy}

In this chapter, we study the massive gravity theory proposed by Arkani-Hamed, Georgi and Schwartz. In this theory, the graviton becomes massive when general covariance is spontaneously broken through the introduction of a field that links two metrics, one of the which will eventually decouple. The excitation of this ``link" field acts like a Goldstone boson
in giving mass to the graviton. By means of gauge fixing terms similar to the renormalizability gauges used in gauge theories, we gives a two-parameter class of graviton and Goldstone boson propagators. We show that for all of those gauges, except for the unitary gauge, the massive graviton propagator approaches that of general relativity in the massless limit. With these massive propagators, we calculate the lowest order tree-level interaction between two external energy momentum tensors. The result is independent of gauge parameters, but is different from the prediction of massless gravity theory, \textit{i.e.,} general relativity in the limit of vanishing graviton mass. This difference is just the van Dam-Veltman-Zakharov (vDVZ) discontinuity. In the end, we also proposed a new massive gravity theory that is free of the vDVZ discontinuity. 
The key to the absence of the discontinuity is to introduce an extra scalar field with 
negative kinetic sign. This type of ghost field has been proposed before to explain the acceleration of our universe.

\ \\

\section{vDVZ Discontinuity}
\label{sec:VDVZD}
It is well known that there is a discontinuity between massive and massless graviton theory \citec{vanDam:1970vg}. Here we will briefly review the properties
of the massive graviton propagator and explain why it does not give the same
physical predictions as in general relativity when the graviton mass goes
to zero. The linearized action of a general massive gravity theory is given by \citec{vanDam:1970vg},
\ba \el{Lagmassivenosource}
S & = & -M_{Pl}^2\int d^4 x \Big[\frac{1}{4}({h_{\mu\nu}}^{,\lambda}{h^{\mu\nu}}_{,\lambda}-h^{,\lambda}h_{,\lambda}-2{h^{\mu\lambda}}_{,\lambda}{h_{\mu\rho}}^{,\rho}+2h^{,\mu}{h_{\mu\lambda}}^{,\lambda})\nn
& & + \frac 1 4 M^2 (h_{\mu\nu}h^{\mu\nu}-\zeta h^2)\Big],
\ea
The choice of $\zeta = 1$ gives the so-called Fierz-Pauli mass term. As we shall see, this is the only choice that can avoid any possible ghosts or tachyons. With this choice, the massive graviton propagator can be obtained (explained below in subsection~\ref{subsec:PoMG}),
\begin{equation}\label{massivegravitonp} {G^{\mu\nu}(p)}_{\alpha\beta}=\frac {\sum_{i=1}^5 {e^i}^{\mu\nu}{e^i}_{\alpha\beta}} {p^2-M^2}\end{equation}
where,
\begin{eqnarray} \sum_{i=1}^5 {e^i}^{\mu\nu}{e^i}_{\alpha\beta}=\frac 1 2 ({\delta^\mu}_\alpha {\delta^\nu}_\beta + {\delta^\nu}_\alpha {\delta^\mu}_\beta-\eta^{\mu\nu}\eta_{\alpha\beta})\nonumber\\
-{\frac 1 2} (\frac {{\delta^\mu}_\alpha p^\nu p_\beta} {M^2}+\frac {{\delta^\nu}_\alpha p^\mu p_\beta} {M^2}+\frac {{\delta^\mu}_\beta p^\nu p_\alpha} {M^2}+\frac {{\delta^\nu}_\beta p^\mu p_\alpha} {M^2})\nonumber\\
+{\frac 2 3}({\frac 1 2}\eta^{\mu\nu}+\frac {p^\mu p^\nu} {M^2})({\frac 1 2}\eta_{\alpha\beta}+\frac {p_\alpha p_\beta} {M^2})
\end{eqnarray}
Here we use the metric with $(+\, -\, -\, -)$ signature which is \emph{different} from the convention in ref.\cite{Arkani-Hamed:2002sp}. Its massless limit will be
\begin{equation}\label{masslesslimit}{G^{\mu\nu}(p)}_{\alpha\beta}=\frac {\frac{1}{2}({\delta^\mu}_\alpha {\delta^\nu}_\beta + {\delta^\nu}_\alpha {\delta^\mu}_\beta-\frac 2 3 \eta^{\mu\nu}\eta_{\alpha\beta})} {p^2}.
\end{equation}
On the other hand, the massless propagator is given by,
\begin{equation}\label{massless}{G^{\mu\nu}(p)}_{\alpha\beta}=\frac {\frac{1}{2}({\delta^\mu}_\alpha {\delta^\nu}_\beta + {\delta^\nu}_\alpha {\delta^\mu}_\beta-\eta^{\mu\nu}\eta_{\alpha\beta})} {p^2}.
\end{equation}
To investigate the vDVZ discontinuity, we introduce external matter sources characterized by two energy momentum tensors $T_{\mu\nu} = T^{a}_{\mu\nu} + T^{b}_{\mu\nu}$ to the Lagrangian. 
The massive Lagrangian (with source) at the linearized level is given by
\ba \el{Lagmassive}
S & = & -M_{Pl}^2\int d^4 x \Big[\frac{1}{4}({h_{\mu\nu}}^{,\lambda}{h^{\mu\nu}}_{,\lambda}-h^{,\lambda}h_{,\lambda}-2{h^{\mu\lambda}}_{,\lambda}{h_{\mu\rho}}^{,\rho}+2h^{,\mu}{h_{\mu\lambda}}^{,\lambda})\nn
& & + \frac 1 4 M^2 (h_{\mu\nu}h^{\mu\nu}-h^2)\Big]+ \int d^4 x( h^{\mu\nu} T^{a}_{\mu\nu} + h^{\mu\nu} T^{b}_{\mu\nu}),
\ea
where the Planck mass is defined as $M_{Pl}^2 = 1/8\pi G$. 
Here, $T^{a}_{\mu\nu}$ and $T^{b}_{\mu\nu}$ are localized at two different points in the position space. From the Hamiltonian (obtained from \eqref{Lagmassive}), the interaction term with these source terms is given by $h^{\mu\nu} (T^{a}_{\mu\nu}+T^{b}_{\mu\nu})$. The value of $h_{\mu\nu}$ is obtained from the equation of motion with source and is of the form of 
\[h_{\mu\nu}(x) = \int d^4 x' G_{\mu\nu}{}^{\xa\xb}(x,x') (T^{a}_{\xa\xb}(x')+ T^{b}_{\xa\xb}(x')).\] So the two-body interacting energy between the sources is given by the product of 
\[\int d^4 x d^4 x' T^{a\, \mu\nu}(x) G_{\mu\nu}{}^{\xa\xb}(x,x') T^{b}_{\xa\xb}(x')\]
or in momentum space,
\[\int d^4 p T^{a\, \mu\nu} (p) G_{\mu\nu}{}^{\xa\xb}(p) T^{b}_{\xa\xb}(-p).\]

For a non-relativistic system with only $T^a_{00}, T^b_{00}\ne 0$, the interaction terms are (following from \er{masslesslimit} and \er{massless})
\ba\frac 2 3 G  \frac {T^a_{00}T^b_{00}} {p^2 + i\xe},&\quad \zt{massless limit}, \\
\frac 1 2 G  \frac {T^a_{00}T^b_{00}} {p^2 + i\xe},&\quad \zt{massless}.
\ea
respectively. To give the same result, we need to choose $G_{\zt{massive}} = \frac 3 4 G_{\zt{massless}}$. Note that they are not equal {\it a priori}. 
Now we can consider the interaction between a non-relativistic source and
an electromagnetic source. The latter has a vanishing trace. As a result,
the difference between the two propagators, \ie the difference between last terms in \er{masslesslimit} and \er{massless} does not contribute to the
interaction term. However, with the choice of $G_{\zt{massive}} = \frac 3 4 G_{\zt{massless}}$, the interaction strengths are different. In a word,
we can fit either the perihelion procession, which is the interaction between two non-relativistic sources or the bending of light, which is the interaction
between a non-relativistic source and a relativistic source, but not both.
Obviously, this implies that massive gravity (with the Fierz-Pauli mass term) cannot be the physical theory that describes our world no matter how small the mass is.

\subsection{Propagator of a Massive Graviton}
\label{subsec:PoMG}
Before we move on, let us make a comment on the massive graviton propagator \er{massivegravitonp}. The massive graviton propagator people often use is in fact not really the Green's function of the equation of motion. However when acting on the conserved source, namely the conserved energy momentum tensor, they will be equivalent.

Usually, the kinetic term and interaction term in the Lagrangian with a general field $\phi$ (with some general indices) will be something like 
\begin{equation}\frac 1 2\phi K \phi + \phi J\end{equation}
and a propagator will satisfy $K G =1$ (in momentum space). We suppress indices for simplicity. In fact the graviton propagator, as we shall see, is not the inverse of the kinetic metric since it can not produce the identity operator. However it is necessary to have an ``effective" propagator
$G'$ which only satisfies $K G' J=J$ for some conserved source $J$.

For a massive graviton, the propagator \eqref{massivegravitonp} will be equivalent to 
\begin{equation}G'=\frac {\frac 1 2({\delta^\mu}_\alpha {\delta^\nu}_\beta + {\delta^\nu}_\alpha {\delta^\mu}_\beta)-\frac{1}{3}\eta^{\mu\nu}\eta_{\alpha\beta}+\frac{1}{3}\frac {p^\mu p^\nu} {M^2}\eta_{\alpha\beta}} {p^2-M^2}\end{equation}
when acting on a source that satisfies $p^\mu T_{\mu \nu} =0$. It is just \eqref{massivegravitonp} after dropping those terms with $p_\alpha$ or $p_\beta$.

Now we show that it is equivalent to the propagator. Following
from \er{Lagmassive}, one has $K$ as,
\begin{eqnarray}
K=p^2 I - \frac 1 2 ({{\delta^\mu}_\alpha p^\nu p_\beta }+{{\delta^\nu}_\alpha p^\mu p_\beta }+{{\delta^\mu}_\beta p^\nu p_\alpha }+{{\delta^\nu}_\beta p^\mu p_\alpha})+\eta^{\mu\nu}{p_\alpha p_\beta} \nonumber\\
+{p^\mu p^\nu}\eta_{\alpha\beta}-p^2 \eta^{\mu\nu}\eta_{\alpha\beta}- M^2(I-\eta^{\mu\nu}\eta_{\alpha\beta})
\end{eqnarray}
where $I =\frac 1 2({\delta^\mu}_\alpha {\delta^\nu}_\beta + {\delta^\nu}_\alpha {\delta^\mu}_\beta)$. Now the product of $(p^2 - \mu^2)K$ and the propagator is,
\begin{eqnarray}
(p^2 - \mu^2) KG' = p^2 (I-\frac{1}{3}\eta^{\mu\nu}\eta_{\alpha\beta}+\frac{1}{3}{p^\mu p^\nu \over M^2}\eta_{\alpha\beta})+2(\frac{1}{3}{p^\mu p^\nu} \eta_{\alpha\beta}-\frac{1}{3}{p^\mu p^\nu \over M^2}p^2 \eta_{\alpha\beta}) \nonumber \\
+\eta^{\mu\nu}(-\frac{1}{3}p^2 \eta_{\alpha\beta})+\frac{1}{3}{p^4\eta^{\mu\nu}\over M^2}\eta_{\alpha\beta}+({p^\mu p^\nu}-p^2\eta^{\mu\nu})(\eta_{\alpha\beta}-4/3\eta_{\alpha\beta}+{p^2\eta_{\alpha\beta}\over 3 M^2}) \nonumber\\
-M^2(I-\frac{1}{3}\eta^{\mu\nu}\eta_{\alpha\beta}-\eta^{\mu\nu}(\eta_{\alpha\beta}-4/3\eta_{\alpha\beta}))+M^2(-{p^\mu p^\nu\over 3 M^2}\eta_{\alpha\beta}+{p^2\eta^{\mu\nu}\over 3 M^2}\eta_{\alpha\beta})\\
=(p^2-M^2)I \nonumber
\end{eqnarray}

\section{The AGS Theory of Massive Gravity}
\label{sec:aCMGT}
However one might wonder whether such discontinuity might disappear in a theory where graviton gains mass through some spontaneous symmetry breaking mechanism. After all, the massive gauge vector boson propagator in unitary gauge also appears not to have a continuous limit to its massless counter part since the part proportional to $\frac {k_\mu k_\nu} {M^2}$ will blow up.

Such a covariant massive gravity theory is proposed by Arkani-Hamed \textit{et al.} \citec{Arkani-Hamed:2002sp} (henceforth referred as AGS theory). In this scenario, the general covariance group (diffeomorphism) is now double to $GC \times GC_0$. The theory is formulated
in a pair of coordinates $x^\mu$ and $x^a_0$. Two independent coordinate
transformation $x^\mu \to y^\mu$, $x_0^a \to y_0^a$ can be applied to them respectively. Another metric $g_0$ is also introduced to keep the sector with coordinate
$x_0^a$ to be covariant. To relate this two sectors, one need to introduce
a link field $Y^a(x^\mu)$, which relates points $x^\mu$ to $x_0^a$ as, 
\[x_0^a= Y^a(x^\mu).\]
This field transforms under both $GC$ and $GC_0$ and converts fields that only transform in $GC_0$ to those that only transform in $GC$. For
example, one can define 
\[G_{\mu\nu} = \frac {\p Y^a}{\p x^\mu} (x^\mu) \frac {\p Y^b}{\p x^\nu}
(x^\mu) g_{0ab}(Y^a(x^\mu)),\]
which only transforms under $GC$ (but not $GC_0$). Geometrically, this is nothing special other than a pull back.

The covariant massive gravity theory can be formulated by adding a mass term 
\be\el{covamass}\CL = \sqrt{g}\Big[a\, g^{\mu\nu}(g_{\mu\nu}-G_{\mu\nu})g^{\rho\xs}(g_{\rho\xs}-G_{\rho\xs})
+b\,g^{\mu\rho}(g_{\mu\nu}-G_{\mu\nu})g^{\nu\xs}(g_{\rho\xs}-G_{\rho\xs})\Big]\ee
The vacuum expectation value of the link field $Y^a$ (like $Y^a = \xd^a{}_\mu
x^\mu$) breaks the group $GC \times GC_0$ down to a certain subgroup. For $Y^a = \xd^a{}_\mu x^\mu$ the remaining subgroup is the diagonal of $GC \times GC_0$. After $M_{0}$, the Planck mass of the $g_0$-sector, is taken to be infinite,
the metric $g_{0ab}$ decouples and remains in the ground state $\eta_{ab}$. In
this case (with the unitary gauge of $Y^a$), \er{covamass} becomes our familiar
form,
\be\CL = \sqrt{g}\Big[a\, h^2 +b\,h^{\mu\nu}h_{\mu\nu})\Big],\ee
where $h_{\mu\nu} = g_{\mu\nu} - \eta_{\mu\nu},\ h = g^{\mu\nu}h_{\mu\nu}$.
Of course, one can restore the link field and maintain the general covariance.
In this sense, this formalism is not much different from the effective field
theories of Goldstone bosons in gauge theories. In both cases, Goldstone
bosons are also realized as the transformation of the broken gauge group.

\subsection{Tree-level Interaction in the AGS Theory}
\label{subsec:TIinMGT}
In this and the next subsections, we review the tree-level interaction in the AGS theory. We have worked out explicitly the graviton propagator derivation
that is not given in the original paper \citec{Arkani-Hamed:2002sp}. The final results agree with those given by the non-covariant theory reviewed in Sec~\ref{sec:VDVZD}. In other words, the vDVZ discontinuity remains.

In the AGS theory, one has the action \citec{Arkani-Hamed:2002sp},
\begin{equation}\label{Lag} S_{grav+mass} = \int d^4 x \sqrt{-g}(-{M_{Pl}}^2 R[g])+ \int d^4 x \sqrt{-g}(aH H+bH_{\mu\nu}H^{\mu\nu})\end{equation}
where the second term gives a mass to graviton. We have already taken the limit in which the other metric in the bi-metric theory decouples. At linearized level, we have,
\begin{equation}H_{\mu\nu}=h_{\mu\nu}+\pi_{\mu,\nu}+\pi_{\nu,\mu},\end{equation}
where the $\pi_\mu$ is the Goldstone field from the linearized link field introduced above,
\[Y^a =\xd^a{}_\mu (x^\mu+\pi^\mu).\]
At quadratic level of $h_{\mu\nu}$, the first term of the Lagrangian is
\begin{eqnarray} \label{2} 
\sqrt{-g}R=-\frac{1}{4}({h_{\mu\nu}}^{,\lambda}{h^{\mu\nu}}_{,\lambda}-h^{,\lambda}h_{,\lambda}-2{h^{\mu\lambda}}_{,\lambda}{h_{\mu\rho}}^{,\rho}+2h^{,\mu}{h_{\mu\lambda}}^{,\lambda})
\end{eqnarray}
Under an infinitesimal coordinate transformation (gauge transformation), $h_{\mu\nu}$ and the Goldstone field $\pi_{\mu}$ transform as
\begin{equation} \label{gauge} h_{\mu\nu} \to h_{\mu\nu} + \xi_{\mu,\nu}+\xi_{\nu,\mu}, \quad \pi_\mu \to \pi_\mu - \xi_\mu .\end{equation} 
One can see that $H_{\mu\nu}$ is invariant under such a gauge transformation. One can obtain the graviton propagator and study its form in the limit when the mass parameters a,b go to zero. One of the obstacles finding the propagator is the mixing term between $h_{\mu\nu}$ and $\pi_{\mu}$. Of course the same kind of mixing occurs in gauge theory, where the mixing is removed by a proper gauge fixing. We will do the same thing soon. But first, it is convenient to write $\pi_{\mu}$ as 
\begin{equation}\pi_\mu= A_\mu+\partial_\mu \phi,\end{equation}
which introduces a new artificial gauge symmetry,
\begin{equation}\label{gauge1} A_\alpha \to A_\alpha+\partial_\alpha \Lambda,\quad \phi \to \phi - \Lambda .\end{equation}
The mass term (second term in the Lagrangian \eqref{Lag}) will give terms of the form, 
\begin{equation}\int d^4 x 4a \phi_{,\mu ,\nu}\phi^{,\mu ,\nu}+ 4b \Box \phi \Box \phi = \int d^4 x (a+b)\Box \phi \Box \phi.\end{equation}
One needs to have $a+b=0$ to avoid the pathological kinetic term with four derivatives which would lead to a tachyon or a ghost. This requirement will lead to the Fierz-Pauli mass term\cite{Fierz-Pauli}.
\begin{equation}\label{5}\frac 1 4 f^4 (h_{\mu\nu}h^{\mu\nu}-h^2)\end{equation}
where $f^4$ is a dimensionful constant that defines the graviton mass,
\begin{equation}-{m_g}^2={f^4 \over {M_{Pl}}^2}.\end{equation}
But the second term of \eqref{2} also gives, in addition to the Fierz-Pauli mass term \eqref{5}, other terms including mixing terms involving $h_{\mu\nu}$, $\phi$ and $A_\mu$ which we will consider next. The $A_\mu$ field has an appropriate kinetic term. Moreover because $a+b=0$, there is no mixing between $A_\mu$ and $\phi$,
\begin{equation}\int d^4 x 4a A_{\mu,\nu}\phi^{,\mu,\nu}+ 4b {A^\mu}_{,\mu} \Box \phi = \int d^4 x (a+b){A^\mu}_{,\mu} \Box \phi =0\end{equation}
But there is a mixing term between $A_\mu$ and $h_{\mu\nu}$:
\begin{equation}f^4 (A_{\mu,\nu}h^{\mu\nu}-h{A^\mu}_{,\mu}),\end{equation}
and a mixing term between $\phi$ and $h_{\mu\nu}$:
\begin{equation}\label{4} f^4 (\phi_{,\mu,\nu}h^{\mu\nu}-h\Box \phi)\end{equation}
As in \citec{Arkani-Hamed:2002sp}, we make the following redefinition 
\begin{equation}\label{redef} h_{\mu\nu}={\widetilde{h}}_{\mu\nu}-{m_g}^2\phi \eta_{\mu\nu}, \end{equation}
to remove the mixing term \eqref{4}. As we shall see in the next subsection,
however this is not enough to remove all the mixing terms between $\phi$ and $h_{\mu\nu}$. We will present the remaining necessary terms to complete this gauge fixing procedure since they are not explicitly given in \citec{Arkani-Hamed:2002sp}.

\subsection{Graviton Propagator in the AGS Theory}
\label{subsec:GPitAGST}
The redefinition \eqref{redef} will remove the mixing term \eqref{4}, but it will also introduce other mixing terms. For whatever reason, this problem was ignored in \citec{Arkani-Hamed:2002sp}. Though no new result will come out, we still feel that it is worthwhile to fill up the necessary gap. In this subsection, we carry out the missing calculation to show that the gauge fixing to remove all the mixing term can indeed be done, and that the result is exactly the same as in \citec{Arkani-Hamed:2002sp}.

Under the transformation \eqref{redef}, the kinetic part \eqref{2} of the graviton Lagrangian will give kinetic terms,
\begin{equation}\label{from2}f^4 (\phi_{,\mu,\nu}{\widetilde{h}}^{\mu\nu}-{\widetilde{h}}\Box \phi) + \frac{3}{2}  f^4 {m_g}^2\phi_{,\mu}\phi^{,\mu}.\end{equation}
(The sign is correct since there is a minus sign in the first term of the definition of Lagrangian \eqref{Lag}.)
The mass term $f^4 (H_{\mu\nu}H^{\mu\nu}-H^2)$ under \eqref{redef} gives
\begin{equation}\frac 3 2 {m_g}^2 f^4 \phi {H'}-3 {m_g}^4 f^4 \phi^2 - f^4 (\phi_{,\mu,\nu}{\tilde h}^{\mu\nu}-\tilde h\Box \phi) -3 f^4 {m_g}^2 \phi_{,\mu}\phi^{,\mu}, \end{equation}
where ${H'}_{\mu\nu} \equiv {\tilde h}_{\mu\nu}+A_{\mu,\nu}+A_{\nu,\mu}$ and $H' = {H'}_{\mu\nu} \eta^{\mu\nu}$. So we still have some mixing. Combining these two terms one will have the kinetic term for $\phi$
\begin{equation}\label{scalar-K} -\frac{3}{2} f^4 {m_g}^2\phi_{,\mu}\phi^{,\mu}.\end{equation}
All together, \eqref{redef} leads to the following terms in the Lagrangian \eqref{Lag},
\begin{equation}\label{psiHmix}\frac 3 2 {m_g}^2 f^4 \phi {H'}-3 {m_g}^4 f^4 \phi^2 -\frac{3}{2} f^4 {m_g}^2\phi_{,\mu}\phi^{,\mu}.\end{equation}
The Lagrangian is invariant under the transformation of \eqref{gauge1},
\begin{equation}\phi \to \phi + \frac {2\omega} {{m_g}^2},\end{equation} 
\begin{equation}A_\mu \to A_\mu - 2{\omega_{,\mu} \over {m_g}^2 },\end{equation}
\begin{equation}\label{Weyl2} {\tilde h}_{\mu\nu} \to {\tilde h}_{\mu\nu}+2\omega \eta_{\mu\nu}.\end{equation} 
To get the propagators we can deal with the gauge freedom by introducing a gauge-fixing term, which, for convenience, we choose to remove the mixing of $H'$ and $\phi$ in the Lagrangian. We choose the following gauge fixing term:
\begin{equation}\label{GF2} \frac{3}{2} f^4 {(\beta {H'} - {{m_g}^2 \phi \over 2\beta})}^2.\end{equation} 
When added to \eqref{psiHmix}, it will remove the mixing term of $\phi$ and $H'$ regardless of the choice of $\beta$ and will give two terms,
\begin{equation}\frac{3}{2} f^4 \beta^2 {H'}^2 + \frac{3}{8} f^4 {m_g}^4 {\phi^2 \over \beta^2} .\end{equation} 

In order to get the canonical form of the kinetic term of $\phi$, let
\begin{equation}\el{cannorm}\phi = {1 \over M_{Pl} {m_g}^2} \phi_C.\end{equation}
So the part of the gauge fixed action involving $\phi$ is going to be
\begin{equation}\label{Psi-sector} \frac{3}{2} ({\phi_C}_{,\mu}{\phi_C}^{,\mu}+2 {m_g}^2 {\phi_C}^2 - {{m_g}^2 \over 4 \beta^2} {\phi_C}^2)\end{equation} 
Since ${\tilde h}_{\mu\nu}$ and $A_\mu$ are combined to form ${H'}_{\mu\nu}$, the Lagrangian is still gauge invariant under the infinitesimal coordinate transformation \eqref{gauge}. We use this symmetry to remove the mixing between $A_\mu$ and ${\tilde h}_{\mu\nu}$,
\begin{equation}\label{3a} 1/2(\zeta{M_{Pl}}({{\tilde h}_{\mu\nu}}^{,\nu}-(1-6\beta^2) {\tilde h}_{,\mu})+{M_{Pl}{m_g}^2 \over \zeta}A_\mu)^2,\end{equation}
where $\zeta$ is a second gauge fixing parameter we introduce. (For more information on gauge fixing and measure in standard gravitational theory, see \textit{e.g.} \citec{Fradkin:1974df}.) This gauge fixing term will give the following terms that only contain the graviton field ${\tilde h}_{\mu\nu}$,
\begin{equation} \label{3} 1/2\zeta^2{M_{Pl}}^2({{\tilde h}^{\mu\nu}}_{,\nu}-(1-6\beta^2){\tilde h}_{,\mu})^2.\end{equation}
Putting together the linearized terms of graviton and scalar field contributions to the Lagrangian and adding the gauge fixing term, we have the Lagrangian of relevant terms (omitting the gauge vector $A_\mu$),
\begin{eqnarray}
\label{G-sector}
\mathcal{L} & = & \frac{1}{4} M_{Pl}^2 ({{\tilde h}_{\mu\nu}}^{,\lambda}{{\tilde h}^{\mu\nu}}_{,\lambda}-{\tilde h}^{,\lambda}{\tilde h}_{,\lambda}-2{{\tilde h}^{\mu\lambda}}_{,\lambda}{{\tilde h}_{\mu\lambda}}^{,\lambda}+2{\tilde h}^{,\mu}{{\tilde h}_{\mu\lambda}}^{,\lambda})\nn
& & -{\frac{1}{2} \zeta^2} M_{Pl}^2 ({{\tilde h}^{\mu\nu}}_{,\nu}-(1-6\beta^2) {\tilde h},\mu)^2  -{m_g}^2 M_{Pl}^2 ({\tilde h}_{\mu\nu}{\tilde h}^{\mu\nu}-(1-6\beta^2) {\tilde h}^2) \nn
& & + \frac{3}{2} ({\phi_C}_{,\mu}{\phi_C}^{,\mu}+2 {m_g}^2 {\phi_C}^2 - {{m_g}^2 \over 4 \beta^2} {\phi_C}^2).
\end{eqnarray}

Finally, let us give the propagator of the graviton and scalar fields and check that this theory indeed gives a gauge-independent result, namely that $\beta$ will drop out from our final result. 
With some effort one can show that the graviton propagator from the Lagrangian \eqref{G-sector} is,
\begin{equation}\label{betagp} \frac {\frac{1}{2}{\delta^\mu}_\alpha {\delta^\nu}_\beta + \frac{1}{2}{\delta^\nu}_\alpha {\delta^\mu}_\beta} {k^2-{m_g}^2}+
{({m_g}^2-6(k^2+{m_g}^2)\beta^2) {\eta^{\mu\nu}\eta_{\alpha\beta}} \over 3(k^2-{m_g}^2)[4k^2 \beta^2 + {m_g}^2 (-1+8 \beta^2)]}. \end{equation}
This returns to the massless graviton propagator (in the form of Eq.(28) in \citec{vanDam:1970vg}) when $m_g \to 0$. The scalar $\phi_C$ has a $\beta$ dependent propagator
\begin{equation}\label{scalarprop}\frac{1}{3} {1 \over k^2 - {m_g}^2(-2+{1\over 4 \beta^2})}.\end{equation}
When $\beta \to 0$ and $m_g$ is finite, \eqref{betagp} will go to the Fierz-Pauli massive graviton propagator while \eqref{scalarprop} vanishes, which is the usual Fierz-Pauli theory. The redefinition \er{redef} introduces coupling between the scalar field and the energy momentum tensor. So effectively, the scalar field gives a propagator of $\frac 1 2  \eta^{\mu\nu}\eta_{\alpha\beta}$ times \eqref{scalarprop} when we consider the interaction between two energy momentum tensors. The sum of these two terms \eqref{betagp} and \eqref{scalarprop} is 
\begin{equation}\frac {\frac{1}{2}{\delta^\mu}_\alpha {\delta^\nu}_\beta + \frac{1}{2}{\delta^\nu}_\alpha {\delta^\mu}_\beta} {k^2-{m_g}^2}-\frac{1}{3} {{\eta^{\mu\nu}\eta_{\alpha\beta}} \over k^2 - {m_g}^2}, \end{equation}
Therefore the interaction between two conserved external sources is independent of the gauge parameter $\beta$ and as noted, exhibits the vDVZ discontinuity.

\subsection{Strong Coupling Behavior of Massive Gravity}
\label{subsec:SCBoMG}
As discussed in subsections~\ref{subsec:TIinMGT} and \ref{subsec:GPitAGST}, even in a covariant Massive gravity theory like the AGS theory, at least at tree level, the vDVZ discontinuity persists. When we try to go beyond tree-level, a big problem, the so-called strong coupling behavior arises. It spells the end of the effective theory because the perturbative calculation discussed previously becomes unreliable above a certain energy scale. In this subsection, we will review this strong coupling behavior, mainly following \citec{Arkani-Hamed:2002sp}.

To see the strong coupling behavior, let us consider the coupling constants of terms of higher powers in $\p^2 \phi_C$. Note that the non-canonical field $\phi$ arises in the form of the massless quantity $\p^2 \phi$ from the terms with derivatives of the Goldstone field, $\p \pi$. Powers of $\p^2 \phi$ appear in the graviton mass term. Since it is massless, we can have any power of $\p^2 \phi$. The first few terms with the lowest powers of $\phi$ are (indices suppressed) \citec{Arkani-Hamed:2002sp},
\[f^4[(\p^2\phi)^2 + (\p^2 \phi)^4 + \p^2 \phi \p A \p A].\]
After the canonical normalization \er{cannorm}, the terms above become,
\be\el{strongcoup} \frac 1 {m_g^4 M_{Pl}} (\p^2\phi)^2 + \frac 1 {m_g^6 M_{Pl}^2} (\p^2 \phi)^4 + \frac 1 {m_g^2 M_{Pl}} \p^2 \phi \p A \p A.\ee
The coupling constants of negative dimensions in mass imply nonrenormalizability.
In the context of effective field theory, the appearance of such coupling constants signals the breakdown of the effective field theory at some cutoff.
This cutoff is determined by the scale of the coupling constants of the nonrenormalizable terms. In the example here, \er{strongcoup} gives three different scales,
the lowest of which is 
\[\xL_5 \sim (m_g^4 M_{Pl})^{1/5}.\]
So the theory will no longer be valid above this scale. As a comparison,
the massless gravity theory has a cutoff scale of $M_{Pl}$, which is much
higher than $\xL_5$.

This $\xL_5$ (or rather $\iv {\xL_5}$) is called the Vainshtein radius \citec{Vainshtein:1972sx}.
Below this length scale, the field theory description we use is no longer valid. The problem is that this scale is generally very large. For a graviton
mass of $m_g \sim 10^{28} \zt{cm}^{-1}$, $\iv {\xL_5} \sim 10^{13} \zt{cm}$.
So the field theory can not be used even to describe the solar system. In
other words, we can not tell whether there is any difference between the massive gravity theory and GR. However, this is not necessarily good news because essentially we cannot tell anything at all about the massive gravity at this scale unless the fundamental theory is known. Moreover, when the mass $m_g$ goes to $0$, so does the energy scale $\xL_5$, and the effective field theory will break down at all scales.

This strong coupling behavior seems to exist in other massive gravity theories
like the DGP model \citec{Luty:2003vm} \ft{DGP model was introduced in \citec{Dvali:2000hr}.}. There are attempts to restore continuity \citec{Vainshtein:1972sx} \citec{Deffayet:2001uk}, but they are based mainly on the classical non-linear equation of motion. It remains unclear whether we can get around the strong coupling behavior to make any reliable prediction. In general, it is unlikely that the underlying fundamental theory gives the same predictions as GR.

\ \\

\section{Ghost and the Absence of vDVZ Discontinuity}
\label{sec:GatAovD}
Now we turn to the simplest form of a theory of a massive graviton
consistent with observation. We work with the covariant massive gravity 
theory proposed by Arkani-Hamed \textit{et al.} \citec{Arkani-Hamed:2002sp}. 
See subsections~\ref{subsec:TIinMGT} and \ref{subsec:GPitAGST} for a review of this theory and the results we are going to use later. As before, we introduce external matter sources characterized by two energy momentum tensors $T_{\mu\nu} = T^{a}_{\mu\nu} + T^{b}_{\mu\nu}$ to the Lagrangian \eqref{G-sector}, 
\begin{eqnarray}
\label{G-sector+}
\mathcal{L} & = & \frac{1}{4} M_{Pl}^2 ({{\tilde h}_{\mu\nu}}^{,\lambda}{{\tilde h}^{\mu\nu}}_{,\lambda}-{\tilde h}^{,\lambda}{\tilde h}_{,\lambda}-2{{\tilde h}^{\mu\lambda}}_{,\lambda}{{\tilde h}_{\mu\lambda}}^{,\lambda}+2{\tilde h}^{,\mu}{{\tilde h}_{\mu\lambda}}^{,\lambda})-{\frac{1}{2} \zeta^2} M_{Pl}^2 ({{\tilde h}^{\mu\nu}}_{,\nu}-(1-6\beta^2) {\tilde h},\mu)^2\nonumber \\
& & -{m_g}^2 M_{Pl}^2 ({\tilde h}_{\mu\nu}{\tilde h}^{\mu\nu}-(1-6\beta^2) {\tilde h}^2) + \frac{3}{2} ({\phi_C}_{,\mu}{\phi_C}^{,\mu}+2 {m_g}^2 {\phi_C}^2 - {{m_g}^2 \over 4 \beta^2} {\phi_C}^2) \nn
& & + h^{\mu\nu} T^{a}_{\mu\nu} + h^{\mu\nu} T^{b}_{\mu\nu}.
\end{eqnarray}
The redefinition \eqref{redef} will also produce an interaction term between $\phi$ and the energy momentum tensor ${T^{\mu}}_\nu$ given by
\begin{equation}\label{sourceT} {T^{\mu}}_\nu{h^{\nu}}_\mu={T^{\mu}}_\nu({{\widetilde{h}}^{\nu}}_\mu-{m_g}^2\phi{\delta^{\nu}}_\mu)={T^{\mu}}_\nu{{\widetilde{h}}^{\nu}}_\mu-{1 \over M_{Pl}} \phi_C T ,\end{equation}
where $T = T^{\mu}{}_{\nu} \delta^{\nu}{}_{\mu}$.

Again, one can read off the interaction between these two sources by looking at those terms that contain a product of $T^{a}_{\mu\nu}$ and $T^{b}_{\mu\nu}$. The Goldstone scalar $\phi$ will provide an extra contribution to the interaction between two sources and this contribution will not go away as the mass of the graviton $m_g$ goes to zero. The extra contribution from this scalar mode in the massless limit $m_g \to 0$ is 
\begin{equation}\label{7} T^{a} T^{b} \over 6 k^2. \end{equation}
Thus, although the graviton propagator coming from \eqref{G-sector} goes to the same massless form as in GR in this limit, the contribution \eqref{7} of the scalar mode leads to exactly the vDVZ discontinuity. The combined contribution of the scalar mode and graviton, when the mass is nonzero is given by 
\begin{equation}\label{intmasgra}\frac {\frac{1}{2}{\delta^\mu}_\alpha {\delta^\nu}_\beta + \frac{1}{2}{\delta^\nu}_\alpha {\delta^\mu}_\beta} {k^2-{m_g}^2}-\frac{1}{3} {{\eta^{\mu\nu}\eta_{\alpha\beta}} \over k^2 - {m_g}^2}, \end{equation}
where $m_g$ is the mass of the graviton. For comparison, we give the massless propagator here
\begin{equation}\frac {\frac{1}{2}{\delta^\mu}_\alpha {\delta^\nu}_\beta + \frac{1}{2}{\delta^\nu}_\alpha {\delta^\mu}_\beta} {k^2}-\frac{1}{2} {{\eta^{\mu\nu}\eta_{\alpha\beta}} \over k^2}. \end{equation}

We can introduce an extra scalar field $\Phi$ with the same coupling as \er{sourceT} but with a kinetic term of negative sign (ghost) and make the vDVZ discontinuity to disappear. More explicitly, this new scalar field has to couple to the matter in the same way as the Goldstone scalar,
\begin{equation}\el{ghostcoupling}-{1 \over M_{Pl}} \Phi T.\end{equation} 
Moreover its kinetic term has to opposite to \eqref{Psi-sector},
\begin{equation}\el{ghostkinetic}-\frac{3}{2} ({\Phi}_{,\mu}{\Phi}^{,\mu} + m_{\Phi}^2 \Phi^2)\end{equation} 
so that its contribution to the interaction between two energy momentum tensors, normaly
\begin{equation}\label{intphantom}\frac {-T^{a} T^{b}} {6 (k^2-m_\Phi^2)},\end{equation} 
which is of opposite sign to \eqref{7}. This field has ghost-like feature. This conclusion is consistent with the fact that a massive gravity theory with a non Fierz-Pauli mass term contains ghosts or tachyons.

Intuitively this ghost field cancels the contribution of the Goldstone scalar so that we can regain the continuous limit. Note that we assume we have an underlying theory whose effective Lagrangian is the combination of a massive gravity sector \eqref{G-sector} and a ghost with the kinetic term \er{ghostkinetic} and the coupling \er{ghostcoupling}.

In fact, similar type of field is introduced in \citec{Caldwell:1999ew} for completely different reasons and the stability problem of this so-called ``phantom" field is discussed in \citec{Carroll:2003st} and \citec{Cline:2003gs}. We found that the phantom field can also be used to cancel the Goldstone scalar to make the massive gravity theory free of vDVZ discontinuity. With the assumption that the phantom sector and the massive graviton sectors come from the same underlying theory, we expect $m_g$ to be related to $m_\Phi$, the mass of the phantom field, which is roughly $10^{-30}M_{Pl}$ \citec{Carroll:2003st} in order to account for the cosmological acceleration. However, there is no strong observational constraints on the ratio of their masses. As long as both of them are small (for example, $10^{-30}M_{Pl}
\sim 10^{-3}$ eV, as mentioned above), we are going to have agreement with any observation at macroscopic scale. The lowest order correction to GR should be proportional to $m_g^2/k^2$ or $m_\Phi/k^2$ as we can see by expanding \eqref{intmasgra} around $m_g =0$ (or expanding \eqref{intphantom} around $m_\Phi =0$).

Of course, the natural guess is that they are of the same order since they are supposed to come from one single underlying theory. This leads to a cutoff scale of the massive gravity theory \citec{Arkani-Hamed:2002sp},
\begin{equation}\label{cutoff3}\Lambda \sim {({m_g}^2 M_{Pl})}^{1/3}\end{equation}
of roughly $100$ MeV. This is surprisingly in agreement with the cutoff scale of the phantom theory obtained in \citec{Carroll:2003st}. This cutoff is obtained by requiring the decay rate of a ghost (at rest) to ghosts and gravitons (more exactly, $\Phi \to 2h 3 \Phi$) to be small compared to the age of the
universe $\iv {H_0}$. However, as pointed out in \citec{Cline:2003gs}, this cutoff for the phantom field theory of $100$ MeV is not correct. Depending on whether one take a Lorentz invariant cutoff or not, the cutoff is about $3$ MeV (Lorentz invariance violating cutoff) or $10^{-3}$ eV (for Lorentz invariant cutoff). Basically their idea is to consider the process of vacuum decay to two photons and two phantoms. The amplitude is given by the vertex of gravitational couplings of 
\[h^{\mu\nu}\p_\mu \Phi \p_\nu \Phi,\]
and
\[h^{\mu\nu} \eta^{\rho \sigma} F_{\mu\rho} F_{\nu \sigma}.\]
The phase space of this vacuum decay amplitude is infinite due to the negative energy of the phantom, and thus one must impose a constraint on the cutoff of the integration in order to get a finite result. By comparing the flux of the photons created via this vacuum decay process and the observed value, Cline \textit{et al.} \citec{Cline:2003gs} obtained the cutoffs mentioned above.

Of course, if we are only interested in the large scale phenomena, then there is no contradiction either. The perturbative theory is valid at the scale of solar system or above and the massive gravity theory with this extra phantom field behaves pretty much the same as our ordinary GR. However, the Lorentz invariant cutoff scale (length scale) is above the length scale at which
we have probed gravity and thus leads to a contradiction that is hard to get around. In other words, a simple combination of massive graviton and phantom field is ruled out by experiment.

\thispagestyle{fancy}
\chapter{Summary and Conclusions}
\thispagestyle{fancy}
\pagestyle{fancy}

\section{LHC Phenomenology}
\label{sec:clpart1}

In Chapter~\ref{CLIC}, we have explored the discovery potential of the
proposed $e^+e^-$ and $\gamma\gamma$ colliders to unmask string
resonances. We  have studied the direct production of Regge
excitations, focusing on the first excited level of open strings
localized on the worldvolume of D-branes. In such a D-brane
construction the resonant parts of the relevant string theory
amplitudes are {\it universal} to leading order in the gauge
coupling. Therefore, it is feasible to extract genuine string effects
that are independent of the compactification scheme. Among the various
processes, we found that, in the minimal extension of the SM,  the $\gamma \gamma \to e^+ e^-$ scattering proceeds only through a spin-2 Regge state.  Our detailed phenomenological studies suggest that for this specific channel, string scales as high as 4~TeV can  be unmasked at the 11$\sigma$ level with the first fb$^{-1}$ of data collected at $\sqrt{s} \approx 5$~TeV. We have also investigated intermediate Regge states of $e^+e^- \to F \bar F$ and we have shown that string theory predicts the {\em precise} value, equal 1/3, of the relative weight of spin 2 and spin 1 contributions. The potential benefit of this striking result becomes evident when analyzing the dimuon angular distribution, which has a pronounced forward-backward asymmetry, providing a very distinct signal of the underlying string physics.

In Chapter~\ref{LHCPheno}, we have extended the work in
Refs.~\citec{Hassanain:2009at} and \citec{Perelstein:2009qi} on an
approximate calculation of string amplitudes in the RS geometry to
include the $J=0$ contribution to bosonic 4-point functions. We have
carried out a phenomenological analysis of the resonant contributions
to dijet production at the LHC, and found that for an integrated
luminosity of 100~fb$^{-1}$, discovery of the resonant signal at
signal-to-noise of 5$\sigma$ is possible for resonant masses of up to
nearly 5~TeV.  However, it should be noted that this is possible only
for the Higgsless model: For the model with the Higgs on or near the
IR brane, the requirement $\Lambda_{\rm IR}\ge $ 1~TeV combined with
the relation $\mu\simeq 5\Lambda_{\rm IR}$ implies $\mu> $ 5~TeV,
greatly narrowing the possible region of discovery.

In Chapter~\ref{zprime}, we have  considered a low-mass string compactification in which the SM gauge multiplets originate in open strings ending on 3 D-branes. For the non-abelian $SU(3)$ and $SU(2)$ groups the D-brane construct requires the existence of two additional $U(1)$ bosons coupled to baryon number and to the trace of the $SU(2)$ multiplets, respectively.  One linear combination of the three $U(1)$ gauge bosons is identified as the the hypercharge $Y$ field, coupled to the anomaly free hypercharge current. The two remaining linear combinations ($Y', Y''$) of the three $U(1)$'s are coupled to anomalous currents, and grow masses in accord with the Green-Schwarz mechanism. After electroweak breaking, mixing with the third component of isospin results in the three observable gauge bosons, where with small mixing $Z'\simeq Y', \, Z''\simeq Y''$.

For a fixed $M_{Z'}$, the model contain several free parameters -- a single mixing angle and a gauge coupling constant unconstrained by the data -- which are chosen to supress the branching of $Z'$ decay into leptons and to accommodate the UA2 90\%CL data on $p \bar p \to jj X$. The remaining two parameters -- $\tan \beta$ and $M_{Z''}$ -- serve to limit the mass shift (due to mixing) of the electroweak $Z$ to conform with LEP observations. The heavier neutral gauge boson $Z''$ is within the reach of LHC.

\ \\

\section{Anomaly Puzzle in N = 1}
\label{sec:clpart2}

In Part II, we have tried to elucidate and settle three problems that are related to the anomaly puzzle in $\CN = 1$ SYM. First, we study the properties of the current operator $R_\mu'$ that is in the same super-multiplet as the stress tensor. We show explicitly
that $R_\mu'$ is not the same as the (anomalous) current $R_\mu$ which transforms the fields according to the charge ratios $1:\frac 2 3: - \frac 1 3$. Only the anomaly of the latter current is of one loop order and satisfies the Adler--Bardeen theorem, while the anomaly of $R_\mu'$ is proportional to the $\xb$ function. By explicit calculation, we show that $R_\mu'$ is a mixing of the $R$-current, $R_\mu$, and the Konishi current. Moreover, we show that the term $- \frac 1 8 \sum_f \xg_f \bar D^2 (\bar \Phi e^V \Phi)$ that appears in the anomaly equation in \citec{Shifman:1986zi} gives the same mixed current $R_\mu'$
and therefore supports the existence of two different ``supercurrents," even though only one supercurrent was proposed in \citec{Shifman:1986zi}. 
 \eml{We then use supersymmetric
QCD at the infrared fixed point, as an example,
to show how the difference between $R_\mu'$ and $R_\mu$ can naturally be explained in terms of two-supercurrents.} 

Secondly, we show that non-local terms must be included for consistency when using the equations of motion in \citec{Grisaru:1985yk} and \citec{Ensign:1987wy}.  This is necessary because 
the equations of motion are used there with the assumption
that their expectation values trivially vanish, while they actually vanish only when the non-local contributions to the expectation values are included.

Finally, we compared the two different calculations of the NSVZ $\xb$ function in \citec{Shifman:1986zi} and \citec{ArkaniHamed:1997mj}.  The second method, which is based on the Jacobian arising from field strength rescaling, seems independent of the infrared behavior of the theory, while the first method seems to depend only on the infrared behavior. 
We resolve this apparent contradiction by showing that the infrared modes are also crucial in getting the multi-loop corrections to the $\xb$ function in the second method.  The reason, as we show, is that the contributions from modes above any arbitrary nonzero scale, $\xL$, to the rescaling Jacobian are proportional to non-renormalizable terms and therefore do not contribute to the $\xb$ function.

\ \\

\section{Hermiticity of the Dirac Hamiltonian in Curved Spacetime}
\label{sec:clpart3}
In part III, we have revisited the quantum mechanics of a one-electron atom 
in an arbitrary curved background.  We addressed the following problem.
The operator $\hat H$, which appears on the rhs of \er{Shro}, $i\partial_t\psi(x) = \hat H\psi(x)$, is not
hermitian with respect to the curved-spacetime scalar product.
But \er{Shro} was obtained directly from the Dirac equation in curved spacetime, so why is $\hat H$ not Hermitian? We resolved this apparent paradox in the following way.
We started from the fundamental Schr{\"o}dinger equation
$ i\partial_t \st{\psi} = {\cal H} \st{\psi}$ of \er{ShroH},
where the operator ${\cal H}$ is Hermitian.
From the completeness relation, \er{comple}, we showed that the eigenstates of position that span the Hilbert space must depend on time as well as spatial position: $\st{\vec{x}, t}$.
The wave function $\psi(x)$ is defined as $\br \vec{x}, t \st{\psi}$.
By applying $\br{\vec{x}, t}|$ from the left to Eq.~(\ref{ShroH}), we found that the position-space representation of ${\cal H}$ is given by \er{xrep}. 
The differential operator
$H$ that appears in this representation is Hermitian with respect
to the curved-spacetime scalar product. 
However, the time derivative of $\br{\vec{x}, t}|$ in the wave function
gives an additional terms in $H$ that does not appear in 
$\hat{H}$.  Thus, we see why \er{Shro} is correct, but does not
involve the Hermitian operator $H$.  We have also discussed
the circumstances in which $\hat H$ is effectively hermitian and can
be used to do perturbation theory to find shifts in energy levels
and transition rates.

\ \\

\section{Massive Gravitons}
\label{sec:clpart4}
In part IV, we showed that the introduction of the ghost field $\Phi$ can effectively remove the vDVZ discontinuity and thus recover GR from a massive gravity theory. Moreover, this ghost field can be regarded as the phantom field previously proposed as an alternate explanation of the cosmological acceleration. We conjecture that there is a single underlying theory that gives this effective theory of massive gravity and ghost. Under this assumption, the mass of the graviton and the ghost are supposed to be at the same order. However, the cutoff of the phantom field is too low to agree with experiments. So this simple model does not really work and further modification is needed. 

\thispagestyle{fancy}
\newpage
\thispagestyle{fancy}
\pagestyle{fancy}

\newpage

\addcontentsline{toc}{chapter}{Appendix}
\label{Appdix}

\thispagestyle{fancy}
\begin{appendix}
\chapter{Basics of Supersymmetry}
\thispagestyle{fancy}
\pagestyle{fancy}
\label{chap:Susy}
\section{Conventions}
We use Weyl spinors $\xt^\xa$, $\bar \xt^{\dot \xa}, \xa = 1,2; \dot \xa
= 1,2$, which transform as
$(\frac 1 2,0)$ and $(0, \frac 1 2)$ representations of the Lorentz group.
The indices are raised and lowered by the antisymmetric tensor $\xe_{\xa\xb} =
\xe_{\dot \xa \dot \xb}$ and its inverse $\xe^{\xa\xb}$ ($\xe^{12} = 1$).
The choice of metric signature in Appendix~\ref{chap:Susy}
is $(-1\,+1\,+1\,+1)$ and $\xs^\mu$ is chosen to be
\be\xs^\mu = (\xs^0, \xs^i),\quad \xs^0 = - 1,\ee
where $\xs^i$ are the usual Pauli matrices.

\ \\

\section{Supersymmetry Algebra and Superconformal Algebra}

Supersymmetry is a generalization of the Poincare algebra. Coleman and Mandula
prove a theorem \citec{Coleman:1967ad} that forbids any extra spacetime symmetries,
\ie generators that do not commute with Poincare group other than those already in the Poincare group \ft{For massless theories, the symmetry group can be
expanded to the conformal group.}. However, this
theorem does not apply to transformations with Grassmann parameters.
So one can add fermionic generators and construct the supersymmetry algebra.
The $\CN = 1$ supersymmetry algebra is given by,
\be
\{Q_\xa,\bar Q_{\dot \xa} \}= 2 \xs^\mu{}_{\xa \dot \xa} P_\mu,\quad \{Q_\xa,Q_\xb\} = \{\bar Q_{\dot \xa},\bar Q_{\dot \xb}\} = 0.\ee
\be\el{algLQ} [L_{\mu\nu}, Q_\xa] = -\frac 1 2 (\xs_{\mu\nu})_\xa{}^\xb Q_\xb,\ee
\be\el{algLbarQ} [L_{\mu\nu}, \bar Q_{\dot \xa}] = \frac 1 2 (\bar \xs_{\mu\nu})_{\dot \xa}{}^{\dot \xb} \bar Q_{\dot \xb},\ee
\be \bar Q_{\dot \xa} = (Q_\xa)^\dagger.\ee
\be[Q_\xa,P_\mu] = 0,\quad [\bar Q_{\dot \xa},P_\mu] = 0.\ee
This is called $\CN = 1$ because it has the minimum number (four) of generators in $4$ dimension.

For massless theories, the scaling symmetry $D$ can be added to the Poincare
group. It does not commute with the translation. So another generator $K_\mu$ needs to be added. Together, they form the conformal algebra with the following new nonvanishing commutators (compared to the Poincare algebra),
\be\el{ApA259}[D,P_\mu] = i P_\mu,\quad [D,K_\mu] = -i K_\mu,\quad [D,L_{\mu\nu}] = 0,\ee
\be[L_{\mu\nu}, K_\xg] = i (\eta_{\nu \xg} K_\mu - \eta_{\mu\xg} K_\nu),\ee
\be\el{ApA261}[P_\mu, K_\nu] = -2i(\eta_{\mu\nu} D + L_{\mu\nu}), \quad [K_\mu,K_\nu]
= 0.\ee

The elements $K_\mu$ and $D$ do not commutes with $Q_\xa$ and therefore we
need to introduce another super-generator $S_\xa$ to close out the algebra. What we get is then the superconformal algebra, which
(in addition to \er{ApA259}-\er{ApA261})
also contains the following extra algebraic relations,
\be[S_\xa, \bar S_{\dot \xa}] = 2 \xs^\mu_{\xa \dot \xa} K_\mu,\ee
\be[Q_\xa, S_\xb] = - (\xs^{\mu\nu})_{\xa\xb} L_{\mu\nu} + 2 i \xe_{\xa \xb}
D - 3 \xe_{\xa\xb} R\ee
\be[Q_\xa, K_\mu] = -(\xs_\mu)_{\xa \dot \xa} \bar S^{\dot \xa},\quad [\bar
Q_{\dot \xa}, K_\mu] = (\xs_\mu)_{\xa \dot \xa} S^{\xa},\ee
\be [S_\xa, P_\mu] = -(\xs_\mu)_{\xa \dot \xa} \bar Q^{\dot \xa},\quad [\bar
S_{\dot \xa}, K_\mu] = (\xs_\mu)_{\xa \dot \xa} Q^{\xa}.\ee

\be[D,Q] = \frac i 2 Q,\quad  [D,\bar Q] = \frac i 2 \bar Q, \quad [D, S] = -\frac
i 2 S,\quad [D,\bar S] = -\frac i 2 \bar S\ee 
\be[R,Q] = -\frac 1 2 Q,\quad [R,\bar Q] = \frac 1 2 \bar Q, \quad [R, S] = \frac 1 2 S,\quad [R,\bar S] = -\frac 1 2 \bar S\ee 

\ \\

\section{Superspace and Superfield}
Minkowski space can be understood as a coset space $G/H$. More explicitly,
one can define an equivalence class 
\be g_1 \sim g_2: g_1 = g_1 h,\quad h \in H.\ee
The coset $G/H$ is just this equivalence class. For Minkowski space, $G = \{L_{\mu\nu}, P_\mu\}$ is the Poincare group and $H$ is the Lorentz group, \be\frac {\{L_{\mu\nu}, P_\mu\}}{\{L_{\mu\nu}\}} = R^4.\ee
The coset space can be parameterized by the translation parameters $x^\mu$.

Superspace is a generalization of the construction above (see \eg \citec{Galperin:2001uw}). Let us first talk about
the real superspace. In this case, $G$ is
the supergroup $\{L_{\mu\nu}, P_\mu, Q_\xa, \bar Q_{\dot \xa}\}$ and $H$
remains the Lorentz group. So the coset is
\be R^{4|4} = \frac {\{L_{\mu\nu}, P_\mu, Q_\xa, \bar Q_{\dot \xa}\}}{\{L_{\mu\nu}\}}
= (x^\mu, \xt^\xa, \bar \xt^{\dot \xa}) \equiv (X^A).\ee
So the supersymmetry transformation parameters become the coordinate of the
superspace (just like $x^\mu$ for Minkowski space). An equivalence class can be
described by one of its element. In this case, one can choose,
\be\xO(x,\theta,\bar \theta) \equiv e^{-i x^\mu P_\mu +i \xt^\xa Q_\xa +i \bar \xt_{\dot \xa} \bar Q^{\dot \xa}}.\ee
Now the action of the group $G$ can be realized as a diffeomorphism on the superspace,
\be g \xO(x,\theta,\bar \theta) = \xO(x',\theta',\bar \theta')h(g,X^A).\ee
From the algebra \er{algLQ} and \er{algLbarQ}, it is clear that under the Lorentz group, $\xt, \bar \xt$ transform as Weyl spinors. Under the susy
transformation,
\be\sQ = e^{i(\xi^\xa Q_\xa + \bar \xi_{\dot \xa} \bar Q^{\dot \xa})},\ee
the parameters change like,
\be\el{susydiffeo} \xd \xt^\xa = \xi^\xa,\quad \xd \bar \xt^{\dot \xa} = \bar \xi^{\dot \xa},\quad \xd x^\mu =i( \xi\xs^\mu \bar \xt - \xt \xs^\mu \bar \xi).\ee
The susy generators are diffeomorphisms and so they can be expressed as tangent vector
fields,
\be \el{SuperT:WessQ} Q_\xa = \frac{\p}{\p \xt^\xa}  - i\p_{\xa \dot \xa} \bar \xt^{\dot \xa},\ee
and
\be \el{susytangent}\bar Q^{\dot \xa} = \frac{\p}{\p \xt_{\dot \xa}} - i\xt^{\xa} \p_{\xa \dot \xb} \xe^{\dot \xb \dot \xa}, \quad \bar Q_{\dot \xa} = -\frac{\p}{\p \xt^{\dot \xa}} + i\xt^{\xa} \p_{\xa \dot \xa}.\ee
In the second line, we use 
\be\frac{\p}{\p \xt_\xa} = -\xe^{\xa \xb} \frac{\p}{\p \xt^\xb}.\ee
Note that we express the vector in terms of spinor coordinate using \er{SuperT:spincor}. The appearance of $\xe^{\dot \xb \dot \xa}$ becomes convenient when we consider the susy transformation $\xi Q + \bar \xi \bar Q$. These vector
fields satisfy the following Lie algebra,
\be\{Q_\xa,\bar Q_{\dot \xa} \}= 2i\p_{\xa \dot \xa},\quad \{Q,Q\} = \{\bar Q,\bar Q\} = 0.\ee

From \er{susydiffeo}, it is natural to understand susy transformations as translations in
the superspace. But these ``translations" are non-Abelian and the parameter of the product of two transformations is not the sum of their parameters.

As in an ordinary QFT, one can define superfields on the superspace as fields that transform under $g$ as a certain representation of $H$. Note that in both cases (Minkowski and $R^{4|4}$), $H$ is the Lorentz group,
\be g: \Phi \to \Phi',\ee 
\be\Phi'^a (x',\theta',\bar \theta') = [h(g,X^A)]^a{}_b\Phi^b(x,\theta,\bar \theta),\ee 
where we use some abstract indices $a,b$ to describe the representations of the Lorentz group. From the analysis above, it is obvious that only those elements in the Lorentz subgroup correspond to nontrivial but $X^A$-independent matrices $h = g$. For an element $g$ in
the form of $\sQ$, the group action is a diffeomorphism of scalar functions. This
can be realized by the vectors \er{SuperT:WessQ} and \er{susytangent}. In operator language, the susy transformation is expressed by
\be\sQ(x,\theta,\bar \theta) \Phi \equiv {\hat \sQ}^{-1}(x,\theta,\bar \theta) \Phi \hat \sQ (x,\theta,\bar \theta),\ee 
where $\hat \sQ$ is an operator that can act on states.
A superfield can be expanded in powers of $\xt$ and $\bar \xt$. This only has a finite number of terms because of the Grassmann nature. Of course,
one can study the transformations of components by expanding (see \eg \er{susycomps}) the
change of the superfield,
\be\label{susytrans1} \delta_\xi \Phi(x,\theta,\bar \theta)  = \Phi' (x,\theta,\bar \theta)- \Phi (x,\theta,\bar \theta).\ee

The superderivatives $D, \bar D$ are defined as the covariant derivatives
in the superspace and they map one superfield to another. They are explicitly given by,
\be\el{SuperT:supd} D_\xa = \frac{\p}{\p \xt^\xa}  + i\p_{\xa \dot \xa} \bar \xt^{\dot \xa},\quad \bar D_{\dot \xa} = -\frac{\p}{\p \xt^{\dot \xa}} - i\xt^{\xa} \p_{\xa \dot \xa}.\ee
They are different from $Q, \bar Q$ by a sign flip of the second term. The commutation relationship is given by,
\be\{D_\xa,\bar D_{\dot \xa} \}= -2i\p_{\xa \dot \xa}.\ee
Moreover, both superderivatives anti-commute with each of $Q,\bar Q$: 
\be\{Q(\zt{or}\ \bar Q), D(\zt{or}\ \bar D)\} = 0.\ee

\subsection{Chiral Superfield}
\com{\sto{Chiral Superfield}}
The definition of chiral \lab{Chiralsfi}superfield is given by the constraint,
\be\label{Chiralsf}{\bar D}_{\dot \xa} \Phi = 0\ee
It can be defined in a fancier way as a function on the
chiral superspace, which we will not discuss
here. The super-derivative of a chiral superfield vanishes. In some sense, a chiral superfield $\Phi$ can be expressed as a function of only $x_L^\mu, \theta$ (not $\bar \theta$),
where $x^\mu_L$ is defined by 
\be\el{Chiralsfi:2} x^\mu_L = x^\mu + i \theta \xs^\mu \bar \theta\ee
Pay attention that ${\bar D}_{\dot \xa} x^\mu_L = 0$. So \eqref{Chiralsf} is manifest. In the coordinates $x_L^\mu, \xt, \bar \xt$, the explicit forms of $D$ and $\bar D$ are given by, 
\be D_\xa = \frac{\p}{\p \xt^\xa} + 2 i \p_{\xa \dot \xa} \bar \xt^{\dot \xa}, \quad \bar D_{\dot \xa} = - \frac{\p}{\p \bar \xt^{\dot \xa}}.\ee
The superderivative $D$ can also be used to obtain the components of a chiral superfield 
\be\Phi = A + \sqrt 2 \xt \psi + \xt^2 F,\ee
\be\psi_\xa = \frac 1 {\sqrt 2} D_\xa \Phi \at{\xt = \bar \xt = 0}.\ee 

The generators $Q$ and $\tilde Q$ in terms of $x_L$ and $\xt, \bar \xt$ are given by,
\be Q_\xa = \frac{\p}{\p \xt^\xa},\quad Q_{\dot \xa} = -\frac{\p}{\p \bar \xt^{\dot \xa}} + i  2 \xt^\xa \xs^\mu_{\xa\dot \xa} \frac{\p}{\p y^\mu}.\ee
The effect of the $\xi Q$ on a chiral field is a transformation of the higher components to the lower components (see \eg \er{susycomps}) while the $\bar \xi \bar Q$ transforms the derivative of the lower components to higher components (see \eg \er{susycomps2}).

Note that $D \Phi$ is no longer a chiral superfield. The reason is that $D$ does not commute with $\bar D$. Neither is $\p \Phi$. On the other hand, $\bar D \bar D U$ for any superfield $U$ is a chiral superfield,
which simply follows from $\bar D^3 = 0$.

One reason the chiral superfield is called chiral is that the CPT counterpart of the states in the supermultiplet (corresponding
to components of the chiral superfield) do not appear in the same superfield as those states. For example, a chiral superfield $\Phi$ describes a multiplet with a scalar and a spinor (helicity $1/2$) but not their CPT conjugate. 

\ \\

\section{Useful Results in Spinor Algebra}
We use the spinor notation to express the vectors. This is defined as,
\be \el{SuperT:spincor} A_{\xa \dot \xa} = \xs^\mu_{\xa \dot \xa} A_\mu,\quad A^\mu = -\frac{1}{2}A_{\xa \dot \xb} {(\bar \xs^\mu)}^{\dot \xb \xa},\ee
where $\bar \xs^\mu$ is defined by,
\be\bar \xs^\mu{}_{\dot \xb \xa} = \xs^\mu{}_{\xa \dot \xb},\ee
So they are related by a transpose. Of course, more often than not, we are dealing with $\bar \xs^\mu{}^{\dot \xb \xa}$, which is $\bar \xs^\mu{}_{\dot \xb \xa}$ with indices raised by the $\xe$. Moreover, we have (indices suppressed),
\be\xs_\mu \bar \xs^\mu = -2 \xd \xd.\ee 
Equivalently, we have
\be\el{Paulimatrices:1} (\bar \xs^i) {}^{A B} = -(\xs^i) {}_{A B}, \quad (\bar \xs^0) {}^{A B} = (\xs^0) {}_{A B}. \ee
where we change the indices to avoid the confusion by the dotted and undotted indices. This is also true even when $\xs^0 = i$ as in the case of Euclidean theory. By the way, we also have 
\be(\bar \xs^\mu) {}^{A B} = -(\xs^{\mu\dagger}){}_{A B}.\ee
In Euclidean theory, we have
\be\el{SuperT:Pauli2} \Tr (\xs_\nu \bar \xs^\mu) = 2 \xd^\mu_\nu.\ee

The volume element in the superspace is defined by,
\be d^2 \xt \equiv -\frac 1 2 \xe_{\xa \xb} d\xt^\xa d \xt^\xb .\ee
As a result, we have
\be\el{xtintegration}\int \xt \xt d^2 \xt = 2.\ee
This is actually easy to prove by writing out the component explicitly. Both $\xt \xt$ and $\xe_{\xa \xb} d\xt^\xa d \xt^\xb$ provide two terms and each combination gives the same contribution. So we have a factor of $4$. 

Another useful result is,
\be -D^2 (\xt \xt) = \xe^{\xa \xb} \frac \p {\p \xt^\xa} \frac \p {\p \xt^\xb} \xt \xt = 4,\ee
This agrees with the fact that the integration of a Grassmann number is like a derivative.

Here are some other useful results:
\be\psi \chi \equiv \psi^\xa \chi_\xa = \chi^\xa \psi_\xa = \chi \psi = -
\psi_\xa \chi^\xa.\ee
\be(\psi \chi)^\dagger = \bar \chi \bar \psi = \bar \chi \bar \psi \equiv \bar \chi_{\dot \xa} \bar \psi^{\dot \xa}.\ee
\be\psi_\xa \psi_\xb = \frac 1 2 \xe_{\xa \xb}\psi^2.\ee
\be\psi^\xa \psi^\xb = -\frac 1 2 \xe^{\xa \xb}\psi^2.\ee

\ \\

\section{Supercurrent}

In this section, we will review some properties of the supercurrent. We mostly use the convention in \citec{Wess:1992cp} including the choice of $\xs^\mu$ matrices and superderivatives. The only two different choices are the form of vector superfield and the integration of Grassmann variables (see Sec~\ref{sec:RAP}). First of all, let us put down the transformation rules for components of
a chiral field $\Phi$ that appears in the Wess-Zumino model with the Lagrangian,
\be\el{supercurrent:1} \CL = \frac 1 4 \int d^2 \xt d^2 \bar \xt \Phi \bar \Phi + \frac 1 2 \int d^2 \xt g \Phi^3.\ee
The chiral superfield can be written in the component form,
\be\el{MuSu2:chifield} \Phi = A + \sqrt 2 \xt \psi + \xt^2 F,\ee 
where the factor of $\sqrt 2$ is to make $\psi$ canonically normalized. The supersymmetry transformation (parameterized by $\xi, \bar \xi$) of the components is given by ($g = 0$ for simplicity),
\be \el{susycomps} \xd A = \sqrt 2 \xi \psi,\quad \xd A^* = \sqrt 2 \bar \xi \bar \psi,\ee
\be \el{susycomps2}\xd \psi = i \sqrt 2 \xs^\mu \bar \xi \p_{\mu} A + \sqrt 2 \xi F,\quad \xd \bar \psi = -i \sqrt 2 \xi \xs^\mu \p_{\mu} A^* + \sqrt 2 \bar \xi F^*\ee
\be\el{MuSu2:susytran} \xd F = i \sqrt 2 \bar \xi \bar \xs^\mu \p_\mu \psi,\quad \xd F^* = i \sqrt 2 \xi \xs^\mu \p_\mu \bar \psi\ee
With this transformation, we can work out the current from Lagrangian,
\ba \el{MuSu2:Lag}\CL & = & i \p_n \bar \psi \bar \xs^n \psi + A^* \Box A + F^* F\nn
& = & \frac i 2 \p_\mu \bar \psi \bar \xs^\mu \psi - \frac i 2  \bar \psi \bar \xs^\mu \p_\mu \psi -\p_\mu A^* \p^\mu A + F^* F,
\ea
which is just \er{supercurrent:1} in component fields.

For simplicity, we set $F = F^* = 0$. Considering only the transformation generated by $\xi$, we have
\ba \el{MuSu2:current} \xd_\xi \CL & = &\sqrt 2\Big[ - \p_\mu A^* \p^\mu (\xi \psi) + \frac i 2 \p_\mu (- i \xi \xs^\nu \p_\nu A^*) \bar \xs^\mu \psi -\frac i 2 (- i \xi \xs^\nu \p_\nu A^*) \bar \xs_\mu \p^\mu \psi \Big]\nn
& = & \sqrt 2 \Big[- \p_\mu A^* \p^\mu (\xi \psi) + \frac 1 2(\p_\mu \xi )\xs^\nu \p_\nu A^* \bar \xs^\mu \psi+ \xi \xs^\nu \p_\mu \p_\nu A^* \bar \xs^\mu \psi - \frac 1 2 \xi \p_\mu (\xs^\nu \p_\nu A^* \bar \xs^\mu \psi) \Big]\nn
& = & \sqrt 2 \Big[- (\p^\mu \xi )\p_\mu A^* \psi -\xi \p_\mu (\p^\mu A^* \psi)+ \frac 1 2( \p_\mu \xi)\xs^\nu \p_\nu A^* \bar \xs^\mu \psi - \frac 1 2 \xi \p_\mu (\xs^\nu \p_\nu A^* \bar \xs^\mu \psi)\Big].\nonumber 
\ea
The current is then given by,
\be \el{MuSu2:current1} J_{\mu} = \sqrt 2 \p_\nu A^* \xs^\nu \bar \xs_\mu \psi. \ee
For a general theory, we need to keep the auxiliary fields when we do the variation. For a free theory, we can see that \er{MuSu2:Lag} implies no $F$ in $J$. The variation of the kinetic term $\bar \psi \xs^\mu\p_\mu \psi$ contains
something proportional to $\p_\mu (\xi F)$. The part with $(\p_\mu F)\xi$ is combined
with another term to form a total derivative $\p_\mu K^\mu$ but this $K^\mu$
does not appear in the conserved current since it has to be the same as the
part (from the variation of $\bar \psi \xs^\mu\p_\mu \psi$) proportional
to $(\p_\mu \xi) F$. 

Despite that, we can use equation of motion in the interacting theory to get $F$ in $J_\mu$, the current following Noether method does not contain $F$. None of the interacting terms containing $F$ have derivatives, and they
therefore cancel identically. The extra terms (proportional to the coupling constant) in $J_\mu$ involves variation of $F$ though. 

Note that the charge generated by this current $Q \equiv \int J^0$ does generate
the correct supersymmetry transformation on component fields. This is obvious for
$\psi$ and $A$. One subtlety is the transformation on $\dot A$. It gives
\[\xd (\dot A) = -i\sqrt 2 \xi [\p_i A^* \xs^i \bar \xs_0 \psi, \dot A] =\sqrt 2 \xi \xs_0 \bar
\xs^i \p_i \psi = -\sqrt 2\xi \xs_0 \bar \xs^0 \p_t \psi = \sqrt 2\xi\dot \psi,\]
where we have used the equation of motion.

However, we actually use the current $J_\mu$ given by, 
\be \el{MuSu2:supcur}J_\mu = \sqrt 2 \Big[\p_\nu A^* \xs^\nu \bar \xs_\mu \psi + \frac {4} 3 \xs_{\mu \nu} \p^\nu (A^* \psi)\Big].\ee
This is the ``improved" supersymmetry current of \citec{Ferrara:1974pz}, which gives the same charge and is also conserved.
Note that the second term does not have any contribution to the supersymmetry charge $Q$ because we have $\xs^{00} = 0$. Although $\xs^{0i} \ne 0$, the spatial derivatives do not contribute to the charge since they only give boundary terms when integrated over a spatial slice (volume). 

The supercurrent is given by,
\be \el{MuSu2:supcWZ} \CJ_\mu \equiv \frac 2 3 \Big[ i \Phi \stackrel {\leftrightarrow}{\p_\mu} \bar \Phi-\frac 1 4 D^\xa \Phi (\xs_{\mu})_{\xa \dot \xa} \bar D^{\dot \xa} \bar \Phi \Big]= \frac {2i} 3 \Phi \stackrel {\leftrightarrow}{\p_{\xa \dot \xa}} \bar \Phi+\frac 1 3 D_\xa \Phi  \bar D_{\dot \xa} \bar \Phi.\ee

The operator $\CJ_\mu$ \er{MuSu2:supcWZ} can be expressed in the form of \er{APN1SYM:supercurrent}. Explicitly, we have
\be\el{MuSu2:rcurrentmatter} C_{\xa \dot \xa} =\frac {2i} 3 A {\overleftrightarrow \p}_{\xa \dot \xa} A^* + \frac 2 3 \psi_\xa \bar \psi_{\dot \xa}.\ee
This indicates a correct $2:-1$ charge ratio for bosonic and fermionic fields. The lowest component $C_\mu$ is related to the R-current,
\[R_\mu = C_\mu.\]
Note that with the introduction of both a mass term $m \Phi^2$ and a cubic term $g \Phi^3$, the $U(1)$ transformation
is no longer a symmetry. A charge given by $R_\mu$ above does not give another supercharge $Q$ when we take their commutator $[R,Q]$.

The $\xt$ component $\chi_\mu$ of this current $\CJ_\mu$
is not
the supersymmetry current $J_\mu$. Instead, they are related by,
\be\el{MuSu2:supsymcur} J_\mu = i(\chi_\mu + \xs_\mu \bar \xs^\nu \chi_\nu).\ee
Now let us show this. Both second terms in \er{MuSu2:supcur} and \er{MuSu2:supsymcur}
give currents that are in the same equivalence class (second term on the rhs) as \er{MuSu2:current1}. 

\ba \el{chicomp}\chi_\mu & = & \frac 2 3 \Big[i \sqrt 2 \psi \stackrel {\leftrightarrow}{\p_\mu} A^* - \frac i {\sqrt 2} \p_\nu A^* \xs^\nu \bar \xs_\mu \psi + \sqrt 2 \xe_{\xa \xb} F \bar \psi_{\dot \xa}\Big]\nn
& = & \frac 2 3 \Big[i \sqrt 2 \psi \stackrel {\leftrightarrow}{\p_\mu} A^* - \frac i {\sqrt 2} \p_\nu A^* \xs^\nu \bar \xs_\mu \psi - \frac {\sqrt 2} 2 \xs_\mu \bar \psi F \Big]\nn
& = & \frac {2\sqrt 2} 3\Big[- i \p_\mu \psi A^*- i (\frac 1 2 \p_\nu A^* \xs_\mu \bar \xs^\nu \psi + \p_\nu A^* \xs^\nu \bar \xs_\mu \psi)\Big].\ea
Note that we use $D = \p_\xt$, $\bar D = - \p_{\bar
\xt} -i 2\xt \p$. Now we have
\ba
J_\mu & = & i(\chi_\mu + \xs_\mu \bar \xs^\nu \chi_\nu)\nn
& = & \frac {2\sqrt 2} 3 i\Big[-(i \p_\mu \psi A^* + i \xs_\mu \bar \xs^\nu \p_\nu \psi A^*) - i (\frac 1 2 \p_\nu A^* \xs_\mu \bar \xs^\nu \psi + \p_\nu A^* \xs^\nu \bar \xs_\mu \psi)\nn
& & - i (\frac 1 2 \p_\nu A^* \xs_\mu \bar \xs^\xl\xs_\xl \bar \xs^\nu \psi + \p_\nu A^* \xs_\mu \bar \xs^\xl \xs^\nu \bar \xs_\xl \psi) \Big]\nn
& = & \sqrt 2 \Big[\p_\nu A^* \xs^\nu \bar \xs_\mu \psi +\frac 4 3 \xs_{\mu \nu} \p^\nu( \psi A^*)\Big], 
\ea
which is exactly \er{MuSu2:supcur}. In the third line, we use $\xs_\mu \bar \xs^\mu = -4$ and $\bar \xs_\nu \xs_\mu \bar \xs^\nu = 2 \xs_\mu$. 

The commutators of $Q$  with the component fields of $\Phi$ give the supersymmetry transformation of the fields when the equation 
of motion holds.

The $M_{\xa \dot \xa}$ for this supercurrent (see \eg \er{APN1SYM:supercurrent})
in fact vanishes. The $\xt^2$
component of $\Phi$ vanishes ($F = 0$) and therefore the first term in $\CJ_\mu$ does not give any contribution. On the other hand, $D\Phi$ has neither
 $\xt^2$ nor $\bar \xt^2$ component. So to get a term involving $\xt^2$, we need a $\xt$ from each factor ($D\Phi$ and $\bar D \bar \Phi$); $D\Phi$ only has a nonvanishing $\bar \xt$\ component. The $\xt$ component is just $F$ and therefore vanishes.
 Similarly, the $\bar \xt$ component of $\bar D \Phi$ vanishes and we have
 a vanishing $\bar M_\mu$.

The bosonic part of the $\xt \bar \xt$ component of the supercurrent can be worked out following from \er{MuSu2:supcWZ},
\be\el{xtbarxtcom}
\frac 2 3 \Big[2\p_{\xa\dot \xb}A\, \bar \xt^{\dot \xb} \xt^\xb\p_{\xb \dot \xa} A^* -  \xt^\xb \bar \xt^{\dot \xb} (\p_{\xb \dot \xb} A \stackrel{\leftrightarrow}{\p_{\xa
\dot \xa}} A^* - A \stackrel{\leftrightarrow}{\p_{\xa\dot \xa}}\p_{\xb \dot \xb} A^*)\Big]
\ee

We use $v_{\mu\nu}$ to denote the $\xt \bar \xt$ component and $t_{\mu\nu}
\equiv \frac 1 2 v_{(\mu\nu)}$ as one half of the symmetric part of the $\xt \bar \xt$ component. With a $t_{\mu\nu}$ defined as
\be \el{supercurtbos} t_{\mu\nu} = \frac 2 3 \Big[-2\p_{(\mu}A \p_{\nu)} A^* +\frac 1 2 \eta_{\mu\nu} \p_\rho A \p^\rho A^*+ \frac 1 2 \p_\mu \p_\nu A\, A^* + \frac 1 2 A\p_\mu \p_\nu A^*\Big],\ee
we can obtain the $\xt \bar \xt$ component explicitly,
\ba
& & 2 t_{\xa \dot \xa \xb \dot \xb} + \frac i 2 \xe_{\xa\xb} \xe_{\dot
\xb \dot \xg} (\p^{\xg \dot \xg} R_{\xg \dot \xa}) -\frac i 2 \xe_{\xb \xg} \xe_{\dot \xa \dot \xb}  \p^{\xg\dot \xg} R_{\xa \dot \xg}\nn
& = & \frac 2 3 \Big[-\p_{\xa\dot \xa} A\ \p_{\xb \dot \xb} A^* - \p_{\xb \dot \xb} A\ \p_{\xa \dot \xa} A^*  - 2 \p_{\xa \dot \xb} A \p_{\xb \dot \xa} A^*  \nn
& & + \p_{\xa \dot \xa} \p_{\xb \dot \xb} A\, A^* + A\p_{\xa \dot \xa} \p_{\xb \dot \xb} A^*\Big].
\ea
Note that this agrees with \er{xtbarxtcom}.

Consider now the relationship between $t_{\mu\nu}$ and the stress tensor.
The stress tensor can be written in the form,
\ba
T_{\mu\nu} & = & - \p_{\mu} A \p_{\nu} A^* - \p_{\nu} A \p_{\mu} A^* + \eta_{\mu\nu} \p_a A \p^a A^* -\frac i 4 (\bar \psi \bar \xs_\mu \p_\nu \psi + \bar \psi \bar \xs_\nu \p_\mu \psi) \nn 
& &  - \frac i 4 (\psi \xs_\mu \p_\nu \bar \psi + \psi \xs_\nu \p_\mu \bar \psi) + \eta_{\mu\nu} (\frac i 2 \bar \psi \bar \xs^\rho \p_\rho \psi
+ \frac i 2 \psi \xs^\rho \p_\rho \bar \psi).
\ea

Of course, we usually use $\vartheta_{\mu\nu}$, which is defined as the improved stress tensor of $T_{\mu\nu}$,
\ba
\vartheta_{\mu\nu} & = & T_{\mu\nu} + \frac 1 3 (\p_\mu\p_\nu - \eta_{\mu\nu}\Box)
AA^*\nn
& = & -\frac 2 3 \p_{\mu}A \p_{\nu} A^* -\frac 2 3 \p_{\nu}A \p_{\mu} A^*+\frac
1 3 \eta_{\mu\nu} \p_{\rho}A \p_{\rho} A^* \nn
& & + \frac 1 3 \p_\mu \p_\nu A\, A^* + \frac 1 3 A\p_\mu \p_\nu A^* - \frac 1 3 \eta_{\mu\nu} (A\Box A^*+\Box
A\, A^*).
\ea
The extra term added to $T_{\mu\nu}$ gives no contribution to the charges $P_\mu$ and certainly does not affect the conservation. Note that we have $t_\rho{}^\rho = \frac 1 3 (A\Box A^*+\Box A\, A^*)$, and therefore
\ba
\el{tcompandvarxt}
t_{\mu\nu} - \eta_{\mu\nu} t_\rho{}^\rho & = & -\frac 4 3 \p_{(\mu}A \p_{\nu)} A^* +\frac 1 3 \eta_{\mu\nu} \p_\rho A \p^\rho A^*+ \frac 1 3 \p_\mu \p_\nu A\, A^* + \frac 1 3 A\p_\mu \p_\nu A^*  \nn
& & - \frac 1 3 \eta_{\mu\nu} (A\Box A^*+\Box A\, A^*)\nn
& = & \vartheta_{\mu\nu}.
\ea
Of course, following \er{tcompandvarxt}, we have $\vartheta_\mu{}^\mu$ proportional to $A\Box A^* + \Box A\, A^*$.
A similar conclusion should hold for improved supersymmetry current $J_\mu$.

Note that this difference between $t_{\mu\nu}$ and $\vartheta_{\mu\nu}$
follows from the difference between $\chi_\mu$ and $J_\mu$ \er{MuSu2:supsymcur}.
This statement is actually only true when $t_{\mu\nu}$ is symmetric. With \er{tcompandvarxt} the variation of $J_\mu$ (under a supersymmetry transformation
generated by $\bar \xi$) is proportional to $\vartheta_{\mu\nu}$ while that of $\chi_{\mu}$ is proportional to $t_{\mu\nu}$.

Let us now include the fermionic fields in our considerations. The $\xt \bar \xt$ component, $v_{\mu\nu} (\xt \xs^\nu \bar \xt) $, of the supercurrent is (following from \er{MuSu2:supcWZ}),
\ba \el{vcompWZ}
v_{\mu\nu} (\xt \xs^\nu \bar \xt) & = & \frac {2i} 3 \Phi \stackrel {\leftrightarrow}{\p_{\xa \dot \xa}} \bar \Phi - \frac 1 6 \bar \xs_\mu^{\dot \xa \xa} D_\xa \Phi  \bar D_{\dot \xa} \bar \Phi \nn
& = & \frac {4 i} 3 (\xt \psi) \stackrel{\leftrightarrow}{\p_\mu} (\bar \xt \bar \psi) - \frac 1 3 \psi(y) \xs_\mu \bar \psi (y^+) \nn
& &- \frac 1 3 \Big[-i 2 (\bar \xt \bar \xs^\nu)^\xa \p_\nu (\xt \psi) (\xs_\mu \bar \psi)_\xa + i 2 (\psi \xs_\mu)_{\dot \xa} (\bar \xs^\nu \xt)^{\dot \xa} \p_\nu (\bar \xt \bar \psi)\Big]\nn
& = & (\xt \xs^\nu \bar \xt)\frac 2 3 \Big[-i \psi \xs_\nu \stackrel{\leftrightarrow}{\p_\mu}\bar \psi - \frac i 2  \p_\nu \psi \xs_\mu \bar \psi + \frac i 2 \psi \xs_\mu \p_\nu \bar \psi - \frac i 2 \p_\rho \psi \xs_\nu \bar \xs^\rho \xs_\mu \bar \psi \nn
& & - \frac i 2 \p_\rho \bar \psi \bar \xs_\nu \xs^\rho \bar \xs_\mu \psi\Big] \nn
& = & (\xt \xs^\nu \bar \xt) \frac 2 3 \Big\{ -i \psi \xs_\nu \stackrel{\leftrightarrow}{\p_\mu}\bar \psi - \frac i 2  \p_\nu \psi \xs_\mu \bar \psi + \frac i 2 \psi \xs_\mu \p_\nu \bar \psi \nn
& & +\Big[ \frac 1 2  \xe_{\nu \rho \mu a} \p^\rho \psi \xs^a \bar \psi - \frac i 2 \p_\rho \psi (\eta_{\nu \mu} \bar \xs^\rho - \xd^\rho_\mu \bar \xs_\nu - \xd^\rho_\nu \bar \xs_\mu) \bar \psi + \zt{H.c.}\Big]\Big\}
\ea

We would like to show that the antisymmetric part
of $v_{\mu\nu}$ satisfies the following expression on-shell,
\be\el{antisymv} v_{[\mu\nu]} = \frac 1 2 \xe_{\mu\nu a b} \p^a R^b .\ee

It is not hard to rewrite $\frac 1 2 \xe_{\mu\nu a b} \p^a R^b$ in the form given by,
\be\el{antiRdspinor} \frac 1 2 \xe_{\mu\nu a b} \p^a R^b = \frac i 2 \xe_{\xa \xb} \p^\xg {}_{(\dot \xb} R_{\xg \dot \xa)} + \zt{H.c.}\ee
Using the relation, 
\[\xs^{[\mu}_{\xa \dot \xa} \xs^{\nu]}_{\xb \dot \xb} = (\xs^{\mu\nu}\xe)_{\xa
\xb} \xe_{\dot \xa \dot \xb} + (\xe\bar \xs^{\mu\nu})_{\dot \xa
\dot \xb} \xe_{\xa \xb}.\]
we have
\ba
-\frac 1 2 \xs^\mu_{\xa \dot \xa} \xs^\nu_{\xb \dot \xb} \xe_{\mu\nu a b} \p^a R^b & = & -\frac 1 2 [(\xs^{\mu\nu}\xe)_{\xa
\xb} \xe_{\dot \xa \dot \xb} + (\xe\bar \xs^{\mu\nu})_{\dot \xa
\dot \xb} \xe_{\xa \xb}] \xe_{\mu\nu a b} \p^a R^b \nn
& = & [i(\xs^{a b}\xe)_{\xa \xb} \xe_{\dot \xa \dot \xb} - i (\xe\bar \xs^{ab})_{\dot \xa \dot \xb} \xe_{\xa \xb}]  \p_a R_b \nn
& = &\frac 1 2 [i \p_{(\xa \dot \xg} R^{\xg \dot \xg} \xe_{\xg \xb)} \xe_{\dot \xa \dot \xb} - i \p^{\xg} {}_{(\dot \xa} R_{\xg \dot \xb)}\xe_{\xa \xb}].
\ea

Now we have
\ba \el{antisymvspin}\frac 1 2 \xe_{\xa \xb} v_{\mu\nu}\xs^\mu_{\xg (\dot \xa} \xs^\nu_{\xd \dot \xb)}\xe^{\xd \xg} & = & i\xe_{\xa \xb} \psi_\xd \p_{\xg (\dot \xa} \bar \psi_{\dot \xb)} \xe^{\xd \xg}  - i \p^{\xg} {}_{(\dot \xa} R_{\xg \dot \xb)}\xe_{\xa \xb}\nn
& = & i \xe_{\xa \xb} \p^\xg{}_{(\dot \xa} (\psi_\xg \bar \psi_{\dot \xb)}) - i \p^{\xg} {}_{(\dot \xa} R_{\xg \dot \xb)}\xe_{\xa \xb}\nn
& = & \frac 1 2 i \xe_{\xa \xb} \p^\xg{}_{(\dot \xa} R_{\xg \dot \xb)}.
\ea
This is the part antisymmetric in $\xa,\xb$ and symmetric in $\dot \xa, \dot \xb$ and corresponds to the self-imaginary dual component of the $\tau_{[\mu\nu]}$.
The Hermitian conjugate gives another term,
\[-\frac 1 2 i \xe_{\dot \xa \dot \xb} \p_{(\xa}{}^{\dot \xg} R_{\xb) \dot \xg}\]
With the use of \er{antisymvspin} and \er{antiRdspinor}, we get \er{antisymv}.

Note that in the derivation above, we did use the equation of motion. In fact, \er{antisymv}
is true on-shell. It follows from 
\be\el{conservationsuperc} D^\xa \CJ_{\xa \dot \xa} = 0,\ee
which can be derived using the explicit form \er{MuSu2:supcWZ} and the equation of motion (and its conjugate)
\[D^2 \Phi = 0,\]
and the commutator,
\[[\bar D_{\dot \xa}, D^2 ] = 4 i D^\xa \xs^\mu_{\xa \dot \xa} \p_\mu.\]

The 
$\bar \xt$ component of eq.\er{conservationsuperc} implies
\ba i (\xs^\nu \bar \xt)^\xa \p_\nu R_{\xa \dot \xa} + (\xs^\nu\bar \xt)^\xa v_{\xa \dot \xa \nu} & = & i\p_\nu R_\mu \bar \xt \xe \bar \xs^\nu \xs^\mu + v_{\mu\nu} \bar \xt \xe \bar \xs^\nu \xs^\mu \nn
& =& \p_a R_b \xe^{a b \mu \nu} \bar \xt (\xe \bar \xs_{\mu \nu}) +2 v_{[\mu\nu]}
\bar \xt (\xe \bar \xs^{\nu\mu}) = 0.
\ea
This is equivalent to \[v_{[\mu\nu]} = \frac 1 2 \xe^{\mu \nu a b } \p_a R_b,\]
which is exactly \er{antisymv}.

Let us now consider SQED, whose supercurrent is given by 
\[\CJ_{\xa \dot \xa} \equiv -\frac 2 {e^2} W_\xa \bar W_{\dot \xa}.\]
We have $R_{\xa \dot \xa} = -\frac 2 {e^2}\xl_\xa \bar \xl_{\dot \xa}$. The $\xt \bar \xt$ component is decomposed in the same way as \er{APN1SYM:supercurrent}
(with only fermionic fields considered),
\ba \el{xtbarxtdecom}2\xt^\xb \bar \xt^{\dot \xb} \frac 1 {e^2}(i \xl_\xa \p_{\xb \dot \xb}\bar \xl_{\dot \xa} - i \p_{\xb \dot \xb} \xl_\xa \bar \xl_{\dot \xa}) & = & 2\xt^\xb \bar \xt^{\dot \xb} \frac 1 {e^2}(i \xl_{(\xa} \p_{\xb) \dot \xb} \bar \xl_{\dot \xa} - i \p_{\xb (\dot \xb} \xl_\xa \bar \xl_{\dot \xa)})\nn
& & +2 \xt^\xb \bar \xt^{\dot \xb} \frac 1 {e^2}(\frac i 2 \xe_{\xa \xb} \xl^\xg \p_{\xg \dot \xb}\bar \xl_{\dot \xa} - \frac i 2 \xe_{\dot \xa \dot \xb} \p_{\xb \dot \xg} \xl_\xa \bar \xl^{\dot \xg}) \nn
& \equiv & 2\xt^\xb \bar \xt^{\dot \xb} \tau_{\xa \dot \xa \xb \dot \xb} + \frac 1 2 \xt_\xa \bar \xt_{\dot \xb} ( i \p^{\xg \dot \xb} R_{\xg \dot \xa} + \zt{H.c}),\ea
where $\tau_{\mu\nu}$ is given by,
\ba\el{supercurvcom} 
\tau_{\mu\nu} & = &  \frac 1 4 \bar \xs_\mu^{\dot \xa \xa} \bar \xs_\nu^{\dot \xb \xb} \frac 1 {e^2}\Big[\frac 1 2 ( i \xl_\xa \p_{\xb \dot \xb} \bar \xl_{\dot \xa} + i \xl_\xb \p_{\xa \dot \xb} \bar \xl_{\dot \xa}  - i \p_{\xb \dot \xb} \xl_\xa \bar \xl_{\dot \xa} - i \p_{\xb \dot \xa} \xl_\xa \bar \xl_{\dot \xb} )
\nn 
& & - (\frac i 2 \xe_{\xa \xb} \p_{\xg \dot \xb} \xl^\xg \bar \xl_{\dot \xa} + \zt{H.c})
\Big]\nn
& = &  -\frac 1 {4 e^2}( i\xl \p_\nu \xs_\mu \bar \xl - \frac i 2 \xl \xs_\nu \bar \xs^\rho \xs_\mu \p_\rho \bar \xl +\zt{H.c}) -\frac 1 {e^2} (\frac i 8 \p_\rho \xl \xs^\rho \bar \xs_\nu \xs_\mu \bar \xl + \zt{H.c}) \nn
& = & -\frac 1 {e^2}(\frac i 4 \xl \p_\nu \xs_\mu \bar \xl - \frac i 8  \xl \xs_\nu \bar \xs^\rho \xs_\mu \p_\rho \bar \xl) - \frac 1 {e^2}(\frac i 8 \xl \xs_\mu \bar \xs^\rho
\xs_\nu \p_\rho \bar \xl + \frac i 4 \xl \xs_\mu \p_\nu \bar \xl) + \zt{H.c}\nn
& = & -\frac 1 {e^2}(\frac i 2 \xl \p_\nu \xs_\mu \bar \xl +\zt{H.c})-\frac 1 4 \xe_{\mu\nu b a} \p^b (\frac 1 {e^2}\xl \xs^a \bar \xl).
\ea
Because $\tau_{\mu\nu}$ includes the term $-\frac 1 4 \xe_{\mu\nu a b} \p^a R^b$, it is not symmetric. Following a similar derivation as in the Wess-Zumino model, we can show that $\tau_{[\mu\nu]} = 0$ and certainly, the symmetric part produces the stress
tensor. This symmetric part $t_{\mu\nu} \equiv \tau_{(\mu\nu)}$ is in fact
related to the stress tensor in the same way as \er{tcompandvarxt},
\be\el{xvtt} \xvt_{\mu\nu} = t_{\mu\nu} - \eta_{\mu\nu} t_\xl{}^\xl.\ee

\goodbreak

\subsection{Conservation of Stress Tensor and SUSY Current}
When the rhs of the anomaly equation (of $\bar D^{\dot \xa} \CJ_{\xa \dot \xa}$) is either the derivative of a chiral superfield,
\be\el{CSTSC:1} \bar D^{\dot \xa} \CJ_{\xa \dot \xa} = D_\xa S,\quad \bar D_{\dot \xa} S = 0,\ee
or a linear superfield,
\be \el{CSTSC:2}\bar D^{\dot \xa} \CJ_{\xa \dot \xa} = \bar D^2 D_{\xa} T, \quad T = \bar T,\ee
the conservation of stress tensor and the susy current is guaranteed. The
two situations are in fact different. For the second case, we have
\ba 
2i\p^{\xa \dot \xa} \CJ_{\xa \dot \xa} & = & \{D^\xa,\bar D^{\dot \xa}\} \CJ_{\xa \dot \xa} \nn
& = & D^\xa \bar D^2 D_{\xa} T + \bar D^{\dot \xa} D^2 \bar D_{\dot
\xa} T \nn
& = & ([D^{\xa},\bar D^2] D_\xa + \bar D^{\dot \xa}[D^2 ,\bar D_{\dot \xa}])T
\nn
& = & -4i \xe^{\xa \xb}\p_{\xb \dot \xa}\bar D^{\dot \xa} D_{\xa} T- 4i \p_{\xa \dot \xa}\bar D^{\dot \xa} D^{\xa} T= 0.
\ea
Note that we use the Hermitian conjugate of \er{CSTSC:2}
\[D^{\xa} \CJ_{\xa \dot \xa} = D^2 \bar D_{\dot \xa} T,\]
in which the key step is $[\bar D (\bar D T)]^\dagger = - D (\bar D T)^\dagger$
(index suppressed for simplicity). $[D^2 ,\bar D_{\dot \xa}] = - 4i D^\xa \p_{\xa \dot \xa}$ can be obtained from a conjugate of $[D_{\xa},\bar D^2]$ \ft{Note that the conjugation does not change the order of the derivatives.}. Anyway, the point is now $\CJ_\mu$ gives conserved currents. The trace
may still be non-zero and therefore trace anomaly and conformal anomaly are
still there.

The case with derivative of the chiral superfield $D_\xa S$ \er{CSTSC:1} has a different character. In this case, it can be shown that the supersymmetric current
($\xt$ and $\bar \xt$ components of $\CJ_{\xa \dot \xa}$) satisfy,
\be\el{CSTSC:3} \p^\mu J_{\xa \mu} = 4 \p_{\xa \dot \xa} \bar \psi^{\dot \xa},\ee
where $\bar \psi$ is the usual $\bar \xt$ component of $\bar S$,
\[\bar S = \bar A + \bar \xt \bar \psi + \bar \xt \bar \xt \bar F.\]
This follows from
\[DD S = (\p_\xt + i 2\bar \xt^{\dot \xa} \p_{\xa \dot \xa})^2 S = 4 i \p_{\xa \dot \xa} \psi^\xa \bar \xt^{\dot \xa} \Rightarrow -i \bar D \bar D \bar S = - 4 \xt^{\xa} \p_{\xa \dot \xa} \bar \psi^{\dot \xa}.\]

Using $\xe^{\xa \xb} J_{\xa \xb \dot \xa} = 4 \bar \psi_{\dot \xa}$, which follows from the other components of \er{CSTSC:1}, we can define
a conserved current,
\be\el{componentchiQ} Q_{\xa \xb \dot \xb} \equiv J_{\xa \xb \dot \xb} + 2 \xe_{\xa \xb} \xe^{\xg \xd}J_{\xg \xd \dot \xb}.\ee
It is easy to see
\[\p^{\xb \dot \xb} Q_{\xa \xb \dot \xb} = 0,\]
where we use \er{SuperT:Pauli2} to rewrite \er{CSTSC:3} in the spinor indices.

However, one has to remember that this $Q_{\xa \mu}$ is not really the classical supersymmetry current with quantum correction. The difference is not proportional to the coupling constant. It is not clear what physical meaning the $Q_{\xa \mu}$ has.

\goodbreak
\section{A Note on Equations Involving Epsilon Dimensional Operators}
\label{subsec:NoExeDO}
\com{\et{A Note on Equations Involving $\xe$ Dimensional Operators}}

Here we prove a few equations involving the $\xe$ dimension. The point is to show how to work in $4-2\xe$ dimensions, especially when spinor indices are used.
Basically, we have the following two equations $\hhat \xd_{\xa \dot \xb}{}^{\xb \dot \xb} = \xe \xd_\xa{}^\xb$, and $\hhat \p = 0$
to work with. Note that we use the conventions in \citec{Gates:1983nr}, \citec{Grisaru:1985ik} and \citec{Ensign:1987wy}.

As a warm-up exercise, let us first show\com{ \rf{GMZ~(5.12)}}\cp{ (see \citec{Grisaru:1985ik})},
\be\p^{\xb \dot \xa} \hhat A_{\xa \dot \xa} = -\xe f^\xb{}_\xa - \hhdb \bar f^{\dot \xa}{}_{\dot \xb}.\ee
The proof is straightforward,
\ba \p^{\xb \dot \xa} \hhat A_{\xa \dot \xa} & = & \hhdg \p^{\xb \dot \xa}  A_{\xg \dot \xg} = \hhdg (\p^{\xb \dot \xa}  A_{\xg \dot \xg} - \p_{\xg \dot \xg}  A^{\xb \dot \xa})\nn
& = & -\hhdg(\xd^\xb_\xg \bar f^{\dot \xa}{}_{\dot
\xg} + \xd^{\dot \xa}_{\dot
\xg} f^\xb{}_\xg)\nn
& = & -\xe f^\xb{}_\xa -\hhdb \bar f^{\dot \xa}{}_{\dot \xb}. 
\ea
In the first line, we replace $\p_\mu A_\nu$
by $F_{\mu\nu}$, using $\hhat \p
= 0$. In the second line, the following relation,\com{ \er{GMZFf}}
\[ F_{ab} = C_{\xa \xb} \bar f_{\dot \xa \dot \xb} + C_{\dot \xa \dot \xb} f_{\xa \xb},\]
is used. 

Now let us move on and prove the following
equation\com{ \rf{E\&M~(2.10)}}\cp{ (see \citec{Ensign:1987wy})},
\ba \el{EM210}
g \bar \cd^\dxa \bar \Phi \hhat \xG_{\xa
\dxa} \Phi = -\xe \bar \cd^2 \cd_\xa (\bar
\Phi \Phi),
\ea
where the covariant derivative is defined by
\[\cd_{\xa \dxa} = \p_{\xa \dxa} - i g \xG_{\xa
\dxa} = i\{\cd_\xa,\bar \cd_\dxa\}.\]
We can rewrite the connection in the $\xe$-dimension
using the $\xe$-dimensional component of
the covariant derivative (because $\hhat \p = 0$). With this replacement, the lhs
side of \er{EM210} can be rewritten as,
\ba
g\bar \cd^\dxa (\bar \Phi \hhat \xG_{\xa
\dxa} \Phi)  & = & \frac 1 2 \bar \cd^\dxa \hhat P_{\xa \dxa} = \frac 1 2 \hhdb \bar \cd^\dxa(i \bar \Phi \overleftrightarrow{\cd}_{\xb
\dxb} \Phi) \nn
& = & \hhdb \bar \cd^\dxa(i \bar \Phi {\cd}_{\xb
\dxb} \Phi) - \frac 1 2 \hhdb \bar \cd^\dxa
{\cd}_{\xb \dxb}(i \bar \Phi \Phi)\nn
& = & \hhdb \Big(- \bar \cd^\dxa
\bar \Phi \bar \cd_\dxb \cd_\xb \Phi - ig \xd^\dxa_\dxb \bar \Phi W_\xb \Phi\Big)\nn
& = & \xe (-i g\bar \Phi W_\xa \Phi)+\hhdb \bar \cd_\dxb (\bar
\cd^\dxa\bar \Phi\cd_\xb \Phi)\nn
& = & \xe (-i g\bar \Phi W_\xa \Phi)-\hhdb \bar \cd_\dxb \cd_\xb(\bar \cd^\dxa\bar \Phi \Phi) - \hhdb \bar \cd_\dxb(-i \cd_\xb{}^{\dxa}\bar \Phi \Phi) \nonumber\ea

This implies
\be - \hhdb \bar \cd^{(\dxa}(i \bar \Phi \overleftarrow{\cd}_{\xb
\dxb)} \Phi) = \frac 1 2 \xe (-i \bar g\Phi W_\xa \Phi)-\frac 1 2 \hhdb \bar \cd_\dxb \cd_\xb(\bar \cd^\dxa\bar \Phi \Phi) \ee
and therefore,
\ba
g\bar \cd^\dxa (\bar \Phi \hhat \xG_{\xa
\dxa} \Phi) & = &  \frac 1 2 \hhdb \bar \cd^\dxa(i \bar \Phi \overleftrightarrow{\cd}_{\xb
\dxb} \Phi) = - \hhdb \bar \cd^{(\dxa}(i \bar \Phi \overleftarrow{\cd}_{\xb
\dxb)} \Phi)- \frac \xe 2 \xd_\xa^\xb \bar \cd^{\dxa}(i \bar \Phi \overleftarrow{\cd}_{\xb
\dxa} \Phi)\nn
& = & \frac \xe 2 (-i g \bar \Phi W_\xa \Phi)-\frac
1 2 \hhdb \bar \cd_\dxb \cd_\xb(\bar \cd^\dxa\bar \Phi \Phi) - \frac \xe 2 \bar \cd^{\dxa}(i \bar \Phi \overleftarrow{\cd}_{\xa
\dxa} \Phi) \nn
& = & \frac \xe 2 (-i g \bar \Phi W_\xa \Phi)
-\hf\xe \bar \cd^2 \cd_\xa (\bar \Phi \Phi)
+\frac i 2 \xe \cd_{\xa \dxa} \bar \cd^\dxa(\bar \Phi \Phi)- \frac \xe 2 \bar \cd^{\dxa}(i \bar \Phi \overleftarrow{\cd}_{\xa
\dxa} \Phi)\nn
& = & \frac \xe 2 (-i g \bar \Phi W_\xa \Phi)
-\hf\xe \bar \cd^2 \cd_\xa (\bar \Phi \Phi)
-\frac \xe 2 (\bar \cd^\dxa\bar \Phi)\bar
\cd_\dxa\cd_\xa \Phi- \frac i 2 \xe ([\bar \cd^{\dxa}, \cd_{\xa\dxa}] \bar \Phi) \Phi\nn
\el{EM210b}& = & -\xe \bar \cd^2 \cd_\xa (\bar \Phi \Phi) \ea
In the third line, we use
\ba \hhdb \bar \cd_\dxb \cd_\xb(\bar \cd^\dxa\bar \Phi \Phi) & = & -\hhdb \cd_\xb \bar \cd_\dxb \bar \cd^\dxa(\bar \Phi \Phi) = \hhdb \xd_\dxb{}^\dxa\cd_\xb\bar \cd^2 (\bar \Phi \Phi)\nn
& = & \xe \bar \cd^2 \cd_\xa (\bar \Phi \Phi)
-i \xe \cd_{\xa \dxa} \bar \cd^\dxa(\bar \Phi \Phi)\ea
In the last line, we use \er{EMC16} and \com{(first eq in \rf{(2.10)}) }\er{EM210a}, which
we are going to prove now,
\be\el{EM210a} \bar \cd^2 \cd_\xa (\bar \Phi \Phi) =( \bar\cd^\dxb \bar \Phi) \bar \cd_\dxb \cd_\xa
\Phi + \bar \Phi [\bar \cd^2, \cd_\xa] \Phi
= (\bar \cd^\dxb \bar \Phi )\bar \cd_\dxb \cd_\xa
\Phi - ig \bar \Phi W_\xa \Phi.\ee
In the last equality of \er{EM210a}, we use\com{ \rf{(C17)}} \er{EMC16} and also the equation of motion\com{ \rf{(2.9)}},
\ba
& (\cd \cdot W)^a = -(\bar \cd \cdot \bar W)^a = -2 i g \bar \Phi T^a \Phi, \nn
\el{EM29}& \cd^2 \Phi  = \bar \cd^2 \bar \Phi = 0.
\ea

Now let us move on to show\com{ \rf{(2.11)}}\cp{ (see \citec{Ensign:1987wy})},
\be\el{eqhhatW} \bar \cd^{\dot \xa} \hhat W_{\xa \dot \xa}
= \frac 1 2 \bar \cd^2 \cd_\xa (\hhat \xG
\cdot \hhat \xG) - 2 \xe \bar \cd^2 \cd_\xa
(\bar \Phi \Phi).\ee
The lhs gives,
\ba \hhdb \bar \cd^{\dot \xa} \bar W_{\dot
\xb} W_\xb & = & \hhdb \xd_\dxb^\dxa \bar \cd_\dxg \bar W^\dxg W_\xb + \hhdb \bar \cd_\dxb \bar W^\dxa W_\xb \nn
& = & \xe\xd_\xa^\xb (2i g \bar \Phi
W_\xb \Phi)+ \hhdb \bar \cd_\dxb \bar W^\dxa W_\xb.
\ea
In the second line, we use\com{ \rf{(2.9)}}
\er{EM29}. The first term on the rhs can be rewritten as,
\ba
\frac 1 2 \bar \cd^2 \cd_\xa (\hhat \xG
\cdot \hhat \xG) & = & \frac 1 2 C_{\xg \xa} \bar \cd^2 (-2 i \hhat \xd_{\xb \dxb}{}^{\xg \dxg} \bar W_{\dxg} \hhat \xG^{\xb \dxb})
\nn
& = & C_{\xg \xa}\Big[ \frac i 2 \hhat \xd_{\xb \dxb}{}^{\xg \dxg} (\bar \cd_\dxd\bar \cd^\dxd \bar W_\dxg) \xG^{\xb \dxb} - i \hhat \xd_{\xb \dxb}{}^{\xg \dxg} \bar \cd_\dxd\bar W_\dxg
(\bar \cd^\dxd \hhat \xG^{\xb \dxb})\Big] \nn
& = &  \frac i 2 C_{\xg \xa}\hhat \xd_{\xb \dxb}{}^{\xg \dxg} (2\bar \cd_\dxg \bar \cd
\cdot \bar W + \bar \cd_\dxd \bar \cd_\dxg
\bar W^\dxd) \xG^{\xb \dxb}+i\xd^\xg_\xa\hhat \xd_{\xg \dxg}{}^{\xb \dxb} \bar \cd_\dxd
\bar W^\dxg \bar \cd^\dxd\hhat \xG_{\xb\dxb} \nn
& = &  2iC_{\xg \xa}\hhat \xd_{\xb \dxb}{}^{\xg \dxg} (\bar \cd_\dxg \bar \cd
\cdot \bar W)  \xG^{\xb \dxb}+ \xd_\xa^\xg\hhat \xd_{\xg \dxg}{}^{\xb \dxb} \bar \cd_\dxb
\bar W^\dxg W_\xb \nn
& = &  -2gC_{\xb \xa}(\bar \cd_\dxb \bar \Phi )\hhat \xG^{\xb \dxb} \Phi+ \hhat \xd_{\xa \dxg}{}^{\xb \dxb} \bar \cd_\dxb
\bar W^\dxg W_\xb\nn
& = & \xe(2i g \bar \Phi
W_\xa \Phi)+2 \xe \bar \cd^2 \cd_\xa \bar \Phi \Phi + \hhdb \bar \cd_\dxb \bar W^\dxa W_\xb.
\ea
In the first line, the following equation\com{ \rf{(C38)}} is used:
\be\el{EMC38}\bar \cd^\dxb \hhat \xG_{\xa\dxa}
= - i \hhat \xd_{\xa \dxa}{}^{\xb \dxb} W_\xb,\quad
\cd^\xb \hhat \xG_{\xa\dxa}
= - i \hhat \xd_{\xa \dxa}{}^{\xb \dxb} \bar
W_\dxb.
\ee
In the 5th line, we use\com{ \rf{(2.11)}},
\[\bar \cd^\dxa W_{\xa \dxa} = 2i g \bar \Phi W_\xa \Phi, \]
which simply follows from \er{EM29}. In the last line, we use\com{ \rf{(2.10)}},
\ba
-2g(\bar \cd_\dxb \bar \Phi ) \hhat \xG^{\xb \dxb} \Phi
& = &-2g \bar \cd_\dxb (\bar \Phi \hhat \xG^{\xb \dxb} \Phi ) +2g \bar \Phi \bar \cd_\dxb \hhat
\xG^{\xb \dxb} \Phi \nn
& = & 2i \xe g \bar \Phi W^\xb \Phi +2 \xe \bar \cd^2 \cd^\xb \bar \Phi \Phi.
\ea
Now we get to \er{eqhhatW}.

Here are some useful results:
\be\el{EMC43} \hhdb \bar \cd^\dxa[\cd_\xb, \bar \cd_\dxb](\zt{gauge singlet}) = -2\xe \bar \cd^2 \cd_\xa (\zt{gauge singlet})\ee
\be\el{EMC16}[\cd_\xb, \cd_{\xa \dxa}] = g\, C_{\xb\xa} \bar W_\dxa,\quad [\bar \cd_\dxb, \cd_{\xa \dxa}] =g\, C_{\dxb\dxa} W_\xa\ee
\be\el{EMC17}[\cd_\xa, \bar \cd^2] = -i \cd_{\xa \dxa}
\bar \cd^\dxa + ig W_\xa = -i \bar \cd^\dxa\cd_{\xa \dxa} - ig W_\xa.\ee

\thispagestyle{fancy}
\chapter{Calculations Relevant to String Phenomenology}
\thispagestyle{fancy}
\pagestyle{fancy}

\section{Some Techniques in the Calculation Scattering Amplitudes}
\com{\et{Calculating Scattering Amplitudes Efficiently}}
\cp{\label{sec:CSAE}}
To calculate the scattering amplitude in QCD effectively\com{\lab{CSAE}}, some special techniques prove to be useful since these calculations turn out to be more complicated than expected\com{ \rf{Dixon~E3}}. \cp{We use these techniques in the main text. So let us review them in this section.} Most of the material in this section is gleaned from \com{\rf{Dixon}}\cp{\citec{Dixon:1996wi}}. We will only focus on the $n$-gluon amplitudes (mainly those at tree-level).\\ 

\sto{Color Management}
In general, these amplitudes can be factorized into a color part and a kinematic
(including helicity) part. More explicitly, any tree diagram for $n$-gluon scattering can be reduced to a sum of single trace terms \com{\rf{(4)}} ($^\clubsuit$ for tree-level)
\ft{Since we mostly work with gluon amplitudes
in this section, to save some space,
we use $A_4^{\clubsuit}(1^{-},2^{+},3^{+}, 4^{+})$ instead of $A_4^{\clubsuit}(g_1^{-},g_2^{+},g_3^{+}, g_4^{+})$.},
\be\el{CSAE:1} \CM_n^{\clubsuit}(\{k_i,\xl_i, a_i\}) = \sum_{\xs = S_n/Z_n} \Tr(T^{a_{\xs(1)}} \dots T^{a_{\xs(n)}}) A_n^{\clubsuit}(\xs(1^{\xl_1}),\dots \xs(n^{\xl_n})).\ee
$A_n^{\clubsuit}(\xs(1^{\xl_1}),\dots \xs(n^{\xl_n}))$ denotes a \eml{partial amplitude}, which contains all
the kinematic information. The number $1$ is used to described momentum $k_1$. A state with $k_1, \xl_1$ is expressed as $1^{\xl_1}$\com{ (\eg \rf{(7)})}. $\xs$ is a map of the permutation \ft{This map is defined as in the following example. For example, under $\xs$, $12345$ is mapped to $23154$, then we can establish a map of $1 \to 2$, $2 \to 3$ etc. Objects in the same position are mapped to each other. In other words, we have $\xs(1) = 2$, $\xs(2) = 3$.}. $S_n$ is the set of all permutations of $n$ objects, while $Z_n$ is the subset of cyclic permutations, which preserves the trace. Note that \eml{$A_n^{\clubsuit}(\xs(1^{\xl_1}),\dots \xs(n^{\xl_n}))$ implicitly contain a sum over the cyclic permutations of
$Z_n$ and therefore are invariant under $Z_n$.}
Let me make a comment on this factorization.
As we know, the contribution (to the amplitude)
from every single diagram can be factorized.
The point is that we are talking about the total amplitude $\CM$, which is the sum of a few diagrams. So it is non-trivial to have factorization.
We will see the proof later. Roughly speaking,
we just group all the diagrams (, the sum of which includes
all the permutations and is invariant under
permutation) according to the color factor.
Each partial amplitude is the sum of diagrams
that have the same color factor.

Because of this factorization, the singularities of the partial amplitudes, poles and (in the loop case) cuts, can only occur in a limited set of momentum channels, those made out of sums of cyclically adjacent momenta. For example, $A_n^{\clubsuit}(1^{\xl_1},2^{\xl_2},3^{\xl_3}, 4^{\xl_4})$ can only have poles in $s_{12}$, $s_{23}$ but not $s_{13}$, where
$s_{ij} \equiv (k_i + k_j)^2$. The idea is that to get a pole, a single propagator
has to be cut to separate diagram into two parts. As we shall see later the color part geometrically
corresponds to a circle. Such a cut also leads to a cut in the ``color circle". In
the color circle, for this order $1234$, gluon $2$ is sandwiched between $1$ and $3$ by two lines and so does $4$ by two different lines. Therefore it is not possible to cut $1$ and $3$ away from $2$, $4$ by cutting just one propagator \ft{Lines in the color
circle correspond to propagators in the Feynman
graph.}, whether it is a gluon propagator or a quark propagator. For example, cutting the former is equivalent to cutting two lines while we need to cut four lines.
On the other hand $1$, $2$ together are connected to the rest by only two
lines, which can be cut (if these two lines are from a single gluon propagator).

\begin{proof}
Now let us give a quick derivation of \er{CSAE:1}. The procedure is easy
to illustrate using diagrams as we shall see. First of all, there are two types of objects with gauge indices $(T^a)_i{}^{\bar j}$ and $f^{abc}$. We can eliminate the structure constants $f^{abc}$ in favor of the $T^a$'s. This corresponds to a diagrammatic change of the first graph of Figure 1 in \citec{Dixon:1996wi}. Secondly, the gluon propagator provides a sum over the adjoint indices $a$.
With the use of\com{ \rf{(2)} (see \iref{ComLA}{Completion of Lie Algebra})}, 
\be{T^a}_i^j {T^a}_k^l = \xd_i^l \xd_k^j - \frac 1 N \xd_i^j \xd_k^l.\ee
the sum over adjoint indices of two $T^a$ can always be written as Kronecker delta with fundamental indices. Note that this corresponds to a diagrammatic change of the second graph of Figure 1 in \citec{Dixon:1996wi}. \eml{Quark propagators} do a sum over the fundamental indices $i,\bar j$ and \eml{provide lines in the graphs}.
Finally, the gauge indices from the external gluons are not summed and therefore we are left with $n$ $T^a$'s. Diagrammatically, we have loops (trace) with curly lines going out (see \eg Figure 2 in \citec{Dixon:1996wi}). \eml{For tree diagrams, it is not hard to see we will only have a single ``gauge loop"} or rather a single trace (see again, Figure 2 in \citec{Dixon:1996wi}). A single trace with $n$ $T^a$'s exactly corresponds to the form in \er{CSAE:1}.
\end{proof}
Similar decomposition of color part and kinematic part can be carried out for loop amplitudes. \com{See \rf{E7-E8}.}

Not all partial amplitudes are independent. For example, using parity (flipping
all helicities) and cyclic ($Z_5$) symmetry, the five-gluon amplitude has only four independent tree-level partial amplitudes\com{ \rf{(7)}}
\ba
A_5^{\clubsuit}(1^{+},2^{+},3^{+}, 4^{+},
5^+) & A_5^{\clubsuit}(1^{-},2^{+},3^{+}, 4^{+}, 5^+) \nn
\label{Dixon7}A_5^{\clubsuit}(1^{-},2^{-},3^{+}, 4^{+},
5^+) & A_5^{\clubsuit}(1^{-},2^{+},3^{-}, 4^{+}, 5^+)
\ea  
Using 4-point partial amplitude as an example,
we have, 
\be 
\el{CSAE:permu}A_4^{\clubsuit}(1^{-},2^{+},3^{+}, 4^{+}) \ne A_4^{\clubsuit}(1^{+},2^{-},3^{+}
, 4^{+}) = A_4^{\clubsuit}(2^{-},3^{+},4^{+}, 1^{+}).\ee
Although they are different, $A_4^{\clubsuit}(2^{-},3^{+},4^{+}, 1^{+})$ (and also $A_4^{\clubsuit}(3^{-},2^{+},4^{+}, 1^{+})$) can be obtained from $A_4^{\clubsuit}(1^{-},2^{+},3^{+}, 4^{+})$ (a function of $4$ momenta) by doing a momentum permutation. So eventually,
\er{Dixon7} gives all the possible inequivalent choices of helicities that give different functions (of the momenta). In fact, we will see that the first two tree partial amplitudes vanish, and there is a group theory relation
\er{Dixon8} between the last two. So there is only one independent
nonvanishing object to calculate.

Let us prove this group theory relation.
Any amplitude containing the extra $U(1)$ photon must vanish.
Hence if we substitute the U(1) generator -- the identity matrix -- into the right-hand-side of \er{CSAE:1} \ft{Note that the partial amplitude is independent of whether the gauge boson is from $SU(N)$ or from $U(1)$.}, and collect the terms with the same remaining color structure \ft{By the way, the sum
of all terms (regardless of the color structure is $0$. Therefore, each partial
sum with different color structure should vanish too.}, that linear combination of partial amplitudes must vanish. We get\com{ \rf{(8)}},
\ba
0 &=& A_n^\tree(1,2,3,\ldots,n) + A_n^\tree(2,1,3,\ldots,n)  + A_n^\tree(2,3,1,\ldots,n) 
\nonumber \\
\label{Dixon8}
&& \hskip 10mm
 + \cdots + A_n^\tree(2,3,\ldots,1,n),   \ea
In the five-point
case, we can use \er{Dixon8} to get,
\ba
A_5^\tree(1^-,2^+,3^-,4^+,5^+) &=& - A_5^\tree(1^-,3^-,2^+,4^+,5^+)
\nonumber \\
&& - A_5^\tree(1^-,3^-,4^+,2^+,5^+)
\nonumber \\
\label{Dixon9}
&& - A_5^\tree(1^-,3^-,4^+,5^+,2^+), 
\ea
which relates the partial amplitude where the two negative helicities are not adjacent to the partial amplitude where they are adjacent. Note that the decoupling of $U(1)$ boson follows from a string calculation in the sense that the necessary process has massive states as the messenger \ft{Note that we do have $gg \to g\xg$.}. Therefore in SM, where only massless states are present, the amplitude with photon vanishes.

Now what remains to be done is the computation of various independent (color-ordered
\ft{\ie with a single trace}) partial amplitudes, which are different functions of external momenta. One can follows this procedure\com{ outlined on \rf{E10}},
\begin{enumerate}
\item Draw all color-ordered graphs, \ie all planar graphs where the cyclic ordering of the external legs matches the ordering of the $T^a$ matrices in the corresponding color structure. Note that we may have contributions
from a few different graphs. Moreover, the Feynman graph to be considered
are those on the lhs of Figure 2 in \citec{Dixon:1996wi} instead of those on the rhs.
\item Evaluate each graph using the color-ordered vertices of Figure 5 in \citec{Dixon:1996wi}.
\end{enumerate}
\ \\

\subsection{Helicity Technique}
\cp{\label{subsec:HeliTech}}
\com{\sto{Helicity Technique}}
We use the notation\com{ \rf{(11)}},
\[\st{i^\pm} = \st{k_i^\pm} = u_\pm(k_i) = v_{\mp}(k_i).\quad \br{i^\pm}|
= \zt{h.c.},\]
to describe spinors with various helicities.
Note that for negative energy solutions, the helicity is the negative of
the chirality or $\xg_5$ eigenvalue. We define the basic spinor products by\com{ \rf{(12)}},
\[\br i\, j \ke \equiv \br i^- | j^+ \ke,\quad [i\, j ] \equiv \br i^+ | j^- \ke.\]
With the explicit form of $\st{i^\pm}$, we can get to the explicit form of
$\br i\, j\ke$ and $[i\, j]$\com{ \rf{(16)} (and \rf{(15)}, \rf{(17})},
\ba
\br i\, j \ke & = & \sqrt{k_i^- k_j^+} e^{i\xvp_{k_i}} - \sqrt{k_i^+ k_j^-} e^{i\xvp_{k_j}} = \sqrt{|s_{ij}|} e^{i\phi_{ij}},\nn
\el{CSAE:explicitij} [i\, j] & = & -\sqrt{k_i^- k_j^+} e^{-i\xvp_{k_i}} + \sqrt{k_i^+ k_j^-} e^{-i\xvp_{k_j}} = \sqrt{|s_{ij}|} e^{-i(\phi_{ij}+\pi)},
\ea
where $s_{ij}\ =\ (k_i+k_j)^2\ =\ 2 k_i\cdot k_j$, and  
\ba \label{Dixon15}
e^{\pm i\varphi_k}\ &\equiv &\ 
  { k^1 \pm ik^2 \over \sqrt{(k^1)^2+(k^2)^2} }
\ =\  { k^1 \pm ik^2 \over \sqrt{k^+k^-} }\ ,
\qquad k^\pm\ =\ k^0 \pm k^3.\\
\label{Dixon17}
\cos\phi_{ij}\ &=&\ { k_i^1 k_j^+ - k_j^1 k_i^+ \over \sqrt{|s_{ij}| k_i^+ k_j^+}}\ , \qquad
\sin\phi_{ij}\ =\ { k_i^2 k_j^+ - k_j^2 k_i^+ \over \sqrt{|s_{ij}| k_i^+ k_j^+}}\ .  
\ea
It is easy to see 
\[(\br i\, j \ke)^\dagger = [j\, i] = \sqrt{|s_{ij}|} e^{-i\phi_{ij}} = \sqrt{|s_{ij}|} e^{-i(\phi_{ji}+\pi)},\]
Note that these equations are consistent with the definition \er{CSAE:explicitij}.
The point is that we have $\phi_{ij} \ne \phi_{ji}$. Instead, the correct
relationship between $\phi_{ij}$ and $\phi_{ji}$ is given by its definition \er{Dixon17}, $\cos \phi_{ij} = - \cos \phi_{ji}$. There are various useful identities\com{ \rf{(18)}-\rf{(24)}},
\ba \label{Dixon18}
& \br i j \ke [ji]
 = \langle i^- | j^+ \rangle \langle j^+ | i^- \rangle
 = \Tr\bigl( \hf (1-\gamma_5) \ksl_i \ksl_j \bigr) = 2 k_i\cdot k_j = s_{ij}.\\
\label{Dixon19} & \langle i^\pm | \gamma^\mu | i^\pm \rangle\ =\ 2 k_i^\mu,\qquad\qquad
  | i^\pm \rangle \langle i^\pm |\ =\ \hf(1\pm\gamma_5)\ksl_i\\
\label{Dixon21} & \langle i^+|\gamma^\mu|j^+\rangle \langle k^+|\gamma_\mu|l^+\rangle
\ =\ 2 \, [ik]\br l j \ke \\
& \label{Dixon22} \langle i^+|\gamma^\mu|j^+\rangle \ =\  \langle j^-|\gamma^\mu|i^-\rangle \\
\label{Dixon23}
& \br ij\ke \br kl\ke\ =\ 
  \br ik\ke \br jl\ke + \br il \ke \br kj\ke.\ea
\be
\label{Dixon24} \sum_{{{i=1} \atop {i\neq j,k}}}^n [ji]\br
ik\ke\ =\ 0.
\ee
Eq.\er{Dixon18} follows from \er{Dixon19}. The first equation in \er{Dixon19} simply from symmetry argument (the only vector being $k_i^\mu$). The second equation follows from the completeness of spinor\com{ (see \eg \rf{Peskin~(3.66)})}. The spinor $\br i^\pm|$  projects out the a certain helicity (since $\br i^+|i^- \ke = [ii] = 0$ from \er{CSAE:explicitij}) and therefore is equivalent to the projection
operator $1\pm \xg_5$. Eq.\er{Dixon21} is straightforward from definition. Eq.\er{Dixon24} follows from momentum conservation and \er{Dixon19}. 

The next step is to introduce a spinor representation for the polarization
vector of a massless gauge boson with definite helicity $\pm$\com{ \rf{(25)}},
\be\el{CSAE:helicity} \xve_\mu^\pm = \pm \frac {\br q^{\mp}|\xg_\mu|k^{\mp}\ke}{\sqrt 2 \br q^\mp|k^\pm\ke}.\ee
A polarization vector defined in this way satisfies the desired properties.
We have\com{ \rf{(26)}}, 
\be\el{Dixon26} k \cdot \xve^{\pm}(k,q) = 0\ee 
following from $\slashed k \st{k^\pm} = 0$, which in turns follows from the EoM. Moreover, under a rotation around $k^\mu$, $\xve^{\pm}(k,q)$ rotates
by an appropriate phase; \ie it is rotated
by a phase twice of that appropriate for a spinor (\eg helicity $\pm\frac 1 2$). Note that only $\st{k^\pm}$ changes (by a phase of $\frac 1 2$ under
this change of coordinate while $\st{q^\pm}$ as a reference state remains fixed. There is a subtlety here. We can transform $\xve_\mu^\pm$ as if they
are vectors. This leads to a transformation on both $\st{k^\pm}$ \ft{The
transformation is again a change of phase.} and $\st{q^\pm}$. As we shall see later any change of the later corresponds to a gauge transformation. Finally, changing the reference momentum $q$ does amount to an on-shell gauge transformation, since $\xve_\mu$ shifts by an amount proportional to $k^\mu$\com{ \rf{(29)}}. Moreover, $\xve^\pm$ also satisfy\com{ \rf{(28)}} $\xve^+ \cdot \xve^+ = 0,\quad \xve^+ \cdot \xve^- = -1$.

For $\xve_i^\pm(q) \equiv \xve(k_i, q_i = q)$ \ft{This means $q_i = q$ while
$k_i$ is arbitrary. Don't be confused.} (vector indices suppressed), we have the following useful identities\com{ \rf{(31)}-\rf{(35)}},
\ba
\el{Dixon31}\xve^\pm_i(q) \cdot q & = & 0,\\
\el{CSAE:helipp}\xve^+_i(q) \cdot \xve_j^+(q) & = & \xve^-_i(q) \cdot \xve_j^-(q)= 0,\\
\el{CSAE:helipm}\xve^+_i(k_j) \cdot \xve_j^-(q) & = & \xve^-_i(k_j) \cdot \xve_j^+(q)= 0,\\
\label{Dixon34}
 {\slashed \xe}_i^+(k_j) | j^+ \rangle &=& 
 {\slashed \xe}_i^-(k_j) | j^- \rangle\ =\ 0, 
\\
\label{Dixon35}\langle j^+ | {\slashed \xe}_i^-(k_j) &=& 
\langle j^- | {\slashed \xe}_i^+(k_j)\ =\ 0. 
\ea
Using\com{ \rf{(20)}} $[i\, i]= \br i\, i \ke = 0$, \er{Dixon21} and \er{Dixon22}, \er{CSAE:helipp} and \er{CSAE:helipm} are
not hard to prove. In particular, it is useful to choose the reference momenta of like-helicity gluons
to be identical, and to equal the external momentum of one of the opposite
helicity set of gluons. For example, consider the 4-point amplitude $A_4^{\clubsuit}(1^-,2^-,3^+,4^+)$. With choice $q_1, q_2 = k_4$ and $q_3,q_4 = k_1$, we can use \er{CSAE:helipm} and \er{Dixon34} to show that only $\xve_2^- \cdot \xve_3^+$ is nonzero among the contractions of $\xve$'s. This fact is used in \com{\iref{TLSA}{Tree-Level $n$-Gluon Superstring Amplitudes}}\cp{Sec~\ref{sec:4pTSA}}. When all helicities are the same, we can choose a single reference momentum for all of them. Note that the reference momentum can not be the same as the momentum of the state, which leads to vanishing denominator ($\propto k \cdot q$) in \er{CSAE:helicity}.

With a proper choice of reference momentum and the use of the relationship \er{Dixon31}-\er{Dixon35}, the (partial) amplitudes can be simplified. Let us consider the $n$-gluon amplitude with all helicities being the same or one being opposite. They all vanish\com{ \rf{(38)}},
\be\el{Dixon38} A_n^{\clubsuit}(1^\pm,2^+,3^+,\dots n^+) = 0\ee  
Similarly, for amplitudes with two quarks\com{ \rf{(39)}}, we have \ft{Note that we switch
back to our normal convention.},
\be\el{Dixon39} A_n^{\clubsuit}(\bar q^\pm_{1},q^+_2,g_3^+,\dots g_n^+) = 0\ee  
This is the reason why we study MHV as the simplest case. Anyway, let us sketch a proof. Each non-Abelian vertex can contribute at most one momentum vector $k_i$ to the numerator of the graph, and there are at most $n- 2$ vertices ($I-V+1 = L, 3V = 2 I + N$ \ft{We only consider 3-point vertex since 4-point vertex leads to fewer number of vertices.}, $I$ being internal lines and $N$ being external). Thus there are at most most $n-2$ momentum vectors available to contract with the $n$ polarization vectors and therefore there is at least one $\xve_i \cdot \xve_j$. With the choice of reference momentum mentioned above, $\xve_i \cdot \xve_j = 0$ when there is at most one different helicity. For the case with two quarks (and $n-2$ gluons), the same graphical argument holds and we have at most $n-2$ vertices. Now we need at least one for the quarks, which offers no momentum to contract with $\xve$. There is at least one remaining $\xve$ that has to be contracted with the spinors. Note that the reference momentum (for all the $+$ gluons) can be chosen to be $k_1$. So from \er{Dixon35}, we know the contraction of $\xve$ with $\st{1^-}$ vanishes and therefore we get to \er{Dixon39}. Let us also consider the case of $A_4^{\clubsuit}(q^\pm, \bar q^\pm, g, g)$. In this case, the two quarks are at different vertices and there is no momentum to contract with. The contraction of $\xve$ with $\st{1^-}$ vanishes once the reference momenta are all chosen to be $k_1$. So the amplitude vanishes. From the viewpoint of string theory, this conclusion of $A_4^{\clubsuit}(q^\pm, \bar q^\pm, g, g) =0$ is obvious from explicit calculation \er{LST535} (with the two spinors $u_3,u_4$ having different helicities).

\ \\

\section{Four-point Tree-Level String Amplitudes}
\label{sec:4pTSA}
\com{\et{Four-point Tree-Level String Amplitudes}}
Now let us review the calculation of four-point tree-Level string amplitudes. For more details, one can refer to \citec{Lust:2008qc} and any
textbook of string theory. A scattering amplitude of strings (all in some particular states) can be calculated from the conformal field theory on the string worldsheet\com{\lab{4pTSA}}. The amplitude is equal to an $n$-point correlation function. Here we restrict ourselves to the case of 4-point amplitude on a disk, which is given by the following 4-point correlation function of vertex operators, 
\be 
\CM(\Phi^{1},\Phi^{2},\Phi^{3},\Phi^{4})=
V_{\zt{CKG}}^{-1}\int
\lb\prod_{k=1}^4 dz_k\rb\ \br V_{\Phi^{1}}(z_1)\ V_{\Phi^{2}}(z_2)\
V_{\Phi^{3}}(z_3)\ V_{\Phi^{4}}(z_4)\ke\ .
\ee
$V_{\Phi^{i}}(z_i)$ are vertex operators located on the boundary of the disk. Basically they represent the strings that participate the scattering. $V_{\zt{CKG}}$ is the volume of the conformal Killing group (generated by the conformal Killing vectors), which in this case is the M\"obius group $\zt{PSL}(2,R)$. Note that \com{\rf{Pol~E181}}this CKG can take three vertex operators to arbitrary positions, except that it does not change the cyclic ordering of the three. So we have two choices if we choose to fix $z_1, z_3$ and $z_4$:
\be\el{LST54} z_1=0\ \ \ ,\ \ \ z_3=1\ \ \ ,\ \ \ z_4=\infty\ ,\ee
and
\be\el{LST55} z_1=1\ \ \ ,\ \ \ z_3=0\ \ \ ,\ \ \ z_4=\infty\ .\ee
Depending on the value of $z_2$, we can have six partial amplitudes\com{ (see \eg \rf{Pol~I~Fig.6.2})}, each of which corresponds to an inequivalent ordering of the four vertex operators \ft{Of course, $z_1, z_3$ and $z_4$ are fixed in either choice of \er{LST54} or \er{LST55}.}. Of course, as we shall see, partial amplitudes of some ordering vanish because of the color factors.  

Let us now write down the vertex operators. In our model gauge bosons are described by open strings with both ends on the same brane while matter fermions are described by open strings with both ends on different branes. The vertex operator for the gauge boson in the ghost $-1$ picture is,
\be\el{LST57} V_{A^a}^{(-1)}(z,\xi,k) ~=~ g_{A} [T^a]^{\alpha_1}_{\alpha_2}\ e^{-\phi(z)}\ \xi^\mu\ 
\psi_\mu(z)\ e^{ik_\rho X^\rho(z)}\ ,\ee
while in the zero--ghost picture it is:
\be \el{LST58}
V_{A^a}^{(0)}(z,\xi,k)=\frac{g_{A}}{(2\ap)^{1/2}} [T^a]^{\alpha_1}_{\alpha_2}\ \xi_\mu\
[\ i\p X^\mu(z)+2\ap\ (k\psi)\ \psi^\mu(z)\ ]\ e^{ik_\rho X^\rho(z)}\ .
\ee
The open string coupling $g_A$ is related to the gauge coupling $g_{Dp_a}$ (of the low energy gauge theory) by 
\[g_A=(2\ap)^{1/2}\ g_{Dp_a}\]
Moreover, $\xi^\mu$ is the (4-dimensional) polarization vector. $X^\mu$ and $\psi^\mu$ are the bosonic and fermionic fields on worldsheet. $[T^a]^{\alpha_1}_{\alpha_2}$ is the Chan-Paton factor. The fundamental indices $\alpha_1, \alpha_2$ denote the branes on which the open string ends. Note that they are branes in the same stack. The adjoint index $a$ denotes the element in Lie algebra this gauge field corresponds to. We won't get into too much details about the different pictures of vertex operators. The ghost number can be identified as the power of $e^\phi$ and the total ghost number is restricted to $-2$. 

The vertex operators for the quarks and leptons are given by 
\ba
V^{(-1/2)}_{\psi^{\alpha}_{\beta}}(z,u,k)&=g_\psi [T^{\alpha}_{\beta}]_{\alpha_1}^{\beta_1}e^{-\phi(z)/2}\ u^{\lambda}
S_{\lambda}(z)\ \Xi^{\scriptscriptstyle a\cap b}(z)\ e^{ik_\rho X^\rho(z)}\ ,\\
V^{(-1/2)}_{\bar\psi^{\beta}_{\alpha}}(z,\bar u,k)&=g_\psi [T_{\alpha}^{\beta}]^{\alpha_1}_{\beta_1}
e^{-\phi(z)/2}\ \bar u_{\dot\lambda} S^{\dot\lambda}(z)\ \overline\Xi^{\scriptscriptstyle a\cap b}(z)\ e^{ik_\rho X^\rho(z)}\ .
\ea
$S_\xl$ is the spin field (vertex operator for the Ramond ground state). Note that the vertex operator for the ground state is always the product of a few twisted fields, each of which is determined by the boundary condition of a pair of dimensions. This operator $S_\xl$ is constructed as if there is only four dimensions. In other words, the only fermionic fields
(or rather their bosonization) involved are $\psi^\mu$ ($\mu = 0,\dots, 3$). The remaining worldsheet fermionic fields $\psi^m$ are used to construct the fermionic twisted fields $s$. This field $s$, together with the bosonic twisted fields $\xs$, forms the boundary changing operator $\Xi^{\scriptscriptstyle a\cap b}$,
\be \Xi^{\scriptscriptstyle a\cap b}=\prod_{j=1}^3\sigma_{\theta^j_{ba}}\ s_{\theta^j_{ba}}\ \ \ ,\ \ \
\overline\Xi^{\scriptscriptstyle a\cap b}=\prod_{j=1}^3\sigma_{-\theta_{ba}^j}\ s_{-\theta_{ba}^j}\ .\ee
The twisted field $s$ is again the vertex operator for the Ramond ground state. But now the open string is extended between two intersecting branes $a,b$ characterized by three angles $\xt_{ba}^j$. Each angle determines the boundary conditions of two pairs of fields
$\psi^m$ and $X^m$ ($m = 4,\dots 9$). The twisted fields $s$ can be bosonized as usual, 
\be s_{\theta^j}=e^{i(\theta^j-\hf)H^j}\ \ \ ,\ \ \
s_{-\theta^j}=e^{-i(\theta^j-\hf)H^j}, \ee
and their correlation functions are not hard to figure out. Anyway, the explicit forms of the correlators of the fermionic twist and bosonic twists can be found in \citec{Klebanov:2003my}. 

Now we are ready to evaluate the $n$-point correlation functions of vertex operators. In general we can compute the correlation functions of various fields that compose the vertex operators. One can refer to standard textbook of string theory for more details. Here we will go through some of the steps of the calculation of 4-gluon amplitude just
to give the reader a general taste. 

\ \\

\com{\sto{Helicity Form Factor}}
\subsection{Helicity Form Factor}
\label{subsec:HFF}
We shall see that the stringy 4-point {\it partial} amplitude for a maximally helicity violation (MHV) configuration indeed has the following form factor\com{ \rf{(2)}},
\be\el{heliformMHV} A(1^-,2^-, 3^+, 4^+) \sim \frac {\br 12\ke^4}{\br 12 \ke\br23\ke\br34\ke\br 41\ke}.\ee
The stringy 4-gluon amplitude is given by the string vertex operators \com{\rf{(12.3.45)} (or \rf{(13)} and \rf{(14)})}\cp{\er{LST57} and \er{LST58}} \ft{We use the convention $2\xa' = 1$.},
\[\br c(z_1)e^{-\phi(z_1)}\psi^\mu(z_1)c(z_2)[\p X^\nu (z_2)+i k_2 \cdot \psi \psi^\nu(z_2)]c(z_3)[\p X^\rho (z_3)+i k_3 \cdot \psi \psi^\rho (z_3)]e^{-\phi(z_4)}\psi^\xs(z_4)\ke\]
Now we use the following correlation functions of the various fields\com{ \rf{(12.4.2a)}}
\[\br c(z_1) c(z_2) c(z_3)\ke = z_{12}z_{13} z_{23},\]
and\com{ \rf{(12.4.2b)}}
\[\br e^{-\phi(z_1)} e^{-\phi(z_4)} \ke = \frac 1 {z_{14}}.\]
Both simply follow from their conformal dimensions.

There are quite a few terms to evaluate. With the choice of $q_1 = q_2 = k_4$ and $q_3 = q_4 = k_1$ and the help of \er{Dixon31}-\er{Dixon35},
we can simplify the amplitude by a lot. When we have both $\p X$, we have $\eta^{\mu\xs}$ from the contraction of the two $\psi$'s and this implies,
\[\br\psi^\mu(z_1)\p X^\nu (z_2)\p X^\rho (z_3)\psi^\xs(z_4)\ke \propto \eta^{\mu \xs} \sim \xve_1^- \cdot \xve_4^+ = 0.\]
So this term does not give an contribution to the MHV amplitude. When there
is only one $\p X$, we have
\[\br\psi^\mu(z_1)[k_{2}\cdot \psi \psi^\nu(z_2)]\p X^\rho (z_3)\psi^\xs(z_4)\ke \propto k_2^\mu\eta^{\nu \xs} \sim \xve_2^- \cdot \xve_4^+ = 0.\]
Note that only $k_2 \cdot \psi$ is contracted with $\psi^\mu$ in the above
expression. Otherwise we will have a $\eta^{\mu\nu}$, which gives to $0$ since $\xve_1^-
\cdot \xve_2^- = 0$.
The contribution from the term
\[\br \psi^\mu(z_1)\p X^\nu (z_2)[k_{3}\cdot \psi \psi^\rho(z_3)]\psi^\xs(z_4)\ke\]
also vanishes. The only nonvanishing term is,
\[\psi^\mu(z_1)[k_2 \cdot \psi \psi^\nu(z_2)][k_3 \cdot \psi \psi^\rho (z_3)]\psi^\xs(z_4)\propto
k_2^\mu \eta^{\nu \rho} k_3^\xs \sim k_2 \cdot \xve_1^- (\xve_2^- \cdot \xve_3^+)(k_3
\cdot \xve_4^+).\]
The formalism we introduced in \er{subsec:HeliTech} implies,
\ba \xe_2^- \cdot \xe^+_3 & = & \lb -\frac {\br 4^+ |\xg^\mu|2^+\ke}{\sqrt 2 \br 4^+| 2^-\ke} \rb\lb \frac {\br 1^- |\xg_\mu|3^-\ke}{\sqrt 2 \br 1^-| 3^+\ke} \rb\nn
& = & -\frac {[43]\br 1 2 \ke}{[42]\br 1 3 \ke},\ea
which follows from \er{Dixon21} and \er{Dixon22},
and
\ba \xe_1^- \cdot k_2 & = & \lb -\frac {\br 4^+ |\xg^\mu|1+\ke}{\sqrt 2 \br 4^+|1-\ke} \rb\lb \frac {\br 2^- |\xg_\mu|2^-\ke}{2} \rb\nn
& = & -\frac {[42]\br 2 1 \ke}{\sqrt 2 [41]},\ea
which follows from \er{Dixon19} and \er{Dixon21}. Then the form factor can be expressed as\com{ \rf{Dixon~(40)}},
\ba
\lb \xve_2^-\cdot\xve_3^+ \rb
   \lb \xve_1^-\cdot k_2 \rb
   \lb \xve_4^+\cdot k_3 \rb &=& 
  \left(-{2\over2}{[43]\br 12 \ke \over [42]\br 13\ke}\right) 
  \left(-{[42]\br21\ke\over\sqrt{2}[41]}\right)
  \left(+{\br 1 3\ke [34]\over\sqrt{2}\br 14\ke}\right)
\nonumber \\
\label{Dixon40}
&=& \frac {s_{12}} 2 \, \frac{\br 12 \ke{[34]}^2} {[12]\br 14\ke [14]}.
\ea
\er{Dixon40} can be further simplified to be \er{heliformMHV}\com{ \rf{Dixon~(42)}}.

\ \\

\subsection{Explicit Forms of Four-point Tree-Level String Amplitudes}
\cp{\label{subsec:EF4TLSA}}
\com{\sto{Explicit Forms of Four-point Tree-Level String Amplitudes}}
String amplitude can be expressed as a sum of partial amplitudes \com{\rf{S\&Tb~(1)}}\cp{\cite{Stieberger:2006te}} just as in field theory. Each partial amplitude can be understood as a particular order of the vertices on the boundary of the disk.
As in the case of field theory, it is easier to calculate the string amplitude with a particular choice of helicities. For $4$-gluon amplitude, if there is at most one different helicity, the amplitude vanishes. This follows from the result of field theory\com{ \iref{CSAE}{Calculating Scattering Amplitudes Efficiently}}\cp{ (Sec~\ref{sec:CSAE})} since string partial amplitude is proportional to that of the field theory\com{ \iref{TLSA}{Tree-Level $n$-Gluon Superstring Amplitudes}}\cp{ \citec{Stieberger:2006te}}. For amplitudes with two quarks and all gluons of the same helicity, the field theory amplitude vanishes. It is not clear whether the same proportionality (between string and field theory amplitudes) holds. But we can get
to this conclusion from the kinetic factor \er{LST535}. So we are left with MHV amplitudes.

Let us start with 4-gluon amplitude. The MHV partial amplitude is given by
\ft{For the derivation of the helicity form factor, see Appendix~\ref{subsec:HFF}.},
\be A(g^-_1,g^-_2, g^+_3, g^+_4) ~=~ 4\, g^2\, {\rm Tr}
 \, (\, T^{a_1}T^{a_2}T^{a_3}T^{a_4})\ {\langle 12\rangle^4\over
   \langle 12\rangle\langle 23\rangle\langle 34\rangle\langle
   41\rangle}\ \hat V(k_1,k_2,k_3,k_4)\ , \ee
where the Veneziano formfactor is defined by \er{formf}. It follows from the integration of the unfixed coordinate $z_2$. The total amplitude is the sum of the partial amplitudes, 
\be{ {\cal M}(g^-_1,g^-_2,g^+_3,g^+_4)
 =4\,g^{2}\langle 12\rangle^4 \sum_{\sigma\in S_4/Z_4 } { {\rm Tr} \, (\,
   T^{a_{1_{\sigma}}}T^{a_{2_{\sigma}}}T^{a_{3_{\sigma}}}T^{a_{4_{\sigma}}})\
   \hat V(k_{1_{\sigma}},k_{2_{\sigma}},k_{3_{\sigma}},k_{4_{\sigma}})\over\langle
   1_{\sigma}2_{\sigma} \rangle\langle
   2_{\sigma}3_{\sigma}\rangle\langle 3_{\sigma}4_{\sigma}\rangle \langle
   4_{\sigma}1_{\sigma}\rangle }\ .}\ee
The explicit expression is,
\ba \el{TotalAmp531}
{\cal M}(g^-_1,g^-_2, g^+_3, g^+_4) = 8\, g^2\langle 12\rangle^4\, \times \quad & \nn
\left\{\  {\hat V_t\over\langle 12\rangle\langle
     23\rangle\langle 34\rangle\langle 41\rangle} \rc \lsb d^{a_1a_2a_3a_4}+{1\over 12}\ \lb f^{a_1a_4n}f^{a_2a_3n}-
f^{a_1a_2n}f^{a_3a_4n} \rb \rsb & \nn
 +{\hat V_s\over\langle 14\rangle\langle
     42\rangle\langle 23\rangle\langle
     31\rangle}\ \lsb\ d^{a_1a_2a_3a_4}+{1\over 12}\ ( f^{a_2a_4n}f^{a_3a_1n}-
f^{a_2a_3n}f^{a_1a_4n} )\ \rsb & \nn
  +{\hat V_u\over\langle 13\rangle\langle
     34\rangle\langle 42\rangle\langle
     21\rangle}\ \lc \lsb\ d^{a_1a_2a_3a_4}+{1\over 12}\ ( f^{a_3a_4n}f^{a_1a_2n}-
f^{a_3a_1n}f^{a_2a_4n} )\ \rsb\ \right\} &,
\ea
where the totally symmetric symbols $d^{a_1a_2a_3}$ and $d^{a_1a_2a_3a_4}$ are defined by 
\ba
d^{a_1a_2a_3} & = & \zt{STr}(T^{a_1}T^{a_2} T^{a_3}) \equiv \frac 1 {2} \lsb \Tr (T^{a_1}T^{a_2} T^{a_3}) + \Tr (T^{a_1} T^{a_3} T^{a_2})\rsb,\\
d^{a_1a_2a_3 a_4} & = &\zt{STr}(T^{a_1}T^{a_2} T^{a_3} T^{a_4}) \nn
& \equiv & \frac 1 {6} \lsb \Tr (T^{a_1}T^{a_2} T^{a_3} T^{a_4}) + \zt{permutation of } a_2,\ a_3,\ a_4\rsb.
\ea
Note that $d^{abc}$ is related to $C^{abc}$
in Chapter~\ref{LHCPheno} by $d^{abc} = \frac 1 4 C^{abc}$. \ft{$d^{abc}$ is defined as the totally symmetric trace and it goes with the $1/n!$ just like the usual bracket of ``[]" and $\{\}$. So effectively, $d^{a\dots b}$ only has one term (because of the averaging). On the other hand, $C^{abc}$, although being totally symmetric, have effectively, $4$ terms (a factor $2$ in the definition included). This difference leads to a factor of $4$ mentioned above. 

By the way, $\CC^{abcd}$ in \er{groupfacs} is not really totally symmetric.} We simplify the Veneziano
form factor \er{formf} by \er{formfsym}. The traces of matrices $T^a$ are simplified using
\ba
{\rm
   Tr}(T^{a_1}T^{a_2}T^{a_3}T^{a_4})\ ~= & d^{a_1a_2a_3a_4}+{i\over 2}(d^{a_1a_4n}f^{a_2a_3n}-
d^{a_2a_3n}f^{a_1a_4n})\nn & +{1\over 12}( f^{a_1a_4n}f^{a_2a_3n}-f^{a_1a_2n}f^{a_3a_4n}).
\ea
Note that $d^{abc}$ follows from the anti-com of two $T^a$ while $f^{abc}$ is the usual structure constant. The amplitude $\CM$, which is the sum of all partial amplitudes (with group factors), is \eml{invariant under any permutation.} The other amplitudes can be obtained by permutation and complex conjugate (corresponding to the amplitude with all helicities reversed). For example, the amplitude $\CM(g_1^-,g_2^+,g_3^-,g_4^+) = \CM(g_1^-,g_3^-,g_2^+,g_4^+)$ follows from \er{TotalAmp531} with $2\leftrightarrow 3$ \com{(see \eg \rf{DHHS~(2.15)} \rf{DHHS~(2.18)} although they are amplitudes near the pole)}. Note that on the other hand, the partial amplitude is only invariant under the cyclic permutation. So there are two independent 4-point partial amplitudes: \com{\rf{PS~(8)}} $A(g^+_1,g^-_2, g^-_3, g^+_4)$ and $A(g^+_1,g^-_2, g^+_3, g^-_4)$. 

The amplitude with two gluons and two fermions can be evaluated following the same procedure. We can consider the 4-point function \ft{Let us digress a little
bit to discuss the Chan-Paton like $[T^{\alpha}_{\beta}]_{\alpha_1}^{\beta_1}$
(in $V^{(-1/2)}_{\psi^{\alpha}_{\beta}}$) when there are two or more stacks of branes. For example, if we have $N_a$ branes in stack $a$ and $N_b$ in stack $b$. The Chan-Paton factor should be a $N_a + N_b$ dimensional matrix. This is in fact obvious when the two stacks overlap. The $U(N_a)$ gauge field has a Chan-Paton factor of a rank-$(N_a + N_b)$ matrix with only a rank-$N_a$ submatrix nonzero. An immediate conclusion is that a disk amplitude with $X$ ($U(1)$ of stack $b$) and $C$ ($U(1)$ of stack $c$) insertion is $0$.}
\[\br {V_{A^{x}}^{(0)}(z_1,\xi_1,k_1)\ V_{A^{y}}^{(-1)}(z_2,\xi_2,k_2)\
V^{(-1/2)}_{\psi^{\alpha_3}_{\beta_3}}(z_3,u_3,k_3) \ V^{(-1/2)}_{\bar\psi^{\beta_4}_{\alpha_4} }(z_4,\bar u_4,k_4)}\ke\ .\]
Let us only consider the case in which both gauge bosons are associated with one stack ($(x,y) = (a_1, a_2)$. In this case, the only possible (inequivalent) choice of $z_1, z_3, z_4$ is \er{LST54}. There are two allowed ranges for $z_2$: $z_2<0$ or $0<z_2<1$.
\ba
& \CM[A^{a_1}(\xi_1,k_1)A^{a_2}(\xi_2,k_2)\psi^{\alpha_3}_{\beta_3}(k_3,u_3)
\bar\psi^{\beta_4}_{\alpha_4}(k_4,\bar u_4)]=-2\ \ap\ g_{Dp_a}^2\ \CK\nn
& \times \lsb\Tr(T^{a_1}T^{a_2}T^{\alpha_3}_{\beta_3}
T^{\beta_4}_{\alpha_4})\ B(s,u)+\Tr(T^{a_2}T^{a_1}T^{\alpha_3}_{\beta_3}
T^{\beta_4}_{\alpha_4})\ \frac{t}{u}\ B(s,t)\rsb\ ,
\ea
where the kinematic factor:
\ba
\CK=\left\{\lsb k_{1\rho}\ (\xi_1\xi_2)-\xi_{1\rho}\ (\xi_2k_1)+\xi_{2\rho}\
(\xi_1k_2)-\frac{s}{t}\ \xi_{2\rho}\ (\xi_1k_3)\rsb\,
(u_3\sigma^{\rho} \overline u_4)\rc\nn
\el{LST535}\hskip2cm\lc-\hf\frac{s}{t}\ k_{1\lambda}\ \xi_{1\mu}\
\xi_{2\rho}\ (u_3\sigma^\lambda\overline\sigma^{\mu}\sigma^\rho \overline u_4)\right\}\ .
\ea
We have shown\com{ \iref{CSAE}{Calculating Scattering Amplitudes Efficiently}}\cp{ in Sec~\ref{sec:CSAE}} that the two gluons have to be opposite in helicities. \eml{The helicities of the two fermions have to be opposite too.} This follows from the conservation of twist charges\com{ \rf{LST~E29b}}. We can also understand this from the view point of field theory, in which the contraction of the two spinors has to be proportional to $\bar u_+(3) \xg^\mu v_+(4) = 0$ (for $q_3^+,\bar q_4^+$ outgoing). The only non-vanishing amplitude is $\CM(g^{-}_1,g^{+}_2,q_3^-,\bar q_4^+)$. After some extra work, we can figure out its explicit form\com{ \rf{(5.39)}},
\be
\CM(g^{-}_1,g^{+}_2,q_3^-,\bar q_4^+)~=~
2\, g^2\, \delta^{\beta_4}_{\beta_3}\,{\langle 13\rangle^2\over \langle
23\rangle\langle 24\rangle}\
\Big[(T^{a_1}T^{a_2})^{\alpha_3}_{\alpha_4}\ {t\over s}\ \hat V_t+
(T^{a_2}T^{a_1})^{\alpha_3}_{\alpha_4}\ {u\over s}\ \hat V_u\Big]\ ,
\ee

\ \\

\subsection{Resonance Scattering Amplitudes}
\com{\sto{Resonance Scattering Amplitudes}}
The helicity form factor like \er{mhvsapp} appears to be annoying. Usually we want something that can be expressed in terms of the Mandelstam variables. In fact, it is claimed that the partial amplitudes can be expressed in $u,t,s$ \com{\rf{PS~(8)} (or \rf{DHHS~(2.15)})},
\be A(g^+_1,g^-_2, g^-_3, g^+_4) = -4 \frac t s,\quad A(g^+_1,g^-_2, g^+_3, g^-_4) = - 4 \frac {u^2}{s t}.\ee
So let us take one more step prove that is also the case for the total amplitude \er{TotalAmp531}. The point is that \eml{all three terms in \er{TotalAmp531} have the same phase.} The phase is more of a gauge choice and can not be determined from the momentum invariants. The first line in \er{TotalAmp531} can be simplified as
\[\frac {\br 12\ke^4}{\br 12 \ke\br23\ke\br34\ke\br 41\ke} = \frac {-s_{12}\br 12\ke^2}{[12]\br23\ke\br34\ke\br 41\ke} = -\frac {s_{12}\br 12\ke^2}{s_{14}\br34\ke^2 } = \frac {s\br 12\ke^2}{u\br34\ke^2 }\]
We multiply numerator and denominator by $[12]$ and use \com{\rf{Dixon~(24)}} $[12]\br23\ke = -[14]\br 4 3 \ke$. The second line can be written as,
\[\frac {\br 12\ke^4}{\br 14 \ke\br42\ke\br23\ke\br 31\ke} = \frac {s_{12}^2\br 12\ke^2}{\br 14 \ke\br43\ke[31][41]\br43\ke\br 31\ke} = -\frac {s^2 \br 12\ke^2}{u t\br34\ke^2 }\]
The last line can be evaluated similarly. One can see all of them are proportional to the phase $\br 12 \ke^2/\br 34\ke^2$, which as an overall phase has no physical effect.

The resonance amplitudes $\CM$ for 4-gluon scattering are given by \com{\rf{AGT~(6)}}, 
\begin{equation}
{\cal
    M}(g^-_1,g^-_2,g^+_3, g^+_4) \to 4\, g^2\,\makebox{Tr}(\{T^{a_1},T^{a_2}\}\{T^{a_3},T^{a_4}\})\frac{\langle
  12\rangle^4}{\langle 12\rangle\langle
      23\rangle\langle 34\rangle\langle
  41\rangle}\frac{u}{s-M^2}\ ,\label{mhvsapp}
  \end{equation}
and \com{\rf{AGT~(6)}}
\begin{equation}
{\cal
    M}(g^-_1,g^+_2,g^+_3, g^-_4) \to 4\, g^2\,\makebox{Tr}(\{T^{a_1},T^{a_2}\}\{T^{a_3},T^{a_4}\})\frac{\langle
  14\rangle^4}{\langle 12\rangle\langle
      23\rangle\langle 34\rangle\langle
  41\rangle}\frac{u}{s-M^2}\ .\label{mhvoapp}\end{equation}
The resonance amplitude for 2-gluon, 2-quark scattering is given by \com{\rf{AGT~(8)}}
\begin{equation}
{\cal
    M}(q^-_1,\bar q^+_2,g^-_3, g^+_4) \to
2\,g^2 \{T^{a_3},T^{a_4}\}_{\alpha_1\alpha_2}\,
\frac{\langle 13\rangle^2}{\langle 14\rangle\langle 24\rangle}\frac{tu}{M^2(s-M^2)}\ .
\label{qlimapp}\end{equation}

In \er{mhvo}, we can see that the gauge factor $\Tr(\{T^{a_1},T^{a_2}\}\{T^{a_3},
T^{a_4}\})$ and the last factor $u/(s-M^2)$ do not follow from the permutation
$2 \leftrightarrow 4$ of \er{mhvsapp}. In fact, only total amplitudes like \com{\rf{AGT~(1)}} \er{TotalAmp531}
are related by the permutation. The resonance amplitude \er{mhvsapp} of $\CM(g_1^-, g_2^-, g_3^+, g_4^+)$ comes from the first and third line of \er{TotalAmp531}
and the gauge factor is
\ba\el{gaugefactor} \Tr(T^{a_1} T^{a_2} T^{a_3}
T^{a_4} + T^{a_2} T^{a_1} T^{a_4}
T^{a_3} + T^{a_2} T^{a_1} T^{a_3}
T^{a_4} + T^{a_1} T^{a_2} T^{a_4}
T^{a_3}) \nn
= \Tr(\{T^{a_1},T^{a_2}\}\{T^{a_3},
T^{a_4}\}).\ea
On the other hand, $\CM(g_1^-,  g_4^-, g_3^+, g_2^+) = \CM(g_1^-,g_2^+, g_3^+, g_4^-)$ (whose resonance amplitude is \er{mhvo}) can be obtained from
the permutation. The total amplitude is obtained by the permutation (under $s \leftrightarrow u$, $2 \leftrightarrow 4$) of \er{TotalAmp531} \com{(or rather the permutation of \rf{DHHS~(2.15)})}, 
\ba \CM(g_1^-,  g_4^-, g_3^+, g_2^+) & = & 4 g^2 \left[\frac u s V_t \Tr (T^{a_1} T^{a_4} T^{a_3}
T^{a_2} + T^{a_4} T^{a_1} T^{a_2}
T^{a_3}) + \frac u t V_s \Tr (T^{a_4} T^{a_1} T^{a_3}
T^{a_2} \rc \nn
& + & \left.  T^{a_1} T^{a_4} T^{a_2}
T^{a_3}) + \frac {u^2}{t s} V_u \Tr(T^{a_1} T^{a_3} T^{a_4}
T^{a_2} + T^{a_3} T^{a_1} T^{a_2}
T^{a_4}) \rsb
\ea
Near the pole, the contribution is from the first and third line.
The gauge factor becomes,
\ba \Tr(T^{a_1} T^{a_4} T^{a_3}
T^{a_2} + T^{a_4} T^{a_1} T^{a_2}
T^{a_3} + T^{a_1} T^{a_3} T^{a_4}
T^{a_2} + T^{a_3} T^{a_1} T^{a_2}
T^{a_4}) \nn
= \Tr(\{T^{a_1},T^{a_2}\}\{T^{a_3},
T^{a_4}\}).\ea
The rest of the resonance amplitude (like the factor of $u/(s-M^2))$ can also
be worked out without too much difficulty. Note that near the pole, we have
\begin{equation}
V_t \to \frac{u}{s-M^2}~,\qquad V_u \to \frac{t}{s-M^2}~.
\end{equation}

As explained above, the resonance scattering amplitudes can be further simplified as \com{\rf{PS~(9)} \rf{PS~(10)}},
\be \CM(g_1^-, g_2^-, g_3^+, g_4^+) = - 2 g^2 \frac {s}{s- M_s^2} \CC^{1234}\ee
\be \CM(g_1^-, g_2^+, g_3^-, g_4^+) = - 2 g^2 \frac {u^2} s \frac 1 {s- M_s^2} \CC^{1234}\ee
There is another independent amplitude $\CM(g_1^-, g_2^+, g_3^-, g_4^+)$, which is related to $\CM(g_1^-, g_2^+, g_3^+, g_4^-)$ by a permutation of $u$ and $t$. This can be easily seen from the $t^4+u^4$ factor in total square amplitude \er{ggggpole}.

Note that even for higher resonances, the pattern remains the same. We have contributions from two of the three terms in $\CM$ and the momentum factors (depending on $u,t,s$) are the same for these two terms. The gauge factor can be combined into a form similar to \er{gaugefactor} \ft{Gauge factors for different amplitudes do not follow from permutation.}. It appears that for even levels (\eg $n=2$) we have commutator instead of the anti-commutator in \er{gaugefactor}. \com{See \eg \rf{DHSS~(2.15)}.}Moreover, resonance amplitudes from $\CM(g_1^-,  g_4^-, g_3^+, g_2^+)$ and those from $\CM(g_1^-, g_2^+, g_3^-, g_4^+)$ are related by $t \leftrightarrow u$ \ft{As we can see, the gauge factor is invariant under $3 \leftrightarrow 4$.}.

\ \\

\subsection{Cross Sections}
\com{\sto{Cross Sections}}
To obtain the cross section, we need to square the amplitude (and sum over the colors). For later convenience, we present the color sum here. We now derive the cross section\com{ \rf{(6.14)}},
\ba \el{cross4gluon}
|{\cal
   M}(g^-_1,g^-_2,\ g^+_3, g^+_4)|^2 =~  g^4(N^2-1)\,s^4\, \nn 
    \times \Bigg[2N^2\left({V^2_t\over s^2u^2}+ {V^2_s\over t^2u^2}
+{V^2_u\over s^2t^2}\right)
+{4(-N^2+3)\over N^2}\left({V_t\over s\,u}
+ {V_s\over tu}+{V_u\over s\,t}\right)^{\! 2}\Bigg]\ea
from \er{TotalAmp531} using the following useful results for the color factors,
\ba
\sum_{a_1,a_2,a_3}d^{a_1a_2a_3}d^{a_1a_2a_3}&  = & {(N^2-1)(N^2-4)\over 16 N}\ ,\\
\sum_{a_1,a_2,a_3,a_4}d^{a_1a_2a_3a_4}d^{a_1a_2a_3a_4}& = & {(N^2-1)(N^4-6N^2+18)\over 96 N^2}\ ,\\
\el{coloridn3}\sum_{a_1,a_2}f^{i_1a_1a_2}f^{i_2a_1a_2}&  = &  N\ \delta^{i_1i_2}\
,\\
\sum_{a_1,a_2,a_3}f^{i_1a_1a_2}f^{i_2a_2a_3}f^{i_3a_3a_1}& = & {N\over 2}f^{i_1i_2i_3}\
,
\ea
Obviously the contraction between $d^{abcd}$ and $f^{abc}$ vanishes. The contraction between $f$'s in the first line and those in the second line (, which is the coefficient of $\hat V_t \hat V_s$ \com{\ft{(?) No hat in \rf{LST} though. This is just a convention difference.}}) is 
\[-\frac 1 {12^2} \lb f^{a_1 a_4 n} f^{a_2 a_3 n} - f^{a_1 a_2 n} f^{a_3 a_4 n}\rb f^{a_2 a_3 m} f^{a_1 a_4 m} \equiv -\frac 1 {12^2} (A + B),\]
where
\[A = f^{a_1 a_4 n} f^{a_2 a_3 n} f^{a_2 a_3 m} f^{a_1 a_4 m} = N^2 (N^2-1),\]
where we use the equation \er{coloridn3}. Similarly, we have
\[B = - f^{a_1 a_2 n} f^{a_3 a_4 n} f^{a_2 a_3 m} f^{a_1 a_4 m} =  f^{a_1 a_2 n} f^{a_4 n a_3} f^{ m a_3 a_2 } f^{a_1 a_4 m}  = \frac N 2  f^{a_1 a_4 m} f^{a_1 a_4 m} = \frac {N^2(N^2-1)} 2.\]
The contraction between the $f$'s in the first line and itself (corresponding to coefficient of $\hat V_t^2$) is
\[\frac 2 {12^2} \lb f^{a_1 a_4 n} f^{a_2 a_3 n} f^{a_1 a_4 m} f^{a_2 a_3 m} + f^{a_1 a_4 n} f^{a_2 a_3 n}f^{a_1 a_2 m} f^{a_3 a_4 m} \rb= \frac 2 {12^2} (A + B).\]
The coefficient of the second term in \er{cross4gluon} (following from the coefficient of the cross product $\hat V_t \hat V_s$) is \com{\ft{For the last line, see mathnb \rf{SU(N) invariants.nb}.}},
\ba 64\lb d^{abcd} d^{abcd} + \frac 1 {12^2} (A + B) \rb &= & 64 \lb\frac {(N^2 - 1)(N^2 - 6N^2 + 18)}{96N^2} + \frac {3N^2(N^2-1)} {2 \cdot 144}\rb \nn
& = & - \frac{4(3+N^4 - 4 N^2)}{N^2},\ea
which is exactly what is in \er{cross4gluon}. Note that we are left with terms $\hat V_t^2/(s^2 u^2) + \dots$ with the coefficient
\[64 \lb\frac 3 {12^2} (A + B) \rb= 2 N^2(N^2-1),\]
which is the coefficient of the first term in \er{cross4gluon}.

\newpage
\thispagestyle{fancy}
\chapter{Calculations Relevant to String Amplitudes in Randall-Sundrum Background}
\label{sec:CalcRS}
\thispagestyle{fancy}
\pagestyle{fancy}

\section{Four Dimensional Spin-0 Sector from
Five Dimensional Spin-2 Field}
\label{sec:spinzero}
\com{\et{Notes on $J=0$ Sector of $B_{MN}$}}
We have a 5d scalar and a 5d spin-2 \ie $(1,1)$ rep of $\zt{SO}(4) \cong \zt{SU}(2) \times \zt{SU}(2)$ in the 5d LEEA of string
theory. Now the vertex of the lowest massive excitation in the $-1$ picture\com{ (see \eg \iref{vertexpicture}{Vertex Operators in Different Pictures})}
is given by \com{\rf{Liu~(2.2a)}}\cp{\citec{Liu:1987tb}}, 
\be V^{-1}_{(1)} = (\xe_{\mu\nu\rho} \psi^\mu
\psi^\nu \psi^\rho + \xS_{\mu\nu} \p X^\mu
\psi^\nu ) e^{i k \cdot X},\ee
where $k^\nu \xS_{\mu\nu} = 0$ and $\xS_\mu{}^\mu
=0$, and $k^\mu \xe_{\mu\nu\rho} + \frac
1 2 (\xS_{\nu\rho}-\xS_{\rho\nu}) = 0$.
Both the 5d scalar and 5d spin-2 tensor can be described by the string excitation $S_{MN} \p X^M \psi^N$ \ft{$i,j$ range from $1$ to $4$.}, where $S_{MN}$ is the symmetric polarization tensor. The spin-2 sector gives a 4d scalar under dimensional reduction. The corresponding $S_{MN}$ (a $10\times 10$ tensor) for this state (4d scalar) is given by
\[g^*_{2,0}:\quad S_{MN}^{J=2} = \frac {\sqrt 3}  2 \bay{cccccc} 0 & 0& 0& 0 & 0 &\\ 0& \frac 1 3 & 0 & 0 & 0 & \\0& 0 & \frac 1 3 & 0 & 0 & \\0& 0 & 0 & \frac 1 3 & 0 & \\0& 0 & 0 & 0 & -1 & \\ & & & & & 0\eay \]
Note that it is traceless. On the other hand, the $S_{MN}$ corresponding to the 5d scalar is the diagonal matrix
\[g^*_{0,0}:\quad S_{MN}^{J=0} = \bay{ccc} 0 &
0 & 0 \\ 0 & \frac {\sqrt{5}} 6 I_4 & 0 \\ 0 & 0 & -\frac 2 {3\sqrt 5} I_5\eay.\]

Let us now consider the amplitude of $gg\to g^*_{0,0}$ and $gg\to g^*_{2,0}$. We restrict the gluons to have only 4-momentum and the polarization vector is also 4d ($\xe^4 = 0$). Note that from the calculation in\com{ \iref{TLSA}{Tree-Level $n$-Gluon Superstring Amplitudes}}\cp{ Sec~\ref{sec:4pTSA}}, it is clear that $S_{44}$ does not give any contribution. So the amplitude is determined by the 3d part ($i,j = 1,2,3$). Both $S_{ij}$ are diagonal in that part. The ratio is $\sqrt{\frac 5 { 3}}$. In other words, we should have
\[|\CM(gg\to g^*_{0,0})|^2 = \frac 5 3 |\CM(gg\to g^*_{2,0})|^2.\]
Moreover, a 4d scalar generated by a $S_{MN} = \zt{diag} (0, \frac {\sqrt 2} 3 I_3, -\frac
1 {3\sqrt 2} I_6)$ with the only nonvanishing components in the range of $i,j = 1,2,3$ is supposed to be responsible for the total $J = 0$ contribution to $gg \to gg$. Note that following the same analysis, we can reach the conclusion that
\[|\CM(\zt{4d scalar})|^2 =\frac 8 3 |\CM(g^*_{2,0})|^2 = |\CM(g^*_{2,0})|^2 + |\CM(g^*_{0,0})|^2.\] 

The contribution to the $gg \to gg$ amplitude by $g^*_{2,0}$ can be read off from the propagator (see Appendix~\ref{subsec:propMF5d}) of the 5d field $B_{MN}$ and the interaction term \er{gF2}\com{
\rf{PS~(49)}}. The 4d components of the 5d propagator includes both the 4d graviton propagator $G_{\mu\nu, \xa \xb}$ and the contribution from $g^*_{2,0}$. However, the contribution by $g^*_{2,0}$ is proportional to $\eta_{\mu\nu} \eta_{\xa \xb}$. Since $T_\mu{}^\mu = 0$, we can ignore the 4d contribution from $g^*_{2,0}$ and focus on its contribution to $G_{44,44}$. Note that the only contribution to $G_{44,44}$ is from $g^*_{2,0}$. More explicitly, we have
\[G_{44,44} \sim S_{44} S_{44} =\frac 3 4.\]
This leads to a $\frac 3 {64} F^2{F'}^2$ term in the $gg \to gg$ amplitude. In the 4d point of view, this contribution (from $g^*_{2,0}$) is from a scalar. Note that we still have the mismatch of
a factor of two. In order to get the $M^8$ term in \er{ggggpole}, we will need a total contribution of $\frac 1 4 F^2 {F'}^2$ (or equivalently $\frac 1 4 \xS_\xa{}^a \xS_\xa'{}^a$) while
we are getting $\frac 1 8 F^2 {F'}^2$.
The other half may come from the contribution
of the pseudoscalar. Note that the scalar
and pseudo scalar couple to the gluon as
\ft{$A$ is a the scalar and $B$ is the pseudoscalar.},
\[A F^2 + B F \tilde F \]
or equivalently,
\[\phi (F + i \tilde F)^2 + \zt{h.c.},\]
where $\phi \equiv A+ i B$. So in principle,
the scalar and pseudo scalar contributes in the same way to the overall amplitude. 
\ \\

\subsection{Coupling of Gluon and Its n = 1 Regge Excitation}
\com{\sto{String Amplitude of $gg \to g^*$}}
\label{subsec:SAgggstar}
Let us derive the coupling of gluon $g$ and its $n = 1$ Regge excitation
$g^*$. Following the procedure as in\com{ \iref{TLSA}{Tree-Level $n$-Gluon Superstring Amplitudes}}\cp{ Sec~\ref{sec:4pTSA}}, we have the $g-g-g^*$ 3-point correlation function as
\[\br \Big[ (k_2 \cdot \xe_1) \xe_2^\nu \p X^\mu - (\xe_1 \cdot \xe_2) k_2^\nu \p X^\mu + \xe_1^\mu \xe_2^\rho \p X_\rho(z_2) \p X^\nu(z_3)  \Big]S_{\mu\nu}\ke \frac{z_{13} z_{23}} {z_{12}}\]
Now we can choose $z_1 = 0, z_3 = 1, z_2 =\infty$. Note that the $\p X^\nu \p X^\mu$ term will bring down a $z_2$ and therefore make the amplitude vanishing. Similarly, we only consider the $\frac {k_1^\mu}{z_{13}}$ term from $\p X^\mu$. This gives us,
\[\lb (k_2 \cdot \xe_1) \xe_2^\nu k_1^\mu - (\xe_1 \cdot \xe_2) k_2^\nu k_1^\mu \rb S_{\mu\nu} \]
For a polarization $S_{ij}$ and 4d momenta in the lab frame, we have
\[\xe_2^\nu k_1^\mu S_{\mu\nu} = \xe_2^i k_1^j \frac {\sqrt 2} {3}\xd_{ij} = 0,\quad k_2^\nu k_1^\mu S_{\mu\nu} = k_2^i k_1^j \frac {\sqrt 2} { 3}\xd_{ij} = \frac {M^2} {6 \sqrt 2}.\]
It appears that setting $z_1 = \infty$ works just fine. In that case, the result is 
\[(k_2 \cdot \xe_1)e_2^\nu S_{\mu\nu} k_2^\nu- (\xe_2 \cdot \xe_1)k_2^\nu S_{\mu\nu} k_2^\mu + (k_1 \cdot \xe_2)k_1^\mu S_{\mu\nu} \xe_1^\nu,\]
which is \er{gF2} in the momentum space.


\ \\

\subsection{Propagator of a Massive Spin-2 Field in Five Dimensions}
\label{subsec:propMF5d}
Let us evaluate the propagator of a massive spin-2 field in 5d. Note that polarization tensor $e_{\mu\nu}$ is traceless and divergenceless $k^\nu e_{\mu\nu} = 0$. The propagator, as in 4d can be expressed in the form of\com{ \rf{Vain'72~(21)}} (similar to \er{Lagmassivenosource}), 
\be\el{massivegraviton5p} {G^{\mu\nu}(p)}_{\alpha\beta}=\frac {\sum_{i=1}^9 {e^i}^{\mu\nu}{e^i}_{\alpha\beta}} {p^2-M^2}\ee
where,
\ba \sum_{i=1}^9 {e^i}^{\mu\nu}{e^i}_{\alpha\beta}=\frac 1 2 ({\delta^\mu}_\alpha {\delta^\nu}_\beta + {\delta^\nu}_\alpha {\delta^\mu}_\beta)- A\eta^{\mu\nu}\eta_{\alpha\beta}\nonumber\\
-{\frac 1 2} (\frac {{\delta^\mu}_\alpha p^\nu p_\beta} {M^2}+\frac {{\delta^\nu}_\alpha p^\mu p_\beta} {M^2}+\frac {{\delta^\mu}_\beta p^\nu p_\alpha} {M^2}+\frac {{\delta^\nu}_\beta p^\mu p_\alpha} {M^2})\nonumber\\
+C(B\eta^{\mu\nu}+\frac {p^\mu p^\nu} {M^2})(B\eta_{\alpha\beta}+\frac {p_\alpha p_\beta} {M^2})
\ea
Contract with $p^\mu$ and focus on the $p_\nu \xd_{\xa \xb}$ term, we have
\[-A +C(B+1)B = 0\]
Note that we use $p^2 = - M^2$. Similarly, the $p_\nu p_\xa p_\xb$ tells us that
\[1 - C(B+1) = 0.\]
These implies $A = B$. 

Now we impose the traceless constraint. From $\xd_{\xa \xb}$ term, we have
\[C(d B + 1)B = (1 - d B),\]
where $d$ is the dimension. From $p_\xa p_\xb$ term, we have
\[2-C(d B + 1) = 0.\]
These two equations \ft{We can replace one of them by $1 - C(B+1) = 0$ above too.} can be solved to give
\[B = \frac 1 {d - 2}\]
and
\[C = \frac {d - 2}{d-1}.\]
When acting on a conserved source, 
\ba \el{massivespin2ddimen} \sum_{i=1}^9 {e^i}^{\mu\nu}{e^i}_{\alpha\beta} & = & \frac 1 2 ({\delta^\mu}_\alpha {\delta^\nu}_\beta + {\delta^\nu}_\alpha {\delta^\mu}_\beta)- A\eta^{\mu\nu}\eta_{\alpha\beta} + C B^2 \eta^{\mu\nu}\eta_{\alpha\beta}\nn
& = & \frac 1 2 ({\delta^\mu}_\alpha {\delta^\nu}_\beta + {\delta^\nu}_\alpha {\delta^\mu}_\beta) - \frac 1 {d-1} \eta^{\mu\nu}\eta_{\alpha\beta} 
\ea
Obviously, when $d = 4$, this agrees with \er{masslesslimit}\com{ \rf{(28)}}. In the rest frame of the particle, the polarization tensors \com{\rf{(22)} }\cp{(see \eg eq.(22)
in \citec{vanDam:1970vg})} are equivalently rank-$(d-1)$ matrix and form a rep of $\zt{SO}(d-1)$. Note that they satisfy\com{ \rf{(23)}}, 
\be\el{VV23}
\sum_{i=1}^5 {e^i}^{\mu\nu}{e^i}_{\alpha\beta}=\left\{\begin{array}{ll} \frac 1 2 ({\delta^\mu}_\alpha {\delta^\nu}_\beta + {\delta^\nu}_\alpha {\delta^\mu}_\beta)- \frac 1 3 \eta^{\mu\nu}\eta_{\alpha\beta}
& \zt{if}\quad \xa,\xb,\mu,\nu \ne 0\nn 0 & \zt{otherwise}\end{array}\rc
\ee
which is very similar to \er{massless}\com{ \rf{(28)}}. However, they are still different when $\xa,\xb = 0$. The interesting part is that $e^1_{\mu\nu}$ and $e^2_{\mu\nu}$ (in the 4d example) under dimensional reduction (and the corresponding $\zt{SO}(3) \to \zt{SO}(2)$) form an irrep of the $\zt{SO}(2)$ group. Acting on the spatial 2d subspace, it behaves like a \emp{massive} spin-2 propagator\com{ \rf{(24)}},
\be\el{VV24}
{e^1}_{\mu\nu}{e^1}_{\alpha\beta} + {e^2}_{\mu\nu}{e^2}_{\alpha\beta} =\left\{\begin{array}{ll} \frac 1 2 ({\delta^\mu}_\alpha {\delta^\nu}_\beta + {\delta^\nu}_\alpha {\delta^\mu}_\beta - \eta^{\mu\nu}\eta_{\alpha\beta})
& \zt{if}\quad \xa,\xb,\mu,\nu = 1,2\nn 0 & \zt{otherwise}\end{array}\rc
\ee
Note that \er{VV24} is \er{massivespin2ddimen} with $d = 3$ \ft{The dimension is a little misleading, note that the $\xa, \xb =0$ components vanish for this polarization tensors but that is not the case for the covariant propagator.}. In other words, under dimensional reduction, a propagator can be decomposed as a sum of the contributions from each irrep. In this case, we decompose the propagator as a sum of \er{VV24}, which is from the 3d spin-2 and $e^5_{\mu\nu} e^5_{\xa\xb}$, which follows from a 3d scalar. There are other contributions from the 3d vector part.

There is some subtlety. This decomposition is for the sum of polarization tensor in the rest frame, which is not covariant. We can see the difference between \er{VV23} and the covariant propagator \er{masslesslimit}. But after the decomposition, we can covariantize each term.

\ \\

\section{Supermultiplet in Randall-Sundrum Background}
\label{supermultipletRS}
A 5d massive vector field has the following action,
\[S = \int d^5 x \sqrt{g}\lsb -\frac 1 4 F^{MN} F_{MN} + \frac 1 2 m^2(A_M + \p_M \xa) (A^M + \p^M \xa)\rsb,\]
where we follows the approach in \citec{Perelstein:2009qi} (see \eg eq.(32) there) to introduce an extra field $\xa$. This $\xa$ introduces an artificial gauge symmetry. 
The Lagrangian is (after splitting fields in 4d and in the warp direction),
\ba\sqrt g \el{Lagdenmasvec} \CL & = & -\frac 1 4 F^{\mu\nu} F_{\mu\nu} - \frac 1 2 A_{\mu}\p_y (e^{-2k|y|} \p_y A^\mu) + \frac 1 2 e^{-2k|y|}m^2 A_\mu
A^\mu \nn
& & +\frac 1 2 e^{-2k|y|}\p_\mu A_5 \p^\mu A_5 + e^{-2k|y|}(\p_y
\p_\mu A^\mu) A_5 - \frac 1 2 e^{-4k|y|}m^2 A_5^2\\
& &+e^{-4k|y|} m^2 A^M \p_M \xa +e^{-2k|y|} m^2 \lsb \frac 1 2 \p_\mu \xa \p^\mu
\xa - \frac 1 2 e^{-2 k |y|} (\p_y \xa)^2 \rsb \nonumber\ea 

It turns out that the following gauge fixing term
is useful: 
\[-\frac 1 2 (\p_\mu A^\mu - m^2 e^{-2k|y|} \xa - \p_y e^{-2k|y|} A_5)^2.\]
After adding this gauge fixing term, we get a new Lagrangian,
\ba\sqrt g \el{Lagdenmasvecgf} \CL_{\zt{gf}} & = & -\frac 1 4 F^{\mu\nu} F_{\mu\nu} - \frac 1 2 A_{\mu}\p_y (e^{-2k|y|} \p_y A^\mu) + \frac 1 2 e^{-2k|y|}m^2 A_\mu
A^\mu - \frac 1 2 (\p_\mu A^\mu)^2\nn
& & +\frac 1 2 e^{-2k|y|}\p_\mu A_5 \p^\mu A_5  - \frac 1 2 e^{-4k|y|}m^2 A_5^2 - \frac 1 2 (\p_y e^{-2k|y|} A_5)^2\nn
& &-2 e^{-4k|y|} m^2 k\, \sy A_5  \xa - \frac 1 2 m^4 e^{-4 k |y|}\xa^2\nn
& & +e^{-2k|y|} m^2 \lsb \frac 1 2 \p_\mu \xa \p^\mu \xa - \frac 1 2 e^{-2 k |y|} (\p_y \xa)^2 \rsb \ea 

The mixing terms that have one $A_\mu$ and either one of $A_5, \xa$ are removed. But we do need to separate
$A_5$ from $\xa$. First of all, let us consider $A_5$. We have the equation of motion,
\[ \p_\mu \p^\mu A_5 -\p_y^2 e^{-2k|y|} A_5+2e^{-2k|y|} m^2 k\, \sy \xa+ m^2e^{-2k|y|} A_5 = 0,\]


Now let us consider the EoM of $\xa$. Note that the derivative $\p_4 \xa$ couples to the gluon in the same way as $A_5$. The equation of motion of $\xa$ is
\[\p^\mu \p_\mu \xa +2e^{-2k|y|} k\, \sy A -e^{2k|y|} \p_y e^{-4 k|y|} \p_y \xa + m^2e^{-2k|y|} \xa = 0.\]

These two equations can be combined to give (with $A \equiv m \xa$),
\be\lsb \Box- \p_y^2 + 4 k\, \sy \p_y + \bay{cc} -4k^2 + m^2 &
2 k m\,\sy \\ 2 k m\,\sy & m^2\eay\rsb\bay{c} A_5 \\ A\eay  = 0\ee
One can perform the following orthogonal transformation,
\[\bay{c} A_5 \\ A \eay = \bay{cc} \frac {\sy \xD_+}{\sqrt{1+\xD_+^2}}& \frac {\sy \xD_-}{\sqrt{1+\xD_-^2}} \\ -\frac {1}{\sqrt{1+\xD_+^2}} & \frac {1}{\sqrt{1+\xD_-^2}} \eay \bay{c} A_+ \\ A_- \eay,\]
where $\xD_\pm = \frac k m (\sqrt{1+\mathfrak m^2}\pm 1)$. This transformation diagonalize
the mass term,
\be\lsb \Box- \p_y^2 + 4 k\, \sy \p_y + \bay{cc} m^2- 2k m \xD_+ & 0 \\ 0 & m^2+ 2k m \xD_-\eay\rsb\bay{c} A_+ \\ A_-\eay  = 0.\ee
The field $A_\pm$ can be decomposed in the usual way,
\[A_\pm(x,y) = \frac {\sy} {\sqrt {\pi r_c}} \sum_{n=0}^\infty A^{(n)}_\pm(x) \xi^{(n)}_\pm(y), \]
We then have the equations for the mode functions,
\be - \p_y^2 \xi^{(n)}_\pm + 4 k\, \sy \p_y \xi^{(n)}_\pm+(m^2 \mp 2k m \xD_\pm) \xi^{(n)}_\pm = e^{2k|y|} (\mu_\pm^{(n)})^2 \xi^{(n)}_\pm.\ee
This can be rewritten in the variable $u = \frac 1 k e^{k|y|}$ as,
\[u^2 \xi^{(n)}_\pm{}''- 3 u \xi^{(n)}_\pm{}' + [(\mu_\pm^{(n)})^2 u^2 - ({\mathfrak m}^2 \mp 2\mathfrak m \xD_\pm)] \xi^{(n)}_\pm = 0.\]
The solution is
\[\xi^{(n)}_\pm(u) = \frac {1} {N_{\xi\pm}} \lsb e^{2k|y|} J_{\nu_\pm}(\mu_\pm^{(n)}  u) + C e^{2k|y|} J_{-\nu_\pm}(\mu_\pm^{(n)} u) \rsb,\]
where $\nu_\pm = \sqrt {{\mathfrak m}^2 \mp 2\mathfrak m \xD_\pm+ 4} = \sqrt{\mathfrak
m^2 + 1} \mp 1$. The masses $\mu_\pm^{(n)}$ can be obtained in a similar
way (as in the case of $B_{\mu\nu}$) by imposing proper boundary conditions. Let us
cheat a little bit and assume $\mu_+^{(n)} = \mu_-^{(n)} \equiv \mu_5^{(n)}$.
Now we only have one mass parameter and therefore
only need one boundary condition. So we require
the gauge invariant combination $\tilde A_5
\equiv A_5 + \p_y \xa$ to have Neumann boundary condition. Now we have,
\ba \tilde A_5 & = & \frac {\sy \xD_+
A_+}{\sqrt{1+\xD_+^2}}
 -\frac {\p_y A_+}{m\sqrt{1+\xD_+^2}} +\frac {\sy \xD_- A_-}{\sqrt{1+\xD_-^2}}
 +\frac {\p_y A_-}{m\sqrt{1+\xD_-^2}} 
\nn
& = & \frac 1 {\sqrt {\pi r_c}} \sum_{n=0}^\infty\left\{ A^{(n)}_+(x)\frac {e^{2k|y|}} {N_{\xi+}} \lsb \frac {(\xD_+- \frac {2k}
 m )J_{\nu_+}(\mu_5^{(n)}u)- \frac k m \mu_5^{(n)} u J_{\nu_+}{}'(\mu_5^{(n)}u)}{\sqrt{1+\xD_+^2}}\rsb
 \rc\nn
& &\left. + A^{(n)}_-(x)\frac {e^{2k|y|}} {N_{\xi-}} \lsb \frac {(\xD_-+ \frac {2k}
 m )J_{\nu_-}(\mu_5^{(n)}u)+ \frac k m \mu_5^{(n)} u J_{\nu_-}{}'(\mu_5^{(n)}u)}{\sqrt{1+\xD_-^2}}\rsb
\right\}\nn
& = & \frac 1 {\sqrt {\pi r_c}} \sum_{n=0}^\infty \lb \frac{A^{(n)}_+(x)} {N_{\xi+}\sqrt{1+\xD_+^2}} + \frac{A^{(n)}_-(x)} {N_{\xi-}\sqrt{1+\xD_-^2}}\rb e^{3k|y|}\frac {\mu_5^{(n)}} m
 J_{\sqrt{\mathfrak
m^2 + 1}} (\mu_5^{(n)}u).\hspace{1cm}
\ea

\ba
(\xD_+- \frac {2k}
 m )J_{\nu_+}- \frac k m \mu_5^{(n)}u J_{\nu_+}{}' & = & \frac k m[\nu_+
 J_{\nu_+}-\mu_5^{(n)} u  J_{\nu_+}{}']\nn
& = & \frac k m \mu_5^{(n)} u  J_{\nu_++1} = \frac k m \mu_5^{(n)} u  J_{\sqrt{\mathfrak
m^2 + 1}}.
\ea
So one can see the mode functions of $\tilde A_5$ are described by the same Bessel function
as in the previous version (massivevectorKK3), which is also the same as the Bessel function for $A_\mu$. So if the boundary condition for $A_\mu$ and $\tilde A_5$ are the same, they are going to have the same mass. Of course, there could be boundary mass-like terms for both $A_\mu$ and $\tilde A_5$, which change the boundary condition. The boundary terms in principle can be obtained from the susy invariance of 5d Lagrangian \citec{Gherghetta:2000qt}.

First let us give the particle contents and
bulk masses for a multiplet containing the massive spin-2 particle. States in $AdS_5$ can be characterized by three numbers $(E_0, s_+,s_-)$, where the pair $(s_+, s_-)$ describes the representation of $SU(2)\times SU(2) \cong SO(4)$. It is essentially the spin in 5d. We have $8$ susy and the total number of states in the 5d graviton multiplet
is $2^4 \times 4 = 64$ (assuming the ground state is a 5d vector $(\frac 1 2, \frac 1 2)$). Half of these states are bosonic. More explicitly, the $32$ bosonic states have the following quantum number $(E_0,s_+, s_-)$.
\ba
& \xD+2 & (\frac 1 2, \frac 1 2) \nn
& \xD+1 & (\frac 1 2, \frac 1 2),\quad (1, 1),\quad (1, 0),\quad (0,1),\quad (0,0),\quad (\frac 1 2, \frac 1 2) \nn
& \xD & (\frac 1 2, \frac 1 2)  
\ea
The state at the lowest level (with $E_0 = \xD$) is the ``ground state" vector $(\frac 1 2,\frac 1 2)$. Let us denote it by $A_M^1$. Those at the second level, with the weight $\xD+1$ are given by applying a pair of susy generators on the ground state. Note that the susy generators have the eigenvalue of $E_0
= \frac 1 2$. $B_{MN}$ is at this level too. So is the 5d scalar $\phi$ (corresponding to $(0,0)$). Similarly, the one at the top (with $E_0 = \xD + 2$) is obtained by stacking $4$ susy generators on the ground state. Let us denote it by
$A_M^2$.

As we can see, there are $4$ different 5d vectors. Let us see which one gives the axion in the same multiplet as $\byy$. Under a dimensional reduction to 4d, we have two $\CN = 2$ multiplets. one of them has $5$ scalars and $1$ vector. The other has a spin-2 ($5$) \ft{The number of degrees of freedom is included so that one can see this multiplet has $24$ bosonic states.}, $6$ vectors ($18$) and $1$ real scalar. In terms of the unbroken $\CN = 1$ susy, the first $\CN = 2$ multiplet gives a vector multiplet and two chiral multiplets. The second $\CN =2$ multiplet gives a graviton multiplet (graviton, vector), two gravitino multiplets (each with $2$ vectors, one of them being
either $A_\mu^1$ or $A_\mu^2$), and a vector multiplet ($1$ vector, $1$ real scalar). Anyway, the second $\CN =2$ multiplet eventually only has one real scalar and can not have the axion. Out of the five scalars in the first $\CN = 2$ multiplet, one has $E_0 = \xD+2$ and the other has $E_0 = \xD$. The remaining all have $E_0 = \xD +1$. The real and imaginary part of the complex scalar in the chiral multiplet is related to each other by two susy generators. So their eigenvalues of $E_0$ are different by $1$.
We have already known that $\phi$ and $\byy$ have $E_0 = \xD+1$. So $A_5$ can either be $A_5^2$ (with $E_0 = \xD+2$) or $A_5^1$ ($E_0 = \xD$). The
other two 5d vectors (denoted by $B^{1,2}_M$) at the same level $E_0 = \xD+1$ in fact comes from $A_M^1$ with a combination of two susy generators which transform as $(0,0)$. In other words, $A_5^1$ is going to be transformed into $B^{1,2}_5$. So at least we know the real scalar in the $\CN=2$ graviton
multiplet (the second $\CN=2$ multiplet) is coming from the linear combination
of $\byy$ and $\phi$. The natural guess is that the linear combination
of $\phi$ and $\byy$ (or maybe either one of them \ft{The best case scenario is a linear
combination, which implies $\phi$ and $\byy$
have the same mass.}) is in the same multiplet as $A_5^1$ or $A_5^2$ while the other combination (orthogonal to the first one) is in the same multiplet as $B_{\mu\nu}$. Only one of them couples to the gluons. Another piece of information we know is that $B_{\mu\nu}$
is in the same $\CN=1$ multiplet as some vector of the same $E_0$ since they
are both created from some gravitino by applying one susy generator.

Now let us consider the mass of $A_5^{1,2}$ and $\phi$ (or $\byy$). The point is that the fields for these states don't have the same mass \citec{Shuster:1999zf} \citec{Gherghetta:2000qt}. The mass of a vector or a scalar is defined by the eigenvalue of the Casimir (of the AdS group $SO(4,2)$), which is related to $(E_0, J)$ by
\[\CC_2 =E_0(E_0-4)+2s_+(s_++1)+2s_-(s_-+1).\] 
For $A_\mu^1$ and $A_5^1$, we have
\[(\CC_2)_{\zt{vector}} = \xD(\xD-4) + 3 = m^2.\]
This implies $\xD = 2 + \sqrt{1 + m^2}$ and the mass of $\phi$ (with $E_0 = \xD+1 = 3+ \sqrt{1 + m^2}$) is
\[m^2_S \equiv (\CC_2)_{\zt{scalar}} = -2+m^2 + 2\sqrt{m^2+1}.\]
Now $\nu = \sqrt{4+m_S^2} = \sqrt{m^2+1} + 1$. So the mode function for $A_\mu^1, A_5^1$ is $J_{\sqrt{m^2+1}}(\mu u)$ while those for $\phi$ (and $B_{yy}$) is $J_{\sqrt{m^2+1} + 1}(\mu u)$. As shown in \citec{Gherghetta:2000qt} (see eq.(20) and eq.(25) there), for even fields (even under the parity $y \to -y$), the boundary condition is expressed in the form \ft{We are a little sloppy here to use a single $J$ to describe both kinds of Bessel functions $J$ and $Y$. Practically, one only need to solve the equation below though.}
of, 
\be\el{bound1} (\frac s 2 - r) J_\nu(\mu) + \mu J_\nu{}'(\mu) = 0,\ee
where $s$ depends on the spin of the fields and $r$ depends on the boundary mass term (measured in $k^2$).
On the other hand, for odd field the boundary condition is 
\be\el{bound2} J_\nu(\mu) =0.\ee
From the property of Bessel function, we know,
\[\nu J_\nu(\mu) \pm\mu J_\nu{}'(\mu) = \mu J_{\nu\mp 1}(\mu).\]
For $\phi$, $s = 4$, in order for $\phi$ and $A_5^1$ to have the same mass, we need $\phi$ and $A_5^1$ to behave differently under the parity. Let us do $\phi$ even and $A_5^1$ odd. This implies $r = 1-\sqrt{m^2+1}$. One can
make a similar argument for $A_5^2$, in this case, $r = 3+\sqrt{m^2+1}$.
The point is that now the real and imaginary part of the complex scalar,
since they are related by susy transformation, are going to have different
weights $E_0$ and therefore different 5d bulk masses. As a result, the $\nu$ in the
Bessel function are different by $1$. In order for the KK modes to
have the same mass, different boundary conditions have to be imposed on the
two fields $A_5$, and $\phi$. Only
when one of them is odd while the other is even, there is a possibility that
the KK masses are the same. Moreover, boundary mass terms need to be added to make sure the boundary conditions \er{bound1} \er{bound2} have the same solution.

Similar analysis can be made for $B_{\mu\nu}$. Since it has the same $E_0$
as the vector in the same multiplet. Their Bessel function is the the same
\ft{If the mass parameter for $B_{\mu\nu}$ is the same as that of a scalar
at the same level and the Lagrangian $B_{\mu\nu}$ is the same (\ie having
the same bulk mass) as given in \citec{Perelstein:2009qi}.}. To get the same mass, one
need to impose the same parity and the same boundary mass term. Btw, if the
real partner of the axion is a linear combination of $\byy$, then it has
the same mass as some vector at this level because they are in the same vector
rep. Although not all vectors have the same boundary mass term (although
they do have the same bulk mass and Bessel function), it is possible that
all vectors have the same mass and therefore $B_{\mu\nu}$ and the axion have
the same mass.

Let us try to figure out the boundary mass
term for $\phi$, which is a scalar at the
level $E_0 = \xD+1$. In fact, the boundary
mass term can be obtained by generalizing
eq.\rf{(37)} in \citec{Gherghetta:2000qt}.
(henceforth, all equation numbers without section numbers refer
to equations in \citec{Gherghetta:2000qt}). Note that the vector multiplet discussed in eq.\rf{(35)} is part of the $\CN=2$ multiplet we discussed above. There
is a vector with $E_0 = \xD$, a symplectic
Majorana spinor $\xl^i$ ($i = 1,2$) \ft{I
believe this is equivalent to a Dirac spinor
in 5d.} with $E_0 = \xD+\frac 1 2$, and a scalar
with $E_0 = \xD + 1$. The vector can be identified
with $A_M^1$ and then the scalar is $\phi$.
To be consistent with the notation in \citec{Gherghetta:2000qt},
let us take $\xD = \frac 3 2 + c$. So the
spinor mass of $\xl^i$ ($\xD + \frac 1 2$) and the scalar
mass are
\ba\el{bulkmassphi} (m_\phi^2)_{\zt{bulk}} & = & c^2 + c - \frac {15} 4,\\
m_{\xl^i}^2 & = & (\CC_2)_{\zt{spinor}} + \frac 5 2 = c^2.\ea
The mass term for the spinor is actually
(see eq.\rf{(35)}) $m_{\xl^{1,2}} = \pm c \xs'$, where $\xs'$ is defined in \rf{(9)}. This agrees with eq.\rf{(45)}
and also eq.\rf{(37)} (the latter only when $c=-\frac 1 2$).
According to \rf{(45)}, the mass term for
$\phi$ should be
\be\el{totalmassphi} m_\phi^2  =  c^2 + c - \frac {15} 4
+ \lb \frac 3 2 - c\rb \xs''.\ee
For later, convenience, let us give the $\nu$
that appears in the Bessel function for $\phi$.
It is given by 
\[\nu_\phi = \sqrt{c^2 + c - \frac {15} 4 + 4} = c + \frac 1 2.\]
Of course, \rf{(45)} is for the hyper-multiplet
and it is not obvious that it applies for the vector multiplet.
So we still need to derive \er{totalmassphi}.
To derive the boundary
mass term (proportional
to $\xs''$), one can first replace the last term in eq.\rf{(36)}
by $-i(\frac 3 2 - c)\xs' \phi (\xs_3)^{ij}
\eta^j$ and $m_{\xS}$ by $m_\phi$ (and also
$\xS$ by $\phi$). The variation \ft{We use
the modified variation $\xd \xl^i$ with the
replacement mentioned above.} of the kinetic term the spinors in \rf{(43)}, \ie
\[i \bar \xl^i \xg^M D_M \xl^i + i m_\xl \bar \xl^i (\xs_3)^{ij}
\xl^j,\]
gives \ft{We will focus on the non-derivative terms.},
\ba K[\xd \xl^i] & = & -(\frac 3 2 -c)i\xs'\phi(\xs_3)^{ik}\lsb i \bar \xl^i \xg^M \lb- \frac {\xs'} 2 \xg_M
(\xs_3)^{ij} \eta^k \rb+ i m_\xl \bar \xl^i
(\xs_3)^{ij} \eta^k \rsb \nn
& = & -(\frac 3 2 -c)i\xs'\phi(\xs_3)^{ik}\lb-\frac
5 2 -c \rb\lb i \bar \xl^i \xs' (\xs_3)^{ij} \eta^k \rb \nn
& = & (m_\phi^2)_{\zt{bulk}}\, \phi \bar \xl^i \eta^i \ea
This will be canceled by the variation of
the mass term of $\phi$, \ie $m_\phi^2 \phi^2$ (with $m_\phi^2$ given by \er{bulkmassphi})
under $\xd \phi = \bar \eta^i \xl^i$. The
modification of the last term in \rf{(36)}
is to compensate the difference of the mass term from \rf{(37)}.

So far, we haven't mentioned anything about the boundary mass term. The point is that the modified term $-i(\frac 3 2 - c)\xs' \phi (\xs_3)^{ij} \eta^j$ in $\xd \xl^i$,
after hit by the derivative $\p_y$ \ft{This happens
when the derivative $\p_y$ in the kinetic terms of $\xl^i$
hits the $\xs'$.}, gives a boundary term
(which we ignore previously) that should
be canceled by the boundary term of $\phi$.
The total mass term can be shown to be exactly \er{totalmassphi}.
One can see from \er{bound1}, \rf{(8)} (definition of $b$), that we have
\[-r + \frac s 2 = c + \frac 1 2 = \nu_\phi.\]
This implies that the boundary condition \er{bound1} for even field (with $E_0 = \xD+1$) exactly 
agrees with that \er{bound2} for odd field (with $E_0 = \xD$). 
In other words, $\phi$ (even) and $A_5^1$ (odd) have the same KK masses.

However, there is a different problem about the orthonormality of the mode functions. The inner product of the two mode functions should be
\be
\frac{1}{\pi r_c} \int_0^{\pi r_c} dy \, e^{-2k|y|}\xi^{(n)}_\pm(\mu_5^{(n)}u)\, \xi^{(m)}_\pm(\mu_5^{(m)}u).
\ee
This agrees with the $e^{-2k|y|}$ in the kinetic term of $A_5$ and $\xa$ (and therefore the kinectic term of $A_\pm$) in the Lagrangian \er{Lagdenmasvec}. However, the $\mu_5^{(n)}$ is obtained by solving equation of the Bessel function $J_{\nu_\pm \pm 1}$ instead of $J_{\nu_\pm}$. So in general, the modes $\xi^{(n)}_\pm(\mu_5^{(n)}u)$ are not orthogonal to each other. This appear to be some sort of inconsistency. There is a possible solution. In fact, if one imposes $Z_2$ odd boundary condition on $\tilde A_5$, \ie $\tilde A_5 (-y) = - \tilde A_5(y)$, $J_{\sqrt{\mathfrak
m^2 + 1}} (\mu_5^{(n)}u)$ have to be vanishing at the boundary. In this case, the modes $\xi^{(n)}_\pm(\mu_5^{(n)}u)$ are orthogonal and everything is fine. \eml{However, it appears that in order to couple to the gluon, $\tilde A_5$ has to be even under the $Z_2$ parity. The coupling to the gluon is like
\[A_\mu \xe^{\mu\nu\rho\xs\xg} F_{\nu \rho}^g F_{\xs \xg}^g,\]
where $F_{\nu \rho}^g$ is the field strength of the gluon, which is $Z_2$ even. The only way to gain a nonvanishing coupling (after the integration over $y$) is to require $\xe^{\mu\nu\rho\xs\xg}$ to be odd. It is not clear whether one can do that. The parity condition that people generally impose on the metric (see \eg \citec{Altendorfer:2000rr}) leads to an even $\xe^{\mu\nu\rho\xs\xg}$.}

\end{appendix}

\newpage

\addcontentsline{toc}{chapter}{Curriculum Vitae}
\label{CV}

\begin{center}
{\LARGE\bf CURRICULUM VITAE} \\
\vskip 2em
\textbf{Xing Huang}
\end{center}
\vskip 2em

\begin{tabbing}
\textbf{EDUCATION}\\
Sep. 2004---present,                    \hskip 1.65cm University of Wisconsin-Milwaukee\\
Bachelor of Science (2004),     \hskip 5mm University of Science and Technology of China

\end{tabbing}

\vskip 1em

\begin{tabbing}
\textbf{AWARDS}\\

2010 \hskip 1.5cm Papastamatiou Scholarship Award (UWM)\\
2008 \hskip 1.5cm Dissertation Fellowship (UWM)\\
2007 \hskip 1.5cm Graduate School Fellowship (UWM)\\
2004-2010 \hskip 5mm Chancellor Fellowship (UWM)

\end{tabbing}

\vskip 1em

\begin{tabbing}
\textbf{PUBLICATIONS}\\

\\

Stringy origin of Tevatron $Wjj$ anomaly\\
Luis A. Anchordoqui, Haim Goldberg, Xing Huang, Dieter L\"ust, Tomasz R. Taylor,\\
e-Print:arXiv:1104.2302\\
\ \\
Searching for string resonances in $e^+e^-$ and $\gamma \gamma$ collisions\\
Luis A. Anchordoqui, Wan-Zhe Feng, Haim Goldberg, Xing Huang, Tomasz R. Taylor,\\
Phys.\ Rev.\  D  (to be published) e-Print: arXiv:1012.3466\\
\ \\

LHC Phenomenology of Lowest Massive Regge Recurrences in the Randall-Sundrum Orbifold\\
Luis A. Anchordoqui, Haim Goldberg, Xing Huang, Tomasz R. Taylor,\\
Phys.\ Rev.\  D {\bf 82}, 106010 (2010)\\
\ \\

Clarifying Some Remaining Questions in the Anomaly Puzzle\\
Xing Huang, Leonard Parker\\
Eur.\ Phys.\ J.\  C {\bf 71}, 1570 (2011)\\
\ \\

Hermiticity of the Dirac Hamiltonian in Curved Spacetime\\
Xing Huang, Leonard Parker\\
Phys.\ Rev.\  D {\bf 79}, 024020 (2009)\\
\ \\

Quasi-equilibrium models for triaxially deformed rotating compact stars\\
Xing Huang, Charalampos Markakis, Noriyuki Sugiyama, Koji Uryu\\
Phys.\ Rev.\ D {\bf 78}, 124023, (2008)

\end{tabbing}

\vskip 1em

\begin{tabbing}
\textbf{PRESENTATIONS}\\

Feb 2010 \ Clarifying Some Remaining Questions in the Anomaly Puzzle in N=1 SYM Theory\\
Center for Gravitation and Cosmology seminar, UW-Milwaukee\\

\ \\

Nov 2010 \      LHC Phenomenology of Regge Excitations in the Randall-Sundrum Model\\
Center for Gravitation and Cosmology seminar, UW-Milwaukee\\

\ \\

May 2011 \      String Physics at CLIC\\
parallel talk at Phenomenology 2011 Symposium,	UW-Madison\\

\end{tabbing}
\ \\

\vskip 10em

\begin{center}

\hrule \vskip 0.2em Co-Major Professor \hfill Date\ \ \ \ \ \ \ \ \ \\
\end{center}

\vskip 4em
\begin{center}

\hrule \vskip 0.2em Co-Major Professor \hfill Date\ \ \ \ \ \ \ \ \ \\
\end{center}

\end{document}